\documentclass[12pt]{article}
\usepackage[utf8]{inputenc}
\usepackage{amsmath}
\usepackage{amssymb}
\usepackage{enumitem}
\usepackage{amsfonts}
\usepackage{eurosym}
\usepackage{makeidx}
\usepackage{times}
\usepackage{anysize}
\usepackage{hyperref}
\usepackage{tikz}
\usepackage{cite}
\usepackage{mathrsfs}
\usepackage{pgfplots}

\makeatletter
\marginsize{2cm}{2cm}{1cm}{1cm}
\newtheorem{theorem}{Theorem}[section]

\newtheorem{assumption}{Assumption}

\newtheorem{condition}[theorem]{Condition}

\newtheorem{corollary}[theorem]{Corollary}

\newtheorem{definition}[theorem]{Definition}

\newtheorem{lemma}[theorem]{Lemma}

\newtheorem{proposition}[theorem]{Proposition}
\newtheorem{remark}[theorem]{Remark}

\newenvironment{proof}[1][Proof]{\noindent\textbf{#1.} }{\ \rule{0.5em}{0.5em}}
\makeatother

\begin{document}

\title{Non-Closing Double-Commutator Flows and the Small-Coupling Limit of
Spin-Boson Models}
\author{J.-B. Bru \and N. Metraud  \and W. de Siqueira Pedra}
\date{\today }
\maketitle

\begin{abstract}
Spin-boson models are paradigmatic examples of open quantum systems and
serve as the theoretical paradigm for current quantum computers based on
single-ion trap technology. Despite their ubiquity, a complete spectral
diagonalization of these models remains an open problem, except in highly
singular regimes. This paper establishes a rigorous framework for the
approximate diagonalization of generalized spin-boson systems. Our approach
is inspired by the Brockett-Wegner double-commutator flow, which is a
non-linear differential equation governing the evolution of (here unbounded)
operators. Unlike relatively recent applications of this flow to quadratic
Hamiltonians in quantum field theory, it does not close in the spin-boson
context. We overcome this fundamental obstruction by performing a detailed
analysis of the resulting non-closed algebraic structure, allowing us to
explicitly bound the higher-order error term with respect to the spin-boson
coupling strength. Consequently, this work provides the first mathematically
rigorous justification for several heuristic diagonalization techniques
widely employed in the theoretical physics literature for small-coupling
regimes, in the simplest non-trivial cases. More broadly, our framework
renders a flow-based algorithm feasible for systematic higher-order
diagonalization and self-energy renormalization. This strategy is
conceptually akin to multi-scale analysis, or, much more recently, to the
iterative, local Lie-Schwinger block-diagonalization method by Fr\"{o}hlich
and Pizzo.\bigskip {}

\bigskip

\end{abstract}

\tableofcontents

\section{Introduction}

\noindent \textbf{Trapped-ion quantum computers.} In the last two decades of
the twentieth century, numerous experimental breakthroughs have been made,
enabling ever more sophisticated control of quantum systems, such as
ultra-cold gases or ions in traps. This paved the way for the possibility of
quantum computers, which have now become a reality, despite major problems
still to be solved in order to make the technological leap. In this race for
quantum computers, several physical platforms exist. Among others, we can
mention IBM's superconductor-based technology and AQT\footnote{%
Alpine Quantum Technologies GmbH, a start-up company at Innsbruck, in
Austria.}'s (lesser-known) single-ion trap technology.

The idea of using a set of closely trapped single ions as quantum bits (as
in AQT) with collective quantized motions produced by laser beams, to
realize quantum logic gates, started with Cirac-Zoller's seminal paper
published in 1995 \cite{Zoller}. This was rapidly implemented experimentally
in \cite{Trapion1,Trapion2} for a single ion, which serves as a so-called
qubit, a two-state quantum system. This has been the basis of trapped-ion
quantum computers and it is still strongly developing as a core research
area. See, e.g., \cite{lastBlatt,reviewQcomput} and references therein.

\begin{figure}[tbp]
\centering
\begin{tikzpicture}[scale=1.5, every node/.style={font=\small}]

\foreach \i in {0, 1, 2} {
	
	\draw[thick] (\i*2,0) rectangle ++(1.5,1.5);
	\node at (0+ 0.75, -0.3) {ion $k-1$};
	\node at (2 + 0.75, -0.3) {ion $k$};
	\node at (4 + 0.75, -0.3) {ion $k+1$};

	\draw[thick] (\i*2 + 0.2,0.5) -- ++(0.6,0); 
	\draw[thick] (\i*2 + 0.2,1.1) -- ++(0.6,0);

	\node[right] at (\i*2 + 0.8,0.5) {\(|\downarrow\rangle\)};
	\node[right] at (\i*2 + 0.8,1.1) {\(|\uparrow\rangle\)};
}

\foreach \i in {0,1,2} {
	\draw[thick] (\i*2+0.5,2.4) rectangle ++(0.4,0.4);
	\node at (0.2+0.5,3) {Laser $k-1$};
	\node at (2.2+0.5,3) {Laser $k$};
	\node at (4.2+0.5,3) {Laser $k+1$};
}

\draw[decorate, decoration={snake, amplitude=1.8pt, segment length=8pt}, -, thick]
(2.7,2.4) -- (2.7,1.6);
\draw[->, thick] (2.7,1.6) -- (2.7,1.52);
\node at (3,1.95) {\(k\)};

\draw[->] (2.5,0.55) -- (2.5,1.05); 
\draw[->] (2.7,1.05) -- (2.7,0.55);

\end{tikzpicture}
\caption{Scheme of a two-level ion-system driven by lasers when the $k$-th
laser is triggered.}
\label{FigProc}
\end{figure}
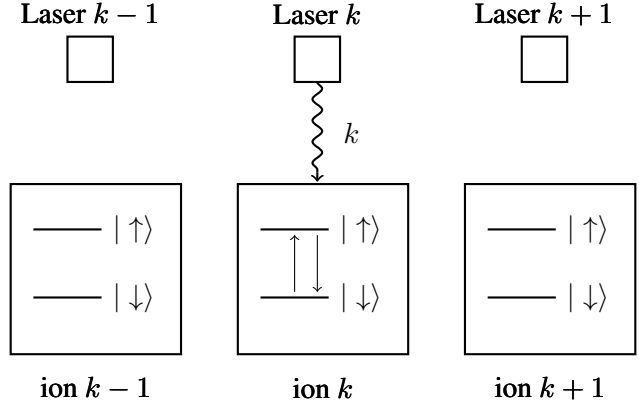

Cirac-Zoller's scheme, as summed up in Figure \ref{FigProc} in the simplest
case of a family of two-level systems, uses a class of effective
interactions acting on $\mathfrak{H}_{N}\otimes \mathcal{F}_{+}$ of the form%
\footnote{%
Here, $c_{k}=c$ and $a_{k}=a$ could be taken as $k$-independent, depending
on the physical situations one wants to consider.} 
\begin{equation}
\frac{1}{\sqrt{N}}\sum_{k=1}^{N}\left( c_{k}P_{k}\otimes a_{k}+\bar{c}%
_{k}P_{k}^{\ast }\otimes a_{k}^{\ast }\right)  \label{form}
\end{equation}%
between $N$ ions on a lattice (typically 1D), modeled as $N$ qubits with
Hilbert space 
\begin{equation}
\mathfrak{H}_{N}\equiv \bigotimes_{k=1}^{N}\mathfrak{h}_{k},\qquad \mathfrak{%
h}_{k}=\mathbb{C}^{2},  \label{form2}
\end{equation}%
and collective quantized motions via the annihilations $a_{k}$ and creations 
$a_{k}^{\ast }$ of photons induced by $n\leq N$ laser beams on each trapped
ions\footnote{%
One can turn on only one laser beam on the $q$-th ion, meaning in this way
that $c_{k}=0$, except for $k=q$.}, where $c_{k}\in \mathbb{C}$\ and $%
P_{k}\in \mathcal{B}\left( \mathfrak{h}_{k}\right) $ is a projector while $%
\mathcal{F}_{+}$ is the bosonic Fock space over some underlying one-particle
Hilbert space $\mathcal{H}$. E.g., 
\begin{equation*}
\mathcal{H}=\bigoplus_{k=1}^{N}\mathcal{H}_{k},
\end{equation*}%
where the family of Hilbert spaces $(\mathcal{H})_{k\in \{1,\cdots ,N\}}$
takes into account $N$ laser beams and/or $N$ possible types of photons%
\footnote{%
Using the bosonic Fock space $\mathcal{F}_{+}\left( \mathcal{H}_{k}\right) $
constructed from $\mathcal{H}_{k}$, note the well-known canonical unitary
isomorphism $\bigotimes_{k=1}^{N}\mathcal{F}_{+}\left( \mathcal{H}%
_{k}\right) \equiv \mathcal{F}_{+}(\mathcal{H})$, where $\mathcal{H}%
=\bigoplus_{k=1}^{N}\mathcal{H}_{k}$.}. Quantum algorithms can then be
implemented by sequentially switching laser beams on different ions,
producing each time some elementary unitary transformation (cf. so-called
quantum channels) driven by the corresponding ion-photon Hamiltonian. An
interaction like (\ref{form}) means that $n\doteq |k:\{c_{k}\neq 0\}|\leq N$
laser beams are turned on at the same time.

As described in \cite{Zoller,Blatt-2003}, the coefficients in (\ref{form})
can be precisely defined via physical parameters of the system, like the
Rabi frequency, the laser phase, and the Lamb-Dicke parameter. Note indeed
that (\ref{form}) is the so-called Lamb-Dicke approximation (see, e.g., \cite%
{Blatt-2003,Jonathan}) of a more general Hamiltonian for which the creation
and annihilation operators $a_{k}$ and $a_{k}^{\ast }$ are replaced by a
single Weyl operator of the form $\exp (i\eta (a_{k}\mathrm{e}^{-i\omega
t}+a_{k}^{\ast }\mathrm{e}^{i\omega t}))$ for each $k$, the constants $\eta $
and $\omega $ being the Lamb-Dicke parameter and the trap frequency,
respectively. For more details, see also \cite{Blatt2008}.

The Lamb-Dicke approximation is justified from a heuristic point of view
when the strength of the coupling is \emph{weak}, in which case it suffices
to consider the linear part of the coupling, leading in this way to an
effective interaction of the form (\ref{form}), which was and is still
widely used in quantum physics \cite%
{lastBlatt,reviewQcomput,forLambDicke,Legget}. We do not deal with the
rigorous justification of the Lamb-Dicke approximation here, but our
results, while general, become particularly relevant in the case of \emph{%
weak} coupling. In addition, it is worth noting that in the class of models
we are rigorously studying, the qubit can be replaced by higher-dimensional
systems such as qudits, i.e., $d$-state quantum systems ($d\in \mathbb{N}$),
or even infinite-dimensional ones. Qudits are being very actively explored
in quantum physics across various platforms (see, e.g., \cite%
{qudits,MorelloQdit}). For instance, for single-ion traps technology \cite%
{qudits} it can be advantageous to limit the number of ions while increasing
the full dimension of the quantum computer via the use of $d $-levels of
each ions to get more efficient quantum algorithms and improved information
encoding.\bigskip

\noindent \textbf{Generalized spin-boson models.} In mathematics, the
Hamiltonians for one single qubit described above within the Lamb-Dicke
approximation, e.g., in \cite{Zoller,Blatt-2003}, are known as spin-boson
Hamiltonians. They were studied in the context of open quantum systems for
which the bosonic environment implies a dissipative behavior on the qubit
(or quantum spin system), even in quantum physics \cite{Legget}. The model
we propose to study here belongs to the so-called generalized spin-boson
models, a class of models for which the interaction term may not necessarily
be linear in the bosonic field and that can include multiple qubits. They
have been studied in many papers, see \cite%
{Arai,Arai2,Arai3,Falconi,Takaesu,Teranishi,Teranishi2,spinboson1,Lonigro,UltravioletRenorm,LillLonigro}%
. In all these papers, their self-adjointness and certain properties such as
ground-state characteristics are mainly studied. However, a mathematically
rigorous framework for the exact diagonalization of these systems has
remained noticeably absent from the literature.

This work addresses this gap by establishing an approximate diagonalization
scheme for a highly general class of spin-boson models that lies outside the
scope of \cite%
{Arai,Arai2,Arai3,Falconi,Takaesu,Teranishi,Teranishi2,spinboson1,Lonigro,UltravioletRenorm,LillLonigro}%
. Our formulation is directly motivated by a generalized Cirac-Zoller
design, extended here to accommodate a potentially infinite collection of
quantum systems, each of which may be infinite-dimensional. While practical
quantum computation requires a scale significantly exceeding current
technological capabilities -- specifically, $N\gg 1000$ qubits -- the regime
of a large but finite number of degrees of freedom is structurally
asymptotic to the infinite-dimensional limit. Consequently, the case of $%
N=\infty $ is treated herein within the same mathematical framework, as this
limiting regime preserves tractability without introducing prohibitive
algebraic complexity.

Consequently, we use a countable collection $\{\mathfrak{h}_{k}\}_{k\in 
\mathbb{N}}$ of (possibly infinite-dimensional) Hilbert spaces to encode
each quantum system. For instance, for $N$ qudits, $\mathfrak{h}_{1},\ldots ,%
\mathfrak{h}_{N}$ are all equals to $\mathbb{C}^{d}$, whereas $\mathfrak{h}%
_{k\geq N}$ can be taken as $\{0\}$. It leads to the Hilbert space 
\begin{equation*}
\mathfrak{H}\equiv \bigotimes_{k\in \mathbb{N}}\mathfrak{h}_{k},
\end{equation*}%
as rigorously defined in Section \ref{not separable Hilbert space S}.
Compare with (\ref{form2}). Then, the Hilbert space of the whole system is
described by the tensor product $\mathfrak{F}\doteq \mathfrak{H}\otimes 
\mathcal{F}_{+}$, recalling that $\mathcal{F}_{+}$ is the bosonic Fock space
over some underlying one-particle Hilbert space $\mathcal{H}$. We then
propose to study an operator of the form%
\begin{equation}
\mathrm{H}_{0}\doteq \sum_{k\in \mathbb{N}}\left( h_{k,0}\otimes \mathbf{1}_{%
\mathcal{F}_{+}}+\mu _{k,0}\otimes a_{k}^{\ast }a_{k}+X_{k,0}\otimes
a_{k}+X_{k,0}^{\ast }\otimes a_{k}^{\ast }\right) ,  \label{H1}
\end{equation}%
with operator-valued coefficients $h_{k,0}=h_{k,0}^{\ast }$, $\mu _{k,0}=\mu
_{k,0}^{\ast }$ and $X_{k,0}$ in $\mathcal{B}(\mathfrak{h}_{k})$ for each $%
k\in \mathbb{N}$. Here, for each $k\in \mathbb{N}$, $a_{k}$ ($a_{k}^{\ast }$%
) is a bosonic annihilation (creation) operator acting on the bosonic Fock
space $\mathcal{F}_{+}$ and $\mathbf{1}_{\mathcal{F}_{+}}$ is the identity
operator acting also on $\mathcal{F}_{+}$. Compare the interaction in (\ref%
{H1}) with (\ref{form}).\bigskip

\noindent \textbf{General assumptions.} We always assume that 
\begin{equation*}
\left\Vert X_{0}\right\Vert _{\mathrm{op},2}^{2}\doteq \sum_{k\in \mathbb{N}%
}\left\Vert X_{k,0}\right\Vert _{\mathrm{op}}^{2}<\infty ,
\end{equation*}%
$\Vert \cdot \Vert _{\mathrm{op}}$ being the operator norm. This condition
is the analogue for an infinite number of quantum systems of the rescaling%
\footnote{%
This factor $1/\sqrt{N}$ results from physical arguments in relation to the M%
\"{o}ssbauer effect \cite{Zoller}.} $1/\sqrt{N}$ in (\ref{form}), since, in
this case, 
\begin{equation*}
\left\Vert X_{0}\right\Vert _{\mathrm{op},2}^{2}=\sum_{k=1}^{N}\left\Vert
X_{k,0}\right\Vert _{\mathrm{op}}^{2}=\frac{1}{N}\sum_{k=1}^{N}\left\vert
c_{k}\right\vert ^{2},\qquad \qquad X_{k,0}=c_{k}P_{k}/\sqrt{N}.
\end{equation*}%
Observe that the scaling factor $1/\sqrt{N}$ in (\ref{form}) and used just
above is physically motivated by a classical treatment of the M\"{o}ssbauer
effect \cite{Zoller}. More generally, the summability condition imposed on
the operator norm reflects the spatial localization of the driving laser
fields. For instance, this accommodates coupling (operator-valued)
coefficients satisfying a Gaussian decay profile of the form $\Vert
X_{k,0}\Vert _{\mathrm{op}}=\mathcal{O}(e^{-\left\vert k-q\right\vert
^{2}/(2\sigma ^{2})})$ for a fixed mean $q$ and standard-deviation $\sigma
>0 $. Mathematically, this summability condition together with another
summability assumptions on $h_{k,0}$ (Assumption \ref{Assumption S}) are
important to define $\mathrm{H}_{0}$ as a self-adjoint and lower semibounded
operator acting on $\mathfrak{F}\doteq \mathfrak{H}\otimes \mathcal{F}_{+}$.
For more details, see Theorem \ref{HamilSA copy(1)}.

Other conditions are used to be able to study the Hamiltonian $\mathrm{H}%
_{0} $. A first one is the fact that $[\mu _{k,0},X_{k,0}]=0$ for all $k\in 
\mathbb{N}$. A gap is also taken in the following sense: 
\begin{equation*}
\mu _{k,0}\pm 2h_{k,0}\geq \alpha _{k,\pm }\mathbf{1}_{\mathfrak{h}_{k}}
\end{equation*}%
for some constants $\alpha _{k,+},\alpha _{k,-}\in \mathbb{R}$ that are
uniformly bounded from below and such that 
\begin{equation*}
\alpha _{k}\doteq \frac{1}{2}\left( \alpha _{k,+}+\alpha _{k,-}\right) \geq 0
\end{equation*}
at any $k\in \mathbb{N}$. This implies in particular that $2\mu _{k,0}\geq
\alpha _{k}\mathbf{1}_{\mathfrak{h}_{k}}$. We assume a uniform gap in the
sense that $\alpha _{k}\geq \alpha >0$ for all $k\in \mathbb{N}$. The above
uniform gap condition could be relaxed, but it would be a priori a highly
non-trivial task, similar to what is done in \cite{bach-bru-memo}.

These are natural conditions for possible applications to the physical
systems described above: The laser beams are all characterized by a fixed
(non-zero) wavelength and are obviously not supposed to be correlated with
the quantum systems (qubits, qudits, etc.), which leads to the canonical
choice $\mu _{k,0}=\varepsilon _{k}\mathbf{1}_{\mathfrak{h}_{k}}$ with $%
\varepsilon _{k}>0$ and even $\varepsilon _{k}=\alpha >0$ if the laser beams
are all the same, implying photons of same energy. In fact, in all physical
examples introduced here, $\mu _{k,0}\in \mathbb{R}\mathbf{1}_{\mathfrak{h}%
_{k}}$ for all $k\in \mathbb{N}$. Observe also that our study also includes
the case of a finite number of quantum systems interacting with a unique
bosonic field (like a monochromatic radiation): In this case, take $%
\mathfrak{h}_{1}$ as the full Hilbert space of this finite number of quantum
systems and $\mathfrak{h}_{k\geq 2}=\{0\}$. Our theoretical framework is
therefore very general, in view of possible applications in physics.\bigskip

\noindent \textbf{Flows driven by double commutators.} One of the major
difficulties in studying the Hamiltonian (\ref{H1}) arises from the fact
that it does not preserve the number of bosons. Indeed, a particle can be
absorbed or emitted by the system. This is reflected by the fact that $[%
\mathrm{N},\mathrm{H}_{0}]\neq 0$, where $\mathrm{N}$ is the boson number
operator. Any self-adjoint operator $A$ acting on $\mathfrak{F}$ is named $%
\mathrm{N}$\emph{-diagonal} when it commutes with $\mathrm{N}$, i.e.,
formally, 
\begin{equation*}
\left[ A,\mathrm{N}\right] \doteq A\mathrm{N}-\mathrm{N}A=0.
\end{equation*}%
While models within this class are amenable to standard analytical
techniques, our primary objective is to construct a unitary transformation $%
\mathrm{U}$ such that $\mathrm{U\mathrm{H}}_{0}\mathrm{U}^{\ast }$ is
rendered $\mathrm{N}$-diagonal.

In general, models governed by $\mathrm{\mathrm{H}}_{0}$ admit no known
closed-form solution and their spectral properties remain completely open.
The rare exceptions for which an exact diagonalization is tractable are
highly specific cases, namely when $X_{k,0}$ commutes with $h_{k,0}$. A
canonical example of this commuting regime is the standard \textquotedblleft
pure-dephasing\textquotedblright\ spin-boson model \cite[Eq. (28)]%
{LillLonigro}, which can be decomposed in two Van-Hove Hamiltonians \cite%
{VanHove}. For more details on this model in relation to the traditional
spin-boson model, see Section \ref{Illustration}.

To $\mathrm{N}$-diagonalize model of the form (\ref{H1}), we use the
Brockett-Wegner flow \cite{Brockett1,Wegner1,bach-bru} with the particle
number operator $\mathrm{N}$, as it has been successfully used to $\mathrm{N}
$-diagonalize fermionic and bosonic quadratic Hamiltonians \cite%
{bach-bru-memo,QuadraHamilFermio}. More precisely, we examine the formal
equation 
\begin{equation}
\partial _{t}\mathrm{H}_{t}=\left[ \mathrm{H}_{t},\left[ \mathrm{H}_{t},%
\mathrm{N}\right] \right] ,\qquad \mathrm{H}_{t=0}=\mathrm{\mathrm{H}}_{0},
\label{flow}
\end{equation}%
for all times $t\in \mathbb{R}_{0}^{+}$. Here, $\left[ A,B\right] \doteq
AB-BA$. See again Section \ref{Illustration} for an illustration of the
method. This approach was already tried in the theoretical physics
literature for the usual spin-boson model \cite{SpinBosonflowphysics}. See
also \cite{Kehrein-livre}. However, all these attempts are far from being
mathematically rigorous. In addition, in contrast to quadratic Hamiltonians,
it is known that the solution $(\mathrm{H}_{t})_{t\in \mathbb{R}_{0}^{+}}$
to the (non-linear) double-commutator flow (\ref{flow}) cannot be of the
form (\ref{H1}), i.e., the flow does not conserve the form of the original
expression. For example, if $\mathrm{H}_{t}$ has components that are
quadratic in terms of bosonic fields, a formal calculation of the double
commutator in (\ref{flow}) leads to third-order terms in the bosonic fields.
In this situation, we say that the flow does not close, in contrast to the
fermionic \cite{QuadraHamilFermio} and bosonic \cite{bach-bru-memo} cases.
This introduces a profound technical challenge, substantially elevating the
analytical complexity beyond that of the only two previous instances \cite%
{bach-bru-memo,QuadraHamilFermio} where the Brockett-Wegner flow has been
successfully implemented for \emph{unbounded} Hamiltonians.\bigskip

\noindent \textbf{Main results.} Fortunately, from a perturbative
perspective, i.e., when $\lambda \doteq \Vert X_{0}\Vert _{\mathrm{op},2}\ll
1$ with not only bounded but also Hilbert-Schmidt operators $X_{k,0}$, $k\in 
\mathbb{N}$, these higher-order terms appear to be negligible, in some
sense. While this constitutes the central physical argument in \cite%
{SpinBosonflowphysics} for a specific spin-boson model, it remains a purely
heuristic assertion lacking a rigorous mathematical foundation. In this
work, we establish this result rigorously for the much broader class of
models introduced above. In particular, we perform an approximate $\mathrm{N}
$-diagonalization here, with a controlled error of order $\mathcal{O}%
(\lambda ^{3})$, and thus \textquotedblleft decouple\footnote{%
It means here that there is no direct energetic exchange through the
creation or absorption of bosonic fields.}\textquotedblright\ the dynamics
of the system. It refers to the proof of the existence of a unitary $U$ such
that 
\begin{equation}
\mathrm{U\mathrm{H}}_{0}\mathrm{U}^{\ast }=\sum_{k\in \mathbb{N}}\left(
h_{k,\infty }\otimes \mathbf{1}_{\mathcal{F}_{+}}+\mu _{k,\infty }\otimes
a_{k}^{\ast }a_{k}\right) +\mathcal{O}(\Vert X_{0}\Vert _{\mathrm{op},2}^{3})
\label{sdsdsdsdsdsd1}
\end{equation}%
with operator-valued coefficients $h_{k,\infty }=h_{k,\infty }^{\ast }$ and $%
\mu _{k,\infty }=\mu _{k,\infty }^{\ast }$ in $\mathcal{B}(\mathfrak{h}_{k})$
for each $k\in \mathbb{N}$. This statement is properly expressed in Theorem %
\ref{MainDiago}, which is our main result. This provides a rigorous
mathematical foundation for the \textquotedblleft small
coupling\textquotedblright\ approximation frequently employed in the physics
literature, thereby facilitating the systematic investigation of the
system's spectral and dynamical properties within the physical contexts
outlined above.

Note finally that our framework has a more restricted coupling factor,
compared with the paper \cite{Lonigro}, but a more general kinetic part
which has additionally led to the development of an extended second
quantization formalism. See Section \ref{extendedquantization} and also
Theorem \ref{MainDisplacement}, which establishes the counterpart of the
Bogoliubov transformation in this situation with a so-called \emph{%
pseudo-displacement transformation}. Displacement transformations are also
named \emph{Weyl transformations}. They are for instance used to obtain some
renormalization results, as shows in the very recent paper \cite%
{UltravioletRenorm}. Notice that in \cite{UltravioletRenorm}, the
assumptions about normality and 2-nilpotency are reminiscent of the
non-commutativity of operator-valued coefficients, yielding higher degree
terms in the fields -- as occurs in our setting -- that are particularly
difficult to control.

More broadly, the analysis presented here enables a flow-based algorithm for
systematic, higher-order diagonalization and self-energy renormalization.
This strategy, detailed in Section \ref{sectionrenorm}, is conceptually
aligned with multi-scale analysis and the iterative, local Lie-Schwinger
block-diagonalization method introduced by Fr\"{o}hlich and Pizzo in 2020 
\cite{Lie-Schwinger0}.\bigskip

\noindent \textbf{Structure of the paper. }To summarize, Theorems \ref%
{MainDiago}--\ref{MainDisplacement} constitute the main results of the
paper, which is divided into two main parts:

\begin{itemize}
\item Section \ref{SectionGSBMain} presents the mathematical framework
required for a rigorous definition of the models, while Section \ref%
{Brockett-Wegner Flow} states our main results, outlines the flow-based
algorithm, and provides some elementary examples.

\item The remaining parts of the paper comprise the technical sections.
Section \ref{Fiberflowsection} studies systems of operator-valued
differential equations within each elementary quantum system (e.g., each
qudit). Section \ref{Foundations} provides the rigorous arguments needed to
define generalized spin-boson models as self-adjoint operators while
establishing crucial bounds. Finally, Section \ref{Unitary Flow}
investigates the time-dependent, unitarily equivalent spin-boson models and
derives their asymptotics, which yields the existence of a unitary operator $%
U$ such that (\ref{sdsdsdsdsdsd1}) holds. The arguments developed in this
section resolve the principal obstruction caused by the non-closedness of
the flow and provide the key ingredients for the proofs of Theorems \ref%
{MainDiago}--\ref{MainDisplacement}.\bigskip
\end{itemize}

\noindent \textbf{General notation. }We denote the set of non-negative real
numbers by $\mathbb{R}_{0}^{+}\doteq \lbrack 0,\infty )$, while $\mathbb{R}%
^{+}\doteq (0,\infty )$. Also, $\mathbb{N}_{0}\doteq \mathbb{N}\cup
\{0\}=\{0,1,2,\ldots \}$. The standard notation $A\lesssim B$ denotes $A\leq
DB$ for a certain constant $D$. We write $A\lesssim _{n}B$ to indicate that
the implicit constant $D=D_{n}$ depends on the parameter $n$. The constant $%
D $ is also used as a generic constant, which is not necessarily the same
from one equation to another. If $s,t\in \mathbb{R}$ are real numbers then $%
s\wedge t$\ and $s\vee t$ stand respectively for the minimum and the maximum
of $s,t$.

If $\mathcal{X}$ is a Banach space, its norm is written by default $%
\left\Vert \cdot \right\Vert _{\mathcal{X}}$, unless another notation is
introduced (see below). If $\mathcal{X}$ is a Hilbert space then its scalar
product is denoted by $\langle \cdot ,\cdot \rangle _{\mathcal{X}}$ and it
is by definition anti-linear in its first argument and linear in its second.
Here, $\mathbf{1\equiv 1}_{\mathcal{X}}\ $is the identity operator on $%
\mathcal{X}$. For any $n\in \mathbb{N}$, note that the $n$--fold tensor
product of the same vector space $\mathcal{X}$ is denoted by $\mathcal{X}%
^{\otimes n}$. By definition, $\mathcal{X}^{\otimes 0}\doteq \mathbb{C}$.
Also, $\vee ^{0}\mathcal{X}\doteq \mathbb{C}$ and, for $n\in \mathbb{N}$, $%
\vee ^{n}\mathcal{X}$ is the subspace of totally symmetric $n$--particle
wave functions in $\mathcal{X}^{\otimes n}$.

The set of (linear) operators acting on a Banach space $\mathcal{X}$ is
denoted by $\mathcal{L}(\mathcal{X})$. The commutator between two operators
is denoted $\left[ A,B\right] \doteq AB-BA$, as usual. For any $A\in 
\mathcal{L}(\mathcal{X})$, $\mathcal{D}(A)$ denotes the corresponding domain
and 
\begin{equation*}
\left\Vert A\right\Vert _{\mathrm{op}}\doteq \sup_{f\in \mathcal{X}%
:\left\Vert f\right\Vert _{\mathcal{X}}=1}\left\Vert Af\right\Vert _{%
\mathcal{X}}
\end{equation*}%
if the supremum is well-defined (i.e., finite), and $\Vert A\Vert _{\mathrm{%
op}}\doteq \infty $ otherwise. It leads to a Banach algebra 
\begin{equation*}
\mathcal{B}\left( \mathcal{X}\right) \doteq \left\{ A\in \mathcal{L}\left( 
\mathcal{X}\right) :\left\Vert A\right\Vert _{\mathrm{op}}<\infty \right\}
\subseteq \mathcal{L}\left( \mathcal{X}\right)
\end{equation*}%
of bounded operators acting on a Banach space $\mathcal{X}$, endowed with
the operator norm topology. It is (trivially) a von Neumann algebra when $%
\mathcal{X}$ is a Hilbert space. We also use the notation%
\begin{equation*}
\mathcal{L}^{2}\left( \mathcal{X}\right) \doteq \left\{ A\in \mathcal{B}%
\left( \mathcal{X}\right) :\left\Vert A\right\Vert _{2}\doteq \sqrt{\mathrm{%
tr}\left( A^{\ast }A\right) }<\infty \right\}
\end{equation*}%
for the Hilbert space of Hilbert-Schmidt operators acting on a separable
Hilbert space $\mathcal{X}$. Here $\Vert \cdot \Vert _{2}\equiv \Vert \cdot
\Vert _{\mathcal{L}^{2}\left( \mathcal{X}\right) }$ and $\mathrm{tr}(\cdot )$
denotes the usual trace for operators. We also use the standard notation $%
\mathcal{L}^{1}(\mathfrak{h})$\ for the space of trace-class operators, with
norm denoted by $\Vert \cdot \Vert _{1}\equiv \Vert \cdot \Vert _{\mathcal{L}%
^{1}\left( \mathcal{X}\right) }$.

\section{Generalized Spin-Boson Models\label{SectionGSBMain}}

Although the construction and rigorous definition of the generalized
spin-boson model as a self-adjoint operator is not, per se, our main result,
it nevertheless constitutes a nontrivial task. In this section, we present
the mathematical framework required for a rigorous definition of the model.

\subsection{Generalized Quantum Spin Systems $\mathcal{S}$\label{not
separable Hilbert space S}}

This system is defined by means of infinite tensor products. The general
mathematical construction was originally constructed by von Neumann \cite%
{Neumann}. It is defined here with a countable collection $\{\mathfrak{h}%
_{k}\}_{k\in \mathbb{N}}$ of (possibly infinite-dimensional) Hilbert spaces.
Then, pick a fixed normalized vector $\varsigma _{k}\in \mathfrak{h}_{k}$
for each $k\in \mathbb{N}$. Consider the pre-Hilbert space 
\begin{equation}
\mathfrak{H}_{\mathrm{pre}}\doteq \mathrm{span}\left\{ \otimes _{k\in 
\mathbb{N}}\varphi _{k}:\varphi _{k}\in \mathfrak{h}_{k}\text{ for }k\in 
\mathbb{N}\text{ and }\exists \Lambda \subseteq \mathbb{N}\text{ such that }%
\left\vert \Lambda \right\vert <\infty \text{, }\varphi _{k}=\varsigma _{k}%
\text{ for }k\in \mathbb{N}\backslash \Lambda \right\}  \label{pre-hilbert}
\end{equation}%
endowed with the inner product uniquely defined by 
\begin{equation*}
\left\langle \otimes _{k\in \mathbb{N}}\varphi _{k},\otimes _{k\in \mathbb{N}%
}\psi _{k}\right\rangle _{\mathfrak{H}}\doteq \prod_{k\in \mathbb{N}%
}\left\langle \varphi _{k},\psi _{k}\right\rangle _{\mathfrak{h}_{k}},\qquad
\varphi _{1},\psi _{1}\in \mathfrak{h}_{1},\ldots ,\varphi _{k},\psi _{k}\in 
\mathfrak{h}_{k}.
\end{equation*}%
By completing this space with respect to the norm $\Vert \cdot \Vert _{%
\mathfrak{H}}$ associated with this inner product, one gets a Hilbert space,
formally written as 
\begin{equation}
\mathfrak{H}\equiv \bigotimes_{k\in \mathbb{N}}\mathfrak{h}_{k}\doteq 
\overline{\mathfrak{H}_{\mathrm{pre}}}^{\ \left\Vert \cdot \right\Vert _{%
\mathfrak{H}}}.  \label{fghfghfghfgh0}
\end{equation}%
This construction depends upon the choice of our family $\{\varsigma
_{k}\}_{k\in \mathbb{N}}$ of \textquotedblleft reference\textquotedblright\
vectors. Note also that the Hilbert space $\mathfrak{H}$ is in general not
separable\footnote{%
Use Cantor's diagonal argument in the case that $\mathfrak{h}_{k}$ has
dimension $2$ for each $k\in \mathbb{N}$. (The set $\{0,1\}^{\mathbb{N}}$ of
infinite binary sequences is uncountable.)}.

On each Hilbert space $\mathfrak{h}_{k}$ we use the von Neumann algebras $%
\mathscr{B}_{k}\subseteq \mathcal{B}\left( \mathfrak{h}_{k}\right) $ and $%
\mathscr{A}_{k}$, which is a $\sigma $-weakly closed two-sided ideal%
\footnote{%
Two-sided ideal: For all $u\in \mathscr{A}_{k}$ and $v,w\in \mathscr{B}_{k}$%
, $uvw\in \mathscr{A}_{k}$. The $\sigma $-weak topology is the weak$^{\ast }$
topology coming from the predual of $\mathscr{B}_{k}$. $\mathscr{A}_{k}$ is $%
\sigma $-weakly closed and, in particular, norm closed.} of $\mathscr{B}_{k}$
(not necessarily proper\footnote{%
I.e., we allow the cases $\mathscr{A}_{k}=\{0\}$ and $\mathscr{A}_{k}=%
\mathscr{B}_{k}$. In fact, if $\mathscr{B}_{k}$ is a factor, then the only $%
\sigma $-weakly closed two-sided ideals are $\{0\}$ and $\mathscr{B}_{k}$,
by \cite[Proposition 3.12]{Tak03I}.}), and do something similar with these
algebras to obtain their infinite tensor products 
\begin{equation}
\mathcal{B}(\mathfrak{H})^{\otimes }\doteq \bigotimes_{k\in \mathbb{N}}%
\mathcal{B}\left( \mathfrak{h}_{k}\right) ,\quad \mathscr{A}^{\otimes
}\doteq \bigotimes_{k\in \mathbb{N}}\mathscr{A}_{k},\quad \mathscr{B}%
^{\otimes }\doteq \bigotimes_{k\in \mathbb{N}}\mathscr{B}_{k},
\label{fghfghfghfgh}
\end{equation}%
which all belong to the von Neumann algebra $\mathcal{B}(\mathfrak{H})$ of
(all) bounded operators acting on $\mathfrak{H}$: First, use the subalgebra
generated by the sets 
\begin{equation*}
\bigcup_{k\in \mathbb{N}}\mathscr{A}_{k}^{\otimes },\qquad \bigcup_{k\in 
\mathbb{N}}\mathscr{B}_{k}^{\otimes },\qquad \bigcup_{k\in \mathbb{N}}%
\mathcal{B}\left( \mathfrak{h}_{k}\right) ^{\otimes },
\end{equation*}%
where, using the notation $\mathbf{1}_{\mathcal{X}}$ for the identity
operator on some space $\mathcal{X}$, 
\begin{equation}
\mathcal{\square }^{\otimes }\doteq \left\{ \bigotimes_{q\in \mathbb{N}%
}A_{q}:A_{k}\in \square \text{ and }A_{q}=\mathbf{1}_{\mathfrak{h}_{q}}\text{
for all }q\in \mathbb{N}\backslash \{k\}\right\} \subseteq \mathcal{B}\left( 
\mathfrak{H}\right) ,\quad k\in \mathbb{N},  \label{fgkljdflkgjdfklgjdfkgj}
\end{equation}%
$\square $ being a symbol representing $\mathcal{B}(\mathfrak{h}_{k})$, $%
\mathscr{B}_{k}$ or $\mathscr{A}_{k}$. Note that, for a given $k\in \mathbb{N%
}$, we can identify $\mathcal{\square }^{\otimes }$ with $\square $, via the
canonical identification 
\begin{equation}
A_{k}\equiv \bigotimes_{q\in \mathbb{N}}\left( \left( 1-\delta _{q,k}\right) 
\mathbf{1}_{\mathfrak{h}_{k}}+\delta _{q,k}A_{q}\right) \in \mathcal{\square 
}^{\otimes },\qquad A_{k}\in \square ,  \label{overloaded notation}
\end{equation}%
where $\delta _{\cdot ,\cdot }$ is the Kronecker symbol, keeping in mind
that $\square $ is a symbol representing $\mathcal{B}(\mathfrak{h}_{k})$, $%
\mathscr{B}_{k}$ or $\mathscr{A}_{k}$. We then take the $\sigma $-weak
closures of the generated subalgebras to obtain von Neumann algebras denoted
by (\ref{fghfghfghfgh}). In particular, by a slight abuse of notation, 
\begin{equation*}
\bigotimes_{k\in \mathbb{N}}\square _{k}\equiv \overline{\bigotimes_{k\in 
\mathbb{N}}\mathcal{\square }_{k}}^{\sigma \text{-weak}}=\left[
\bigotimes_{k\in \mathbb{N}}\square _{k}\right] ^{\prime \prime }
\end{equation*}%
(cf. the bicommutant theorem \cite[Theorem 2.4.11]{BratteliRobinsonI}),
where $\square _{k}$ is a symbol representing $\mathcal{B}(\mathfrak{h}_{k})$%
, $\mathscr{B}_{k}$ or $\mathscr{A}_{k}$. These von Neumann algebras are
highly dependent upon the choice of our family $\{\varsigma _{k}\}_{k\in 
\mathbb{N}}$, as is well-known \cite{Powers,Araki-Woods}.

An important example is given by quantum spins or qudits for which $%
\mathfrak{h}_{k}$ is a copy of the fixed Hilbert space $\mathbb{C}^{d}$ with
fixed dimension $d\in \mathbb{N}$. Note that the term \textquotedblleft
generalized quantum spins\textquotedblright\ simply refers to quantum
degrees of freedom associated with a possibly infinite-dimensional Hilbert
space.

Another example is given by the infinite tensor product space 
\begin{equation}
\mathfrak{H}=\bigotimes_{k\in \mathbb{N}}\mathcal{F}_{k}  \label{case Fock}
\end{equation}%
where, given an orthonormal basis $\left\{ f_{k}\right\} _{k\in \mathbb{N}}$
of some Hilbert space $\mathcal{H}_{\mathcal{S}}$, each $\mathfrak{h}_{k}$
in (\ref{fghfghfghfgh0}) is the infinite-dimensional (one--mode) Fock space 
\begin{equation*}
\mathcal{F}_{k}\doteq \bigoplus_{n\in \mathbb{N}_{0}}\mathrm{span}%
(f_{k})^{\otimes n},
\end{equation*}
together with the algebra of bounded operators over this space. (Recall that 
$\mathcal{X}^{\otimes 0}\doteq \mathbb{C}$ and $\mathcal{X}^{\otimes n}$
denotes the $n$--fold tensor product of the same vector space $\mathcal{X}$%
.) In this situation, we naturally take the fixed normalized vector $%
\varsigma _{k}$ as being the vacuum $(1,0,\ldots )$ of each Fock space. Note
that a Hilbert space $\mathfrak{H}$ defined by (\ref{case Fock}) is again
not separable and in this case there is an uncountable number\footnote{%
However, if the tensor product was finite then $\mathfrak{H}$ would be
unitarly equivalent to the Fock space associated with the finite-dimensional
Hilbert space with basis $\left\{ f_{1},\ldots ,f_{k_{\max }}\right\} $.} of
\textquotedblleft copies\textquotedblright\ of the separable Fock space $%
\mathcal{F(H}_{\mathcal{S}})$, see \cite[Definition 4.5, Lemma A.2]%
{SaschaLill}. See also \cite[Definition 4.6]{SaschaLill} for some fermionic
cases. More generally, the infinite tensor products of Equations (\ref%
{fghfghfghfgh0}) and (\ref{fghfghfghfgh}) can be used to model a countable
number of distinguishable quantum objects.

It now remains to define an Hamiltonian characterizing the generalized
quantum spin system. For this purpose, we use sequences $y\equiv
(y_{k})_{k\in \mathbb{N}}$ of operators in $\mathcal{B}(\mathfrak{h}%
_{k})\equiv \mathcal{B}(\mathfrak{h}_{k})^{\otimes }$ having for some $p\in 
\mathbb{N}$ a finite $\ell _{p}$--operator norm in the following sense:%
\begin{equation}
\left\Vert y\right\Vert _{\mathrm{op},p}\doteq \left( \sum_{k\in \mathbb{N}%
}\left\Vert y_{k}\right\Vert _{\mathrm{op}}^{p}\right) ^{1/p}<\infty .
\label{lp-operator norm}
\end{equation}%
Then, taking a sequence $h_{0}\equiv (h_{k,0})_{k\in \mathbb{N}}$ of
self-adjoint operators where each element of index $k\in \mathbb{N}$ is in $%
\mathcal{B}(\mathfrak{h}_{k})\equiv \mathcal{B}(\mathfrak{h}_{k})^{\otimes }$%
, the Hamiltonian of the quantum system $\mathcal{S}$ is the self-adjoint
operator formally defined by 
\begin{equation}
H_{\mathcal{S}}\doteq \sum_{k\in \mathbb{N}}h_{k,0}.  \label{HS}
\end{equation}%
When $\Vert h_{0}\Vert _{\mathrm{op},1}<\infty $, this infinite sum defines
a self-adjoint bounded operator $H_{\mathcal{S}}\in \mathcal{B}\left( 
\mathfrak{H}\right) $ acting on the infinite tensor product $\mathfrak{H}$
because of the absolute convergence of the above sum in the operator norm of
the Banach space $\mathcal{B}\left( \mathfrak{H}\right) $. Otherwise, $H_{%
\mathcal{S}}$ can be ill-defined or at least an unbounded operator. In this
situation, we consider an alternative hypothesis (other than $\Vert
h_{0}\Vert _{\mathrm{op},1}<\infty $) on the sequence $h_{0}\equiv
(h_{k,0})_{k\in \mathbb{N}}$ of self-adjoint operators in $\mathcal{B}(%
\mathfrak{h}_{k})\equiv \mathcal{B}(\mathfrak{h}_{k})^{\otimes }$:

\begin{assumption}
\label{Assumption S}\mbox{}\newline
Let $u\equiv (u_{k})_{k\in \mathbb{N}}$ with $u_{k}=u_{k}^{\ast }\in 
\mathcal{B}(\mathfrak{h}_{k})\equiv \mathcal{B}(\mathfrak{h}_{k})^{\otimes }$%
. The sequence satisfies the assumption if there is $k_{0}\in \mathbb{N}$
such that $\varsigma _{k}\in \ker u_{k}$ for all natural numbers $k\geq
k_{0} $, and the negative part of the sequence is operator-norm summable: 
\begin{equation*}
\sum_{k\in \mathbb{N}}\left\Vert u_{k}\mathbf{1}\left[ u_{k}\leq 0\right]
\right\Vert _{\mathrm{op}}<\infty ,
\end{equation*}%
where $\mathbf{1}\left[ A\leq 0\right] $ denotes the spectral projection of
a self-adjoint operator $A$ associated with its negative spectrum.
\end{assumption}

\noindent Then, if $h_{0}=u_{0}+v_{0}$ with $u_{0}\equiv (u_{k,0})_{k\in 
\mathbb{N}}$ and $v_{0}\equiv (v_{k,0})_{k\in \mathbb{N}}$ being two
sequences of self-adjoint operators respectively satisfying Assumption \ref%
{Assumption S} and $\Vert v_{0}\Vert _{\mathrm{op},1}<\infty $, then the
operator $H_{\mathcal{S}}$ is well defined by Equation (\ref{HS}) on the
dense domain $\mathfrak{H}_{\mathrm{pre}}$ (\ref{pre-hilbert}). In fact, $H_{%
\mathcal{S}}$ is essentially self-adjoint and lower semibounded, see
Proposition \ref{Proposition self-adjoint copy(1)}. By a slight abuse of
notation, we identify $H_{\mathcal{S}}\equiv H_{\mathcal{S}}^{\ast \ast }$
with its self-adjoint extension and, hence, the pre-Hilbert space $\mathfrak{%
H}_{\mathrm{pre}}$ can be seen as a core of $H_{\mathcal{S}}$.

In the example given by (\ref{case Fock}), for each $k\in \mathbb{N}$, the
normalized vector $\varsigma _{k}=(1,0,\ldots )$ is the vacuum of the Fock
space $\mathcal{F}_{k}$ and $h_{k,0}$ could be taken as equal to $%
h_{k,0}=\varepsilon _{k}a_{k}^{\ast }a_{k}P_{N_{k}}$ for some real parameter 
$\varepsilon _{k}\in \mathbb{R}$ and some natural number $N_{k}\in \mathbb{N}
$, where $a_{k}\doteq a(f_{k})$ ($a_{k}^{\ast }\doteq a(f_{k})^{\ast }$) is
the corresponding annihilation (creation) operator acting on $\mathcal{F}%
_{k} $ while $P_{N_{k}}$ is the orthogonal projection on the subspace 
\begin{equation*}
\bigoplus_{n\in \left\{ 0,\ldots ,N_{k}\right\} }\mathrm{span}%
(f_{k})^{\otimes n}\subseteq \mathcal{F}_{k}
\end{equation*}%
corresponding to wave functions whose number of particles is less than or
equal to $N_{k}$.

\subsection{Quasi-Free Bosonic Fields\label{Bosonic fields}}

Bosonic fields interacting with the system $\mathcal{S}$ are modeled via a
Fock space whose underlying one-particle space is another separable complex
Hilbert space $\mathcal{H}$. This refers to the reservoir in the theory of
open quantum systems. The Hilbert space $\mathcal{H}$ is assumed to have
always infinite dimension, i.e., $\mathrm{dim}(\mathcal{H})=\infty $. In
fact, all our established results hold in the finite-dimensional setting ($%
\mathrm{dim}(\mathcal{H})<\infty $) since it offers a simpler framework.

The bosonic Fock space over $\mathcal{H}$ is defined as usual by 
\begin{equation}
\mathcal{F}_{+}\equiv \mathcal{F}_{+}\left( \mathcal{H}\right) \doteq
\bigoplus_{n\in \mathbb{N}_{0}}\vee ^{n}\mathcal{H},  \label{Fock}
\end{equation}%
where $\vee ^{0}\mathcal{H}\doteq \mathbb{C}$ and, for $n\in \mathbb{N}$, $%
\vee ^{n}\mathcal{H}$ is the subspace of totally symmetric $n$--particle
wave functions in $\mathcal{H}^{\otimes n}$, the $n$--fold tensor product of 
$\mathcal{H}$. Recall that the scalar product $\left\langle \cdot ,\cdot
\right\rangle _{\mathcal{F}_{+}}$ on $\mathcal{F}_{+}$ is the sum over $n\in 
\mathbb{N}$ of each canonical scalar product on the $n$--sector $\vee ^{n}%
\mathcal{H}$.

Given any orthonormal basis $\{\psi _{k}\}_{k\in \mathbb{N}}$ of the
separable Hilbert space $\mathcal{H}$, we use the notation $a_{k}\doteq
a(\psi _{k})$ ($a_{k}^{\ast }\doteq a(\psi _{k})^{\ast }$), $k\in \mathbb{N}$%
, for the corresponding bosonic annihilation (creation) operators acting on $%
\mathcal{F}_{+}$. As is well-known, they are defined on the domain $\mathcal{%
D}(\mathrm{N}^{1/2})$ of the positive operator $\mathrm{N}^{1/2}$, where 
\begin{equation}
\mathrm{N}\doteq \sum_{k\in \mathbb{N}}a_{k}^{\ast }a_{k}  \label{defN}
\end{equation}%
is the so-called particle number operator. Annihilation and creation
operators obey the so-called Canonical Commutation Relation (CCR): 
\begin{equation}
\left[ a_{k},a_{l}\right] =0\qquad \text{and}\qquad \left[ a_{k},a_{l}^{\ast
}\right] =\left\langle \psi _{k},\psi _{l}\right\rangle _{\mathcal{H}}%
\mathbf{1}_{\mathcal{F}_{+}},  \label{CCR}
\end{equation}%
where $\mathbf{1}_{\mathcal{F}_{+}}$ is the identity operator on $\mathcal{F}%
_{+}$ and $\left[ A,B\right] \doteq AB-BA$ is the usual commutator of
operators $A$ and $B$.

We consider a quasi-free bosonic field. This means that the corresponding
Hamiltonian acting on the bosonic Fock space is a quadratic Hamiltonian in
terms of the creation and annihilation operators. We also use a
gauge-invariant bosonic field. In this situation, the Hamiltonian is just
the second quantization $\mathrm{d}\Gamma (\mathrm{h})$ of a certain
one-particle Hamiltonian $\mathrm{h}$ acting on the one-particle Hilbert
space $\mathcal{H}$ (cf. (\ref{Fock})). In this case, we choose an
orthonormal basis of eigenvectors of $\mathrm{h}=\mathrm{h}^{\ast }$ for $%
\{\psi _{k}\}_{k\in \mathbb{N}}\subseteq \mathcal{H}$. Each $\psi _{k}$ is
associated with the eigenvalue $\mu _{k}\in \mathbb{R}$. Then,%
\begin{equation}
\mathrm{d}\Gamma \left( \mathrm{h}\right) \doteq \sum_{k\in \mathbb{N}}\mu
_{k}a_{k}^{\ast }a_{k},  \label{Model second quantized}
\end{equation}%
which well-defines a self-adjoint operator acting on the bosonic Fock space $%
\mathcal{F}_{+}$. See \cite[Section 5.2.1]{BratteliRobinson}.

\subsection{Coupled Systems\label{Coupled Systems}}

The total system encompasses the generalized quantum spin system $\mathcal{S}
$ and quasi-free bosonic fields. This leads us to consider the Hilbert space 
$\mathfrak{F}\doteq \mathfrak{H}\otimes \mathcal{F}_{+}$ endowed with the
usual scalar product uniquely defined by 
\begin{equation*}
\left\langle f\otimes \varphi ,g\otimes \psi \right\rangle _{\mathfrak{F}%
}\doteq \left\langle f,g\right\rangle _{\mathfrak{H}}\left\langle \varphi
,\psi \right\rangle _{\mathcal{F}_{+}},\qquad f,g\in \mathfrak{H},\ \varphi
,\psi \in \mathcal{F}_{+}.
\end{equation*}%
In this situation, Equation (\ref{Model second quantized}) can be extended
to $\mathfrak{F}$ simply by taking a tensor product with $\mathbf{1}_{%
\mathfrak{H}}$, recalling again that $\mathbf{1}_{\mathfrak{H}}$ is the
identity operator on $\mathfrak{H}$. Observe in this case that the
eigenvalues $\{\mu _{k}\}_{k\in \mathbb{N}}\subseteq \mathbb{R}$ in (\ref%
{Model second quantized}) can be seen as a set of self-adjoint operators $%
\mu _{k}\equiv \mu _{k}\mathbf{1}_{\mathfrak{H}}$, $k\in \mathbb{N}$. In
fact, we can take a broader perspective by generalizing $\mathbf{1}_{%
\mathfrak{H}}\otimes \mathrm{d}\Gamma \left( \mathrm{h}\right) $ to
operator-valued coefficients: Fix an orthonormal basis $\{\psi _{k}\}_{k\in 
\mathbb{N}}$ of $\mathcal{H}$ and define the operator 
\begin{equation}
\mathrm{d}\Gamma ^{\otimes }\left( \mu \right) \doteq \sum_{k\in \mathbb{N}%
}\mu _{k}\otimes a_{k}^{\ast }a_{k}  \label{Bosonic Hamilto}
\end{equation}%
acting on $\mathfrak{F}$ for any sequence $\mu \equiv \left( \mu _{k}\right)
_{k\in \mathbb{N}}$ of self-adjoint operators in $\mathcal{B}(\mathfrak{h}%
_{k})\equiv \mathcal{B}(\mathfrak{h}_{k})^{\otimes }$.

In Section \ref{extendedquantization}, we verify with standard arguments
that, for any sequence $\mu \equiv \left( \mu _{k}\right) _{k\in \mathbb{N}}$
of self-adjoint operators in $\mathcal{B}(\mathfrak{h}_{k})\equiv \mathcal{B}%
(\mathfrak{h}_{k})^{\otimes }$, the operator $\mathrm{d}\Gamma ^{\otimes
}\left( \mu \right) $ defined on the dense subset 
\begin{equation}
\mathcal{D}_{\infty }\doteq \mathfrak{H}_{\mathrm{pre}}\otimes \mathrm{span}%
\left( \bigcup_{n\in \mathbb{N}_{0}}\bigvee^{n}\left( \mathrm{span}\{\psi
_{k}\text{ }:\text{ }k\in \mathbb{N}\}\right) \right)  \label{domain D}
\end{equation}%
is essentially self-adjoint, where $\vee ^{0}(\cdot )\doteq \mathbb{C}$ and $%
\vee ^{n}$ for $n\in \mathbb{N}$ means here that we project everything on
totally symmetric ($n$--)particle wave functions. So, Equality (\ref{Bosonic
Hamilto}) has to be understood by means of self-adjoint extensions. Put
another way, we identify the essentially self-adjoint $\mathrm{d}\Gamma
^{\otimes }\left( \mu \right) $ defined on the above domain with its
(unique) self-adjoint extension.

In the situation described by the Hamiltonian (\ref{Bosonic Hamilto}) for $%
\mu _{k}\equiv \mu _{k}\mathbf{1}_{\mathfrak{H}}$, i.e., $\mathbf{1}_{%
\mathfrak{H}}\otimes \mathrm{d}\Gamma \left( \mathrm{h}\right) $ (see (\ref%
{Model second quantized})), the full system is obviously decoupled, i.e.,
the two quantum subsystems are not linked and evolve independently of each
other. Otherwise, the system is said (by us) to be \emph{pseudo-decoupled},
meaning that the two quantum subsystems can still be strongly linked, but
without any direct energetic exchange through the creation or absorption of
bosons.

The total model we consider nevertheless introduces such an energy exchange.
In other words, the system $\mathcal{S}$ can both absorb and emit a boson
within the full model. For example, $\mathcal{S}$ can be reduced to a single
atom (or ion) with $d$ energy levels (i.e., $\mathfrak{H}=\mathbb{C}^{d}$).
A transition between these levels results from absorption of a boson if the
final energy level is higher, or emission if it is lower. It is precisely
what one wants to achieve with trapped ions for quantum computing: the
laser(s) trigger(s) energy level transition(s) of the ions. This case
belongs to a very general class of interactions formally defined by 
\begin{equation}
\sum_{k\in \mathbb{N}}\left( X_{k,0}\otimes a_{k}+X_{k,0}^{\ast }\otimes
a_{k}^{\ast }\right) =\sum_{k\in \mathbb{N}}\left( X_{k,0}\otimes a\left(
\psi _{k}\right) +X_{k,0}^{\ast }\otimes a\left( \psi _{k}\right) ^{\ast
}\right)  \label{Interaction}
\end{equation}%
for some sequence $\left( X_{k,0}\right) _{k\in \mathbb{N}}$ of operators in 
$\mathcal{B}(\mathfrak{h}_{k})\equiv \mathcal{B}(\mathfrak{h}_{k})^{\otimes
} $ with finite $\ell _{2}$--operator norm, i.e., 
\begin{equation*}
\Vert X_{0}\Vert _{\mathrm{op},2}\doteq \left( \sum_{k\in \mathbb{N}%
}\left\Vert X_{k,0}\right\Vert _{\mathrm{op}}^{2}\right) ^{1/2}<\infty .
\end{equation*}%
See (\ref{lp-operator norm}) for $p=2$. In the special case of an $n$-level
atom $S$ described above, the operators $X_{k,0}$ would typically be some
very simple operators of the form $\left\vert f\right\rangle \left\langle
g\right\vert $ for (vector) states $f$ and $g$, using the bra-ket notation
widely used in quantum physics. When $\mathfrak{H}=\mathbb{C}$, $\left(
X_{k,0}\right) _{k\in \mathbb{N}}$ is a sequence of numbers and the coupling
functions $\Psi \doteq \sum_{k\in \mathbb{N}}X_{k,0}\psi _{k}$ in (\ref%
{Interaction}) are known in the literature as form factors. One current
challenge is to make sense of these class of Hamiltonians when $\Psi \notin 
\mathfrak{\mathcal{H}}$, but this is \emph{not} addressed here.

Consequently, bearing in mind (\ref{HS}), (\ref{Bosonic Hamilto}) and (\ref%
{Interaction}), the Hamiltonian of the full system is of the form 
\begin{equation}
\mathrm{H}_{0}\doteq \sum_{k\in \mathbb{N}}\left( h_{k,0}\otimes \mathbf{1}_{%
\mathcal{F}_{+}}+\mu _{k,0}\otimes a_{k}^{\ast }a_{k}+X_{k,0}\otimes
a_{k}+X_{k,0}^{\ast }\otimes a_{k}^{\ast }\right) ,  \label{def_H0}
\end{equation}%
where $(h_{k,0})_{k\in \mathbb{N}}$, $(\mu _{k,0})_{k\in \mathbb{N}}$ and $%
\left( X_{k,0}\right) _{k\in \mathbb{N}}$ are appropriate sequences of
operators for which each element with index $k$ is in $\mathcal{B}(\mathfrak{%
h}_{k})\equiv \mathcal{B}(\mathfrak{h}_{k})^{\otimes }$. These parameters
are subject to several assumptions outlined below, which form the basis of
our mathematical framework.

\begin{condition}
\label{Assumption BWMmainresult}\mbox{}

\begin{itemize}
\item[\emph{BW0}] Let $(\mathfrak{h}_{k})_{k\in \mathbb{N}}$ be a countable
family of separable Hilbert spaces. Let $\mathcal{F}_{+}$ denote the bosonic
Fock space built over an infinite-dimensional separable Hilbert space $%
\mathcal{H}$, with an orthonormal basis given by $\{\psi _{k}\}_{k\in 
\mathbb{N}}$. In addition, $\mathscr{B}_{k}\subseteq \mathcal{B}\left( 
\mathfrak{h}_{k}\right) $ is a von Neumann algebra and $\mathscr{A}_{k}$ is
a $\sigma $-weakly closed two-sided ideal of $\mathscr{B}_{k}$.

\item[\emph{BW1}] Let $h_{0}\equiv (h_{k,0})_{k\in \mathbb{N}}$ be a
sequence of self-adjoint operators $h_{k,0}\in \mathscr{B}_{k}\equiv %
\mathscr{B}_{k}^{\otimes }$ that can be decomposed as $h_{0}=u_{0}+v_{0}$,
where $u_{0}\equiv (u_{k,0})_{k\in \mathbb{N}}$ and $v_{0}\equiv
(v_{k,0})_{k\in \mathbb{N}}$ are two sequences of self-adjoint operators $%
u_{k,0},v_{k,0}\in \mathcal{B}(\mathfrak{h}_{k})\equiv \mathcal{B}(\mathfrak{%
h}_{k})^{\otimes }$, satisfying Assumption \ref{Assumption S} and $\Vert
v_{0}\Vert _{\mathrm{op},1}<\infty $, respectively.

\item[\emph{BW2}] Let $\mu _{0}\equiv (\mu _{k,0})_{k\in \mathbb{N}}$ be a
sequence of self-adjoint operators $\mu _{k,0}\in \mathscr{B}_{k}\cap %
\mathscr{A}_{k}^{\prime }$, satisfying $\mu _{k,0}\pm 2h_{k,0}\geq \alpha
_{k,\pm }\mathbf{1}_{\mathfrak{h}_{k}}$ for some constants $\alpha _{k,\pm
}\in \mathbb{R}$ such that $\alpha _{k,+}+\alpha _{k,-}\geq 0$ at any $k\in 
\mathbb{N}$.

\item[\emph{BW3}] Let $X_{0}\equiv (X_{k,0})_{k\in \mathbb{N}}$ be a
sequence of operators $X_{k,0}\in \mathscr{A}_{k}\equiv \mathscr{A}%
_{k}^{\otimes }$, with $\Vert X_{0}\Vert _{\mathrm{op},2}<\infty $.

\item[\emph{BW4}] Assume that $X_{0}$ is a sequence of Hilbert-Schmidt
operators and that 
\begin{equation*}
\begin{array}{l}
2\sqrt{3}\left\Vert X_{k,0}\right\Vert _{2}<\alpha _{k,+}+\alpha _{k,-}, \\ 
\\ 
\inf \{\alpha _{k,\pm }:k\in \mathbb{N},\;X_{k,0}\neq 0\}>-\infty , \\ 
\\ 
\inf \left\{ \alpha _{k,+}+\alpha _{k,-}:k\in \mathbb{N},\;X_{k,0}\neq
0\right\} \doteq 2\alpha >0.%
\end{array}%
\end{equation*}
\end{itemize}
\end{condition}

Some of these conditions allow $\mathrm{H}_{0}$ in (\ref{def_H0}) to be
defined as an essentially self-adjoint operator on the same domain $\mathcal{%
D}\subseteq \mathfrak{F}$ as $\mathrm{d}\Gamma ^{\otimes }\left( \mu
_{0}\right) $. This refers to the following assertion:

\begin{theorem}[Generalized spin-boson Hamiltonians]
\label{HamilSA copy(1)}\mbox{}\newline
Assume \emph{BW0--BW1} and \emph{BW3}, with $\mathscr{A}_{k}=\mathscr{B}_{k}=%
\mathcal{B}(\mathfrak{h}_{k})$, as well as \emph{BW2} with $\mu _{k,0}\in 
\mathcal{B}(\mathfrak{h}_{k})$ replacing $\mu _{k,0}\in \mathscr{B}_{k}\cap %
\mathscr{A}_{k}^{\prime }$. Moreover, let 
\begin{equation*}
\sum_{k\in \mathbb{N}}\left( \alpha _{k,+}+\alpha _{k,-}\right)
^{-1}\left\Vert X_{k}\right\Vert _{\mathrm{op}}^{2}<\infty
\end{equation*}%
under the convention $0^{-1}\times 0\doteq 0$ when $\alpha _{k,+}+\alpha
_{k,-}=X_{k}=0$. Then, $\mathrm{H}_{0}$ is essentially self-adjoint on the
core $\mathcal{D}_{\infty }\subseteq \mathfrak{F}$ of $\mathrm{d}\Gamma
^{\otimes }\left( \mu _{0}\right) $ and we identify $\mathrm{H}_{0}$ with
its (unique) self-adjoint extension.
\end{theorem}

\begin{proof}
Combine Theorem \ref{HamilSA} and Proposition \ref{Proposition self-adjoint
copy(1)}.
\end{proof}

Note that the von Neumann algebras $\mathscr{B}_{k}\subseteq \mathcal{B}(%
\mathfrak{h}_{k})$ and $\mathscr{A}_{k}$ in Assumptions BW0-BW3 are only
needed when we apply a continuous flow of unitary transformations to an
Hamiltonian of the form (\ref{def_H0}). Furthermore, BW4 is not strictly
required here, although we impose it\footnote{%
BW4 is only a sufficient condition; see Sections \ref{Section diescrete
family} and \ref{Unitary Flow} for more details.} below to establish a
continuous unitary flow of generalized spin-boson Hamiltonians.

This model is a general version of spin-boson Hamiltonians. They were first
introduced in \cite{Arai} under the name of \textit{generalized spin-boson
models}, which now form a class of models for which the interaction term may
not necessarily be linear in the bosonic field. They have been studied in
many papers, see \cite%
{Arai2,Arai3,Falconi,Takaesu,Teranishi,Teranishi2,spinboson1,Lonigro,UltravioletRenorm,LillLonigro}%
. In all these papers, their definition as essentially self-adjoint
operators and certain properties such as ground-state properties are mainly
studied. However, no exact diagonalization scheme has been considered in the
mathematical literature, and our study goes in this direction. Note
additionally that the generalized spin-boson models in \cite%
{Arai2,Arai3,Falconi,Takaesu,Teranishi,Teranishi2,spinboson1,Lonigro,UltravioletRenorm}
have a bosonic Hamiltonian of the form $\mathbf{1}_{\mathfrak{H}}\otimes 
\mathrm{d}\Gamma \left( \mathrm{h}\right) $ (see (\ref{Model second
quantized})), which is a restricted version of the term $\mathrm{d}\Gamma
^{\otimes }\left( \mu _{0}\right) $ (\ref{Bosonic Hamilto}) used in (\ref%
{def_H0}) for general self-adjoint operators $\mu _{k,0}$ in $\mathcal{B}(%
\mathfrak{h}_{k})$ for $k\in \mathbb{N}$.

In fact, even if our main interest is to start from $\mu _{k,0}\equiv \mu
_{k,0}\mathbf{1}_{\mathfrak{H}}$, i.e., from a bosonic Hamiltonian of the
form $\mathbf{1}_{\mathfrak{H}}\otimes \mathrm{d}\Gamma \left( \mathrm{h}%
\right) $, we wish to achieve a pseudo-decoupling of the system via some
unitary transformation, and thus arrive at a model of the form 
\begin{equation}
\sum_{k\in \mathbb{N}}\left( \mu _{k,\infty }\otimes a_{k}^{\ast
}a_{k}+h_{k,\infty }\otimes \mathbf{1}_{\mathcal{F}_{+}}\right)
\label{sdsdsdsdssdsd}
\end{equation}%
for some particular sequences $(h_{k,\infty })_{k\in \mathbb{N}}$ and $(\mu
_{k,\infty })_{k\in \mathbb{N}}$ of self-adjoint operators, where each
element of index $k\in \mathbb{N}$ is in $\mathcal{B}(\mathfrak{h}%
_{k})\equiv \mathcal{B}(\mathfrak{h}_{k})^{\otimes }$. Non-rigorous
calculations in this direction have been used in the past, as in \cite%
{SpinBosonflowphysics,Kehrein-livre} on the usual spin-boson models; see
Section \ref{Illustration}.

Nevertheless, such a result cannot be generally mathematically correct
because of the nature of a unitary transformation canceling an exchange term
like (\ref{Interaction}). Formally, such a transformation would encode the
production of infinitely many bosons, leading to technical difficulties in
giving it a precise mathematical meaning. We can, however, obtain a
perturbative result by controlling exactly the error committed for a weak
exchange term. This is achieved using as a guideline the Brockett-Wegner
flow, which has been successfully implemented to diagonalize quadratic
Hamiltonians in the bosonic \cite{bach-bru-memo} and fermionic \cite%
{QuadraHamilFermio} cases.

\section{Pseudo-Decoupling via Double-Commutator Flows\label{Brockett-Wegner
Flow}}

\subsection{Application of the Brockett-Wegner Flow}

The Brockett-Wegner flow \cite{Brockett1,Wegner1,bach-bru} is a
(quadratically) non-linear first-order differential equation. It is
explained here by considering the bosonic particle number operator $\mathbf{1%
}_{\mathfrak{H}}\otimes \mathrm{N}$ and taking the model (\ref{def_H0}) as
an initial operator. In this case, it refers to the following differential
equation:%
\begin{equation}
\forall t\in \mathbb{R}_{0}^{+}:\qquad \partial _{t}\mathrm{H}_{t}=\left[ 
\mathrm{H}_{t},\left[ \mathrm{H}_{t},\mathbf{1}_{\mathfrak{H}}\otimes 
\mathrm{N}\right] \right] \ ,\qquad \mathrm{H}_{t=0}\doteq \mathrm{H}_{0}=%
\mathrm{H}_{0}^{\ast }\ ,  \label{brockettWegner}
\end{equation}%
for operators $\mathrm{H}_{t} $ acting on the Hilbert space $\mathfrak{F}%
\doteq \mathfrak{H}\otimes \mathcal{F}_{+}$, with $[A,B]\doteq AB-BA$ being
the commutator between $A$ and $B$ (provided it exists on some domain). An
heuristic solution to (\ref{brockettWegner}) is given by unitary conjugation 
\begin{equation}
\mathrm{H}_{t}=\mathrm{U}_{t,s}\mathrm{H}_{s}\mathrm{U}_{t,s}^{\ast },\qquad
s,t\in \mathbb{R}_{0}^{+},  \label{unitary equivalent}
\end{equation}%
of the initial Hamiltonian $\mathrm{H}_{0}$, where $(\mathrm{U}%
_{t,s})_{s,t\in \mathbb{R}_{0}^{+}}\subseteq \mathcal{B}\left( \mathfrak{F}%
\right) $ is a family of unitary operators solving the non-autonomous
(hyperbolic) evolution equation 
\begin{equation}
\forall s,t\in \mathbb{R}_{0}^{+}:\qquad \partial _{t}\mathrm{U}_{t,s}=-i%
\mathrm{G}_{t}\mathrm{U}_{t,s},\qquad \mathrm{U}_{s,s}\doteq \mathbf{1}_{%
\mathfrak{F}},  \label{non-autonomous evolution equation}
\end{equation}%
with self-adjoint generator $\mathrm{G}_{t}\doteq i\left[ \mathbf{1}_{%
\mathfrak{H}}\otimes \mathrm{N},\mathrm{H}_{t}\right] $. Formally, one might
expect $\mathrm{H}_{t}$ and $\mathrm{U}_{t,0}$ to converge in some sense as $%
t\rightarrow \infty $. This asymptotic behavior holds for full-rank matrices 
\cite{bach-bru} as well as for quadratic fermionic \cite{QuadraHamilFermio}
and bosonic \cite{bach-bru-memo} Hamiltonians under certain (sufficient)
conditions. Heuristically, such convergence would lead to a limiting
Hamiltonian $\mathrm{H}_{\infty }=\mathrm{U}_{\infty ,0}\mathrm{\mathrm{H}}%
_{0}\mathrm{U}_{\infty ,0}^{\ast }$, which satisfies $\left[ \mathbf{1}_{%
\mathfrak{H}}\otimes \mathrm{N},\mathrm{H}_{\infty }\right] =0$, yielding an
expression of the form (\ref{sdsdsdsdssdsd}) as \emph{a first approximation}%
. See Sections \ref{Diff flow}--\ref{Main Results} for more details.

\subsection{Differential Flows on Operator-Valued Coefficients\label{Diff
flow}}

The differential equation (\ref{brockettWegner}) raises fundamental
mathematical issues and cannot be used as it stands, given that the
operators involved are all unbounded. To overcome this difficulty, we
consider an approximate (a priori) solution $(\mathrm{H}_{t})_{t\in \mathbb{R%
}_{0}^{+}}$ of the following form: 
\begin{equation}
\mathrm{H}_{t}\doteq \sum_{k\in \mathbb{N}}\left( h_{k,t}\otimes \mathbf{1}_{%
\mathcal{F}_{+}}+\mu _{k,t}\otimes a_{k}^{\ast }a_{k}+X_{k,t}\otimes
a_{k}+X_{k,t}^{\ast }\otimes a_{k}^{\ast }\right)  \label{defHt}
\end{equation}%
for any $t\in \mathbb{R}_{0}^{+}$, where, for each $k\in \mathbb{N}$, the
families $(h_{k,t})_{t\in \mathbb{R}_{0}^{+}}$, $(\mu _{k,t})_{t\in \mathbb{R%
}_{0}^{+}}$ and $(X_{k,t})_{t\in \mathbb{R}_{0}^{+}}$ consist of mappings
from $\mathbb{R}_{0}^{+}$ to $\mathscr{B}_{k}$, $\mathscr{B}_{k}$ and $%
\mathscr{A}_{k}$, respectively (with $\mu _{k,0}\in \mathscr{B}_{k}\cap %
\mathscr{A}_{k}^{\prime }$). Next, we formally evaluate the double
commutator of the Brockett-Wegner flow (\ref{brockettWegner}) using the CCR (%
\ref{CCR}), and equate the coefficients of each monomial up to second order
in the creation and annihilation operators with those of $\partial _{t}%
\mathrm{H}_{t}$. (Neglecting terms of third order and higher in this
commutator calculation is purely heuristic at this stage, but this
truncation is rigorously justified in the sequel.)

Consequently, we find that, for each $k\in \mathbb{N}$, the three operator
families $(h_{k,t})_{t\in \mathbb{R}_{0}^{+}}$, $(\mu _{k,t})_{t\in \mathbb{R%
}_{0}^{+}}$ and $(X_{k,t})_{t\in \mathbb{R}_{0}^{+}}$ are governed by the
same system of non-linear operator-valued differential equations. Written
for three abstract families $(h_{t})_{t\in \mathbb{R}_{0}^{+}}$, $(\mu
_{t})_{t\in \mathbb{R}_{0}^{+}}$ and $(X_{t})_{t\in \mathbb{R}_{0}^{+}}$ of
operators acting on a separable Hilbert space $\mathfrak{h}$, this system
reads as follows: For any time $t\in \mathbb{R}_{0}^{+}$,%
\begin{equation}
\left\{ 
\begin{array}{llll}
\partial _{t}h_{t} & =-2X_{t}X_{t}^{\ast } & , & h_{0}\in \mathscr{B}, \\ 
\partial _{t}\mu _{t} & =2\left[ X_{t}^{\ast },X_{t}\right] =2X_{t}^{\ast
}X_{t}-2X_{t}X_{t}^{\ast } & , & \mu _{0}\in \mathscr{B}, \\ 
\partial _{t}X_{t} & =\left( h_{t}-\mu _{t}/2\right) X_{t}+X_{t}\left(
-h_{t}-\mu _{t}/2\right) & , & X_{0}\in \mathscr{A},%
\end{array}%
\right.  \label{flowoneparticle}
\end{equation}%
where $\mathscr{B}\subseteq \mathcal{B}\left( \mathfrak{h}\right) $ is a
fixed von Neumann algebra and $\mathscr{A}$ is any $\sigma $-weakly closed
two-sided ideal of $\mathscr{B}$. This differential flow on operator-valued
coefficients is an asymmetric counterpart of the hyperbolic and elliptic
flows studied in \cite{bach-bru-memo} and \cite{EllipticFlow}, respectively.

In Theorem \ref{LocalExistenceMainResult}, we give an abstract
local-existence result for a whole class of flow that includes (\ref%
{flowoneparticle}) as well as the hyperbolic and elliptic flows \cite%
{bach-bru-memo,EllipticFlow} in the bounded case. To ensure the global
existence of solutions to (\ref{flowoneparticle}), however, stringent
\textquotedblleft gap conditions\textquotedblright\ are required. These
results, applied to the case of interest of this paper, are gathered in the
following theorem:

\begin{theorem}[Operator-valued coefficient flows]
\label{GlobalConvergence}\mbox{}\newline
Let $\mathfrak{h}$ be a separable Hilbert space, $\mathscr{B}\subseteq 
\mathcal{B}(\mathfrak{h})$ a von Neumann algebra, and $\mathscr{A}$ a
two-sided $\sigma $-weakly closed ideal of $\mathscr{B}$. Assume the initial
data $(h_{0},\mu _{0},X_{0})\in \mathscr{B}^{2}\times (\mathscr{A}\cap 
\mathcal{L}^{2}(\mathfrak{h}))$ satisfies\footnote{%
H2 is always satisfied for sufficiently small $\alpha _{\pm }$.}:

\begin{itemize}
\item[\emph{(H1)}] $h_{0}$ and $\mu _{0}$ are self-adjoint;

\item[\emph{(H2)}] $\mu _{0}\pm 2h_{0}>\alpha _{\pm }\mathbf{1}_{\mathfrak{H}%
}$ for some constants $\alpha _{\pm }\in \mathbb{R}$;

\item[\emph{(H3)}] The mean parameter $\alpha \doteq (\alpha _{+}+\alpha
_{-})/2$ satisfies the small-coupling condition $\alpha >2\sqrt{3}\Vert
X_{0}\Vert _{2}$.
\end{itemize}

\noindent Then the following assertions hold:

\begin{itemize}
\item[(i)] The system (\ref{flowoneparticle}) admits a unique global
solution 
\begin{equation*}
(h,\mu ,X)\in C(\mathbb{R}_{0}^{+},\mathscr{B})^{2}\times C(\mathbb{R}%
_{0}^{+},\mathscr{A}).
\end{equation*}%
Furthermore, for each $t\in \mathbb{R}_{0}^{+}$, the operators $\mu _{t}$
and $h_{t}$ are self-adjoint.

\item[(ii)] $2\mu _{t}\geq \alpha \mathbf{1}$ and the interaction satisfies
the exponential decay bounds: 
\begin{equation*}
\Vert X_{t}\Vert _{2}^{2}\leq \mathrm{e}^{-\alpha t}\Vert X_{0}\Vert
_{2}^{2}\qquad \text{and}\qquad \Vert X_{t}\Vert _{\mathrm{op}}^{2}\leq 
\mathrm{e}^{-\alpha t}\Vert X_{0}\Vert _{\mathrm{op}}^{2},\qquad t\in 
\mathbb{R}_{0}^{+}.
\end{equation*}

\item[(iii)] For any $t\in \lbrack 0,\infty ]$, the solution components
satisfy the integral equations 
\begin{equation*}
\mu _{t}=\mu _{0}+2\int_{0}^{t}[X_{\tau }^{\ast },X_{\tau }]\,\mathrm{d}\tau
\qquad \text{and}\qquad h_{t}=h_{0}-2\int_{0}^{t}X_{\tau }X_{\tau }^{\ast }\,%
\mathrm{d}\tau ,
\end{equation*}%
and the integrals converge in $\mathcal{L}^{1}(\mathfrak{h})$ as $%
t\rightarrow \infty $. In particular, in this limit, $X_{t}\rightarrow 0$ in 
$\mathcal{L}^{2}(\mathfrak{h})$, while $\mu _{t}\rightarrow \mu _{\infty }$
and $h_{t}\rightarrow h_{\infty }$ in $\mathcal{L}^{1}(\mathfrak{h})\cap %
\mathscr{B}$.
\end{itemize}
\end{theorem}

\begin{proof}
See Proposition \ref{controleX}, Corollary \ref{pseudoHS} and Theorem \ref%
{globalexistence}.
\end{proof}

\begin{corollary}[Well-defined time-dependent Hamiltonians]
\label{coro-Ht}\mbox{}\newline
Assume \emph{BW0-BW4} of Condition \ref{Assumption BWMmainresult}. Then,
given a solution $(h_{k,t})_{t\in \mathbb{R}_{0}^{+}}$, $(\mu _{k,t})_{t\in 
\mathbb{R}_{0}^{+}}$ and $(X_{k,t})_{t\in \mathbb{R}_{0}^{+}}$ to (\ref%
{flowoneparticle}) with respective limits $h_{k,\infty }$, $\mu _{k,\infty }$
and $X_{k,\infty }=0$ for any $k\in \mathbb{N}$, the operators $\mathrm{H}%
_{t}$, $t\in \lbrack 0,\infty ]$, defined by (\ref{defHt}) on the domain $%
\mathcal{D}_{\infty }$ (\ref{domain D}) are essentially self-adjoint.
\end{corollary}

\begin{proof}
We apply Theorem \ref{SufiCondAbstrat} to verify that the hypotheses of
Proposition \ref{Lemma Htbis} (i) are satisfied, the latter of which yields
the desired result.
\end{proof}

\noindent As before, under Assumptions BW0-BW4 of Condition \ref{Assumption
BWMmainresult}, we identify $\mathrm{H}_{t}$ with its self-adjoint extension
for each time $t\in \lbrack 0,\infty ]$. Note that the uniform condition $%
\alpha >0$ (cf. BW2 and BW4) could probably be relaxed, but this is
postponed to further studies given the length of the present paper.

\subsection{Main Results\label{Main Results}}

\subsubsection{Small-Coupling Limit of Spin-Boson Models -- Sobolev-like
Estimates\label{Small-Coupling Limit}}

Given an Hamiltonian corresponding to (\ref{def_H0}) we can infer from
Theorem \ref{GlobalConvergence} the existence of families 
\begin{equation*}
\left( h_{k},\mu _{k},X_{k}\right) _{k\in \mathbb{N}}\subseteq \left[
C\left( \mathbb{R}_{0}^{+},\mathcal{B}(\mathfrak{H})^{\otimes }\right) ^{3}%
\right] ^{\mathbb{N}}
\end{equation*}%
such that, for each $k\in \mathbb{N}$, $\left( h_{k},\mu _{k},X_{k}\right) $
is the (unique) solution to (\ref{flowoneparticle}) with initial data $%
\left( h_{k,0},\mu _{k,0},X_{k,0}\right) $. Moreover, by Corollary \ref%
{coro-Ht}, there exists an associated family of self-adjoint Hamiltonians $%
\mathrm{H}_{t}$ for each time $t\in \mathbb{R}_{0}^{+}$ (see also (\ref%
{defHt})). Unfortunately, an algebraic computation of the Brockett-Wegner
flow (\ref{brockettWegner}) shows that 
\begin{equation}
\partial _{t}\mathrm{H}_{t}\neq i\left[ \mathrm{H}_{t},\mathrm{G}_{t}\right]
=\left[ \mathrm{H}_{t},\left[ \mathrm{H}_{t},\mathbf{1}_{\mathfrak{H}%
}\otimes \mathrm{N}\right] \right] \qquad \text{with}\qquad \mathrm{G}%
_{t}\doteq i\left[ \mathbf{1}_{\mathfrak{H}}\otimes \mathrm{N},\mathrm{H}_{t}%
\right] .  \label{brockettWegner2}
\end{equation}%
As a consequence, the family $\left( \mathrm{H}_{t}\right) _{t\in \mathbb{R}%
_{0}^{+}}$ does \emph{not} lead to a formal solution to this flow. Indeed,
by using (\ref{flowoneparticle}) and the CCR (\ref{CCR}), one computes that,
for any $t\in \mathbb{R}_{0}^{+}$, 
\begin{equation}
\mathrm{G}_{t}\doteq i\sum_{k\in \mathbb{N}}\left( X_{k,t}^{\ast }\otimes
a_{k}^{\ast }-X_{k,t}\otimes a_{k}\right)  \label{generator}
\end{equation}%
as well as 
\begin{equation}
\partial _{t}\mathrm{H}_{t}=i\left[ \mathrm{H}_{t},\mathrm{G}_{t}\right] +%
\mathrm{W}_{t}=\left[ \mathrm{H}_{t},\left[ \mathrm{H}_{t},\mathbf{1}_{%
\mathfrak{H}}\otimes \mathrm{N}\right] \right] +\mathrm{W}_{t},
\label{Eq 3rd order}
\end{equation}%
where%
\begin{eqnarray}
\mathrm{W}_{t} &\doteq &\frac{1}{2}\sum_{k\in \mathbb{N}}\left( \left[
X_{k,t},\mu _{k,t}\right] \otimes a_{k}^{\ast }a_{k}a_{k}+\left[ \mu
_{k,t},X_{k,t}^{\ast }\right] \otimes a_{k}^{\ast }a_{k}^{\ast }a_{k}\right)
\notag \\
&&\qquad \qquad +\frac{1}{2}\sum_{k\in \mathbb{N}}\left( \left[ X_{k,t},\mu
_{k,t}\right] \otimes a_{k}a_{k}^{\ast }a_{k}+\left[ \mu
_{k,t},X_{k,t}^{\ast }\right] \otimes a_{k}^{\ast }a_{k}a_{k}^{\ast }\right)
.  \label{ldfj}
\end{eqnarray}%
In other words, we get third-order terms in the bosonic fields that are not
present in the expression (\ref{defHt}) of $\mathrm{H}_{t}$, $t\in \mathbb{R}%
_{0}^{+}$.

So, the Brockett-Wegner flow applied to generalized spin-boson Hamiltonians
around the particle number operator \emph{does not close}. That is to say,
the structure of the initial Hamiltonian $\mathrm{H}_{0}$ given in (\ref%
{defHt}) is not conserved by the flow, in contrast to the quadratic
fermionic \cite{QuadraHamilFermio} and bosonic \cite{bach-bru-memo} cases.
The non-closure represents a major difficulty, rendering the analysis
significantly more complex than the only two previous applications of the
Brockett-Wegner flow to unbounded operators investigated in full detail \cite%
{bach-bru-memo,QuadraHamilFermio}. Consequently, new technical developments
are required to make the deployment of a double-commutator flow meaningful
here.

In the physics literature, this obstacle is overcome by just ignoring
third-order terms as is done in \cite{SpinBosonflowphysics,Kehrein-livre}.
The argument used in \cite{SpinBosonflowphysics} is that the coupling to
terms that are quadratic in bosonic fields should be unimportant for the low
lying excitations of the system, in particular on the ground state and if
the coupling strength is small. See for example \cite[Eq. (2.5)]%
{SpinBosonflowphysics}. It is also considered in the physics book \cite[%
Section 4.3]{Kehrein-livre} as a perturbative approximation in terms of the
coupling given by the $X_{k,0}$'s, without any mathematically rigorous
arguments.

Here, we study the mathematical validity of such a truncation for a far more
general class of Hamiltonians than in \cite%
{SpinBosonflowphysics,Kehrein-livre}. To this end, we need to examine the
non-autonomous evolution equation 
\begin{equation}
\partial _{t}\mathrm{U}_{t,s}=-i\mathrm{G}_{t}\mathrm{U}_{t,s},\qquad s,t\in 
\mathbb{R}_{0}^{+},  \label{flow non auto}
\end{equation}%
where the time-dependent generator $\mathrm{G}_{t}$ is defined by (\ref%
{generator}) and is thus unbounded, but with constant core $\mathcal{D}(%
\mathbf{1}_{\mathfrak{H}}\otimes \mathrm{N}^{1/2})$, thanks to Proposition %
\ref{corollarire-selfadjoint2}. Through \cite[Section VII.1]{bach-bru-memo}
we show that some conditions on the generator imply a unique solution to
this evolution equation as a unitary evolution system, see Theorem \ref%
{propagators}. The mapping $t\mapsto \mathrm{U}_{t,0}\mathrm{H}_{0}\mathrm{U}%
_{t,0}^{\ast }$ solves the following time-dependent Heisenberg equation:%
\begin{equation}
\forall t\in \mathbb{R}_{0}^{+}:\qquad \partial _{t}\mathrm{Y}_{t}=i\left[ 
\mathrm{Y}_{t},\mathrm{G}_{t}\right] \ ,\qquad \mathrm{Y}_{t=0}\doteq 
\mathrm{H}_{0}=\mathrm{H}_{0}^{\ast }.  \label{Heisenberg equation}
\end{equation}%
We then compare this solution with the family of Hamiltonians $\mathrm{H}%
_{t} $, $t\in \mathbb{R}_{0}^{+}$, of Corollary \ref{coro-Ht}. According to
the formal equations (\ref{Eq 3rd order}) and (\ref{Heisenberg equation}),
by the method of variation of parameters, we expect that 
\begin{equation}
\mathrm{H}_{t}=\mathrm{U}_{t,s}\mathrm{H}_{s}\mathrm{U}_{s,t}+\int_{s}^{t}%
\mathrm{U}_{t,\tau }\mathrm{W}_{\tau }\mathrm{U}_{\tau ,t}\mathrm{d}\tau
,\qquad s,t\in \mathbb{R}_{0}^{+},  \label{formaltranfo}
\end{equation}%
with $\mathrm{U}_{t,s}^{\ast }=\mathrm{U}_{s,t}$ for any $s,t\in \mathbb{R}%
_{0}^{+}$ and where the family of the Hamiltonians $\mathrm{W}_{t}$, $t\in 
\mathbb{R}_{0}^{+}$, gather the third-order terms highlighted in Equation (%
\ref{ldfj}). In fact, we prove that Equality (\ref{formaltranfo}) can be
made rigorous as an equality between two self-adjoint operators, thanks to
Theorem \ref{Gluing}. By taking the limit $t\rightarrow \infty $ and $s=0$,
we obtain an approximated pseudo-decoupling of the original model $\mathrm{H}%
_{0}$ via the Hamiltonian 
\begin{equation}
\mathrm{H}_{\infty }\doteq \sum_{k\in \mathbb{N}}\left( h_{k,\infty }\otimes 
\mathbf{1}_{\mathcal{F}_{+}}+\mu _{k,\infty }\otimes a_{k}^{\ast
}a_{k}\right) .  \label{Hinfinit}
\end{equation}%
It refers to the following theorem:

\begin{theorem}[Small-coupling limit of spin-boson models]
\label{MainDiago}\mbox{}\newline
Suppose Condition \ref{Assumption BWMmainresult} holds. Then there exist two
families of unitary transformations $\mathrm{U}_{\infty ,s}$ and closed
symmetric operators $\mathrm{W}_{s}$ for $s\in \mathbb{R}_{0}^{+}$, sharing
the constant core $\mathcal{D}(\mathbf{1}_{\mathfrak{H}}\otimes \mathrm{N}%
^{3/2})$, such that 
\begin{equation*}
\mathrm{H}_{\infty }=\mathrm{U}_{\infty ,0}\mathrm{H}_{0}\mathrm{U}%
_{0,\infty }+\int_{0}^{\infty }\mathrm{U}_{\infty ,\tau }\mathrm{W}_{\tau }%
\mathrm{U}_{\tau ,\infty }\mathrm{d}\tau
\end{equation*}%
holds as an operator equation on the domain $\mathcal{D}(\mathrm{H}_{0})=%
\mathcal{D}(\mathrm{H}_{\infty })$, where $\mathrm{U}_{\tau ,\infty }\equiv 
\mathrm{U}_{\infty ,\tau }^{\ast }$. Furthermore, the error introduced by
the $\mathrm{N}$-diagonal form satisfies the operator norm estimate: 
\begin{equation}
\left\Vert \left( \mathrm{H}_{\infty }-\mathrm{U}_{\infty ,0}\mathrm{H}_{0}%
\mathrm{U}_{0,\infty }\right) \left( \mathbf{1}_{\mathfrak{H}}\otimes \left( 
\mathrm{N}+\mathbf{1}_{\mathcal{F}_{+}}\right) ^{-\frac{3}{2}}\right)
\right\Vert _{\mathrm{op}}\leq D\alpha ^{-2}\Vert X_{0}\Vert _{\mathrm{op}%
,2}^{3}\,\mathrm{e}^{D\alpha ^{-1}\Vert X_{0}\Vert _{\mathrm{op},2}},
\label{sdsdsds}
\end{equation}%
where $D\in \mathbb{R}^{+}$ is a fixed constant.
\end{theorem}

\begin{proof}
The integral written in the above expression is formal on $\mathcal{D}(%
\mathrm{H}_{0})$, but this notation can be justified because it is a true
(Banach-valued) Riemann integral on $\mathfrak{F}$ as soon as $\varphi \in 
\mathcal{D}(\mathbf{1}_{\mathfrak{H}}\otimes \mathrm{N}^{3/2})$. This is
explained via Lemmata \ref{Lemma-self-adjoint1 copy(1)}, \ref%
{TechnicalIntegralWandRWR copy(1)} and Theorem \ref{Uinfinitime}. In fact,
defining $\mathrm{W}_{t}$ by (\ref{ldfj}) and Lemma \ref{Lemma-self-adjoint1
copy(1)} on the domain $\mathcal{D}(\mathbf{1}_{\mathfrak{H}}\otimes \mathrm{%
N}^{3/2})$ and using the families $h_{k}$, $\mu _{k}$ and $X_{k}$ for any $%
k\in \mathbb{N}$ obtained from Theorem \ref{globalexistence} with the
corresponding initial conditions, the result is a direct application of
Lemma \ref{DomHconst} and Theorem \ref{Hinftyunitequiv} combined with
Theorem \ref{SufiCondAbstrat} and Inequality (\ref{sufficient1}).
\end{proof}

Observe that $\mathbf{1}_{\mathfrak{H}}\otimes (\mathrm{N}+\mathbf{1}_{%
\mathcal{F}_{+}})^{-3/2}$ acts here as a smoothing operator. By
self-adjointness of $\mathbf{1}_{\mathfrak{H}}\otimes (\mathrm{N}+\mathbf{1}%
_{\mathcal{F}_{+}})^{-3/2}$, using $\Delta _{\mathrm{H}}\doteq \mathrm{H}%
_{\infty }-\mathrm{U}_{\infty ,0}\mathrm{H}_{0}\mathrm{U}_{0,\infty
}\subseteq \Delta _{\mathrm{H}}^{\ast }$ one has the equality 
\begin{equation*}
\left\Vert \left( \mathbf{1}_{\mathfrak{H}}\otimes \left( \mathrm{N}+\mathbf{%
1}_{\mathcal{F}_{+}}\right) ^{-\frac{3}{2}}\right) \Delta _{\mathrm{H}%
}\right\Vert _{\mathrm{op}}\leq \left\Vert \Delta _{\mathrm{H}}\left( 
\mathbf{1}_{\mathfrak{H}}\otimes \left( \mathrm{N}+\mathbf{1}_{\mathcal{F}%
_{+}}\right) ^{-\frac{3}{2}}\right) \right\Vert _{\mathrm{op}}.
\end{equation*}%
As the operator $\mathbf{1}_{\mathfrak{H}}\otimes (\mathrm{N}+\mathbf{1}_{%
\mathcal{F}_{+}})^{-3/2}$ is injective, note additionally that 
\begin{equation*}
\left\Vert \varphi \right\Vert _{-3/2}\doteq \left\Vert \mathbf{1}_{%
\mathfrak{H}}\otimes \left( \mathrm{N}+\mathbf{1}_{\mathcal{F}_{+}}\right)
^{-\frac{3}{2}}\varphi \right\Vert _{\mathfrak{F}},\qquad \varphi \in 
\mathfrak{F},
\end{equation*}%
defines a new norm on $\mathfrak{F}$. In particular, for a linear operator $%
A $ on $\mathfrak{F}$, 
\begin{equation*}
\left\Vert \mathbf{1}_{\mathfrak{H}}\otimes \left( \mathrm{N}+\mathbf{1}_{%
\mathcal{F}_{+}}\right) ^{-\frac{3}{2}}A\right\Vert _{\mathrm{op}}
\end{equation*}%
is nothing but the operator norm of $A$ seen as a linear mapping from the
original normed space $(\mathfrak{F},\left\Vert \cdot \right\Vert _{%
\mathfrak{F}})$ to the (new) normed space $(\mathfrak{F},\left\Vert \cdot
\right\Vert _{-3/2})$, the completion of which yields an extrapolated
(negative) Sobolev space. Indeed, the new norm $\left\Vert \cdot \right\Vert
_{-3/2}$ is the abstract Sobolev norm of order $-3/2$ from \cite[1.4
Definition]{Nagel}, associated with $-\mathbf{1}_{\mathfrak{H}}\otimes 
\mathrm{N}$ as the generator of a $C_{0}$-semigroup of contractions on $%
\mathfrak{F}$. Hence, the bound (\ref{sdsdsds}) can be understood as a \emph{%
Sobolev-like estimate} for the difference of two linear operators.

One can apply Theorem \ref{MainDiago} when $\mathscr{A}_{k}=\mathscr{B}%
_{k}\doteq \mathcal{B}(\mathfrak{h}_{k})$ and thus $\mathscr{A}_{k}^{\prime
}=\mathbb{C}\mathbf{1}_{\mathfrak{h}_{k}}$ for all $k\in \mathbb{N}$. By the
assumption BW2 of Condition \ref{Assumption BWMmainresult}, it means in this
case that the initial self-adjoint elements $\mu _{k,0}$, $k\in \mathbb{N}$,
can simply be considered as real numbers. This is the main situation of
interest in physics, but many other examples can be produced, with
non-trivial two-sided $\sigma $-weakly closed ideal $\mathscr{A}_{k}\neq 
\mathcal{B}(\mathfrak{h}_{k}),\{0\}$ and commutant $\mathscr{A}_{k}^{\prime
}\neq \mathbb{C}\mathbf{1}_{\mathfrak{h}_{k}}$:\medskip

\noindent \textbf{Elementary example. }Every $\sigma $-weakly closed
two-sided ideal of a von Neumann algebra $\mathfrak{M}$ must be of the form $%
P\mathfrak{M}$, where is a $P$ is a projection that lies in the center $%
\mathfrak{M}\cap \mathfrak{M}^{\prime }$ of $\mathfrak{M}$, see \cite[%
Proposition 3.12]{Tak03I}. Therefore, if we want a non-trivial situation, $%
\mathscr{B}_{k}$ must not be a factor\footnote{%
A von Neumann algebra $\mathfrak{M}$ is a factor if $\mathfrak{M}\cap 
\mathfrak{M}^{\prime }=\mathbb{C}\mathbf{1}$.}. For example, for any $k\in 
\mathbb{N}$, take $d\in \mathbb{N}$, 
\begin{equation}
\mathfrak{h}_{k}=\mathbb{C}^{d}\oplus \mathbb{C}^{d},\quad \mathscr{B}_{k}=%
\mathcal{B}(\mathbb{C}^{d})\oplus \mathcal{B}(\mathbb{C}^{d}),\quad %
\mathscr{A}_{k}=\mathcal{B}(\mathbb{C}^{d})\oplus \left\{ 0\right\} .
\label{ex1}
\end{equation}%
Obviously, $\mathscr{B}_{k}$ is not a factor and $\mathscr{A}_{k}$ is a
non-trivial two-sided $\sigma $-weakly closed ideal of $\mathscr{B}_{k}$
with non-trivial commutant 
\begin{equation}
\mathscr{A}_{k}^{\prime }=\mathbb{C}\mathbf{1}_{\mathbb{C}^{d}}\oplus 
\mathcal{B}(\mathbb{C}^{d}).  \label{ex2}
\end{equation}%
This example can easily be generalized. \medskip

\noindent \textbf{Generic example. }To construct a whole class of examples,
it suffices to use the general theory of direct integrals of measurable
families of Hilbert spaces, operators, and von Neumann algebras \cite%
{Niesen-direct-integrals}. Take the simpler case of constant-fiber spaces or
algebras, as explained in \cite[Section XIII.16]{ReedSimonIV}: Let $(%
\mathcal{X},\mu )$ be any $\sigma $-finite\footnote{%
There exists a strictly positive, integrable measurable function.} measure
space and $\mathcal{Y}$ any separable (possibly infinite-dimensional)
Hilbert space. The constant fiber direct integral\ of $\mathcal{Y}$ over $%
\mathcal{X}$ is, by definition, the Hilbert space 
\begin{equation*}
\int_{\mathcal{X}}^{\oplus }\mathcal{Y}\,\mu \left( \mathrm{d}x\right)
\equiv L^{2}\left( \mathcal{X},\mathcal{Y},\mu \right) \doteq \left\{ F\in 
\mathcal{Y}^{\mathcal{X}}\,:F\text{ measurable with}\,\left\Vert F\left(
\cdot \right) \right\Vert _{\mathcal{Y}}^{2}\in L^{1}\left( \mathcal{X},\mu
\right) \right\}
\end{equation*}%
of equivalence classes of measurable, square-integrable, $\mathcal{Y}$%
--valued functions with scalar product\footnote{%
The scalar product is well-defined, by the polarization identity and the
Cauchy-Schwarz inequality. See for instance \cite[Section 7.3.2]%
{Bru-Pedra-livre}.}%
\begin{equation*}
\left\langle \varphi ,\psi \right\rangle \equiv \left\langle \varphi ,\psi
\right\rangle _{L^{2}\left( \mathcal{X},\mathcal{Y},\mu \right) }\doteq
\int_{\mathcal{X}}\left\langle \varphi \left( x\right) ,\psi \left( x\right)
\right\rangle _{\mathcal{Y}}\,\mu \left( \mathrm{d}x\right) ,\qquad \varphi
,\psi \in L^{2}\left( \mathcal{X},\mathcal{Y},\mu \right) ,
\end{equation*}%
and the pointwise vector space operations 
\begin{equation*}
\left( \varphi +\psi \right) \left( x\right) =\varphi \left( x\right) +\psi
\left( x\right) ,\text{$\qquad $}\left( \text{$\alpha $}\varphi \right)
\left( x\right) =\text{$\alpha $}\varphi \left( x\right) ,\qquad \text{$%
\alpha \in \mathbb{C}$},\ \varphi ,\psi \in L^{2}\left( \mathcal{X},\mathcal{%
Y},\mu \right) .
\end{equation*}%
A mapping $A:\mathcal{X}\rightarrow \mathcal{B}(\mathcal{Y})$ is measurable
whenever the mapping $x\mapsto \left\langle \varphi ,A(x)\psi \right\rangle
_{\mathcal{Y}}$ from $\mathcal{X}$ to $\mathbb{C}$ is measurable for all $%
\varphi ,\psi \in \mathcal{Y}$. Let $L^{\infty }(\mathcal{X},\mathcal{B}(%
\mathcal{Y}),\mu )$ be the $C^{\ast }$-algebra of equivalence classes%
\footnote{$A,B\in \mathcal{Y}^{\mathcal{X}}$ belong to the same equivalence
class if $\Vert A-B\Vert _{\infty }=0$.} of measurable functions $A:\mathcal{%
X}\rightarrow \mathcal{B}(\mathcal{Y})$ with%
\begin{equation}
\left\Vert A\right\Vert _{\infty }\doteq \mathrm{ess-}\sup \left\{
\left\Vert A\left( x\right) \right\Vert _{\mathrm{op}}:x\in \mathcal{X}%
\right\} <\infty .  \label{norm}
\end{equation}%
Here, $\mathrm{ess-}\sup $ denotes the essential supremum. A bounded
operator $D$ on $L^{2}(\mathcal{X},\mathcal{Y},\mu )$ is decomposable if
there is $L^{\infty }(\mathcal{X},\mathcal{B}(\mathcal{Y}),\mu )$ such that,
for all $\psi \in L^{2}(\mathcal{X},\mathcal{Y},\mu )$,%
\begin{equation*}
\left( D\psi \right) \left( x\right) =A\left( x\right) \psi \left( x\right)
\ ,\qquad x\in \mathcal{X}\text{\quad (}\mu \text{-a.e.)}.
\end{equation*}%
If such an $A$ exists, then it is unique. In addition, the above mapping $%
A\mapsto D$ is a $\ast $-homomorphism that is isometric. See \cite[Theorem
XIII.83]{ReedSimonIV}. The operators $A(x)\in \mathcal{B}(\mathcal{Y})$, $%
x\in \mathcal{X}$, are called the fibers of $D$, which is usually written as%
\begin{equation*}
D=\int_{\mathcal{X}}^{\oplus }A\left( x\right) \,\mu \left( \mathrm{d}%
x\right) .
\end{equation*}%
The space of decomposable operators is thus identified with $L^{\infty }(%
\mathcal{X},\mathcal{B}(\mathcal{Y}),\mu )$. Then, for any $k\in \mathbb{N}$%
, take $\mathfrak{h}_{k}=L^{2}\left( \mathcal{X},\mathcal{Y},\mu \right) $, $%
\mathscr{B}_{k}=L^{\infty }(\mathcal{X},\mathcal{B}(\mathcal{Y}),\mu )$ and,
given any fixed borelian $U\subsetneq \mathcal{X}$, 
\begin{equation*}
\mathscr{A}_{k}=\left\{ A\in \mathscr{B}_{k}:A(x)=0\quad \mu \text{-a.e. on }%
U\right\} .
\end{equation*}%
The von Neumann algebra $\mathscr{B}_{k}$ is not a factor since its center
contains the space 
\begin{equation*}
L^{\infty }(\mathcal{X},\mathbb{C},\mu )\mathbf{1}_{\mathcal{Y}}\equiv
\left\{ \int_{\mathcal{X}}^{\oplus }f\left( x\right) \mathbf{1}_{\mathcal{Y}%
}\,\mu \left( \mathrm{d}x\right) :f\in L^{\infty }(\mathcal{X},\mathbb{C}%
,\mu )\right\} ,
\end{equation*}%
while $\mathscr{A}_{k}$ is a two-sided $\sigma $-weakly closed ideal with
non-trivial commutant 
\begin{equation*}
\mathscr{A}_{k}^{\prime }=\left\{ A\in \mathscr{B}_{k}:A(x)\in \mathbb{C}%
\mathbf{1}_{\mathcal{X}}\ \mu \text{-a.e. on }\mathcal{X}\backslash U\text{,}%
\quad A(x)\in \mathcal{B}(\mathcal{Y})\ \mu \text{-a.e. on }U\right\} .
\end{equation*}%
This example generalizes (\ref{ex1})--(\ref{ex2}) to a vast array of
contexts. We can even make the example more sophisticated by using
non-constant Hilbert spaces and von Neumann algebras \cite%
{Niesen-direct-integrals}, while retaining the same idea.

\subsubsection{Pseudo-Displacement Transformation\label%
{SectionmainResultPseudoDisplacement}}

We now study the unitary transformations $\mathrm{U}_{\infty ,s}$, $s\in 
\mathbb{R}_{0}^{+}$, appearing in Theorem \ref{MainDiago}, by their effect
on creation and annihilation operators. In the quadratic case studied in 
\cite{bach-bru-memo,QuadraHamilFermio}, the family of unitary propagators
obtained from a flow like (\ref{flow non auto}) implements a continuous
Bogoliubov transformation. See \cite[Theorem 14]{bach-bru-memo} in the
bosonic case and \cite[Proposition 2.12]{QuadraHamilFermio} in the fermionic
case. Here, the unitary propagator is an evolution system generated by a
strongly continuous family of polynomials of order one in the bosonic
fields. Consequently, it does not implement a continuous Bogoliubov
transformation, but rather a continuous displacement transformation, up to
correction terms that decrease quadratically with the coupling strength.

\begin{theorem}[Pseudo-displacement unitary transformation]
\label{MainDisplacement}\mbox{}\newline
Suppose Condition \ref{Assumption BWMmainresult} holds. Then for any $s\in 
\mathbb{R}_{0}^{+}$ and $k\in \mathbb{N}$, the identity 
\begin{equation}
\mathrm{U}_{\infty ,s}\left( \mathbf{1}_{\mathfrak{H}}\otimes a_{k}\right) 
\mathrm{U}_{s,\infty }^{\ast }=\mathbf{1}_{\mathfrak{H}}\otimes a_{k}+%
\mathbf{g}_{k,\infty ,s}\otimes \mathbf{1}_{\mathcal{F}_{+}}+\mathcal{R}%
_{k,\infty ,s}  \label{sdsdsd}
\end{equation}%
holds as an operator equation on $\mathcal{D}(\mathbf{1}_{\mathfrak{H}%
}\otimes \mathrm{N}^{1/2})$. Here, the displacement coefficient is given by 
\begin{equation*}
\mathbf{g}_{k,\infty ,s}\doteq (\delta _{X_{k,0},0}-1)\int_{s}^{\infty
}X_{k,\tau }^{\ast }\mathrm{d}\tau ,
\end{equation*}%
where $\delta _{i,j}$ denotes the standard Kronecker delta, and the
remainder satisfies the operator norm estimate 
\begin{equation*}
\left\Vert \mathcal{R}_{k,\infty ,s}(\mathrm{N}+\mathbf{1})^{-\frac{1}{2}%
}\right\Vert _{\mathrm{op}}\leq \left( \mathrm{e}^{D\alpha ^{-1}\left\Vert
X_{0}\right\Vert _{\mathrm{op},2}}-1\right) D\alpha ^{-1}\left\Vert
X_{k,0}\right\Vert _{\mathrm{op}}
\end{equation*}%
for some fixed constant $D\in \mathbb{R}^{+}$.
\end{theorem}

\begin{proof}
For $s=0$, combine Proposition \ref{TransfoDisplimit} with Corollary \ref%
{pseudoHS} ($T_{\mathrm{max}}=\infty $), Theorem \ref{SufiCondAbstrat}, and
Inequality (\ref{sufficient1}). Mutatis mutandis for $s\in \mathbb{R}^{+}$,
because of Corollary \ref{pseudoHS}.
\end{proof}

As shown in Equations (\ref{jk0}) and (\ref{jk0bis}), the remainders from
Propositions \ref{TransfoDisp} and \ref{TransfoDisplimit} possess explicit
formulas. One can leverage them alongside Theorems \ref{MainDisplacement}
and \ref{Uinfinitime} to derive the following operator equation on the
domain $\mathcal{D}(\mathbf{1}_{\mathfrak{H}}\otimes \mathrm{N})$: 
\begin{equation*}
\mathrm{U}_{\infty ,s}\left( \mathbf{1}_{\mathfrak{H}}\otimes \mathrm{N}%
\right) \mathrm{U}_{s,\infty }^{\ast }=\mathbf{1}_{\mathfrak{H}}\otimes 
\mathrm{N}+\sum_{k\in \mathbb{N}}\mathbf{g}_{k,\infty ,s}^{\ast }\otimes
a_{k}+\sum_{k\in \mathbb{N}}\mathbf{g}_{k,\infty ,s}\otimes a_{k}^{\ast }+%
\mathcal{R}_{\infty ,s}.
\end{equation*}%
This relation comes with an explicit bound 
\begin{equation*}
\left\Vert \mathcal{R}_{\infty ,s}(\mathrm{N}+\mathbf{1})^{-1}\right\Vert _{%
\mathrm{op}}<\infty .
\end{equation*}%
For example, we note that%
\begin{equation*}
\sum_{k\in \mathbb{N}}\left\Vert (\mathrm{N}+\mathbf{1})^{\frac{n}{2}}%
\mathcal{R}_{k,\infty ,s}(\mathrm{N}+\mathbf{1})^{-\frac{n+1}{2}}\right\Vert
_{\mathrm{op}}^{2}<\infty ,\qquad n\in \mathbb{N},
\end{equation*}%
because $\Vert X_{0}\Vert _{\mathrm{op},2}<\infty $ and all other terms in (%
\ref{sdsdsd}) satisfy the same property. We omit the formal presentation of
these supplementary results to avoid further lengthening the paper.

The pseudo-displacement transformation explained above is reminiscent of a
dressing transformations. See \cite[Definition 2.4]{DressingTransform}. In
fact, there is a strong affinity between our method and (ultraviolet)
renormalization theory; see Section \ref{sectionrenorm}.

\begin{remark}[Commutating case]
\label{Commutating case}\mbox{}\newline
If $[X_{k,t},\mu _{k,t}]=[X_{k,t}^{\ast },\mu _{k,t}]=0$ for all $k\in 
\mathbb{N}$ and all $t\in \mathbb{R}_{0}^{+}$, then $\mathcal{R}_{k,\infty
,s}\equiv 0$ and $\mathrm{W}_{t}=0$ in Theorems \ref{MainDiago} and \ref%
{MainDisplacement}. This holds in particular if $\mathfrak{h}_{k}\equiv 
\mathbb{C}$. More broadly, this behavior is preserved for general Hilbert
spaces $\mathfrak{h}_{k}$ under specific structural assumptions that force $%
\mu _{k,t}\in \mathscr{B}_{k}\cap \mathscr{A}_{k}^{\prime }$ for all $t\in 
\mathbb{R}_{0}^{+}$. See Section \ref{Illustration} for an elementary
example.
\end{remark}

\subsubsection{Toward Multi-Scale Analysis and Self-energy Renormalization
of Spin-Boson Models\label{sectionrenorm}}

Before proceeding to the technical machinery underlying the proof of our
main results, we present a conjectural algorithm suggested by our method and
an illustrative example.

Multi-scale analysis and the renormalization group (RG) are powerful
mathematical frameworks designed to address notoriously difficult problems,
particularly systems with infinitely many degrees of freedom or wildly
differing scales where standard perturbation theory fails. The
renormalization group is generally formulated as a dynamical system in which
the \textquotedblleft time\textquotedblright\ parameter corresponds to a
physical scale, such as energy. Tracking how the system parameters evolve
along this scale yields the RG flow. See, e.g., \cite{Salmhofer-livre} and
references therein. Together with multi-scale analysis, the RG flow has
become an indispensable tool for analyzing rigorously quantum many-body
problems.

In the context of generalized spin-boson models, a renormalization scheme
can be implemented to address UV-diverging self-energy \cite%
{UltravioletRenorm,Benjamin1,Benjamin2}. Similarly, the approach developed
in this paper suggests a continuous formulation of the renormalization
group: The differential flow of the spin-boson model continuously
renormalizes the scale term $\lambda _{t}\doteq \Vert X_{t}\Vert _{\mathrm{op%
},2}$ driving it to zero in the limit $t\rightarrow \infty $. Consequently,
the dominant terms are renormalized into $\mathrm{H}_{\infty }$ (\ref%
{Hinfinit}) at the expense of a higher-order remainder depending on $\lambda
_{0}$ and the creation and annihilation operators; see (\ref{ldfj}) and (\ref%
{sdsdsds}).

The renormalization group has an intrinsic limitation in terms of coupling
strengths and it is very effective to use it inductively for different
energy scales. This requires inductive bounds, scale-by-scale estimates, and
precise error control after each step. A similar strategy can be employed
here by recursively defining a sequence of differential flows, each designed
to cancel the leading-order non-$\mathrm{N}$-diagonal remainder produced by
the previous step. Note, however, that while close in spirit, our approach
is a perturbative refinement in powers of the coupling strength, rather than
a Wilsonian decomposition into physical energy or length scales.
Conceptually, the proposed iterative approach shares meanwhile strong
similarities with the iterative, local Lie-Schwinger block-diagonalization
method -- a powerful approach introduced by Fr\"{o}hlich and Pizzo in 2020 
\cite{Lie-Schwinger0} to diagonalize Hamiltonians subject to small
perturbations. See also \cite%
{Lie-Schwinger1,Lie-Schwinger2,Lie-Schwinger3,Lie-Schwinger4,Lie-Schwinger5}%
.\ 

The flow-based algorithm could proceed as follows: For the second step, one
considers the formal expression 
\begin{equation}
\mathrm{H}_{t}^{(2)}\doteq \sum_{k\in \mathbb{N}}\left( h_{k,t}^{(2)}\otimes 
\mathbf{1}_{\mathcal{F}_{+}}+\mu _{k,t}^{(2,1)}\otimes n_{k}+\mu
_{k,t}^{(2,2)}\otimes n_{k}^{2}+\mu _{k,t}^{(2,3)}\otimes n_{k}^{3}\right) +%
\mathrm{X}_{t}^{(2)},\qquad t\in \mathbb{R}_{0}^{+},  \label{H2}
\end{equation}%
where $n_{k}\doteq a_{k}^{\ast }a_{k}$ for any $k\in \mathbb{N}$ and%
\begin{equation*}
\mathrm{X}_{t}^{(2)}\doteq \frac{1}{2}\sum_{k\in \mathbb{N}%
}X_{k,t}^{(2)}\otimes \left( a_{k}n_{k}+n_{k}a_{k}\right)
+(X_{k,t}^{(2)})^{\ast }\otimes \left( n_{k}a_{k}^{\ast }+a_{k}^{\ast
}n_{k}\right) .
\end{equation*}%
At time $t=0$, assume that the sequences $(h_{k,0}^{(2)})_{k\in \mathbb{N}}$%
, $(\mu _{k,0}^{(2,1)})_{k\in \mathbb{N}}$ and $(X_{k,0}^{(2)})_{k\in 
\mathbb{N}}$ satisfy Condition \ref{Assumption BWMmainresult} and $\mu
_{k,0}^{(2,2)}=\mu _{k,0}^{(2,3)}=0$ for any $k\in \mathbb{N}$. We then
formally define 
\begin{equation*}
\mathrm{G}_{t}^{(2)}\doteq i[\mathbf{1}_{\mathfrak{H}}\otimes \mathrm{N},%
\mathrm{H}_{t}^{(2)}],\qquad t\in \mathbb{R}_{0}^{+}.
\end{equation*}%
A computation using the CCR (\ref{CCR}) yields 
\begin{equation}
\mathrm{G}_{t}^{(2)}=i[\mathbf{1}_{\mathfrak{H}}\otimes \mathrm{N},\mathrm{X}%
_{t}^{(2)}]=\frac{i}{2}\sum_{k\in \mathbb{N}}\left( (X_{k,t}^{(2)})^{\ast
}\otimes \left( n_{k}a_{k}^{\ast }+a_{k}^{\ast }n_{k}\right)
-X_{k,t}^{(2)}\otimes \left( a_{k}n_{k}+n_{k}a_{k}\right) \right)  \label{G2}
\end{equation}%
for any $t\in \mathbb{R}_{0}^{+}$. Comparing this expression with (\ref%
{generator}), we find exactly the same general structure as in the initial
step. We define the new differential flow via the equation 
\begin{equation*}
\partial _{t}\mathrm{H}_{t}^{(2)}=i[\mathrm{H}_{t}^{(2)},\mathrm{G}%
_{t}^{(2)}]+\mathrm{W}_{t}^{(2)}.
\end{equation*}%
We then equate the coefficients of each monomial of $\mathrm{H}_{t}^{(2)}$
in the creation and annihilation operators with those of $\partial _{t}%
\mathrm{H}_{t}$. The resulting system of operator-valued differential
equations is structurally very similar to (\ref{flowoneparticle}) and can be
studied in the same manner. In fact, preliminary investigations, based on
both personal computations and AI-assisted explorations, suggest that the
following system of operator-valued differential equations has to be
analyzed: 
\begin{equation*}
\left\{ 
\begin{array}{llll}
\partial _{t}h_{t} & =-\frac{1}{2}X_{t}X_{t}^{\ast } & , & h_{0}\in %
\mathscr{B}, \\ 
\partial _{t}\mu _{t}^{(1)} & =\frac{1}{2}\left( X_{t}^{\ast
}X_{t}-5X_{t}X_{t}^{\ast }\right) & , & \mu _{0}^{(1)}\in \mathscr{B}, \\ 
\partial _{t}\mu _{t}^{(2)} & =-4X_{t}X_{t}^{\ast }-2X_{t}^{\ast }X_{t} & ,
& \mu _{0}^{(2)}=0, \\ 
\partial _{t}\mu _{t}^{(3)} & =2\left[ X_{t}^{\ast },X_{t}\right] & , & \mu
_{0}^{(3)}=0, \\ 
\partial _{t}X_{t} & =L_{t}X_{t}+X_{t}R_{t} & , & X_{0}\in \mathscr{A},%
\end{array}%
\right.
\end{equation*}%
where $\mathscr{B}\subseteq \mathcal{B}\left( \mathfrak{h}\right) $ is a
fixed von Neumann algebra and $\mathscr{A}$ is any $\sigma $-weakly closed
two-sided ideal of $\mathscr{B}$ while 
\begin{equation*}
L_{t}\doteq h_{t}+\sum_{j=1}^{3}\frac{\left( -1\right) ^{j}}{2^{j}}\mu
_{t}^{(j)}\qquad \text{and}\qquad R_{t}\doteq -h_{t}-\sum_{j=1}^{3}\frac{1}{%
2^{j}}\mu _{t}^{(j)}.
\end{equation*}%
Due to the striking similarity between this system and Equation (\ref%
{flowoneparticle}), one can readily adapt the mathematical framework
developed in Section \ref{Fiberflowsection}. For instance, 
\begin{equation*}
\partial _{t}L_{t}=-\frac{3}{2}\left( X_{t}X_{t}^{\ast }+X_{t}^{\ast
}X_{t}\right) \leq 0\qquad \text{and}\qquad \partial _{t}R_{t}=-\frac{3}{2}%
\left( X_{t}X_{t}^{\ast }-3X_{t}^{\ast }X_{t}\right) .
\end{equation*}%
Consequently, the global convergence of these quantities can be proven.

Meanwhile, consider the non-autonomous evolution equation 
\begin{equation}
\partial _{t}\mathrm{U}_{t,s}^{(2)}=-i\mathrm{G}_{t}^{(2)}\mathrm{U}%
_{t,s}^{(2)},\qquad s,t\in \mathbb{R}_{0}^{+},  \label{ssdsdsdsd}
\end{equation}%
where the time-dependent generator $\mathrm{G}_{t}^{(2)}$ is unbounded, but
well-defined on $\mathcal{D}\!\left( \mathbf{1}_{\mathfrak{H}}\otimes 
\mathrm{N}^{3/2}\right) $, by Lemma\ \ref{Lemma-self-adjoint1 copy(1)}. The
mapping $t\mapsto \mathrm{U}_{t,0}^{(2)}\mathrm{H}_{0}\mathrm{U}_{0,t}^{(2)}$
solves the time-dependent Heisenberg equation%
\begin{equation*}
\forall t\in \mathbb{R}_{0}^{+}:\qquad \partial _{t}\mathrm{Y}_{t}=i[\mathrm{%
Y}_{t},\mathrm{G}_{t}^{(2)}]\ ,\qquad \mathrm{Y}_{t=0}\doteq \mathrm{H}%
_{t}^{(2)}\ .
\end{equation*}%
By the method of variation of parameters, we therefore expect that 
\begin{equation*}
\mathrm{H}_{t}^{(2)}=\mathrm{U}_{t,s}^{(2)}\mathrm{H}_{s}^{(2)}\mathrm{U}%
_{s,t}^{(2)}+\int_{s}^{t}\mathrm{U}_{t,\tau }^{(2)}\mathrm{W}_{\tau }^{(2)}%
\mathrm{U}_{\tau ,t}^{(2)}\mathrm{d}\tau ,\qquad s,t\in \mathbb{R}_{0}^{+},
\end{equation*}%
where $(\mathrm{U}_{t,s}^{(2)})^{\ast }=\mathrm{U}_{s,t}^{(2)}$ for any $%
s,t\in \mathbb{R}_{0}^{+}$, and where the family of Hamiltonians $\mathrm{W}%
_{t}^{(2)}$, $t\in \mathbb{R}_{0}^{+}$, collects the fifth-order terms of
the \emph{non}-$\mathrm{N}$-diagonal terms. Taking the limit $t\rightarrow
\infty $ with $s=0$, exactly as in Theorem~\ref{MainDiago}, we obtain the
Hamiltonian%
\begin{eqnarray*}
\mathrm{H}_{\infty }^{(2)} &\doteq &\sum_{k\in \mathbb{N}}\left( h_{k,\infty
}^{(2)}\otimes \mathbf{1}_{\mathcal{F}_{+}}+\mu _{k,\infty }^{(2,1)}\otimes
n_{k}+\mu _{k,\infty }^{(2,2)}\otimes n_{k}^{2}+\mu _{k,\infty
}^{(2,3)}\otimes n_{k}^{3}\right) \\
&=&\mathrm{U}_{t,0}^{(2)}\mathrm{H}_{0}^{(2)}\mathrm{U}_{0,t}^{(2)}+%
\int_{0}^{\infty }\mathrm{U}_{t,\tau }^{(2)}\mathrm{W}_{\tau }^{(2)}\mathrm{U%
}_{\tau ,t}^{(2)}\mathrm{d}\tau .
\end{eqnarray*}%
Furthermore, the error introduced by the $\mathrm{N}$-diagonal form is
expected to satisfy the operator norm estimate 
\begin{equation*}
\left\Vert \left( \mathrm{H}_{\infty }^{(2)}-\mathrm{U}_{t,0}^{(2)}\mathrm{H}%
_{0}^{(2)}\mathrm{U}_{0,t}^{(2)}\right) \left( \mathbf{1}_{\mathfrak{H}%
}\otimes \left( \mathrm{N}+\mathbf{1}_{\mathcal{F}_{+}}\right)
^{-5/2}\right) \right\Vert _{\mathrm{op}}\lesssim _{\alpha }\Vert X_{0}\Vert
_{\mathrm{op},2}^{5}.
\end{equation*}

To make these observations and the second step mathematically rigorous,
several claims are to be established, including the following:

\begin{itemize}
\item \textbf{Self-adjointness.} One must first show that expression (\ref%
{H2}) defines a well-defined Hamiltonian $\mathrm{H}_{t}^{(2)}$ with core $%
\mathcal{D}_{\infty }$, in direct analogy with Theorem\ \ref{HamilSA copy(1)}
for the initial step. One must likewise verify that the expression (\ref{G2}%
) defines a self-adjoint operator $\mathrm{G}_{t}^{(2)}$ with core $\mathcal{%
D}_{\infty }$, in analogy with Proposition~\ref{corollarire-selfadjoint2}.
This last property is essential for rigorously defining the family $(\mathrm{%
U}_{t,s}^{(2)})_{s,t\in \mathbb{R}_{0}^{+}}\subseteq \mathcal{B}\left( 
\mathfrak{F}\right) $ of unitary operators that solves the non-autonomous
evolution equation (\ref{ssdsdsdsd}).

\item \textbf{Bridging the gap between steps.} While the first point is
relatively standard, the second difficulty is considerably more subtle. The
Hamiltonian produced by the initial step does not take the form (\ref{H2})
at $t=0$; instead, it reads 
\begin{equation*}
\tilde{\mathrm{H}}_{0}^{(2)}=\mathrm{U}_{\infty ,0}\,\mathrm{H}_{0}\,\mathrm{%
U}_{0,\infty }=\mathrm{H}_{\infty }-\int_{0}^{\infty }\mathrm{U}_{\infty
,\tau }\,\mathrm{W}_{\tau }\,\mathrm{U}_{\tau ,\infty }\,\mathrm{d}\tau .
\end{equation*}%
On one hand, Theorem\ \ref{MainDiago} guarantees that this operator is
self-adjoint, which renders the first open problem moot at initial time $t=0$%
. On the other hand, the presence of the unitary conjugations $\mathrm{U}%
_{\infty ,\tau }$ and $\mathrm{U}_{\tau ,\infty }$ inside the integral means
that the expression does not reduce to a pure third-order term in the
creation and annihilation operators. Note furthermore that the domain of $%
\mathrm{H}_{t}^{(2)}$ is strictly smaller than that of $\tilde{\mathrm{H}}%
_{0}^{(2)}$. There is therefore a structural gap that must be bridged before
one can move on to this second stage or to an adapted version.
\end{itemize}

\noindent Assuming these difficulties can be solved inductively at every
step $n\in \mathbb{N}$, one might ultimately obtain an expression of the
form 
\begin{equation*}
\mathrm{UH}_{0}\mathrm{U}=\sum_{k\in \mathbb{N}}\left( h_{k,\infty
}^{(\infty )}\otimes \mathbf{1}_{\mathcal{F}_{+}}+\sum_{j\in \mathbb{N}}\mu
_{k,\infty }^{(\infty ,j)}\otimes n_{k}^{j}\right)
\end{equation*}%
at least on the dense domain $\mathcal{D}_{\infty }\subseteq \mathcal{D}(%
\mathbf{1}_{\mathfrak{H}}\otimes \mathrm{N}^{j})$, $j\in \mathbb{N}$, where $%
\mathrm{U}$ is some unitary operator. This polynomial series in the number
of particles $n_{k}\doteq a_{k}^{\ast }a_{k}$ can be understood from a
perturbative perspective: By combining an arbitrary number of Feynman
diagrams associated with terms of the form $X_{k}\otimes a_{k}+X_{k}^{\ast
}\otimes a_{k}^{\ast }$, the system generates diagrams corresponding to
terms of the form $(X_{k}^{\ast }X_{k})^{j}\otimes n_{k}^{j}$ for all $j\in 
\mathbb{N}$. This operator identity on $\mathcal{D}_{\infty }$, however, is
far from obvious and may not even be valid. Nonetheless, we believe that
this approach may yield new results on generalized spin-boson models, at
least up to some finite inductive step. Recall indeed that an exact
diagonalization of the spin-boson model is generally out of reach, except in
a few very special cases. This is illustrated in the next section.

Last but not least, the flow-based algorithm we propose also resonates with
self-energy renormalization: When the $\ell _{2}$--operator norm of $X_{0}$
is infinite, i.e., $\Vert X_{0}\Vert _{\mathrm{op},2}=+\infty $, it is still
possible to define from Theorem \ref{HamilSA copy(1)} an Hamiltonian of the
form 
\begin{equation}
\mathrm{H}_{0}^{(N)}\doteq \sum_{k\in \mathbb{N}}\left( h_{k,0}\otimes 
\mathbf{1}_{\mathcal{F}_{+}}+\mu _{k,0}\otimes a_{k}^{\ast }a_{k}\right)
+\sum_{k<N}\left( X_{k,0}\otimes a_{k}+X_{k,0}^{\ast }\otimes a_{k}^{\ast
}\right) ,\qquad N\in \mathbb{N},  \label{Hcutoff}
\end{equation}%
and run the flow for each $N\in \mathbb{N}$. In this case, by Theorem \ref%
{GlobalConvergence}, we can arrive at a limit operator $h_{\infty }\equiv
h_{\infty }^{(N)}$ with initial conditions $(h_{0},\mu _{0},X_{0}^{(N)})$,
where $X_{0}^{(N)}\equiv (\mathbf{1}[k<N]X_{k,0})_{k\in \mathbb{N}}$. The $%
\ell _{2}$--operator norms $E_{N}\doteq \Vert h_{\infty }^{(N)}\Vert _{%
\mathrm{op},1}$ for $N\in \mathbb{N}$\ are natural candidates for energy
counterterms. In a suitable topology, we should formally have 
\begin{equation*}
\lim_{N\rightarrow +\infty }\left( \mathrm{H}_{0}^{(N)}-E_{N}\mathbf{1}_{%
\mathfrak{F}}\right) =\mathrm{H}^{(\mathrm{ren})},
\end{equation*}%
where $\mathrm{H}^{(\mathrm{ren})}$ is a self-adjoint Hamiltonian, as in 
\cite{DressingTransform}. In the specific case of the van Hove Hamiltonian
(when $\mathfrak{H}=\mathbb{C}$), we recover exactly the known counter term
used in its renormalization.

Furthermore, integrating this idea with the flow-based algorithm suggests
that the iterative terms $h_{\infty }^{(2)},h_{\infty }^{(3)},\ldots $
effectively correspond to successive higher-order energy counterterms.
Although the models differ, these iterative coefficients can be compared to
the hierarchy of operators that must be controlled in \cite[Section 1.2 and
Equation (1.8)]{Benjamin2}. Incorporating higher-order terms should allow
one to relax the summability hypothesis on $X_{0}$, requiring only that 
\begin{equation*}
\sum_{k\in \mathbb{N}}\mu _{k,0}^{-p}\left\Vert X_{k,0}\right\Vert _{\mathrm{%
op}}<+\infty
\end{equation*}%
for some $p\in (1,+\infty )$. Therefore, Theorem \ref{MainDiago} on
approximate diagonalizations of spin-boson models appears to provide a
natural starting point for such a development, and we hope that it will
stimulate further work in this direction.

\subsection{Example: The Standard Spin-Boson Model\label{Illustration}}

Consider a single qubit interacting with a monochromatic bosonic bath. Thus,
we construct the Fock space $\mathcal{F}_{+}$ over a one-dimensional
underlying one-particle space $\mathcal{H}$. Fix a normalized vector $\psi
\in \mathcal{H}$ and set $a=a(\psi )$ and $a^{\ast }\doteq a^{\ast }(\psi )$%
. In this case, it suffices to take $\mathfrak{H}=\mathfrak{h}_{0}=\mathbb{C}%
^{2}$. We use the usual notation%
\begin{equation*}
\sigma _{x}=\left( 
\begin{array}{cc}
0 & 1 \\ 
1 & 0%
\end{array}%
\right) ,\quad \sigma _{y}=\left( 
\begin{array}{cc}
0 & -i \\ 
i & 0%
\end{array}%
\right) ,\quad \sigma _{z}=\left( 
\begin{array}{cc}
1 & 0 \\ 
0 & -1%
\end{array}%
\right)
\end{equation*}%
for the Pauli matrices. Bosons are taken in a fixed mode of energy $\mu \in 
\mathbb{R}^{+}$. Then, the Hamiltonian of the (pure dephasing) spin-boson
model is equal to 
\begin{equation}
H_{SB}\doteq \Delta \sigma _{x}\otimes \mathbf{1}_{\mathcal{F}_{+}}+\mu 
\mathbf{1}_{\mathbb{C}^{2}}\otimes a^{\ast }a+\lambda \sigma _{x}\otimes
\left( a+a^{\ast }\right) ,  \label{ssssss}
\end{equation}%
where $\lambda \in \mathbb{R}$ is the coupling constant between the bath and
the qubit while $\Delta \in \mathbb{C}$ is the so-called tunneling factor of
the qubit. In this case, the flow (\ref{flowoneparticle}) is elementary to
solve with the initial condition 
\begin{equation*}
X_{0}=\lambda \sigma _{x},\qquad h_{0}=\Delta \sigma _{x}\qquad \text{and}%
\qquad \mu _{0}=\mu \mathbf{1}_{\mathbb{C}^{2}},
\end{equation*}%
because it is decoupled: Let $\lambda :\mathbb{R}_{0}^{+}\mathbb{\rightarrow
R}$ be the unique solution to the Cauchy problem 
\begin{equation*}
\forall t\in \mathbb{R}_{0}^{+}:\qquad \lambda ^{\prime }\left( t\right)
=-\mu \lambda \left( t\right) ,\qquad \lambda \left( 0\right) =\lambda ,
\end{equation*}%
i.e., $\lambda \left( t\right) =\lambda \mathrm{e}^{-\mu t}$ for $t\in 
\mathbb{R}_{0}^{+}$, and $y:\mathbb{R}^{+}\rightarrow \mathbb{C}$ be defined
for all $t\in \mathbb{R}_{0}^{+}$ by 
\begin{equation*}
y(t)\doteq -2\int_{0}^{t}\lambda \left( \tau \right) ^{2}\mathrm{d}\tau =%
\frac{\lambda ^{2}}{\mu }\left( \mathrm{e}^{-2\mu t}-1\right) .
\end{equation*}%
Then, 
\begin{equation*}
t\mapsto \left( \Delta \sigma _{x}+y\left( t\right) \mathbf{1},\mu \mathbf{1}%
,\lambda \left( t\right) \sigma _{x}\right)
\end{equation*}%
is solution to the flow (\ref{flowoneparticle}) applied to this simple
situation. Obviously, $\lambda _{t}\rightarrow 0$ and the decay of $t\mapsto
\lambda _{t}$ is even exponentially fast. As announced in Section \ref%
{sectionrenorm}, observe that $\lim_{t\rightarrow +\infty }y(t)$ corresponds
to the self-energy renormalization term in \cite[Equation (18)]%
{DressingTransform}.

In this case, the flow (\ref{flowoneparticle}) is the \emph{untruncated}
Brockett-Wegner flow, so the diagonalization is \emph{exact}. See Remark \ref%
{Commutating case}. If we change $\lambda \sigma _{x}$ by $\lambda \sigma
_{z}$ in Equation (\ref{ssssss}), i.e., 
\begin{equation*}
\tilde{H}_{SB}\doteq \Delta \sigma _{x}\otimes \mathbf{1}_{\mathcal{F}%
_{+}}+\mu \mathbf{1}_{\mathbb{C}^{2}}\otimes a^{\ast }a+\lambda \sigma
_{z}\otimes \left( a+a^{\ast }\right) ,
\end{equation*}%
then we obtain the standard spin-boson model, as it is commonly referred to
in the literature unless otherwise specified. See \cite[Section 2]%
{LillLonigro} for a zoology of variations of these models. In this case, 
\begin{equation*}
\left[ \sigma _{z},\sigma _{x}\right] =2i\sigma _{y}\neq 0
\end{equation*}%
and therefore, one checks that the Brockett-Wegner flow \emph{does not not
close}. One can perform an approximated pseudo-decoupling by means of
Theorem \ref{MainDiago}. This is similar to what is done in \cite%
{SpinBosonflowphysics,Kehrein-livre}, except that the approximation in \cite%
{SpinBosonflowphysics,Kehrein-livre} is non-rigorous and ignore all the
terms produced in the normal ordering 
\begin{equation*}
:(a^{\ast }+a)(a^{\ast }+a):\qquad \text{and}\qquad :(a^{\ast }-a)(a^{\ast
}-a):\ .
\end{equation*}%
In particular, this produces a term $\mu _{(\cdot )}\equiv \mu \mathbf{1}_{%
\mathbb{C}^{2}}$ that remains constant, which is clearly wrong. As a
consequence, third-order terms do not appear in \cite%
{SpinBosonflowphysics,Kehrein-livre}, as they were indirectly canceled by
their non-rigorous approach. In our method, we account for changes in the
kinetic pre-factor $\mu $, as mathematically required.

Another example is given by the spin-boson model within the so-called
\textquotedblleft rotating wave approximation\textquotedblright . In this
approximation the model has terms of the form%
\begin{equation*}
\hat{H}_{SB}\doteq \Delta \sigma _{x}\otimes \mathbf{1}_{\mathcal{F}%
_{+}}+\mu \mathbf{1}_{\mathbb{C}^{2}}\otimes a^{\ast }a+\lambda \left(
\sigma _{-}\otimes a^{\ast }+\sigma _{+}\otimes a\right)
\end{equation*}%
with $\sigma _{\pm }\doteq (\sigma _{x}\pm \sigma _{y})/2$. Such
off-diagonal terms with $\sigma _{\pm }$ also make the diagonalization of
the Hamiltonian impossible with the usual methods. In particular, the
Brockett-Wegner flow does not close, again.

A general presentation of spin-boson models and their generalization along
the lines of what is done here can be found, for example, in the paper \cite[%
Introduction and Section 2.3]{spinboson1}.

\section{Systems of Operator-Valued Differential Equations\label%
{Fiberflowsection}}

Let $\mathfrak{h}$ be a Hilbert space. In this section, $\mathfrak{h}$ is
not necessarily separable. In fact, the separability is only required in
some specific situations. Recall that $\mathcal{B}(\mathfrak{h})$ is the
Banach space of bounded operators acting on $\mathfrak{h}$, with unit often
denoted for simplicity by $\mathbf{1\equiv 1}_{\mathfrak{h}}$. Let $%
\mathscr{B}\subseteq \mathcal{B}(\mathfrak{h})$ be a von Neumann algebra and 
$\mathscr{A}$ a $\sigma $-weakly closed two-sided ideal of $\mathscr{B} $.
In this section we study the system of non-linear operator-valued
differential equations (\ref{flowoneparticle}), that is, 
\begin{equation}
\left\{ 
\begin{array}{llll}
\partial _{t}h_{t} & =-2X_{t}X_{t}^{\ast } & , & h_{0}\in \mathscr{B}, \\ 
\partial _{t}\mu _{t} & =2\left[ X_{t}^{\ast },X_{t}\right] =2X_{t}^{\ast
}X_{t}-2X_{t}X_{t}^{\ast } & , & \mu _{0}\in \mathscr{B}, \\ 
\partial _{t}X_{t} & =\left( h_{t}-\mu _{t}/2\right) X_{t}+X_{t}\left(
-h_{t}-\mu _{t}/2\right) & , & X_{0}\in \mathscr{A},%
\end{array}%
\right.  \label{goodflow}
\end{equation}%
for three families $(h_{t})_{t\geq 0}$, $(\mu _{t})_{t\geq 0}$ and $%
(X_{t})_{t\geq 0}$ of bounded operators acting on $\mathfrak{h}$. We begin
with a preliminary analysis of some Dyson series.

\subsection{Preliminary Study}

Throughout this subsection, we fix a closed set $I\subseteq \mathbb{R}%
_{0}^{+}$: Either $I\doteq \lbrack 0,T]$ for some strictly positive time $%
T\in \mathbb{R}^{+}$, or $I\doteq \mathbb{R}_{0}^{+}$. Given any
norm-continuous family $\Delta \in C(I,\mathcal{B}(\mathfrak{h}))$ of
bounded operators acting on $\mathfrak{h}$, we set 
\begin{eqnarray}
Z_{t,s}^{g}\left( \Delta \right) &\doteq &\mathbf{1}+\sum_{n\in \mathbb{N}%
}\int_{s}^{t}\mathrm{d}\tau _{1}\cdots \int_{s}^{\tau _{n-1}}\mathrm{d}\tau
_{n}\prod_{k=1}^{n}\Delta _{\tau _{k}},  \label{eq:dysonzg} \\
Z_{t,s}^{d}\left( \Delta \right) &\doteq &\mathbf{1}+\sum_{n\in \mathbb{N}%
}\int_{s}^{t}\mathrm{d}\tau _{1}\cdots \int_{s}^{\tau _{n-1}}\mathrm{d}\tau
_{n}\prod_{k=1}^{n}\Delta _{\tau _{n-k+1}},  \label{dysonzd}
\end{eqnarray}%
for any $s,t\in I$. Observe that 
\begin{equation*}
\left( Z_{t,s}^{g}\left( \Delta \right) \right) _{s,t\in I},\left(
Z_{t,s}^{d}\left( \Delta \right) \right) _{s,t\in I}\in C\left( I^{2},%
\mathcal{B}\left( \mathfrak{h}\right) \right) .
\end{equation*}%
In the following lemma, we give first some elementary properties of these
Dyson series.

\begin{lemma}[Non-autonomous evolutions and boundedness of Dyson series]
\label{lemma-Dyson-Series}\mbox{}\newline
Given $\Delta \in C(I,\mathcal{B}(\mathfrak{h}))$, let $Z_{t,s}^{g}\left(
\Delta \right) $ and $Z_{t,s}^{d}\left( \Delta \right) $ be respectively
defined by (\ref{eq:dysonzg}) and (\ref{dysonzd}) for all $s,t\in I$. Then,
the following hold:

\begin{itemize}
\item[(i)] $(t,s)\mapsto Z_{t,s}^{g}\left( \Delta \right) $ and $%
(t,s)\mapsto Z_{t,s}^{d}\left( \Delta \right) $ are the unique solutions in $%
C(I,\mathcal{B}(\mathfrak{h}))$ to the following non-autonomous evolution
equations: For any $s,t\in I$, 
\begin{equation*}
\partial _{t}Z_{t,s}^{g}\left( \Delta \right) =\Delta _{t}Z_{t,s}^{g}\left(
\Delta \right) ,\quad \partial _{t}Z_{t,s}^{d}\left( \Delta \right)
=Z_{t,s}^{d}\left( \Delta \right) \Delta _{t}\quad \text{with}\quad
Z_{t,t}^{g}\left( \Delta \right) =Z_{t,t}^{d}\left( \Delta \right) =\mathbf{1%
}.
\end{equation*}%
Furthermore, 
\begin{equation*}
\partial _{s}Z_{t,s}^{g}\left( \Delta \right) =-Z_{t,s}^{g}\left( \Delta
\right) \Delta _{s}\qquad \text{and}\qquad \partial _{s}Z_{t,s}^{d}\left(
\Delta \right) =-\Delta _{s}Z_{t,s}^{d}\left( \Delta \right) .
\end{equation*}

\item[(ii)] If there is $\gamma \in \mathbb{R}$ such that $\Delta _{t}\leq
\gamma \mathbf{1}$ for any $t\in I$ then, for any $s,t\in I$, 
\begin{equation*}
Z_{t,s}^{g}\left( \Delta \right) ^{\ast }Z_{t,s}^{g}\left( \Delta \right)
\leq \mathrm{e}^{2\gamma \left( t-s\right) }\mathbf{1}\qquad \text{and}%
\qquad Z_{t,s}^{d}\left( \Delta \right) Z_{t,s}^{d}\left( \Delta \right)
^{\ast }\leq \mathrm{e}^{2\gamma \left( t-s\right) }\mathbf{1}.
\end{equation*}
\end{itemize}
\end{lemma}

\begin{proof}
Assertion (i) is an elementary property of these Dyson series. We omit the
details and refer the reader to a standard textbook, such as \cite[Section
X.12]{ReedSimonII}. Observe from (i) that, for any $s,t\in I$, 
\begin{eqnarray*}
\partial _{t}\left\{ Z_{t,s}^{g}\left( \Delta \right) ^{\ast }\mathrm{e}%
^{-2\gamma \left( t-s\right) }Z_{t,s}^{g}\left( \Delta \right) \right\}
&=&2Z_{t,s}^{g}\left( \Delta \right) ^{\ast }\mathrm{e}^{-\gamma \left(
t-s\right) }\left( \Delta _{t}-\gamma \mathbf{1}\right) \mathrm{e}^{-\gamma
\left( t-s\right) }Z_{t,s}^{g}\left( \Delta \right) , \\
\partial _{t}\left\{ Z_{t,s}^{d}\left( \Delta \right) \mathrm{e}^{-2\gamma
\left( t-s\right) }Z_{t,s}^{d}\left( \Delta \right) ^{\ast }\right\}
&=&2Z_{t,s}^{d}\left( \Delta \right) \mathrm{e}^{-\gamma \left( t-s\right)
}\left( \Delta _{t}-\gamma \mathbf{1}\right) \mathrm{e}^{-\gamma \left(
t-s\right) }Z_{t,s}^{d}\left( \Delta \right) ^{\ast }.
\end{eqnarray*}%
Thus, if $\Delta _{t}\leq \gamma \mathbf{1}$ for any $s,t\in I$ then it
follows by integration that 
\begin{equation*}
Z_{t,s}^{g}\left( \Delta \right) ^{\ast }\mathrm{e}^{-2\gamma \left(
t-s\right) }Z_{t,s}^{g}\left( \Delta \right) \leq \mathbf{1}\qquad \text{and}%
\qquad Z_{t,s}^{d}\left( \Delta \right) \mathrm{e}^{-2\gamma \left(
t-s\right) }Z_{t,s}^{d}\left( \Delta \right) ^{\ast }\leq \mathbf{1}.
\end{equation*}%
This leads to Assertion (ii).
\end{proof}

Note from (\ref{eq:dysonzg}) and (\ref{dysonzd}) combined with the triangle
inequality that, for any $\Delta \in C(I,\mathcal{B}(\mathfrak{h}))$ and $%
s,t\in I$,%
\begin{multline*}
\max \left\{ \left\Vert \Delta _{t}Z_{t,s}^{g}\left( \Delta \right)
\right\Vert _{\mathrm{op}},\left\Vert Z_{t,s}^{g}\left( \Delta \right)
\Delta _{t}\right\Vert _{\mathrm{op}},\left\Vert Z_{t,s}^{d}\left( \Delta
\right) \Delta _{t}\right\Vert _{\mathrm{op}},\left\Vert Z_{t,s}^{d}\left(
\Delta \right) \Delta _{t}\right\Vert _{\mathrm{op}}\right\} \\
\leq \left\Vert \Delta _{t}\right\Vert _{\mathrm{op}}\exp \left(
\int_{s}^{t}\left\Vert \Delta _{t}\right\Vert _{\mathrm{op}}\mathrm{d}\tau
\right) .
\end{multline*}%
This bound is very elementary but not uniform with respect to the operator
norm of the operator family $\Delta \in C(I,\mathcal{B}(\mathfrak{h}))$. In
the sequel we use sequences of such operator families and more uniform
estimates are necessary. We therefore use the same strategy as in the
monograph \cite{bach-bru-memo} for upper semibounded operators to derive a
less trivial upper bound, even if we only handle here families of bounded
operators. This requires an additional continuity property, namely the
Lipschitz continuity of the generator $\Delta $, in the spirit of the theory
of parabolic evolution equations. The arguments used below are somehow
\textquotedblleft known\textquotedblright , but not as elementary as in the
previous lemma.

\begin{lemma}[Uniform estimates on Dyson series]
\label{lemma-Dyson-Series2}\mbox{}\newline
Given $\Delta \in C(I,\mathcal{B}(\mathfrak{h}))$, let $Z_{t,s}^{g}\left(
\Delta \right) $ and $Z_{t,s}^{d}\left( \Delta \right) $ be respectively
defined by (\ref{eq:dysonzg}) and (\ref{dysonzd}) for all $s,t\in I$. Assume
additionally that $\Delta _{0}\leq 0$, $\Delta $ is Lipschitz continuous in
the norm topology, i.e., 
\begin{equation}
\left\Vert \Delta _{t}-\Delta _{s}\right\Vert _{\mathrm{op}}\leq C_{\Delta
}\left\vert t-s\right\vert ,\qquad s,t\in I,  \label{rdef0}
\end{equation}%
for some fixed constant $C_{\Delta }\in \mathbb{R}^{+}$, and 
\begin{equation}
r\doteq \sup_{t\in I}\left\Vert \Delta _{t}-\Delta _{0}\right\Vert _{\mathrm{%
op}}<\infty .  \label{rdef}
\end{equation}%
Then, for any $s,t\in I$ with $t\neq s$, 
\begin{multline*}
\max \left\{ \left\Vert \Delta _{t}Z_{t,s}^{g}\left( \Delta \right)
\right\Vert _{\mathrm{op}},\left\Vert Z_{t,s}^{g}\left( \Delta \right)
\Delta _{s}\right\Vert _{\mathrm{op}},\left\Vert Z_{t,s}^{d}\left( \Delta
\right) \Delta _{t}\right\Vert _{\mathrm{op}},\left\Vert \Delta
_{s}Z_{t,s}^{d}\left( \Delta \right) \right\Vert _{\mathrm{op}}\right\} \\
\leq \mathrm{C}\left( s,t,C_{\Delta },r\right) +\frac{1}{\mathrm{e}%
\left\vert t-s\right\vert },
\end{multline*}%
with the function $\mathrm{C}$ defined by 
\begin{equation}
\mathrm{C}\left( s,t,C_{\Delta },r\right) \doteq \left( 3r+\mathrm{e}%
^{-1}\left( \left( \sqrt{2}\pi +1\right) r^{2}+C_{\Delta }\right) \left\vert
t-s\right\vert \right) \mathrm{e}^{r\left( t-s\right) }+4\sqrt{2}r\mathrm{e}%
^{-1}.  \label{definition de C 1}
\end{equation}
\end{lemma}

\begin{proof}
The proof follows very closely the one given in \cite[Lemma 35]%
{bach-bru-memo}, which is even more complicated because it is done for
semibounded generators. We reproduce it to establish explicit and uniform
bounds and to ensure the paper is self-contained. However we do the proof
only for the Dyson series (\ref{eq:dysonzg}), the arguments for the other
series (\ref{dysonzd}) being basically the same, with a few modifications.
We assume all the assumptions of the lemma and divide the proof into several
steps. To simplify the equations, we use the notation $Z_{t,s}^{g}\equiv
Z_{t,s}^{g}\left( \Delta \right) $.\medskip

\noindent \underline{Step 1:} For any positive operator $X=X^{\ast }\geq 0$, 
$\beta \in \mathbb{R}^{+}$ and $\alpha \in \mathbb{R}_{0}^{+}$, the operator 
$X^{\alpha }\mathrm{e}^{-\beta X}$ is bounded in operator norm by%
\begin{equation}
\left\Vert X^{\alpha }\mathrm{e}^{-\beta X}\right\Vert _{\mathrm{op}}\leq
\sup_{x\in \mathbb{R}_{0}^{+}}\left\{ x^{\alpha }\mathrm{e}^{-\beta
x}\right\} =\left( \frac{\alpha }{\mathrm{e}\beta }\right) ^{\alpha }.
\label{petit inequality new}
\end{equation}%
Let $s,t\in I$. Since $\Delta _{0}=\Delta _{0}^{\ast }\leq 0$, we then apply
this inequality to $X=-\Delta _{0}$, $\alpha =1$ and $\beta =|t-s|$ to
obtain that 
\begin{equation}
\left\Vert \Delta _{0}\mathrm{e}^{\left\vert t-s\right\vert \Delta
_{0}}\right\Vert _{\mathrm{op}}\leq \frac{1}{\mathrm{e}\left\vert
t-s\right\vert },\qquad s,t\in I.  \label{petit inequality newbisbis}
\end{equation}%
Moreover, for any $s,t,u\in I$ with $u\geq t>s$, observe that 
\begin{eqnarray}
\left\Vert \mathrm{e}^{\left( u-s\right) \Delta _{0}}-\mathrm{e}^{\left(
t-s\right) \Delta _{0}}\right\Vert _{\mathrm{op}} &=& \left\Vert \left( 
\mathbf{1}-\mathrm{e}^{\left( u-t\right) \Delta _{0}}\right) \mathrm{e}%
^{\left( t-s\right) \Delta _{0}}\right\Vert _{\mathrm{op}}\leq \sup_{x\in 
\mathbb{R}_{0}^{-}}\left\{ \left( 1-\mathrm{e}^{\left( u-t\right) x}\right) 
\mathrm{e}^{\left( t-s\right) x}\right\}  \notag \\
&\leq &\left( u-t\right) \left( t-s\right) ^{-1},  \label{interpolation1}
\end{eqnarray}%
while, as $\Delta _{0}\leq 0$, 
\begin{equation}
\left\Vert \mathrm{e}^{\left( u-s\right) \Delta _{0}}-\mathrm{e}^{\left(
t-s\right) \Delta _{0}}\right\Vert _{\mathrm{op}}\leq 2.
\label{interpolation2}
\end{equation}%
We can thus interpolate the two above estimates to get 
\begin{equation}
\left\Vert \mathrm{e}^{\left( u-s\right) \Delta _{0}}-\mathrm{e}^{\left(
t-s\right) \Delta _{0}}\right\Vert _{\mathrm{op}}\leq 2^{1-\nu }\left(
u-t\right) ^{\nu }\left( t-s\right) ^{-\nu }  \label{interpolation3}
\end{equation}%
for any $s,t,u\in I$ with $u\geq t>s$ and every $\nu \in \left[ 0,1\right] $%
. \medskip

\noindent \underline{Step 2:} We use now Duhamel's formula to obtain that 
\begin{equation}
Z_{t,s}^{g}=\mathrm{e}^{\left( t-s\right) \Delta _{0}}+\int_{s}^{t}\mathrm{e}%
^{\left( t-\tau \right) \Delta _{0}}\left( \Delta _{\tau }-\Delta
_{0}\right) Z_{\tau ,s}^{g}\mathrm{d}\tau ,\qquad s,t\in I.
\label{definition of W 2}
\end{equation}%
Remark also from Lemma \ref{lemma-Dyson-Series} (ii) combined with the
assumptions $\Delta _{0}\leq 0$ and (\ref{rdef}) that 
\begin{equation}
\left\Vert Z_{t,s}^{g}\right\Vert _{\mathrm{op}}\leq \mathrm{e}^{r\left(
t-s\right) },\qquad s,t\in I.  \label{upper bound for W}
\end{equation}%
Now, we use Equations (\ref{rdef}) and (\ref{interpolation3})--(\ref{upper
bound for W}) together with $\Delta _{0}\leq 0$ and the triangle inequality
to get that, for any $s,t,u\in I$ with $u\geq t>s$ and every $\nu \in
\lbrack 0,1)$,%
\begin{eqnarray}
\left\Vert Z_{u,s}^{g}-Z_{t,s}^{g}\right\Vert _{\mathrm{op}} &\leq &2^{1-\nu
}\left( u-t\right) ^{\nu }\left( t-s\right) ^{-\nu }+r\left( u-t\right) 
\mathrm{e}^{r\left( u-s\right) }  \notag \\
&&+\frac{2^{1-\nu }r}{1-\nu }\mathrm{e}^{r\left( t-s\right) }\left(
u-t\right) ^{\nu }\left( t-s\right) ^{1-\nu }.  \label{interpolation4}
\end{eqnarray}%
This inequality is useful to obtain below integrable estimates. See (\ref%
{estimate integral estimaatebis})--(\ref{estimate integral estimaatebisbis})
below. \medskip

\noindent \underline{Step 3:} Now, fix $s\in I$ with $s<T$ if $I=[0,T]$. For
any sufficiently small $\epsilon \in \mathbb{R}^{+}$, let 
\begin{equation*}
\mathcal{W}_{t,s}^{(\epsilon )}\doteq \int_{s}^{t-\epsilon }\mathrm{e}%
^{\left( t-\tau \right) \Delta _{0}}\left( \Delta _{\tau }-\Delta
_{0}\right) Z_{\tau ,s}^{g}\mathrm{d}\tau ,\qquad t\in I\cap (s+\epsilon
,\infty ).
\end{equation*}%
Compare this definition with the second term on the right-hand side of
Equation (\ref{definition of W 2}). As $\epsilon \rightarrow 0^{+}$, $%
\mathcal{W}_{t,s}^{\left( \epsilon \right) }$ converges to $\mathcal{W}%
_{t,s}^{(0)}$ in the norm topology and $\mathcal{W}_{t,s}^{(\epsilon )}$ is
norm differentiable with respect to $t\in I\cap (s+\epsilon ,\infty )$, with
derivative equal to%
\begin{equation}
\partial _{t}\mathcal{W}_{t,s}^{(\epsilon )}=\mathrm{e}^{\epsilon \Delta
_{0}}\left( \Delta _{t-\epsilon }-\Delta _{0}\right) Z_{t-\epsilon
,s}^{g}+\int_{s}^{t-\epsilon }\Delta _{0}\mathrm{e}^{\left( t-\tau \right)
\Delta _{0}}\left( \Delta _{\tau }-\Delta _{0}\right) Z_{\tau ,s}^{g}\mathrm{%
d}\tau  \label{interpolation4bis}
\end{equation}%
for any $t\in I\cap (s+\epsilon ,\infty )$. Therefore, using (\ref{rdef}), (%
\ref{interpolation2}) and (\ref{upper bound for W}) together with $\Delta
_{0}\leq 0$, the triangle inequality and the equality%
\begin{eqnarray*}
\int_{s}^{t-\epsilon }\Delta _{0}\mathrm{e}^{\left( t-\tau \right) \Delta
_{0}}\left( \Delta _{\tau }-\Delta _{0}\right) Z_{\tau ,s}^{g}\mathrm{d}\tau
&=&\int_{s}^{t-\epsilon }\Delta _{0}\mathrm{e}^{\left( t-\tau \right) \Delta
_{0}}\left( \left( \Delta _{\tau }-\Delta _{0}\right) Z_{\tau ,s}^{g}-\left(
\Delta _{t}-\Delta _{0}\right) Z_{t,s}^{g}\right) \mathrm{d}\tau \\
&&+\left( \mathrm{e}^{\left( t-s\right) \Delta _{0}}-\mathrm{e}^{\epsilon
\Delta _{0}}\right) \left( \Delta _{t}-\Delta _{0}\right) Z_{t,s}^{g},
\end{eqnarray*}%
we find that 
\begin{equation}
\Vert \partial _{t}\mathcal{W}_{t,s}^{(\epsilon )}\Vert _{\mathrm{op}}\leq 3r%
\mathrm{e}^{r\left( t-s\right) }+\int_{s}^{t-\epsilon }\left\Vert \Delta _{0}%
\mathrm{e}^{\left( t-\tau \right) \Delta _{0}}\right\Vert _{\mathrm{op}%
}\left\Vert \left( \Delta _{\tau }-\Delta _{0}\right) Z_{\tau ,s}^{g}-\left(
\Delta _{t}-\Delta _{0}\right) Z_{t,s}^{g}\right\Vert _{\mathrm{op}}\mathrm{d%
}\tau .  \label{interpolation5}
\end{equation}%
We use now (\ref{petit inequality newbisbis}), (\ref{upper bound for W}), (%
\ref{interpolation4}) for $\nu \in (0,1)$, and the Lipschitz continuity (\ref%
{rdef0}) of the operator family $\Delta $ to get 
\begin{eqnarray}
&&\int_{s}^{t-\epsilon }\left\Vert \Delta _{0}\mathrm{e}^{\left( t-\tau
\right) \Delta _{0}}\right\Vert _{\mathrm{op}}\left\Vert \left( \Delta
_{\tau }-\Delta _{0}\right) Z_{\tau ,s}^{g}-\left( \Delta _{t}-\Delta
_{0}\right) Z_{t,s}^{g}\right\Vert _{\mathrm{op}}\mathrm{d}\tau  \notag \\
&\leq &\int_{s}^{t-\epsilon }\frac{r}{\mathrm{e}\left( t-\tau \right) }%
\left\Vert Z_{\tau ,s}^{g}-Z_{t,s}^{g}\right\Vert _{\mathrm{op}}\mathrm{d}%
\tau +\int_{s}^{t-\epsilon }\frac{\mathrm{e}^{r\left( t-s\right) }}{\mathrm{e%
}\left( t-\tau \right) }\left\Vert \Delta _{\tau }-\Delta _{t}\right\Vert _{%
\mathrm{op}}\mathrm{d}\tau  \label{estimate integral estimaatebis} \\
&\leq &\frac{2^{1-\nu }r}{\mathrm{e}}\left( \int_{s}^{t}\left( t-\tau
\right) ^{\nu -1}\left( \tau -s\right) ^{-\nu }\mathrm{d}\tau +\frac{r%
\mathrm{e}^{r\left( t-s\right) }}{\left( 1-\nu \right) }\int_{s}^{t}\left( 
\frac{\tau -s}{t-\tau }\right) ^{1-\nu }\mathrm{d}\tau \right)  \notag \\
&&+\frac{1}{\mathrm{e}}\left( r^{2}+C_{\Delta }\right) \left( t-s\right) 
\mathrm{e}^{r\left( t-s\right) }.  \label{estimate integral estimaatebisbis}
\end{eqnarray}%
Using the change of variable, $u=(\tau -s)(t-\tau )^{-1}$, we remark that 
\begin{equation*}
\int_{s}^{t}\left( \frac{\tau -s}{t-\tau }\right) ^{1-\nu }\mathrm{d}\tau
=\left( t-s\right) \int_{0}^{\infty }\frac{u^{1-\nu }}{\left( u+1\right) ^{2}%
}\mathrm{d}u,
\end{equation*}%
while one easily obtains the upper bound\footnote{%
This integral is actually a beta function and its exact value is $\pi /\sin
(\pi \nu )$ for $\nu \in (0,1)$.} 
\begin{eqnarray*}
\int_{s}^{t}\frac{\left( t-\tau \right) ^{\nu -1}}{\left( \tau -s\right)
^{\nu }}\mathrm{d}\tau &\leq &2^{1-\nu }\left( t-s\right) ^{\nu -1}\int_{s}^{%
\frac{t+s}{2}}\frac{1}{\left( \tau -s\right) ^{\nu }}\mathrm{d}\tau \\
&&+\frac{2^{\nu }}{\left( t-s\right) ^{\nu }}\int_{\frac{t+s}{2}}^{t}\left(
t-\tau \right) ^{\nu -1}\mathrm{d}\tau \\
&=&\frac{1}{1-\nu }+\frac{1}{\nu }.
\end{eqnarray*}
Then, using these last computations, we infer from (\ref{interpolation5})
and (\ref{estimate integral estimaatebisbis}) that, for any $\nu \in (0,1)$, 
$s\in I$ (with $s<T $ if $I=[0,T]$), sufficiently small $\epsilon \in 
\mathbb{R}^{+}$ and all times $t\in I\cap (s+\epsilon ,\infty )$,%
\begin{equation}
\Vert \partial _{t}\mathcal{W}_{t,s}^{(\epsilon )}\Vert _{\mathrm{op}}\leq 
\mathrm{C}_{t,s,\nu }<\infty ,  \label{interpolation6}
\end{equation}%
where $\mathrm{C}_{t,s,\nu }$ is the positive constant defined by 
\begin{multline}
\mathrm{C}_{t,s,\nu }\doteq \left( 3r+\frac{1}{\mathrm{e}}\left( \left( 1+%
\frac{2^{1-\nu }}{1-\nu }\int_{0}^{\infty }\frac{u^{1-\nu }}{\left(
u+1\right) ^{2}}\mathrm{d}u\right) r^{2}+C_{\Delta }\right) \left(
t-s\right) \right) \mathrm{e}^{r\left( t-s\right) }  \notag \\
+\frac{2^{1-\nu }r}{\mathrm{e}}\left( \frac{1}{1-\nu }+\frac{1}{\nu }\right)
.  \label{interpolation6bis}
\end{multline}%
The last bound is uniform with respect to $\epsilon $ and, for any compact
subset $K\subseteq I$, $\mathrm{C}_{t,s,\nu }$ can be bounded by some
constant $C_{\nu }$ not depending upon $s\in K$ and $t\in K\cap (s+\epsilon
,\infty )$. \medskip

\noindent \underline{Step 4:} As we prove (\ref{interpolation6}), using (\ref%
{petit inequality newbisbis}), (\ref{interpolation4}), the Lipschitz
continuity of $\Delta $, and the triangle inequality one can verify that, as 
$\epsilon \rightarrow 0$, $\partial _{t}\mathcal{W}_{t,s}^{(\epsilon )}$
converges in the norm topology to 
\begin{equation}
\mathcal{V}_{t,s}\doteq \left. \partial _{t}\mathcal{W}_{t,s}^{(\epsilon
)}\right\vert _{\epsilon =0}=\left( \Delta _{t}-\Delta _{0}\right)
Z_{t,s}^{g}+\int_{s}^{t}\Delta _{0}\mathrm{e}^{\left( t-\tau \right) \Delta
_{0}}\left( \Delta _{\tau }-\Delta _{0}\right) Z_{\tau ,s}^{g}\mathrm{d}\tau
,  \label{interpolation6+1}
\end{equation}%
for all $s,t\in I$ with $t>s$. By (\ref{interpolation6}), we clearly have%
\begin{equation}
\left\Vert \mathcal{V}_{t,s}\right\Vert _{\mathrm{op}}\leq \mathrm{C}%
_{t,s,\nu }<\infty  \label{petit inequality newbisbisbis}
\end{equation}%
for any $\nu \in (0,1)$ and $s,t\in I$ with $t>s$. The operator $\mathcal{V}%
_{t,s}$ is also norm-continuous in $t>s$, see (\ref{interpolation6+1}).
Since $\mathcal{W}_{t,s}^{(\epsilon )}$ converges to 
\begin{equation*}
\mathcal{W}_{t,s}^{(0)}\doteq \int_{s}^{t}\mathrm{e}^{\left( t-\tau \right)
\Delta _{0}}\left( \Delta _{\tau }-\Delta _{0}\right) Z_{\tau ,s}^{g}\mathrm{%
d}\tau ,\qquad t\in I\cap (s,\infty ),
\end{equation*}%
in the norm topology, we can take the limit $\epsilon \rightarrow 0^{+}$ in
the equation%
\begin{equation*}
\mathcal{W}_{u,s}^{\left( \epsilon \right) }-\mathcal{W}_{t,s}^{\left(
\epsilon \right) }=\int_{t}^{u}\partial _{\tau }\mathcal{W}_{\tau
,s}^{(\epsilon )}\mathrm{d}\tau
\end{equation*}%
and get the equality 
\begin{equation*}
\mathcal{W}_{u,s}^{\left( 0\right) }-\mathcal{W}_{t,s}^{\left( 0\right)
}=\int_{t}^{u}\mathcal{V}_{\tau ,s}\mathrm{d}\tau
\end{equation*}%
for any $s,t,u\in I$ with $u\geq t>s$. Since $\mathcal{V}_{t,s}$ is
norm-continuous in $t>s$, one checks that $\mathcal{V}_{t,s}=\partial _{t}%
\mathcal{W}_{t,s}^{(0)}$, i.e., $\mathcal{W}_{t,s}^{(0)}$\ is norm
differentiable with respect to $t>s$. By (\ref{definition of W 2}), for any $%
t\in I\cap (s,\infty )$,%
\begin{equation}
Z_{t,s}^{g}=\mathrm{e}^{\left( t-s\right) \Delta _{0}}+\mathcal{W}%
_{t,s}^{(0)}.  \label{definition of W 2bis}
\end{equation}%
As a consequence, for any $s,t\in I$ with $t>s$, 
\begin{equation*}
\partial _{t}Z_{t,s}^{g}=\Delta _{0}\mathrm{e}^{\left( t-s\right) \Delta
_{0}}+\mathcal{V}_{t,s}
\end{equation*}%
with respect to the norm topology. Compare this equality with Lemma \ref%
{lemma-Dyson-Series} (i). In particular, by (\ref{petit inequality newbisbis}%
) and (\ref{petit inequality newbisbisbis}), 
\begin{equation}
\left\Vert \partial _{t}Z_{t,s}^{g}\right\Vert _{\mathrm{op}}\leq \frac{1}{%
\mathrm{e}\left( t-s\right) }+\mathrm{C}_{t,s,\nu }<\infty
\label{definition of W 2bisbis}
\end{equation}%
for any $\nu \in (0,1)$ and $s,t\in I$ with $t>s$. For $s,t\in I$ with $t>s$%
, it yields 
\begin{equation*}
\left\Vert \Delta _{t}Z_{t,s}^{g}\left( \Delta \right) \right\Vert _{\mathrm{%
op}}\leq \mathrm{C}\left( s,t,C_{\Delta },r\right) +\frac{1}{\mathrm{e}%
\left\vert t-s\right\vert }
\end{equation*}%
for the Dyson series (\ref{eq:dysonzg}) by taking $\nu =1/2$ and using that 
\begin{equation*}
\int_{0}^{\infty }\frac{u^{1/2}}{\left( u+1\right) ^{2}}\mathrm{d}%
u=2\int_{0}^{\infty }\frac{x^{2}}{\left( x^{2}+1\right) ^{2}}\mathrm{d}x=%
\frac{\pi }{2}.
\end{equation*}%
To get the same assertion when $t<s$, the arguments are exactly the same, up
to trivial modifications. The remaining inequalities are similar. We omit
the details.
\end{proof}

\begin{remark}[Improvement of the bound under negative generators]
\label{remark utiles}\mbox{}\newline
If the operator family $\Delta $ satisfies $\Delta _{t}\leq 0$ for all $t\in
I$, then the function $\mathrm{C}$ can be taken to be equal to 
\begin{equation*}
\mathrm{C}\left( s,t,C_{\Delta },r\right) \doteq r\left( 3+4\sqrt{2}\mathrm{e%
}^{-1}\right) +\mathrm{e}^{-1}\left( \left( \sqrt{2}\pi +1\right)
r^{2}+C_{\Delta }\right) \left\vert t-s\right\vert .
\end{equation*}%
Indeed, the factor $\mathrm{e}^{r\left( t-s\right) }$ in (\ref{definition de
C 1}) arises solely from the estimates of $\Vert Z_{t,s}^{g}(\Delta )\Vert _{%
\mathrm{op}}$ and $\Vert Z_{t,s}^{d}(\Delta )\Vert _{\mathrm{op}}$ deduced
from Lemma \ref{lemma-Dyson-Series} (ii) with $\gamma =r$; see (\ref{upper
bound for W}). However, if $\Delta _{t}\leq 0$, one can instead apply Lemma %
\ref{lemma-Dyson-Series} (ii) with $\gamma =0$, thereby replacing the factor 
$\mathrm{e}^{r\left( t-s\right) }$ with $1$.
\end{remark}

\begin{remark}[Extension to upper semibounded initial data]
\label{remark utiles copy(2)}\mbox{}\newline
Lemmata \ref{lemma-Dyson-Series} and \ref{lemma-Dyson-Series2} can be
extended to possibly unbounded operator families $\Delta $ such that $\Delta
_{0}\leq 0$ and $\left( \Delta _{t}-\Delta _{0}\right) _{t\in I}\in C(I,%
\mathcal{B}(\mathfrak{h}))$. In this setting, the operator norm topology
used throughout the proofs of these lemmata is often replaced by the strong
operator topology; see \cite[Lemma 35]{bach-bru-memo} for further details.
Moreover, one can verify that the mapping $(Z_{t,s}^{g}(\Delta ))_{t\geq
s+\epsilon \geq 0}$ is H\"{o}lder-continuous in the norm topology for every $%
\epsilon >0$. See, for instance, similar arguments used to prove \cite[Chap.
5, Theorem 6.9]{Pazy}.
\end{remark}

\begin{corollary}[Uniform Lipschitz continuity on compacta]
\label{coro uniqueness copy(1)}\mbox{}\newline
Under Conditions of Lemma \ref{lemma-Dyson-Series2}, for each compact subset 
$K\subseteq I$, there are two constants \textrm{C}$_{K},\mathrm{D}%
_{K}<\infty $ such that, for each $s\in K$, any sufficiently small $\epsilon
\in \mathbb{R}^{+}$ and all $t,u\in K\cap \lbrack s+\epsilon ,\infty )$, 
\begin{equation*}
\max \left\{ \left\Vert Z_{u,s}^{g}\left( \Delta \right) -Z_{t,s}^{g}\left(
\Delta \right) \right\Vert _{\mathrm{op}},\left\Vert Z_{u,s}^{d}\left(
\Delta \right) -Z_{t,s}^{d}\left( \Delta \right) \right\Vert _{\mathrm{op}%
}\right\} \leq 2\left( \mathrm{C}_{K}+\mathrm{D}_{K}\epsilon ^{-1}\right)
\left\vert u-t\right\vert .
\end{equation*}%
Here, \textrm{C}$_{K},\mathrm{D}_{K}$ do not depend upon the operator norm
of the initial operator $\Delta _{0}\leq 0$.
\end{corollary}

\begin{proof}
Combine Lemmata \ref{lemma-Dyson-Series} and \ref{lemma-Dyson-Series2}.
\end{proof}

\subsection{Local Existence -- General Statement}

Let $\mathscr{B}\subseteq \mathcal{B}\left( \mathfrak{h}\right) $ be a von
Neumann algebra and $\mathscr{A}$ a two-sided $\sigma $-weakly closed ideal
of $\mathscr{B}$. The local existence of the system of differential
equations (\ref{goodflow}) is a consequence of the well-posedness of the
following system of operator-valued differential equations for some $T\in 
\mathbb{R}^{+}$ and times $t\in \lbrack 0,T]$: 
\begin{equation}
\left\{ 
\begin{array}{llll}
\partial _{t}L_{t} & =f\left( X_{t}\right) & , & L_{0}\in \mathscr{B}, \\ 
\partial _{t}R_{t} & =g\left( X_{t}\right) & , & R_{0}\in \mathscr{B}, \\ 
\partial _{t}X_{t} & =L_{t}X_{t}+X_{t}R_{t} & , & X_{0}\in \mathscr{A},%
\end{array}%
\right.  \label{localexistenceflowgeneral}
\end{equation}%
where $f$ and $g$ are two mappings from $\mathscr{B}$ to itself satisfying $%
f(0)=g\left( 0\right) =0$ and which are locally Lipschitz-continuous around $%
0\in \mathcal{B(\mathfrak{h})}$. Here, a mapping $\xi :\mathcal{B(\mathfrak{h%
})\rightarrow B(\mathfrak{h})}$ is locally Lipschitz-continuous around $C\in 
\mathcal{B(\mathfrak{h})}$ if, for any closed ball $\overline{B_{r}\left(
C\right) }\subseteq \mathcal{B(\mathfrak{h})}$ of center $C$ and radius $%
r\in \mathbb{R}^{+}$, there is a constant $\kappa _{\xi ,r,C}\in \mathbb{R}%
^{+}$ such that%
\begin{equation}
\left\Vert \xi \left( A\right) -\xi \left( B\right) \right\Vert _{\mathrm{op}%
}\leq \kappa _{\xi ,r,C}\left\Vert A-B\right\Vert _{\mathrm{op}},\qquad
A,B\in \overline{B_{r}\left( C\right) }.  \label{kappa}
\end{equation}%
The constant $\kappa _{\xi ,r,C}$ is named the Lipschitz constant of $\xi $
in $\overline{B_{r}\left( C\right) }$.

The strategy for establishing the local existence of a solution to (\ref%
{localexistenceflowgeneral}) is similar to that employed in \cite%
{bach-bru-memo,EllipticFlow}. However, rather than utilizing a single
evolution system as in \cite{bach-bru-memo,EllipticFlow}, our approach
relies on two distinct evolution systems of operators. These systems are
defined via the Dyson series (\ref{eq:dysonzg}) and (\ref{dysonzd}), thereby
accommodating the asymmetric nature of (\ref{localexistenceflowgeneral}).
Indeed, Lemma \ref{lemma-Dyson-Series} (i) implies that, for any $X\in
C([0,T],\mathcal{B}(\mathfrak{h}))$ and two other families $L,R\in C([0,T],%
\mathcal{B}(\mathfrak{h}))$ with $T\in \mathbb{R}^{+}$ satisfying (\ref%
{localexistenceflowgeneral}), the operator family defined by 
\begin{equation}
\tilde{X}_{t}\doteq Z_{t,s}^{g}\left( L\right) X_{s}Z_{t,s}^{d}\left(
R\right) ,\qquad s,t\in \left[ 0,T\right] ,  \label{trytryt1}
\end{equation}%
must\footnote{%
See Lemma \ref{lemma-Dyson-Series} (i) and Equation (\ref{argument1}) below.}
be equal at any time $t\in \lbrack 0,T]$ to $\tilde{X}_{t}=X_{t}$, with 
\begin{equation}
\partial _{t}\tilde{X}_{t}=L_{t}\tilde{X}_{t}+\tilde{X}_{t}R_{t},\qquad t\in %
\left[ 0,T\right] .  \label{trytryt2}
\end{equation}%
Compare this equality with the third equation in ( \ref%
{localexistenceflowgeneral}). In other words, by setting $s=0$ in (\ref%
{trytryt1}), we can construct $(\tilde{X}_{t}=X_{t})_{t\geq 0}$ only using
the initial data $X_{0}\in \mathcal{B}(\mathfrak{h})$ and the two families $%
Z_{\cdot ,0}^{g}(L)$ and $Z_{\cdot ,0}^{d}(R)$. This reduces the local
existence problem for (\ref{localexistenceflowgeneral}) to finding the local
existence of two operator-valued functions $(L_{t})_{t\geq 0}$ and $%
(R_{t})_{t\geq 0}$ for a fixed $X_{0}\in \mathcal{B}(\mathfrak{h})$. This
suggests a natural fixed-point argument, which we set up below:

Given some maximum time $T\in \mathbb{R}^{+}$, we use below a contraction
mapping principle on the Banach space $\mathfrak{C}^{2}$ with the norm 
\begin{equation*}
\left\Vert (U,V)\right\Vert _{\mathfrak{C}^{2}}\doteq \max \left\{
\left\Vert U\right\Vert _{\mathfrak{C}},\left\Vert V\right\Vert _{\mathfrak{C%
}}\right\} ,\qquad U,V\in \mathfrak{C},
\end{equation*}%
where $\mathfrak{C}\doteq C_{b}([0,T],\mathcal{B}(\mathfrak{h}))$ is the
Banach space of continuous mappings from $[0,T]$ to $\mathcal{B}(\mathfrak{h}%
)$ endowed with the supremum norm 
\begin{equation}
\left\Vert U\right\Vert _{\mathfrak{C}}\doteq \sup_{t\in \left[ 0,T\right]
}\left\Vert U_{t}\right\Vert _{\mathrm{op}},\qquad U\in C\left( \left[ 0,T%
\right] ,\mathcal{B}\left( \mathfrak{h}\right) \right) .
\label{supremum norm}
\end{equation}%
We denote by $\mathbf{B}_{r}\equiv \mathbf{B}_{r}\left( 0\right) \subseteq 
\mathfrak{C}^{2}$ the open ball of center $0$ and radius $r\in \mathbb{R}%
^{+} $ in $\mathfrak{C}^{2}$. Our aim is then to find a fixed point of the
mapping $\mathcal{J}:\mathbf{B}_{r}\rightarrow \mathfrak{C}^{2}$ defined at
fixed $X_{0}\in \mathcal{B}\left( \mathfrak{h}\right) $ by%
\begin{equation}
\mathcal{J}\left( 
\begin{array}{c}
U \\ 
V%
\end{array}%
\right) =\left( 
\begin{array}{c}
\tilde{U} \\ 
\tilde{V}%
\end{array}%
\right) ,  \label{defFfrak0}
\end{equation}%
where, for any $t\in \left[ 0,T\right] $, 
\begin{equation}
\left( 
\begin{array}{c}
\tilde{U}_{t} \\ 
\tilde{V}_{t}%
\end{array}%
\right) \doteq \left( 
\begin{array}{c}
\int_{0}^{t}f\left( Z_{\tau ,0}^{g}\left( U\right) X_{0}Z_{\tau
,0}^{d}\left( V\right) \right) \mathrm{d}\tau \\ 
\int_{0}^{t}g\left( Z_{\tau ,0}^{g}\left( U\right) X_{0}Z_{\tau
,0}^{d}\left( V\right) \right) \mathrm{d}\tau%
\end{array}%
\right) .  \label{defFfrak}
\end{equation}%
This is established in the following lemma:

\begin{lemma}[Local existence of a fixed point of $\mathcal{J}$]
\label{localexistencegeneral}\mbox{}\newline
Let $\mathfrak{h}$ be a Hilbert space and $X_{0}\in \mathcal{B}\left( 
\mathfrak{h}\right) $. Let $f$ and $g$ be two mappings $\mathcal{B(\mathfrak{%
h})\rightarrow B(\mathfrak{h})}$ vanishing at $0\in \mathcal{B(\mathfrak{h})}
$ and locally Lipschitz-continuous around $0$. Given $r\in \mathbb{R}^{+}$,
fix $T=1/\left( 2r\right) $, and denote the Lipschitz constants of $f$ and $%
g $ in the closed ball $\overline{B_{r}\left( 0\right) }\subseteq \mathcal{B(%
\mathfrak{h})}$ by $\kappa _{f,r}\equiv \kappa _{f,r,0}$ and $\kappa
_{g,r}\equiv \kappa _{g,r,0}$, respectively; see (\ref{kappa}) with $C=0$
for $\xi =f$ (and $\xi =g$, respectively). For any $r\in \mathbb{R}^{+}$
such that 
\begin{equation}
r^{-2}\max \left\{ \kappa _{f,r},\kappa _{g,r}\right\} \left\Vert
X_{0}\right\Vert _{\mathrm{op}}\leq \frac{1}{8\left( \mathrm{e}%
^{1/2}-1\right) ^{2}}  \label{inequality stupide}
\end{equation}%
(provided it exists), the fixed point problem 
\begin{equation}
\mathcal{J}\left( 
\begin{array}{c}
U \\ 
V%
\end{array}%
\right) =\left( 
\begin{array}{c}
U \\ 
V%
\end{array}%
\right)  \label{fixed point problem}
\end{equation}%
has a unique solution $(L^{\Delta },R^{\Delta })$ in the open ball $\mathbf{B%
}_{r}\subseteq \mathfrak{C}^{2}$. In addition, $L^{\Delta }\in \mathfrak{C}$
and $R^{\Delta }\in \mathfrak{C}$ are Lipschitz continuous in the norm
topology and if $X_{0}\in \mathscr{A}$, $f(\mathscr{B})\subseteq \mathscr{B}$
and $g(\mathscr{B})\subseteq \mathscr{B}$, then the same fixed point problem
has a unique solution in $\mathbf{B}_{r}\cap C([0,T],\mathscr{B})^{2}$.
\end{lemma}

\begin{proof}
For the moment, let $T\in \mathbb{R}^{+}$ and $r\in \mathbb{R}^{+}$ be
arbitrary parameters whose precise values will be determined later. We first
observe that, according to Lemma \ref{lemma-Dyson-Series} (ii), for any $%
T\geq t\geq s\geq 0$ and for any $U\in C([0,T],\mathcal{B}(\mathfrak{h}))$, 
\begin{equation}
\max \left\{ \left\Vert Z_{t,s}^{g}\left( U\right) \right\Vert _{\mathrm{op}%
},\left\Vert Z_{t,s}^{d}\left( U\right) \right\Vert _{\mathrm{op}}\right\}
\leq \exp \left( \left( t-s\right) \sup_{\tau \in \left[ s,t\right]
}\left\Vert U_{\tau }\right\Vert _{\mathrm{op}}\right) .  \label{eq:borneZg}
\end{equation}%
Using $f(0)=g\left( 0\right) =0$, (\ref{kappa}) and (\ref{eq:borneZg}), we
compute from (\ref{defFfrak0})--(\ref{defFfrak}) that, for any $\left(
U,V\right) \in \mathbf{B}_{r}$, 
\begin{equation}
\left\Vert \mathcal{J}\left( 
\begin{array}{c}
U \\ 
V%
\end{array}%
\right) \right\Vert _{\mathfrak{C}^{2}}\leq \frac{\kappa _{r}}{2r}\left( 
\mathrm{e}^{2rT}-1\right) \left\Vert X_{0}\right\Vert _{\mathrm{op}},
\label{sdsdsdsfgfggf1}
\end{equation}%
where 
\begin{equation*}
\kappa _{r}\doteq \max \left\{ \kappa _{f,r},\kappa _{g,r}\right\} \in 
\mathbb{R}^{+}.
\end{equation*}%
Moreover, by (\ref{kappa}) and (\ref{eq:borneZg}), for any $%
(U_{1},V_{1}),(U_{2},V_{2})\in \mathbf{B}_{r}$,%
\begin{eqnarray}
&&\left\Vert \mathcal{J}\left( 
\begin{array}{c}
U_{1} \\ 
V_{1}%
\end{array}%
\right) -\mathcal{J}\left( 
\begin{array}{c}
U_{2} \\ 
V_{2}%
\end{array}%
\right) \right\Vert _{\mathfrak{C}^{2}}  \label{x} \\
&\leq &\frac{\kappa _{r}}{r}\left( \mathrm{e}^{rT}-1\right) \left\Vert
X_{0}\right\Vert _{\mathrm{op}}\left( \sup_{t\in \left[ 0,T\right]
}\left\Vert Z_{t,0}^{g}\left( U_{1}\right) -Z_{t,0}^{g}\left( U_{2}\right)
\right\Vert _{\mathrm{op}}+\sup_{t\in \left[ 0,T\right] }\left\Vert
Z_{t,0}^{d}\left( V_{1}\right) -Z_{t,0}^{d}\left( V_{2}\right) \right\Vert _{%
\mathrm{op}}\right) .  \notag
\end{eqnarray}%
Using now the Dyson series (\ref{eq:dysonzg}) and (\ref{dysonzd}) or,
equivalently, Lemma \ref{lemma-Dyson-Series} (i) written as integral
equations, we note that, for any $t\in \lbrack 0,T]$,%
\begin{multline*}
\sup_{t\in \left[ 0,T\right] }\left\Vert Z_{t,0}^{g}\left( U_{1}\right)
-Z_{t,0}^{g}\left( U_{2}\right) \right\Vert _{\mathrm{op}} \\
\leq \frac{1}{r}\left( \mathrm{e}^{rT}-1\right) \sup_{t\in \left[ 0,T\right]
}\left\Vert U_{1}-U_{2}\right\Vert _{\mathrm{op}}+rT\sup_{t\in \left[ 0,T%
\right] }\left\Vert Z_{t,0}^{g}\left( U_{1}\right) -Z_{t,0}^{g}\left(
U_{2}\right) \right\Vert _{\mathrm{op}},
\end{multline*}%
which in turn implies that 
\begin{equation}
\sup_{t\in \left[ 0,T\right] }\left\Vert Z_{t,0}^{g}\left( U_{1}\right)
-Z_{t,0}^{g}\left( U_{2}\right) \right\Vert _{\mathrm{op}}\leq \frac{\mathrm{%
e}^{rT}-1}{r\left( 1-rT\right) }\sup_{t\in \left[ 0,T\right] }\left\Vert
U_{1}-U_{2}\right\Vert _{\mathrm{op}},  \label{eq:borneZU}
\end{equation}%
provided $rT<1$. In the same way, if $rT<1$ then%
\begin{equation}
\sup_{t\in \left[ 0,T\right] }\left\Vert Z_{t,0}^{d}\left( V_{1}\right)
-Z_{t,0}^{d}\left( V_{2}\right) \right\Vert _{\mathrm{op}}\leq \frac{\mathrm{%
e}^{rT}-1}{r\left( 1-rT\right) }\sup_{t\in \left[ 0,T\right] }\left\Vert
V_{1}-V_{2}\right\Vert _{\mathrm{op}}.  \label{borneZV}
\end{equation}%
As a consequence, by combining the inequality 
\begin{equation*}
\sup_{t\in \left[ 0,T\right] }\left\Vert V_{1}-V_{2}\right\Vert _{\mathrm{op}%
}+\sup_{t\in \left[ 0,T\right] }\left\Vert U_{1}-U_{2}\right\Vert _{\mathrm{%
op}}\leq 2\left\Vert \left( 
\begin{array}{c}
U_{1}-U_{2} \\ 
V_{1}-V_{2}%
\end{array}%
\right) \right\Vert _{\mathfrak{C}^{2}},
\end{equation*}%
with (\ref{eq:borneZU})--(\ref{borneZV}), we infer from Inequality (\ref{x})
that if $rT<1$ then%
\begin{equation}
\left\Vert \mathcal{J}\left( 
\begin{array}{c}
U_{1} \\ 
V_{1}%
\end{array}%
\right) -\mathcal{J}\left( 
\begin{array}{c}
U_{2} \\ 
V_{2}%
\end{array}%
\right) \right\Vert _{\mathfrak{C}^{2}}\leq \frac{2\kappa _{r}\left( \mathrm{%
e}^{rT}-1\right) ^{2}}{r^{2}\left( 1-rT\right) }\left\Vert X_{0}\right\Vert
_{\mathrm{op}}\left\Vert \left( 
\begin{array}{c}
U_{1}-U_{2} \\ 
V_{1}-V_{2}%
\end{array}%
\right) \right\Vert _{\mathfrak{C}^{2}}.  \label{sdsdsdsfgfggf2}
\end{equation}%
We fix now the maximum time $T=1/\left( 2r\right) $. Thus, by a direct
application of (\ref{sdsdsdsfgfggf1}) and (\ref{sdsdsdsfgfggf2}), 
\begin{equation*}
\left\Vert \mathcal{J}\left( 
\begin{array}{c}
U \\ 
V%
\end{array}%
\right) \right\Vert _{\mathfrak{C}^{2}}\leq \frac{\kappa _{r}\left( \mathrm{e%
}-1\right) }{2r}\left\Vert X_{0}\right\Vert _{\mathrm{op}},
\end{equation*}%
as well as 
\begin{equation*}
\left\Vert \mathcal{J}\left( 
\begin{array}{c}
U_{1} \\ 
V_{1}%
\end{array}%
\right) -\mathcal{J}\left( 
\begin{array}{c}
U_{2} \\ 
V_{2}%
\end{array}%
\right) \right\Vert _{\mathfrak{C}^{2}}\leq 4\frac{\kappa _{r}}{r^{2}}\left( 
\mathrm{e}^{1/2}-1\right) ^{2}\left\Vert X_{0}\right\Vert _{\mathrm{op}%
}\left\Vert \left( 
\begin{array}{c}
U_{1}-U_{2} \\ 
V_{1}-V_{2}%
\end{array}%
\right) \right\Vert _{\mathfrak{C}^{2}}.
\end{equation*}%
In particular, if $r\in \mathbb{R}^{+}$ can be chosen such that%
\begin{equation}
\frac{\kappa _{r}\left( \mathrm{e}-1\right) }{2r}\left\Vert X_{0}\right\Vert
_{\mathrm{op}}\leq r\qquad \text{and}\qquad 4\frac{\kappa _{r}}{r^{2}}\left( 
\mathrm{e}^{1/2}-1\right) ^{2}\left\Vert X_{0}\right\Vert _{\mathrm{op}}\leq 
\frac{1}{2},  \label{condition stupide}
\end{equation}%
the mapping $\mathcal{J}:\overline{\mathbf{B}_{r}}\rightarrow \overline{%
\mathbf{B}_{r}}$ is a contraction. In this case, the contraction mapping
principle on the closed set defined by $\overline{\mathbf{B}_{r}}\subseteq 
\mathfrak{C}^{2}$ yields a unique fixed point $\left( L^{\Delta },R^{\Delta
}\right) =\mathcal{J}\left( L^{\Delta },R^{\Delta }\right) \in \overline{%
\mathbf{B}_{r}}$. Note that (\ref{condition stupide}) for $r\in \mathbb{R}%
^{+}$ is equivalent to 
\begin{equation*}
\frac{\kappa _{r}}{r^{2}}\left\Vert X_{0}\right\Vert _{\mathrm{op}}\leq 
\frac{1}{8\left( \mathrm{e}^{1/2}-1\right) ^{2}}<\frac{2}{\mathrm{e}-1}.
\end{equation*}%
In particular, if (\ref{condition stupide}) holds, $\mathcal{J}:\overline{%
\mathbf{B}_{r}}\rightarrow \mathbf{B}_{r}$. Additionally, since$\left\Vert
(L,R)\right\Vert _{\mathfrak{C}^{2}}\leq r$, by (\ref{eq:borneZg}), it is
clear that $L\in \mathfrak{C}$ and $R\in \mathfrak{C}$ are Lipschitz
continuous in the norm topology on $\left[ 0,T\right] $. Finally, for any $%
(U,V)\in \mathbf{B}_{r}\cap C([0,T],\mathscr{B})$ and any $s,t\in \mathbb{R}%
_{0}^{+}$, 
\begin{equation*}
Z_{t,s}^{g}(U)X_{0}Z_{t,s}^{d}(V)\in \mathscr{A}\subseteq \mathscr{B},
\end{equation*}%
by ($\sigma $-weak) closedness of $\mathscr{B}$ and because $\mathscr{A}$ is
a two-sided ideal. As a consequence, if $X_{0}\in \mathscr{A}$, $f(%
\mathscr{B})\subseteq \mathscr{B}$ and $g(\mathscr{B})\subseteq \mathscr{B}$%
, then $\mathcal{J}(\mathbf{B}_{r}\cap C([0,T],\mathscr{B}))\subseteq 
\mathbf{B}_{r}\cap C([0,T],\mathscr{B})$ and by the Banach contraction
principle, we can achieve the fixed point $\left( L^{\Delta },R^{\Delta
}\right) $ by taking an arbitrary $(U,V)\in \mathbf{B}_{r}\cap C([0,T],%
\mathscr{B})$ and infinitely iterating $\mathcal{J}$. Using the closedness
of $\mathscr{B}$, $\left( L^{\Delta },R^{\Delta }\right) $ must belong in
this particular case to $\mathbf{B}_{r}\cap C([0,T],\mathscr{B})^{2}$.
\end{proof}

Lemma \ref{localexistencegeneral} can then be used to deduce the local
existence of a unique solution to the general system of non-linear
operator-valued differential equations (\ref{localexistenceflowgeneral}):

\begin{theorem}[Local existence and uniqueness]
\label{LocalExistenceMainResult}\mbox{}\newline
Let $\mathfrak{h}$ be a Hilbert space and $X_{0}\in \mathcal{B}\left( 
\mathfrak{h}\right) $. Let $f$ and $g$ be two mappings $\mathcal{B(\mathfrak{%
h})\rightarrow B(\mathfrak{h})}$ vanishing at $0\in \mathcal{B(\mathfrak{h})}
$ and locally Lipschitz-continuous around $0$. Given $r\in \mathbb{R}^{+}$,
fix $T=1/\left( 2r\right) $, and denote the Lipschitz constants of $f$ and $%
g $ in the closed ball $\overline{B_{r}\left( 0\right) }\subseteq \mathcal{B(%
\mathfrak{h})}$ by $\kappa _{f,r}\equiv \kappa _{f,r,0}$ and $\kappa
_{g,r}\equiv \kappa _{g,r,0}$, respectively. For any $r\in \mathbb{R}^{+}$
such that (\ref{inequality stupide}) holds, there is a unique solution $%
(L,R,X)\in \mathbf{B}_{r}((L_{0},R_{0}))\times \mathfrak{C}$ to (\ref%
{localexistenceflowgeneral}). In addition, $L,R,X$ are Lipschitz continuous
in the norm topology and if $X_{0}\in \mathscr{A}$, $f(\mathscr{B})\subseteq %
\mathscr{B}$, $g(\mathscr{B})\subseteq \mathscr{B}$ and $L_{0},R_{0}\in %
\mathscr{B}$, then the solution belongs to 
\begin{equation*}
\left( \mathbf{B}_{r}((L_{0},R_{0}))\cap C([0,T],\mathscr{B})^{2}\right)
\times \left( \mathfrak{C}\cap C([0,T],\mathscr{A})\right) .
\end{equation*}
\end{theorem}

\begin{proof}
Assume the existence of some parameter $r\in \mathbb{R}^{+}$ such that (\ref%
{inequality stupide}) holds true. Then, Lemma \ref{localexistencegeneral}
provides us with a unique fixed point $(L^{\Delta },R^{\Delta })\in 
\mathfrak{C}^{2}$, corresponding to two Lipschitz continuous operator-valued
mappings. We thus define 
\begin{equation*}
L\doteq L_{0}+L^{\Delta }\qquad \text{and}\qquad R\doteq R_{0}+R^{\Delta }.
\end{equation*}%
Then we define the evolution systems 
\begin{equation*}
\left( Z_{t,s}^{g}\left( L\right) \right) _{T\geq t,s\geq 0},\left(
Z_{t,s}^{d}\left( R\right) \right) _{T\geq t,s\geq 0}\in C\left( \left[ 0,T%
\right] ^{2},\mathcal{B}\left( \mathfrak{h}\right) \right)
\end{equation*}%
according to the Dyson series (\ref{eq:dysonzg}) and (\ref{dysonzd})
respectively, as well as the family $X\in C([0,T],\mathcal{B}(\mathfrak{h}))$
by Equation (\ref{trytryt1}). $X$ is clearly Lipschitz continuous and even
of class $C^{1}$ by (\ref{eq:dysonzg}) and (\ref{dysonzd}) combined with the
Lipschitz continuity of $L$ and $R$ in the corresponding ball of radius $%
r\in \mathbb{R}^{+}$. Then, by Equations (\ref{trytryt2}) and (\ref%
{defFfrak0})--(\ref{defFfrak}) together with Lemma \ref%
{localexistencegeneral} there is a solution $(L,R,X)\in \mathbf{B}%
_{r}((L_{0},R_{0}))\times \mathfrak{C}$ to (\ref{localexistenceflowgeneral}%
). Now take any other solution $(\tilde{L},\tilde{R},\tilde{X})\in \mathbf{B}%
_{r}((L_{0},R_{0}))\times \mathfrak{C}$. Then, observe from Lemma \ref%
{lemma-Dyson-Series} (i) that, for any $s,t\in \lbrack 0,T]$, 
\begin{equation}
\partial _{s}\left( Z_{t,s}^{g}(\tilde{L})\tilde{X}_{s}Z_{t,s}^{d}(\tilde{R}%
)\right) =0,  \label{argument1}
\end{equation}%
which yields by integration that 
\begin{equation}
\Tilde{X}_{t}=Z_{t,0}^{g}(\tilde{L})X_{0}Z_{t,0}^{d}(\tilde{R}),\qquad t\in %
\left[ 0,T\right] .  \label{argument2}
\end{equation}%
Therefore, $(\tilde{L}-L_{0})$ and $(\tilde{R}-R_{0})$\ are solutions to (%
\ref{fixed point problem}). By Lemma \ref{localexistencegeneral}, it follows
that $\tilde{L}=L$ and $\tilde{R}=R$, which in turn imply that $X=\Tilde{X}$%
. The last statement follows from the fact that the fixed point provided by
Lemma \ref{localexistencegeneral} gives an operator family which is in $%
\mathscr{B}^{2}$ at all time $t\in \lbrack 0,T]$. It is thus also the case
for $(L,R)$. Note additionally that $X\in C([0,T],\mathscr{A})$ because $%
\mathscr{B}$ is a von Neumann algebra and $\mathscr{A}\subseteq \mathscr{B}$
is a two-sided ideal.
\end{proof}

The extension of the local solution to (\ref{localexistenceflowgeneral})
given by Theorem \ref{LocalExistenceMainResult} to a global solution (i.e.,
for all times $t\in \mathbb{R}_{0}^{+}$) cannot really be carried out as it
is. Indeed, we lack information on this general system of operator-valued
differential equations. An important issue concerns the limited control over
the parameters $r\in \mathbb{R}^{+}$ satisfying (\ref{inequality stupide}),
assuming in the first place such parameter exists, which is not obvious. In
fact, it is useful to estimate the maximum time $T=1/(2r)$ in Theorem \ref%
{LocalExistenceMainResult}, as a function of the initial parameters. This
can be studied when the Lipschitz constants $\kappa _{f,r}$ and $\kappa
_{g,r}$ can be expressed explicitly with respect to the radius $r$. This is
established in the following corollary, which analyzes the specific system (%
\ref{goodflow}) by providing an explicit radius $r$ that satisfies (\ref%
{inequality stupide}):

\begin{corollary}[Local well-posedness of the differential flow]
\label{coro uniqueness}\mbox{}\newline
Given a Hilbert space $\mathfrak{h}$, let $h_{0},\mu _{0},X_{0}\in %
\mathscr{B}^{2}\times \mathscr{A}$ with $h_{0}=h_{0}^{\ast }$ and $\mu
_{0}=\mu _{0}^{\ast }$. If 
\begin{equation*}
T=\left( 2\times 8^{2}\left( \mathrm{e}^{1/2}-1\right) ^{2}\left\Vert
X_{0}\right\Vert _{\mathrm{op}}\right) ^{-1},
\end{equation*}%
then there is a solution $(h,\mu ,X)\in \mathfrak{C}^{3}$ to (\ref{goodflow}%
). Additionally, $\mu _{t}$ and $h_{t}$ are self-adjoint, and $(\mu
_{t},h_{t},X_{t})\in \mathscr{B}^{2}\times \mathscr{A}$ for any time $t\in
\lbrack 0,T]$.
\end{corollary}

\begin{proof}
Let $f$ and $g$ be two mappings on $\mathcal{B(\mathfrak{h})}$ defined by 
\begin{equation}
f\left( A\right) \doteq -AA^{\ast }-A^{\ast }A\qquad \text{and}\qquad
g\left( A\right) \doteq 3AA^{\ast }-A^{\ast }A  \label{defined mapping}
\end{equation}%
for any $A\in \mathcal{B}(\mathfrak{h})$. Clearly, $f\left( 0\right)
=g\left( 0\right) =0$ and $v(\mathscr{B})\subseteq \mathscr{B}$ for all $%
v\in \{f,g\}$. For any closed ball $\overline{B_{r}\left( 0\right) }%
\subseteq \mathcal{B(\mathfrak{h})}$ of center $0$ and radius $r\in \mathbb{R%
}^{+}$,%
\begin{equation*}
\max \left\{ \left\Vert f\left( A\right) -f\left( B\right) \right\Vert _{%
\mathrm{op}},\left\Vert g\left( A\right) -g\left( B\right) \right\Vert _{%
\mathrm{op}}\right\} \leq 8r\left\Vert A-B\right\Vert _{\mathrm{op}},\qquad
A,B\in \overline{B_{r}\left( 0\right) }.
\end{equation*}%
In other words, $f$ and $g$ have $\kappa _{f,r}=\kappa _{g,r}=8r$ as a
Lipschitz constant in $\overline{B_{r}\left( 0\right) }$ for any $r\in 
\mathbb{R}^{+}$. Let $h_{0},\mu _{0},X_{0}\in \mathcal{B}(\mathfrak{h})$. As
soon as%
\begin{equation}
r\geq r_{0}\doteq 8^{2}\left( \mathrm{e}^{1/2}-1\right) ^{2}\left\Vert
X_{0}\right\Vert _{\mathrm{op}},  \label{sdsdssdsddsd}
\end{equation}%
observe that (\ref{inequality stupide}) is satisfied. Then, using all these
parameters as well as the relations%
\begin{equation}
h_{t}=\frac{1}{2}\left( L_{t}-R_{t}\right) \qquad \text{and}\qquad \mu
_{t}=-\left( L_{t}+R_{t}\right) ,  \label{sdsdssdsdd}
\end{equation}%
we can apply Theorem \ref{LocalExistenceMainResult} to find a solution $%
(h,\mu ,X)\in \mathfrak{C}^{3}$ to (\ref{goodflow}) for any $t\in \lbrack
0,T]$ with $T=1/(2r_{0})$. The functions $f$ and $g$ defined by (\ref%
{defined mapping}) map any operator to self-adjoint ones and preserve $%
\mathscr{B}$ as said above. So, if $\mu _{0}$ and $h_{0}$ are self-adjoint
then $\mu _{t}$ and $h_{t}$ are also self-adjoint and $(\mu
_{t},h_{t},X_{t})\in \mathscr{B}^{2}\times \mathscr{A}$ for any $t\in
\lbrack 0,T]$.
\end{proof}

\subsection{Existence, Uniqueness and Long-Time Behavior of the Flow}

We define $T_{\max }$ to be the (possibly infinite)\ maximal time for which
the system (\ref{goodflow}) of operator-valued differential equations has a
solution:%
\begin{equation}
T_{\max }\doteq \sup \left\{ T\geq 0:\exists \mathrm{\mathrm{\ }}\text{a
solution}\mathrm{\ }(h,\mu ,X)\in \mathfrak{C}^{3}\ \text{to\ (\ref{goodflow}%
)}\right\} \in (0,\infty ]\ ,  \label{deftmax}
\end{equation}%
where we recall that $\mathfrak{C\equiv C}_{T}=C([0,T],\mathcal{B}(\mathfrak{%
h}))$ is the Banach space endowed with the supremum norm (\ref{supremum norm}%
) for any $T\in \mathbb{R}^{+}$. Note that $T_{\max }>0$ because of\
Corollary \ref{coro uniqueness}. We first show that the solution to (\ref%
{goodflow}) is unique.

\begin{lemma}[Uniqueness of the solution to the system of differential
equations]
\label{uniqueness}\mbox{}\newline
Given a Hilbert space $\mathfrak{h}$, let $h_{0},\mu _{0},X_{0}\in \mathcal{B%
}(\mathfrak{h})$ with $h_{0}=h_{0}^{\ast }$ and $\mu _{0}=\mu _{0}^{\ast }$.
For all $T\in (0,T_{\max })$, there is a unique solution $(h,\mu ,X)\in 
\mathfrak{C}^{3}$ to (\ref{goodflow}) with initial condition $(h_{0},\mu
_{0},X_{0})$.
\end{lemma}

\begin{proof}
Let $T\in (0,T_{\max })$. If $(h,\mu ,X)\in \mathfrak{C}^{3}$ is a solution
to the initial value problem (\ref{goodflow}) then, using essentially the
same arguments as those used around (\ref{argument1})--(\ref{argument2}) we
deduce that $X$ must be of the form (\ref{trytryt1}) with 
\begin{equation}
L_{t}=h_{t}-\mu _{t}/2\qquad \text{and}\qquad R_{t}=-h_{t}-\mu _{t}/2.
\label{relation bis0}
\end{equation}%
As a consequence, if $(h,\mu ,X)\in \mathfrak{C}^{3}$ is a solution to (\ref%
{goodflow}) then we deduce that, for any $x\in \lbrack 0,T)$, $\delta \in
(0,T-x]$ and $t\in \lbrack 0,\delta ]$, the equalities 
\begin{equation}
L_{t}=h_{x+t}-h_{x}+\left( \mu _{x}-\mu _{x+t}\right) /2\qquad \text{and}%
\qquad R_{t}=h_{x}-h_{x+t}+\left( \mu _{x}-\mu _{x+t}\right) /2,
\label{relation bis}
\end{equation}%
leads to a solution $(L,R)$ to the integral equation (\ref{fixed point
problem}) for all times $t\in \lbrack 0,\delta ]$, with $f$ and $g$ defined
by (\ref{defined mapping}). Thus, two (non-zero) solutions $(h^{\left(
1\right) },\mu ^{\left( 1\right) },X^{\left( 1\right) }),(h^{\left( 2\right)
},\mu ^{\left( 2\right) },X^{\left( 2\right) })\in \mathfrak{C}^{3}$ to (\ref%
{goodflow}) give also two solutions $(L^{\left( 1\right) },R^{\left(
1\right) })$ and $(L^{\left( 2\right) },R^{\left( 2\right) })$ to the same
integral equations for $t\in \lbrack 0,\delta ]$. Note meanwhile that these
solutions belong to some closed ball of center $0$, provided its radius $r$
is sufficiently large. Because of (\ref{sdsdssdsddsd}) and Lemma \ref%
{localexistencegeneral}, there is a unique solution in this (closed) ball
for $\delta =1/(2r)\in (0,T-x]$. It means that $L^{\left( 1\right)
}=L^{\left( 2\right) }$ and $R^{\left( 1\right) }=R^{\left( 2\right) }$,
which in turn imply that%
\begin{equation*}
h_{t+x}^{\left( 1\right) }-h_{x}^{\left( 1\right) }=h_{t+x}^{\left( 2\right)
}-h_{x}^{\left( 2\right) }\qquad \text{and}\qquad \mu _{t+x}^{\left(
1\right) }-\mu _{t}^{\left( 1\right) }=\mu _{t+x}^{\left( 2\right) }-\mu
_{t}^{\left( 2\right) }
\end{equation*}%
for any $x\in \lbrack 0,T)$, $\delta \in (0,T-x]$ and $t\in \lbrack 0,\delta
]$, $\delta =1/(2r)$ being some sufficiently small constant. Remark that,
given any solution $(L,R)$ to the integral equation (\ref{fixed point
problem}), the mappings $t\mapsto \Vert L\Vert _{\mathrm{op}}$ and $t\mapsto
\Vert R\Vert _{\mathrm{op}}$ are always continuous, and so bounded on any
compact set (where the solution exists). As a consequence, since $T\in
(0,T_{\max })$, using Lemma \ref{lemma-Dyson-Series} (ii) and the explicit
form (\ref{trytryt1}) of $X$, one checks that 
\begin{multline*}
\sup_{t\in \left[ 0,T\right] }\left\Vert X_{t}\right\Vert _{\mathrm{op}}\leq
\left\Vert X_{0}\right\Vert _{\mathrm{op}}\sup_{t\in \left[ 0,T\right]
}\left\{ \left\Vert Z_{t,0}^{g}\left( L\right) \right\Vert _{\mathrm{op}%
}\left\Vert Z_{t,0}^{d}\left( R\right) \right\Vert _{\mathrm{op}}\right\} \\
\leq \left\Vert X_{0}\right\Vert _{\mathrm{op}}\mathrm{e}^{T\sup_{t\in \left[
0,T\right] }\Vert L_{t}\Vert _{\mathrm{op}}}\mathrm{e}^{T\sup_{t\in \left[
0,T\right] }\Vert R_{t}\Vert _{\mathrm{op}}}<\infty .
\end{multline*}%
Hence, the radius 
\begin{equation}
r\geq 8^{2}\left( \mathrm{e}^{1/2}-1\right) ^{2}\sup_{t\in \left[ 0,T\right]
}\left\Vert X_{t}\right\Vert _{\mathrm{op}}  \label{79}
\end{equation}
(cf. (\ref{sdsdssdsddsd})) and so, the constant $\delta =1/(2r)\in (0,T]$
can be taken independently of the choice of the initial time $x\in \lbrack
0,T]$. It follows that, for any $T\in (0,T_{\max })$ and $t\in \lbrack 0,T]$%
, 
\begin{equation}
h_{t}^{\left( 1\right) }=h_{t}^{\left( 2\right) }\qquad \text{and}\qquad \mu
_{t}^{\left( 1\right) }=\mu _{t}^{\left( 2\right) }.  \label{poipoipoi}
\end{equation}%
Because any solution $X\ $must be of the form (\ref{trytryt1}) with $L,R$
defined by (\ref{relation bis}), we infer from (\ref{poipoipoi}) that $%
X^{\left( 1\right) }=X^{\left( 2\right) }$.
\end{proof}

The next step is to show the global existence of a (unique) solution to the
system (\ref{goodflow}) of differential equations. In other words, we study
the maximum time $T_{\max }$ (\ref{deftmax}) to show that, under some
sufficient conditions, it is infinite, i.e., $T_{\max }=\infty $. We first
prove that this maximum time of existence of a solution to (\ref{goodflow})
can only be finite if the operator norm of the component $X\equiv
(X_{t})_{t\geq 0}$ of the solution blows up in the limit $t\rightarrow
T_{\max }$. This refers to the following assertion:

\begin{lemma}[The blow--up alternative]
\label{blowup}\mbox{}\newline
Given a Hilbert space $\mathfrak{h}$, let $h_{0},\mu _{0},X_{0}\in \mathcal{B%
}(\mathfrak{h})$ with $h_{0}=h_{0}^{\ast }$ and $\mu _{0}=\mu _{0}^{\ast }$
and consider (\ref{goodflow}) with initial condition $(h_{0},\mu _{0},X_{0})$%
. Then, either $T_{\max }=\infty $ and we have a unique global solution $%
(h,\mu ,X)\in \mathfrak{C}^{3}\ $to\ (\ref{goodflow}) or $T_{\max }<\infty $
and 
\begin{equation*}
\lim_{t\rightarrow T_{\max }^{-}}\left\Vert X_{t}\right\Vert _{\mathrm{op}%
}=\infty .
\end{equation*}
\end{lemma}

\begin{proof}
Let $(h,\mu ,X)\in \mathfrak{C}^{3}$ be a solution to the initial value
problem (\ref{goodflow}) for any time $t\in \lbrack 0,T_{\max })$. By
contradiction, assume $T_{\max }<\infty $ and the existence of a sequence $%
(t_{n})_{n\in \mathbb{N}}\subseteq \lbrack 0,T_{\max })$ converging to $%
T_{\max }$ such that 
\begin{equation}
\varkappa \doteq \sup_{n\in \mathbb{N}}\left\Vert X_{t_{n}}\right\Vert _{%
\mathrm{op}}<\infty .  \label{toto3bisassumption}
\end{equation}%
Fix any radius $r\geq 8^{2}\left( \mathrm{e}^{1/2}-1\right) ^{2}\varkappa $
(cf. (\ref{sdsdssdsddsd}) and (\ref{79})) and $\delta =1/(2r)$. Then, there
is $n_{0}\in \mathbb{N}$ such that, for all natural numbers $n\geq n_{0}$, $%
\delta +t_{n}>T_{\max }$. Furthermore, by using Theorem \ref%
{LocalExistenceMainResult} together with arguments used in Lemma \ref%
{uniqueness} (see, e.g., (\ref{relation bis}) and (\ref{79})), we obtain a
solution$\mathrm{\ }(h,\mu ,X)\in \mathfrak{C}^{3}\ $to\ (\ref{goodflow})
until the time $\delta +t_{n_{0}}>T_{\max }$, which contradicts the
definition of $T_{\max }$ as the maximum time until which such a solution
exists; see (\ref{deftmax}).
\end{proof}

We now introduce sufficient conditions to guarantee the global existence of
the system (\ref{goodflow}) of differential equations. To do this, we use a
kind of gap equation that prevents the operator norm of the $X $-part of the
solution to (\ref{goodflow}) from exploding; cf. Lemma \ref{blowup}. This
refers to the following sufficient conditions: Let $\alpha _{+},\alpha
_{-}\in \mathbb{R}$ be such that 
\begin{equation}
\mu _{0}\pm 2h_{0}\geq \alpha _{\pm }\mathbf{1}.  \label{alpha+-}
\end{equation}%
Since the operators $\mu _{0}$ and $h_{0}$ are bounded, these inequalities
automatically hold for sufficiently small choices of $\alpha _{\pm }$. For
any strictly positive parameter $\varepsilon \in \mathbb{R}^{+}$, we then
define the maximum time 
\begin{equation}
T_{\varepsilon }\doteq \sup \left\{ t\in \left[ 0,T_{\max }\right) :\forall
s\in \left[ 0,t\right] ,\ \mu _{s}\pm 2h_{s}\geq \left( \alpha _{\pm
}-\varepsilon \right) \mathbf{1}\right\} \in (0,\infty ].  \label{Tespsilon}
\end{equation}%
When Inequalities (\ref{alpha+-}) is satisfied, $T_{\varepsilon }$ is
well-defined\ as a strictly positive time, by norm continuity of the
solution to (\ref{goodflow}). Then, using (\ref{alpha+-})--(\ref{Tespsilon})
we obtain an exponential decay of the Hilbert-Schmidt norm of the $X$-part
of the differential system. Given a separable Hilbert space $\mathfrak{h}$,
recall that $\mathcal{L}^{1}(\mathfrak{h})$\ and $\mathcal{L}^{2}(\mathfrak{h%
})$ are the spaces of trace-class and Hilbert-Schmidt operators,
respectively. The norms of $\mathcal{L}^{1}(\mathfrak{h})$\ and $\mathcal{L}%
^{2}(\mathfrak{h})$ are respectively denoted by $\Vert \cdot \Vert _{1}$ and 
$\Vert \cdot \Vert _{2}$.

\begin{proposition}[Hilbert-Schmidt exponential decay of the $X$-component]
\label{controleX}\mbox{}\newline
Given a separable Hilbert space $\mathfrak{h}$, take $h_{0}=h_{0}^{\ast
},\mu _{0}=\mu _{0}^{\ast }\in \mathcal{B}(\mathfrak{h})$ and $X_{0}\in 
\mathcal{L}^{2}(\mathfrak{h})$ satisfying (\ref{alpha+-}) and 
\begin{equation}
\alpha \doteq \frac{1}{2}\left( \alpha _{+}+\alpha _{-}\right) >\sqrt{3}%
\left\Vert X_{0}\right\Vert _{2}.  \label{fgdfgdfgfdgdfgdgfdfgd}
\end{equation}%
Then, $T_{\alpha /2}=T_{\max }$. Furthermore, for any $t\in \lbrack
0,T_{\max })$, 
\begin{equation*}
2\mu _{t}\geq\alpha \mathbf{1}\qquad \text{and}\qquad \left\Vert
X_{t}\right\Vert _{2}^{2}\leq \mathrm{e}^{-\alpha t}\left\Vert
X_{0}\right\Vert _{2}^{2}.
\end{equation*}
\end{proposition}

\begin{proof}
Assume Inequalities (\ref{alpha+-}) and (\ref{fgdfgdfgfdgdfgdgfdfgd})
together with $X_{0}\neq 0$, in order to avoid the trivial case\footnote{%
If $X_{0}=0$ then $X_{t}=0$, $\mu _{t}=\mu _{0}$ and $h_{t}=h_{0}$ for all
times.}. Since $\alpha >0$, we can fix some strictly positive parameter $%
\varepsilon \in (0,\alpha )$. Then, by Equation (\ref{goodflow}) and
cyclicity of the trace, for any $t\in \left[ 0,T_{\varepsilon }\right) $, 
\begin{equation}
\partial _{t}\mathrm{tr}\left( X_{t}X_{t}^{\ast }\right) =-\mathrm{tr}\left(
X_{t}\left( \mu _{t}+2h_{t}\right) X_{t}^{\ast }\right) -\mathrm{tr}\left(
X_{t}^{\ast }\left( \mu _{t}-2h_{t}\right) X_{t}\right) \leq -2\left( \alpha
-\varepsilon \right) \mathrm{tr}\left( X_{t}X_{t}^{\ast }\right) ,
\label{sdsdsdsdklklkl}
\end{equation}%
the time $T_{\varepsilon }$ being defined by (\ref{Tespsilon}). Invoking Gr%
\"{o}nwall's lemma, it follows that, for any $u,t\in \left[ 0,T_{\varepsilon
}\right) $, 
\begin{equation}
\mathrm{tr}\left( X_{t}X_{t}^{\ast }\right) \leq \mathrm{e}^{-2\left( \alpha
-\varepsilon \right) \left( t-u\right) }\mathrm{tr}\left( X_{u}X_{u}^{\ast
}\right) .  \label{sdsdsdsd}
\end{equation}%
In particular, setting $u=0$ and integrating both sides from $0$ to $t$
yields 
\begin{equation}
2\int_{0}^{t}\mathrm{tr}\left( X_{s}X_{s}^{\ast }\right) \mathrm{d}s\leq
\left( \alpha -\varepsilon \right) ^{-1}\left\Vert X_{0}\right\Vert
_{2}^{2},\qquad t\in \left[ 0,T_{\varepsilon }\right) .  \label{sdsdsdsds}
\end{equation}%
Using again the flow (\ref{goodflow}), we easily compute that 
\begin{equation*}
\partial _{t}\left( \mu _{t}+2h_{t}\right) =2X_{t}^{\ast
}X_{t}-6X_{t}X_{t}^{\ast }\ ,\qquad t\in \left[ 0,T_{\max }\right) .
\end{equation*}%
We thus combine this last computation with Equation (\ref{sdsdsdsds}) to
obtain that, for any $t\in \left[ 0,T_{\varepsilon }\right) $, 
\begin{equation}
\mu _{t}+2h_{t}=\mu _{0}+2h_{0}+2\int_{0}^{t}X_{s}^{\ast }X_{s}\mathrm{d}%
s-6\int_{0}^{t}X_{s}X_{s}^{\ast }\mathrm{d}s\geq \left( \alpha
_{+}-6\int_{0}^{t}\mathrm{tr}\left( X_{s}X_{s}^{\ast }\right) \mathrm{d}%
s\right) \mathbf{1}\geq \tilde{\alpha}_{+,\varepsilon }\mathbf{1}
\label{eq1}
\end{equation}%
with 
\begin{equation}
\tilde{\alpha}_{+,\varepsilon }\doteq \alpha _{+}-\frac{3}{\alpha
-\varepsilon }\left\Vert X_{0}\right\Vert _{2}^{2}\ .  \label{alphatildeplus}
\end{equation}%
So, if there are $r\in \mathbb{R}^{+}$ and $\varepsilon \in \lbrack 0,\alpha
)$ such that 
\begin{equation}
3\left( 1+r\right) \left\Vert X_{0}\right\Vert _{2}^{2}<\varepsilon \left(
\alpha -\varepsilon \right) ,  \label{sdsssdsdsdsd}
\end{equation}%
then 
\begin{equation*}
\tilde{\alpha}_{+,\varepsilon }>\alpha _{+}-\frac{\varepsilon }{1+r}.
\end{equation*}%
Therefore, by performing the limit $t\rightarrow T_{\varepsilon }$ we obtain
that 
\begin{equation}
\lim_{t\rightarrow T_{\varepsilon }}\left( \mu _{t}+2h_{t}\right) \geq
\left( \alpha _{+}-\frac{\varepsilon }{1+r}\right) \mathbf{1}>\left( \alpha
_{+}-\varepsilon \right) \mathbf{1}.  \label{contiargument}
\end{equation}%
Either $T_{\mathrm{max}}=T_{\varepsilon }$ or $T_{\mathrm{max}%
}>T_{\varepsilon }$. In the second case, by continuity, (\ref{contiargument}%
) yields the existence of a strictly positive parameter $\delta _{+}\in
(0,T_{\max }-T_{\varepsilon })$ such that 
\begin{equation}
\mu _{t}+2h_{t}\geq \left( \alpha _{+}-\varepsilon \right) \mathbf{1},\quad
t\in \lbrack T_{\varepsilon },T_{\varepsilon }+\delta _{+}).
\label{alpha+avecepsilon}
\end{equation}%
Moreover, integrating the respective differential system (\ref{goodflow})
and using the positivity of $X_{s}X_{s}^{\ast }$ and $X_{s}X_{s}^{\ast }$
for all $s\in \lbrack 0,T_{\mathrm{max}})$ imply that, for any $t\in \left[
0,T_{\varepsilon }\right) $, 
\begin{equation}
\mu _{t}-2h_{t}=\mu _{0}-2h_{0}+2\int_{0}^{t}\left( X_{s}^{\ast
}X_{s}+X_{s}X_{s}^{\ast }\right) \mathrm{d}s\geq \alpha _{-}\mathbf{1}.
\label{eq2}
\end{equation}%
Therefore, if there are $r\in \mathbb{R}^{+}$ and $\varepsilon \in \lbrack
0,\alpha )$ such that (\ref{sdsssdsdsdsd}) holds, then the only possible
alternative is $T_{\varepsilon }=T_{\max }$. This yields in particular the
assertion $T_{\alpha /2}=T_{\max }$ with (\ref{sdsdsdsd}) for $\varepsilon
=\alpha /2$. Finally, remark that the inequalities 
\begin{equation}
\mu _{t}\pm 2h_{t}\geq \left( \alpha _{\pm }-\alpha /2\right) \mathbf{1}
\label{LowBoundmupmh}
\end{equation}%
implies that%
\begin{equation*}
\left( \alpha _{+}-\alpha /2\right) \mathbf{1}-\mu _{t}\leq 2h_{t}\leq \mu
_{t}-\left( \alpha _{-}-\alpha /2\right) \mathbf{1},
\end{equation*}%
i.e., $2\mu _{t}\geq \alpha \mathbf{1}$.
\end{proof}

\begin{corollary}[Operator-norm exponential decay of the $X$-component]
\label{pseudoHS}\mbox{}\newline
Given a separable Hilbert space $\mathfrak{h}$, take $h_{0}=h_{0}^{\ast
},\mu _{0}=\mu _{0}^{\ast }\in \mathcal{B}(\mathfrak{h})$ and $X_{0}\in 
\mathcal{L}^{2}(\mathfrak{h})$ satisfying Inequalities (\ref{alpha+-}) and (%
\ref{fgdfgdfgfdgdfgdgfdfgd}). Then, the solution $(h,\mu ,X)$ to the
differential system (\ref{goodflow}) satisfies 
\begin{equation*}
\left\Vert X_{t}\right\Vert _{\mathrm{op}}^{2}\leq \mathrm{e}^{-\alpha
t}\left\Vert X_{0}\right\Vert _{\mathrm{op}}^{2},\qquad t\in \lbrack 0,T_{%
\mathrm{max}}).
\end{equation*}
\end{corollary}

\begin{proof}
As explained in (\ref{argument1})--(\ref{argument2}), any solution $X\ $must
be of the form (\ref{trytryt1}). In other words, 
\begin{equation}
X_{t}=Z_{t,s}^{g}\left( L\right) X_{s}Z_{t,s}^{d}\left( R\right) ,\qquad
s,t\in \mathbb{R}_{0}^{+},  \label{equality trivial2}
\end{equation}%
where, for any $t\in \lbrack 0,T_{\mathrm{max}})$, 
\begin{equation}
L_{t}=h_{t}-\mu _{t}/2\qquad \text{and}\qquad R_{t}=-h_{t}-\mu _{t}/2.
\label{relation bis02}
\end{equation}%
Hence, a direct application of Lemma \ref{lemma-Dyson-Series} (ii) and (\ref%
{LowBoundmupmh}) yields the asserted inequality.
\end{proof}

We conclude this section with the main theorem of Section \ref%
{Fiberflowsection}, which establishes the global existence, uniqueness, and
long-time asymptotic behavior of the differential flow:

\begin{theorem}[Global well-posedness and asymptotics of the flow]
\label{globalexistence}\mbox{}\newline
Let $\mathfrak{h}$ be a separable Hilbert space, $\mathscr{B}\subseteq 
\mathcal{B}\left( \mathfrak{h}\right) $ a von Neumann algebra and $%
\mathscr{A}$ a two-sided $\sigma $-weakly closed ideal of $\mathscr{B}$.
Assume the initial data $h_{0}=h_{0}^{\ast },\mu _{0}=\mu _{0}^{\ast }\in %
\mathscr{B}\subseteq \mathcal{B}(\mathfrak{h})$ and $X_{0}\in \mathcal{L}%
^{2}(\mathfrak{h})\cap \mathscr{A}$ satisfy $\mu _{0}\pm 2h_{0}\geq \alpha
_{\pm }\mathbf{1}$, where $\alpha _{+}+\alpha _{-}>2\sqrt{3}\left\Vert
X_{0}\right\Vert _{2}$. Then a solution $(h,\mu ,X)\in \mathfrak{C}^{3}$ to
the system (\ref{goodflow}) exists globally ($T_{\max }=\infty $) and is
unique. For all $t\in \mathbb{R}_{0}^{+}$, $(\mu _{t},h_{t},X_{t})\in %
\mathscr{B}^{2}\times \mathscr{A}$, and 
\begin{equation*}
\mu _{t}=\mu _{0}+2\int_{0}^{t}\left[ X_{\tau }^{\ast },X_{\tau }\right] 
\mathrm{d}\tau ,\qquad h_{t}=h_{0}-2\int_{0}^{t}X_{\tau }X_{\tau }^{\ast }%
\mathrm{d}\tau .
\end{equation*}%
Furthermore, these integrals converge absolutely in $\mathcal{L}^{1}(%
\mathfrak{h})$ as $t\rightarrow \infty $. In this limit, $X_{t}\rightarrow 0$
in $\mathcal{L}^{2}(\mathfrak{h})$, while $\mu _{t}\rightarrow \mu _{\infty
} $ and $h_{t}\rightarrow h_{\infty }$ in $\mathcal{L}^{1}(\mathfrak{h})$
with $(\mu _{\infty },h_{\infty })\in \mathscr{B}^{2}$.
\end{theorem}

\begin{proof}
Assume Inequality (\ref{fgdfgdfgfdgdfgdgfdfgd}). Then, the assertion $%
T_{\max }=\infty $ results from Lemma \ref{blowup} and\ Proposition \ref%
{controleX} together with the inequality 
\begin{equation}
\left\Vert A\right\Vert _{\mathrm{op}}\leq \left\Vert A\right\Vert
_{2},\qquad A\in \mathcal{B}(\mathfrak{h}).  \label{sdsdsdsdsdsdsfdfgfgf}
\end{equation}%
Lemma \ref{uniqueness} guarantees the uniqueness of the solution to the
differential system. Then, the rest of the mathematical statements are
simply a consequence of the exponential decay of the Hilbert-Schmidt norm of
the $X$-part of the differential flow given by Proposition \ref{controleX}
combined with the triangle inequality. We omit the details. If $\mathscr{B}%
\subseteq \mathcal{B}\left( \mathfrak{h}\right) $ is a von Neumann algebra, $%
\mathscr{A}$ is a two-sided $\sigma $-weakly closed ideal of $\mathscr{B}$
and $(\mu _{0},h_{0},X_{0})\in \mathscr{B}^{2}\times \mathscr{A}$, then one
easily checks that $(\mu _{t},h_{t},X_{t})\in \mathscr{B}^{2}\times %
\mathscr{A}$ for all $t\in \mathbb{R}_{0}^{+}$. This property extends to $%
t=\infty $ because $\mathscr{B}$ is a von Neumann algebra (and is therefore
closed under the relevant topology).
\end{proof}

\subsection{Discrete Family of Differential Flows\label{Section diescrete
family}}

We apply the differential flow (\ref{goodflow}) to the operator-valued
coefficients of the initial generalized spin-boson Hamiltonian \textrm{H}$%
_{0}$, using the initial data $(\mu _{k,0},X_{k,0},h_{k,0})_{k\in \mathbb{N}%
} $. This generates time-dependent coefficients that define a family of
time-dependent generalized spin-boson Hamiltonians. To ensure that these
evolving coefficients possess the required structural properties, we assume
certain conditions.

To present them, for a countable collection $\mathfrak{h}\equiv \{\mathfrak{h%
}_{k}\}_{k\in \mathbb{N}}$ of (possibly infinite-dimensional) Hilbert spaces
we use the Banach space 
\begin{equation}
\ell ^{2}\left( \mathcal{B}\left( \mathfrak{h}\right) \right) \equiv \ell
^{2}\left( \mathbb{N},\{\mathcal{B}\left( \mathfrak{h}_{k}\right) \}_{k\in 
\mathbb{N}}\right) \doteq \left\{ \left( X_{k}\right) _{k\in \mathbb{N}%
}:\forall k\in \mathbb{N},\ X_{k}\in \mathcal{B}(\mathfrak{h}_{k})\quad 
\text{and}\quad \left\Vert X\right\Vert _{\mathrm{op},2}<\infty \right\}
\label{L2}
\end{equation}%
endowed with the $\ell _{2}$--operator norm $\Vert \cdot \Vert _{\mathrm{op}%
,2}$, which is defined by Equation (\ref{lp-operator norm}) for $p=2$.
Furthermore, for any subset $I\subseteq \mathbb{R}_{0}^{+}$, $C(I;\ell ^{2}(%
\mathcal{B}(\mathfrak{h})))$ represents the vector space of continuous
mappings from $I$ to $\ell ^{2}(\mathcal{B}(\mathfrak{h}))$. Here, as is
usual, for elements $(A_{k})_{k\in \mathbb{N}}$ and $(B_{k})_{k\in \mathbb{N}%
}$ where $A_{k}\doteq (A_{k,t})_{t\in I}$ and $B_{k}\doteq (B_{k,t})_{t\in
I} $, we define addition and multiplication as:%
\begin{equation*}
A+B\doteq (A_{k}+B_{k})_{k\in \mathbb{N}}\doteq (\left(
A_{k,t}+B_{k,t}\right) _{t\in I})_{k\in \mathbb{N}}\qquad \text{and}\qquad
AB\doteq (A_{k}B_{k})_{k\in \mathbb{N}}\doteq (\left( A_{k,t}B_{k,t}\right)
_{t\in I})_{k\in \mathbb{N}}.
\end{equation*}%
In particular, 
\begin{equation*}
\lbrack A,B]\doteq (\left( \left[ A_{k,t},B_{k,t}\right] \right) _{t\in
I})_{k\in \mathbb{N}},
\end{equation*}%
where we recall that $[\cdot ,\cdot ]$ is the usual commutator for operators.

We are now in a position to present our sufficient conditions:

\begin{condition}[Finite-time horizon hypotheses]
\label{AbstractAssumptions}\mbox{}\newline
Let $\mathfrak{h}\equiv (\mathfrak{h}_{k})_{k\in \mathbb{N}}$ be a countable
family of separable Hilbert spaces. The family $(\mu _{k},X_{k},h_{k})_{k\in 
\mathbb{N}}$ of operator-coefficients is assumed to satisfy the following
properties:

\begin{itemize}
\item[\emph{(i)}] For each $k\in \mathbb{N}$, $(h_{k},\mu _{k},X_{k})\in C(%
\mathbb{R}_{0}^{+},\mathcal{B}(\mathfrak{h}_{k}))^{3}$ is a solution to the
flow (\ref{goodflow}). For all $t\in \mathbb{R}_{0}^{+}$, $h_{k,t}$ and $\mu
_{k,t}$ are self-adjoint, with $\mu _{k,0}\pm 2h_{k,0}\geq \alpha _{k,\pm }%
\mathbf{1}_{\mathfrak{h}_{k}}$ for some constants $\alpha _{k,\pm }\in 
\mathbb{R}$ satisfying $\alpha _{k,+}+\alpha _{k,-}\geq 0$ and 
\begin{equation*}
\inf \{\alpha _{k,\pm }:k\in \mathbb{N},\;X_{k,0}\neq 0\}>-\infty .
\end{equation*}%
Furthermore, for any $T\in \mathbb{R}_{0}^{+}$, there exists a constant $%
D_{T}\in \mathbb{R}^{+}$ such that 
\begin{equation*}
\sup_{t\in \left[ 0,T\right] }\left\Vert X_{k,t}\right\Vert _{\mathrm{op}%
}^{2}\leq D_{T}\left\Vert X_{k,0}\right\Vert _{\mathrm{op}}^{2},\qquad k\in 
\mathbb{N}.
\end{equation*}

\item[\emph{(ii)}] For any $t\in \mathbb{R}_{0}^{+}$, the operator $h_{t}$
decomposes as $h_{t}=u_{t}+v_{t}$, where $u_{t}\equiv (u_{k,t})_{k\in 
\mathbb{N}}$ and $v_{t}\equiv (v_{k,t})_{k\in \mathbb{N}}$ are sequences of
self-adjoint operators in $\mathcal{B}(\mathfrak{h}_{k})\equiv \mathcal{B}(%
\mathfrak{h}_{k})^{\otimes }$. The sequence $u_{t}$ satisfies Assumption \ref%
{Assumption S}, and $v_{t}$ satisfies $\Vert v_{t}\Vert _{\mathrm{op}%
,1}<\infty $.

\item[\emph{(iii)}] For any $t\in \mathbb{R}_{0}^{+}$, $\Vert X_{t}\Vert _{%
\mathrm{op},2}<\infty $ and $\Vert \lbrack X_{t},\mu _{t}]\Vert _{\mathrm{op}%
,2}<\infty $. Moreover, $X,\mu \in C(\mathbb{R}^{+};\ell ^{2}(\mathcal{B}(%
\mathfrak{h})))$ with $X$ being locally-Lipschitz continuous.

\item[\emph{(iv)}] For each $k\in \mathbb{N}$, there exists a constant $%
c_{k}\in \mathbb{R}_{0}^{+}$ such that $\mu _{k,t}\geq c_{k}\mathbf{1}$ for
all $t\in \mathbb{R}_{0}^{+}$, and 
\begin{equation*}
\sum_{k\in \mathbb{N}}c_{k}^{-1}\left\Vert X_{k,t}\right\Vert _{\mathrm{op}%
}^{2}<\infty ,
\end{equation*}%
using the convention $0^{-1}\times 0\equiv 0$ if $c_{k}=\Vert X_{k,t}\Vert _{%
\mathrm{op}}=0$.

\item[\emph{(v)}] The following gap condition holds:\footnote{%
Note that the property (v) combined with $\Vert X_{0}\Vert _{\mathrm{op}%
,2}<+\infty $ implies property (iv) via Corollary \ref{pseudoHS}. Property
(v) is a stronger condition invoked only when strictly necessary.} 
\begin{equation*}
c\doteq \inf \left\{ c_{k}:k\in \mathbb{N},\text{\quad }X_{k,0}\neq 0,\text{%
\quad }\mu _{k,t}\geq c_{k}\mathbf{1}_{\mathfrak{h}_{k}}\right\} >0.
\end{equation*}
\end{itemize}
\end{condition}

\begin{condition}[Infinite-time horizon hypotheses]
\label{Assumptionsasympt}\mbox{}\newline
Suppose Condition \ref{AbstractAssumptions} holds. For each $k\in \mathbb{N}$%
, assume that the triplet $(h_{k,t},\mu _{k,t},X_{k,t})$ converges in the
operator norm as $t\rightarrow \infty $ to a limit $(h_{k,\infty },\mu
_{k,\infty },0)\in \mathcal{B}(\mathfrak{h}_{k})^{3}$, where $h_{k,\infty }$
and $\mu _{k,\infty }$ are self-adjoint. Furthermore, assume that Conditions %
\ref{AbstractAssumptions} (ii)--(v) remain valid at $t=\infty $. Setting $%
x=\Vert X_{0}\Vert _{\mathrm{op},2}$, we impose the following final
requirement:

\begin{itemize}
\item[\emph{(vi)}] For each $p\in \{1,2\}$, the interaction and its
commutator satisfy the integrability bounds: 
\begin{equation*}
\int_{0}^{\infty }\left\Vert X_{t}\right\Vert _{\mathrm{op},2}^{p}\mathrm{d}%
t\leq f_{p}\left( x\right) \qquad \text{and}\qquad \int_{0}^{\infty
}\left\Vert \left[ X_{t},\mu _{t}\right] \right\Vert _{\mathrm{op},2}\mathrm{%
d}t\leq g\left( x\right) ,
\end{equation*}%
where $f_{p},g:\mathbb{R}_{0}^{+}\rightarrow \mathbb{R}$ are continuous,
non-decreasing functions such that $f_{p}(0)=g(0)=0$.
\end{itemize}
\end{condition}

The results of Section \ref{Fiberflowsection} provide sufficient conditions
to ensure that all requirements of Condition \ref{AbstractAssumptions} are
satisfied. As shown in the next assertion, these requirements are perfectly
met under Condition \ref{Assumption BWMmainresult}:

\begin{theorem}[Sufficient conditions]
\label{SufiCondAbstrat}\label{SufiCondAsympt copy(1)}\mbox{}\newline
Condition \ref{Assumption BWMmainresult} (BW1--BW4) implies the existence of
a family $(h_{k},\mu _{k},X_{k})_{k\in \mathbb{N}}$ that satisfies
Conditions \ref{AbstractAssumptions}--\ref{Assumptionsasympt}.
\end{theorem}

\begin{proof}
Theorem \ref{globalexistence} and Condition \ref{Assumption BWMmainresult}
guarantee that, for each $k\in \mathbb{N}$, Equation (\ref{goodflow}) admits
a unique solution $(h_{k},\mu _{k},X_{k})_{k\in \mathbb{N}}\in C(\mathbb{R}%
_{0}^{+},\mathcal{B}(\mathfrak{h}_{k}))^{3}$. As $t\rightarrow \infty $,
this solution converges in the operator norm to some $(h_{k,\infty },\mu
_{k,\infty },0)\in \mathcal{B}(\mathfrak{h})^{3}$, where $h_{k,t}$ and $\mu
_{k,t}$ are self-adjoint for all $t\in \lbrack 0,\infty ]$. Condition \ref%
{Assumption BWMmainresult} (BW2, BW4) says that $\mu _{k,0}\pm 2h_{k,0}\geq
\alpha _{k,\pm }\mathbf{1}_{\mathfrak{h}_{k}}$ for some constants $\alpha
_{k,\pm }\in \mathbb{R}$ satisfying $\alpha _{k,+}+\alpha _{k,-}\geq 0$, as
well as 
\begin{equation*}
\inf \{\alpha _{\pm ,k}:k\in \mathbb{N},\ X_{k,0}\neq 0\}>-\infty .
\end{equation*}%
Furthermore, Condition \ref{Assumption BWMmainresult} (BW4) and Corollary %
\ref{pseudoHS} yield the uniform bound:%
\begin{equation*}
\sup_{t\in \mathbb{R}_{0}^{+}}\left\Vert X_{k,t}\right\Vert _{\mathrm{op}%
}^{2}\leq \left\Vert X_{k,0}\right\Vert _{\mathrm{op}}^{2}.
\end{equation*}%
Condition \ref{AbstractAssumptions} (i) is thus verified.

Condition \ref{Assumption BWMmainresult} (BW1) combined with Corollary \ref%
{pseudoHS} and Theorem \ref{globalexistence} yields Hypothesis (ii) of\
Conditions \ref{AbstractAssumptions} and \ref{Assumptionsasympt}, i.e., for
all times $t\in \lbrack 0,\infty ]$\ the operator $h_{t}$ can be decomposed
as $h_{t}=u_{t}+v_{t}$, where these two terms respectively satisfy
Assumption \ref{Assumption S} and $\Vert v_{t}\Vert _{\mathrm{op},1}<\infty $%
. Indeed, by Corollary \ref{pseudoHS} and Theorem \ref{globalexistence} (cf.
(\ref{goodflow})) together with (\ref{lp-operator norm}), Tonelli's theorem
and the triangle inequality, we get that%
\begin{equation}
\left\Vert h_{t}-h_{0}\right\Vert _{\mathrm{op},1}\leq
2\int_{0}^{t}\left\Vert X_{\tau }\right\Vert _{\mathrm{op},2}^{2}\mathrm{d}%
\tau ,\qquad t\in \left[ 0,\infty \right] ,  \label{sufficient0}
\end{equation}%
where 
\begin{equation}
\int_{0}^{\infty }\left\Vert X_{\tau }\right\Vert _{\mathrm{op},2}^{p}%
\mathrm{d}\tau \leq 2\left( p\alpha \right) ^{-1}\left\Vert X_{0}\right\Vert
_{\mathrm{op},2}^{p},\qquad p\in \{1,2\}.  \label{sufficient1}
\end{equation}%
By Condition \ref{Assumption BWMmainresult} (BW1), $h_{0}$ is decomposed as $%
h_{0}=u_{0}+v_{0}$, where $u_{0}$ and $v_{0}$ are two sequences of bounded
self-adjoint operators satisfying Assumption \ref{Assumption S} and $\Vert
v_{0}\Vert _{\mathrm{op},1}<\infty $, respectively. By Equations (\ref%
{sufficient0})--(\ref{sufficient1}), for each $t\in \lbrack 0,\infty ]$, $%
h_{t}$ can thus be decomposed as $u_{t}+v_{t}$ with $u_{t}\doteq u_{0}$ and $%
v_{t}\doteq h_{t}-h_{0}+v_{0}$. It defines two sequences $u_{t}$ and $v_{t}$
of bounded self-adjoint operators satisfying again Assumption \ref%
{Assumption S} and $\Vert v_{t}\Vert _{\mathrm{op},1}<\infty $,
respectively. Hypothesis (ii) of\ Conditions \ref{AbstractAssumptions} and %
\ref{Assumptionsasympt} is thus verified.

In the same way, since for each $k\in \mathbb{N}$ and $t\in \lbrack 0,\infty
]$, $[\mu _{k,0},X_{k,t}]=0$ under Condition \ref{Assumption BWMmainresult}
(cf. Theorem \ref{globalexistence}), we deduce from (\ref{goodflow}) and (%
\ref{sufficient1}) that 
\begin{equation}
\int_{0}^{\infty }\left\Vert \left[ X_{t},\mu _{t}\right] \right\Vert _{%
\mathrm{op},2}\mathrm{d}t\leq 8\int_{0}^{\infty }\left\Vert X_{t}\right\Vert
_{\mathrm{op},2}\int_{0}^{\tau }\left\Vert X_{\tau }\right\Vert _{\mathrm{op}%
,2}^{2}\mathrm{d}\tau \mathrm{d}t\leq 16\alpha ^{-2}\left\Vert
X_{0}\right\Vert _{\mathrm{op},2}^{3}.  \label{sufficient2}
\end{equation}%
Inequalities (\ref{sufficient1})--(\ref{sufficient2}) clearly yield
Condition \ref{Assumptionsasympt} (vi). Furthermore, by Proposition \ref%
{controleX} and Theorem \ref{globalexistence}, for any $k\in \mathbb{N}$
such that $X_{k,0}\neq 0$, 
\begin{equation*}
2\mu _{k,t}\geq \left( \alpha _{k,+}+\alpha _{k,-}\right) \mathbf{1}_{%
\mathfrak{h}_{k}},\qquad t\in \lbrack 0,\infty ],
\end{equation*}%
with 
\begin{equation*}
\inf \left\{ \alpha _{k,+}+\alpha _{k,-}:k\in \mathbb{N},\ X_{k,0}\neq
0\right\} >0,
\end{equation*}%
thanks to Condition \ref{Assumption BWMmainresult} (BW4). Note that the
components $k\in \mathbb{N}$ for which $X_{k,0}=0$ can be ignored since they
produce a trivial flow. Therefore, Hypothesis (v) of\ Conditions \ref%
{AbstractAssumptions} and \ref{Assumptionsasympt}, i.e., for all times $t\in
\lbrack 0,\infty ]$, holds. This obviously implies Hypothesis (iv) of\
Conditions \ref{AbstractAssumptions} and \ref{Assumptionsasympt}.

It remains to check Hypothesis (iii) of\ Conditions \ref{AbstractAssumptions}
and \ref{Assumptionsasympt}. First, the boundedness part of (iii) follows
easily from Corollary \ref{pseudoHS} and Theorem \ref{globalexistence} and,
in particular, from the fact that $[\mu _{k,0},X_{k,t}]=0$ for all $k\in 
\mathbb{N}$ and $t\in \lbrack 0,\infty ]$. The final step is thus to verify
that the mappings the mappings $t\mapsto X_{t}$ and $t\mapsto \left[
X_{t},\mu _{t}\right] $ are, respectively, locally Lipschitz continuous and
continuous on $\mathbb{R}^{+}$ with respect to the norm $\Vert \cdot \Vert _{%
\mathrm{op},2}$: In fact, for any $k\in \mathbb{N}$\ and $s,t\in \mathbb{R}%
_{0}^{+}$, the operators $L_{k,t}$ and $R_{k,t}$ defined by (\ref{relation
bis02}) satisfies the bounds%
\begin{eqnarray}
\max \left\{ \left\Vert L_{k,t}-L_{k,s}\right\Vert _{\mathrm{op}},\left\Vert
R_{k,t}-R_{k,s}\right\Vert _{\mathrm{op}}\right\} &\leq &\left\Vert
h_{k,t}-h_{k,s}\right\Vert _{\mathrm{op}}+\frac{1}{2}\left\Vert \mu
_{k,t}-\mu _{k,s}\right\Vert _{\mathrm{op}}  \notag \\
&\leq &4\int_{t\wedge s}^{t\vee s}\left\Vert X_{k,\tau }\right\Vert _{%
\mathrm{op}}^{2}\mathrm{d}\tau ,  \label{Equation sup1}
\end{eqnarray}%
using the triangle inequality and Theorem \ref{globalexistence} (cf. (\ref%
{goodflow})). By Corollary \ref{pseudoHS}, it follows that, for any $k\in 
\mathbb{N}$ and $s,t\in \mathbb{R}_{0}^{+}$, 
\begin{equation*}
\max \left\{ \left\Vert L_{k,t}-L_{k,s}\right\Vert _{\mathrm{op}},\left\Vert
R_{k,t}-R_{k,s}\right\Vert _{\mathrm{op}}\right\} \leq 4\alpha
^{-1}\left\Vert X_{k,0}\right\Vert _{\mathrm{op}}^{2}\left( \mathrm{e}%
^{-\alpha \left( t\wedge s\right) }-\mathrm{e}^{-\alpha \left( t\vee
s\right) }\right) .
\end{equation*}%
As a consequence, $L_{k}$ and $R_{k}$ are Lipschitz continuous with $4\Vert
X_{k,0}\Vert _{\mathrm{op}}^{2}$ as Lipschitz constant, for any $k\in 
\mathbb{N}$. Therefore, we infer from Lemma \ref{lemma-Dyson-Series2} that,
for any $k\in \mathbb{N}$ and $s,t\in \mathbb{R}_{0}^{+}$ with $t\neq s$, 
\begin{eqnarray}
&&\max \left\{ \left\Vert L_{k,t}Z_{t,s}^{g}\left( L_{k}\right) \right\Vert
_{\mathrm{op}},\left\Vert Z_{t,s}^{g}\left( L_{k}\right) L_{k,s}\right\Vert
_{\mathrm{op}},\left\Vert Z_{t,s}^{d}\left( R_{k}\right) R_{k,t}\right\Vert
_{\mathrm{op}},\left\Vert R_{k,s}Z_{t,s}^{d}\left( R_{k}\right) \right\Vert
_{\mathrm{op}}\right\}  \notag \\
&\leq &\mathrm{C}\left( s,t,4\left\Vert X_{k,0}\right\Vert _{\mathrm{op}%
}^{2},4\alpha ^{-1}\left\Vert X_{k,0}\right\Vert _{\mathrm{op}}^{2}\right) +%
\frac{1}{\mathrm{e}\left\vert t-s\right\vert }.  \label{inequality penible}
\end{eqnarray}%
Note that Remark \ref{remark utiles} also applies here, by virtue of
Proposition \ref{controleX}. By (\ref{goodflow}) and (\ref{equality trivial2}%
), it follows that, for any $k\in \mathbb{N}$, $\epsilon ,T\in \mathbb{R}%
^{+} $ and $s,t\in \lbrack \epsilon ,T]$, 
\begin{equation}
\sup_{\tau \in \left[ \epsilon ,\infty \right) }\left\{ \left\Vert L_{k,\tau
}X_{k,\tau }\right\Vert _{\mathrm{op}}+\left\Vert X_{k,\tau }R_{k,\tau
}\right\Vert _{\mathrm{op}}\right\} \leq \left\Vert X_{k,0}\right\Vert _{%
\mathrm{op}}\left( \mathrm{D}\left( T,x_{\max },\alpha \right) +2\left( 
\mathrm{e}\epsilon \right) ^{-1}\right)  \label{dfg0}
\end{equation}%
and%
\begin{eqnarray}
\left\Vert X_{k,t}-X_{k,s}\right\Vert _{\mathrm{op}} &\leq &\left\vert
t-s\right\vert \sup_{\tau \in \left[ \epsilon ,\infty \right) }\left\{
\left\Vert L_{k,\tau }X_{k,\tau }\right\Vert _{\mathrm{op}}+\left\Vert
X_{k,\tau }R_{k,\tau }\right\Vert _{\mathrm{op}}\right\}  \notag \\
&\leq &\left\vert t-s\right\vert \left\Vert X_{k,0}\right\Vert _{\mathrm{op}%
}\left( \mathrm{D}\left( T,x_{\max },\alpha \right) +2\left( \mathrm{e}%
\epsilon \right) ^{-1}\right) ,  \label{dfg}
\end{eqnarray}%
where 
\begin{equation}
x_{\max }=\sup_{k\in \mathbb{N}}\left\Vert X_{k,0}\right\Vert _{\mathrm{op}%
}^{2}\leq \sum_{k\in \mathbb{N}}\left\Vert X_{k,t}\right\Vert _{\mathrm{op}%
}^{2}=\Vert X_{0}\Vert _{\mathrm{op},2}^{2}<\infty ,  \label{xmax}
\end{equation}%
while the function $\mathrm{D}$ is defined by 
\begin{equation}
\mathrm{D}\left( y_{1},y_{2},y_{3}\right) =8\mathrm{e}^{-1}y_{2}y_{3}^{-1}%
\left( \left( 3\mathrm{e}+4\sqrt{2}\right) +\left( 4y_{3}^{-1}\left( \sqrt{2}%
\pi +1\right) y_{2}+y_{3}\right) \left\vert y_{1}\right\vert \right)
\label{D}
\end{equation}%
for any $y_{1},y_{2}\in \mathbb{R}$ and $y_{3}\in \mathbb{R}\backslash \{0\}$%
. Inequality (\ref{dfg}) clearly yields the locally Lipschitz continuity of
the mapping $(X_{t})_{t\in \mathbb{R}^{+}}$, in the $\ell _{2}$--operator
norm $\Vert \cdot \Vert _{\mathrm{op},2}$. Now, similar to (\ref{sufficient2}%
), using the triangle inequality, Corollary \ref{pseudoHS} and Theorem \ref%
{globalexistence} and, in particular, the fact that $[\mu _{k,0},X_{k,t}]=0$
for each $k\in \mathbb{N}$ and all $t\in \mathbb{R}_{0}^{+}$, note that, for 
$k\in \mathbb{N}$ and $s,t\in \mathbb{R}_{0}^{+}$,%
\begin{eqnarray*}
\left\Vert \left[ \mu _{k,t},X_{k,t}\right] -\left[ \mu _{k,s},X_{k,s}\right]
\right\Vert _{\mathrm{op}} &\leq &8\left\Vert X_{k,0}\right\Vert _{\mathrm{op%
}}\int_{t\wedge s}^{t\vee s}\left\Vert X_{\tau }\right\Vert _{\mathrm{op}%
,2}^{2}\mathrm{d}\tau \\
&&+8\left\Vert X_{k,t}-X_{k,s}\right\Vert _{\mathrm{op}}\int_{0}^{s}\left%
\Vert X_{\tau }\right\Vert _{\mathrm{op},2}^{2}\mathrm{d}\tau ,
\end{eqnarray*}%
which, combined again with Corollary \ref{pseudoHS}, yields 
\begin{eqnarray*}
\left\Vert \left[ \mu _{k,t},X_{k,t}\right] -\left[ \mu _{k,s},X_{k,s}\right]
\right\Vert _{\mathrm{op}} &\leq &8\alpha ^{-1}\left\Vert X_{k,0}\right\Vert
_{\mathrm{op}}^{2}\left( \mathrm{e}^{-\alpha \left( t\wedge s\right) }-%
\mathrm{e}^{-\alpha \left( t\vee s\right) }\right) \\
&&+8\alpha ^{-1}\left\Vert X_{k,t}-X_{k,s}\right\Vert _{\mathrm{op}%
}\left\Vert X_{k,0}\right\Vert _{\mathrm{op}}^{2}\left( 1-\mathrm{e}%
^{-\alpha s}\right) .
\end{eqnarray*}%
Keeping in mind (\ref{dfg})--(\ref{xmax}), it is then easy to deduce from
this last inequality that the mapping $t\mapsto \left[ X_{t},\mu _{t}\right] 
$ is continuous on $\mathbb{R}^{+}$ with respect to the norm $\Vert \cdot
\Vert _{\mathrm{op},2}$.
\end{proof}

We conclude this section with a pivotal proposition for our subsequent
analysis. Recall that $\ell ^{2}(\mathcal{B}(\mathfrak{h}))$ denotes the
Banach space (\ref{L2}) of sequences $X_{k}\in \mathcal{B}(\mathfrak{h}_{k})$%
, $k\in \mathbb{N}$, equipped with the $\ell ^{2}$--operator norm $\Vert
\cdot \Vert _{\mathrm{op},2}$. Furthermore, for an interval $I\subseteq 
\mathbb{R}_{0}^{+}$, $C(I;\ell ^{2}(\mathcal{B}(\mathfrak{h})))$ is the
Banach space of continuous mappings from $I$ to $\ell ^{2}(\mathcal{B}(%
\mathfrak{h}))$.

\begin{proposition}[Solution to the differential flow in $C(\mathbb{R}%
^{+};\ell ^{2}(\mathcal{B}(\mathfrak{h})))$]
\label{prop useful}\mbox{}\newline
Let $(h_{k},\mu _{k},X_{k})_{k\in \mathbb{N}}$ be a family satisfying
Hypotheses (i) and (iii) of Condition \ref{AbstractAssumptions}. These
families define three mappings $h$, $\mu $ and $X$ in $C(\mathbb{R}%
_{0}^{+};\ell ^{2}(\mathcal{B}(\mathfrak{h})))$. With respect to the norm $%
\Vert \cdot \Vert _{\mathrm{op},2}$, the following differential equations
also hold for any strictly positive time $t\in \mathbb{R}^{+}$:%
\begin{equation*}
\left\{ 
\begin{array}{ll}
\partial _{t}h_{t} & =-2X_{t}X_{t}^{\ast } \\ 
\partial _{t}\mu _{t} & =2X_{t}^{\ast }X_{t}-2X_{t}X_{t}^{\ast } \\ 
\partial _{t}X_{t} & =\left( h_{t}-\mu _{t}/2\right) X_{t}+X_{t}\left(
-h_{t}-\mu _{t}/2\right)%
\end{array}%
\right.
\end{equation*}
\end{proposition}

\begin{proof}
Clearly, for any $T\in \mathbb{R}^{+}$ and $s,t\in \lbrack 0,T]$, there
exists a constant $D_{T}\in \mathbb{R}^{+}$ such that%
\begin{equation}
\left\Vert X_{k,t}-X_{k,s}\right\Vert _{\mathrm{op}}\leq 2D_{T}\left\Vert
X_{k,0}\right\Vert _{\mathrm{op}}^{2},\qquad k\in \mathbb{N},
\label{stupide ine}
\end{equation}%
thanks to the triangle inequality and Condition \ref{AbstractAssumptions}
(i). By Hypotheses (i) and (iii) of\ Condition \ref{AbstractAssumptions}
together with (\ref{stupide ine}) we can invoke Lebesgue's dominated
convergence theorem to deduce that, for any $s\in \mathbb{R}_{0}^{+}$, 
\begin{equation}
\lim_{t\rightarrow s}\left\Vert X_{t}-X_{s}\right\Vert _{\mathrm{op}%
,2}=\lim_{t\rightarrow s}\sum_{k=0}^{\infty }\left\Vert
X_{k,t}-X_{k,s}\right\Vert _{\mathrm{op}}=0.  \label{ert1}
\end{equation}%
In particular, the family $X$ is continuous with respect to $\Vert \cdot
\Vert _{\mathrm{op},2}$ and since any continuous real function is of course
locally bounded, for all $s\in \mathbb{R}_{0}^{+}$, 
\begin{equation}
\lim_{t\rightarrow s}\int_{s}^{t}\left\Vert X_{\tau }\right\Vert _{\mathrm{op%
},2}^{2}\mathrm{d}\tau =0.  \label{ert2}
\end{equation}%
Combined with (\ref{goodflow}), it shows that $h,\mu ,X\in C(\mathbb{R}%
_{0}^{+};\ell ^{2}(\mathcal{B}(\mathfrak{h})))$. In fact, via Lebesgue's
dominated convergence theorem and Hypotheses (i) and (iii) of\ Condition \ref%
{AbstractAssumptions}, one checks that, for all $t\in \mathbb{R}_{0}^{+}$
(including $t=0$), 
\begin{equation}
\partial _{t}h_{t}=-2X_{t}X_{t}^{\ast }\qquad \text{and}\qquad \partial
_{t}\mu _{t}=2X_{t}^{\ast }X_{t}-2X_{t}X_{t}^{\ast }  \label{flow bis1}
\end{equation}%
with respect to the norm $\Vert \cdot \Vert _{\mathrm{op},1}$ (which is even
stronger than $\Vert \cdot \Vert _{\mathrm{op},2}$.) Additionally, by
Condition \ref{AbstractAssumptions} (i) (cf. (\ref{goodflow})), for any $%
k\in \mathbb{N}$, $t\in \mathbb{R}^{+}$ and $\varepsilon \in \mathbb{R}$,%
\begin{equation}
\lim_{\varepsilon \rightarrow 0}\left\Vert \varepsilon ^{-1}\left(
X_{k,(t+\varepsilon )\vee 0}-X_{k,t}\right)
-L_{k,t}X_{k,t}-X_{k,t}R_{k,t}\right\Vert _{\mathrm{op}}=0,
\label{continuity add4}
\end{equation}%
where the operators $L_{k,t}$ and $R_{k,t}$ are defined by (\ref{relation
bis02}) for any $k\in \mathbb{N}$\ and $t\in \mathbb{R}_{0}^{+}$.
Furthermore, by (\ref{goodflow}), (\ref{equality trivial2}) and the triangle
inequality, for any $k\in \mathbb{N}$, $\epsilon ,T\in \mathbb{R}^{+}$, $%
\epsilon <T$ and $t\in \lbrack \epsilon ,T]$, $\varepsilon \in \mathbb{R}$
such that $t+\varepsilon \geq \epsilon $,%
\begin{eqnarray}
&&\left\Vert \varepsilon ^{-1}\left( X_{k,(t+\varepsilon )\vee
0}-X_{k,t}\right) -L_{k,t}X_{k,t}-X_{k,t}R_{k,t}\right\Vert _{\mathrm{op}}
\label{continuity add5} \\
&\leq &2\left\Vert X_{k,0}\right\Vert _{\mathrm{op}}\sup_{\tau \in \left[
((t+\varepsilon )\wedge t)\vee 0,(t+\varepsilon )\vee t\right] }\left\{
\left\Vert L_{k,\tau }Z_{k,\tau ,0}^{g}\right\Vert _{\mathrm{op}}\left\Vert
Z_{k,\tau ,0}^{d}\right\Vert _{\mathrm{op}}+\left\Vert Z_{k,\tau
,0}^{g}\right\Vert _{\mathrm{op}}\left\Vert Z_{k,\tau ,0}^{d}R_{k,\tau
}\right\Vert _{\mathrm{op}}\right\} ,  \notag
\end{eqnarray}%
where $Z_{k,t,s}^{g}\doteq Z_{t,s}^{g}(L_{k})$ and $Z_{k,t,s}^{d}\doteq
Z_{t,s}^{d}(R_{k})$ for any $s,t\in \mathbb{R}_{0}^{+}$ (cf. (\ref%
{eq:dysonzg})--(\ref{dysonzd})). Combining \ref{lemma-Dyson-Series} (ii) and %
\ref{lemma-Dyson-Series2} with Remark \ref{remark utiles}, Equations (\ref%
{relation bis02}), (\ref{flow bis1}) and Condition \ref{AbstractAssumptions}
(i), we find that there exist constants $D_{1},D_{2}\in \mathbb{R}^{+}$ such
that, for all $k\in \mathbb{N}$ with $X_{k,0}\neq 0$, and any $\epsilon
,T,\tau \in \mathbb{R}^{+}$ satisfying $\epsilon <T$ and $\epsilon \leq \tau
\leq T$, 
\begin{equation*}
\left\Vert Z_{k,\tau ,0}^{d}\right\Vert _{\mathrm{op}}\leq \mathrm{e}%
^{-(\alpha _{k,+}/2)\tau }D_{1}
\end{equation*}%
and 
\begin{equation*}
\left\Vert L_{k,\tau }Z_{k,\tau ,0}^{g}\right\Vert _{\mathrm{op}}\leq
\left\Vert \left( L_{k,\tau }+\frac{\alpha _{k,-}}{2}\mathbf{1}\right)
Z_{k,\tau ,0}^{g}\right\Vert _{\mathrm{op}}+\frac{1}{2}\left\vert \alpha
_{k,-}\right\vert \left\Vert Z_{k,\tau ,0}^{g}\right\Vert _{\mathrm{op}}\leq
D_{2}\mathrm{e}^{-\alpha _{k,-}\tau /2}\left( 1+\epsilon ^{-1}\right) .
\end{equation*}%
Indeed, the inequality $\alpha _{k,+}+\alpha _{k,-}\geq 0$ combined with the
lower bound 
\begin{equation*}
\gamma _{\pm }\doteq \inf \{\alpha _{\pm ,k}:k\in \mathbb{N},\ X_{k,0}\neq
0\}>-\infty
\end{equation*}%
(cf. Condition \ref{AbstractAssumptions} (i)) implies that, without loss of
generality, we can assume 
\begin{equation*}
\inf \{\left\vert \alpha _{\pm ,k}\right\vert :k\in \mathbb{N},\ X_{k,0}\neq
0\}<\infty .
\end{equation*}%
Therefore, since $\alpha _{k,+}+\alpha _{k,-}\geq 0$ for all $k\in \mathbb{N}
$, for any $\epsilon ,T\in \mathbb{R}^{+}$ with $\epsilon <T$, there exists
a constant $D_{3}\in \mathbb{R}^{+}$ such that 
\begin{equation*}
\sup_{\tau \in \left[ ((t+\varepsilon )\wedge t)\vee \epsilon
,(t+\varepsilon )\vee t\right] }\left\{ \left\Vert L_{k,\tau }Z_{k,\tau
,0}^{g}\right\Vert _{\mathrm{op}}\left\Vert Z_{k,\tau ,0}^{d}\right\Vert _{%
\mathrm{op}}\right\} \leq D_{3}\left( 1+\epsilon ^{-1}\right)
\end{equation*}%
for each $k\in \mathbb{N}$ satisfying $X_{k,0}\neq 0$, and for any $t\in
\lbrack \epsilon ,T]$ and $\varepsilon \in \mathbb{R}$ such that $%
t+\varepsilon \geq \epsilon $. Mutatis mutandis for 
\begin{equation*}
\sup_{\tau \in \left[ ((t+\varepsilon )\wedge t)\vee \epsilon
,(t+\varepsilon )\vee t\right] }\left\{ \left\Vert Z_{k,\tau
,0}^{g}\right\Vert _{\mathrm{op}}\left\Vert Z_{k,\tau ,0}^{d}R_{k,\tau
}\right\Vert _{\mathrm{op}}\right\} .
\end{equation*}%
In particular, for any $\epsilon ,T\in \mathbb{R}^{+}$ with $\epsilon <T$,
there is a constant $D_{4}\in \mathbb{R}^{+}$ such that 
\begin{equation*}
\sup_{k\in \mathbb{N}}\sup_{\tau \in \left[ ((t+\varepsilon )\wedge t)\vee
\epsilon ,(t+\varepsilon )\vee t\right] }\left\{ \left\Vert L_{k,\tau
}Z_{k,\tau ,0}^{g}\right\Vert _{\mathrm{op}}\left\Vert Z_{k,\tau
,0}^{d}\right\Vert _{\mathrm{op}}+\left\Vert Z_{k,\tau ,0}^{g}\right\Vert _{%
\mathrm{op}}\left\Vert Z_{k,\tau ,0}^{d}R_{k,\tau }\right\Vert _{\mathrm{op}%
}\right\} \leq D_{4}\left( 1+\epsilon ^{-1}\right)
\end{equation*}%
for each $k\in \mathbb{N}$ satisfying $X_{k,0}\neq 0$, and for any $t\in
\lbrack \epsilon ,T]$ and $\varepsilon \in \mathbb{R}$ such that $%
t+\varepsilon \geq \epsilon $. Therefore, we can invoke Lebesgue's dominated
convergence theorem to obtain from (\ref{continuity add4})--(\ref{continuity
add5}) and $\Vert X_{0}\Vert _{\mathrm{op},2}<\infty $ (cf. Condition \ref%
{AbstractAssumptions} (iii)) that, for any strictly positive time $t\in 
\mathbb{R}^{+}$, 
\begin{eqnarray*}
&&\lim_{\varepsilon \rightarrow 0}\left\Vert \varepsilon ^{-1}\left(
X_{k,(t+\varepsilon )\vee 0}-X_{k,t}\right)
-L_{k,t}X_{k,t}-X_{k,t}R_{k,t}\right\Vert _{\mathrm{op},2}^{2} \\
&=&\lim_{\varepsilon \rightarrow 0}\sum_{k\in \mathbb{N}}\left\Vert
\varepsilon ^{-1}\left( X_{k,(t+\varepsilon )\vee 0}-X_{k,t}\right)
-L_{k,t}X_{k,t}-X_{k,t}R_{k,t}\right\Vert _{\mathrm{op}}^{2}=0.
\end{eqnarray*}%
Together with (\ref{flow bis1}), this implies the asserted system of
differential equations.
\end{proof}

\begin{remark}[Finite case]
\label{remark utiles copy(1)}\mbox{}\newline
If the set $\{k\in \mathbb{N},\;X_{k,0}\neq 0\}$ has finite cardinality,
then the differential equations (\ref{flow bis1}) remain valid at $t=0$. In
this case, Lemma \ref{lemma-Dyson-Series2} and Remark \ref{remark utiles},
which are employed in the proof of Proposition \ref{prop useful}, are no
longer necessary.
\end{remark}

\section{Foundations of Spin-Boson Models\label{Foundations}}

Recall that 
\begin{equation*}
\mathfrak{H}\equiv \bigotimes_{k\in \mathbb{N}}\mathfrak{h}_{k}
\end{equation*}%
is the (generally non-separable) Hilbert space defined by (\ref%
{fghfghfghfgh0}), which refers to the generalized quantum spin system $%
\mathcal{S}$. See Section \ref{not separable Hilbert space S} for more
details. The second Hilbert space $\mathcal{F}_{+}$ is the bosonic Fock
space (\ref{Fock}), i.e., 
\begin{equation*}
\mathcal{F}_{+}\equiv \mathcal{F}_{+}\left( \mathcal{H}\right) \doteq
\bigoplus_{n\in \mathbb{N}_{0}}\vee ^{n}\mathcal{H},
\end{equation*}%
which is constructed over a fixed, infinite-dimensional, separable complex
Hilbert space $\mathcal{H}$. See Section \ref{Bosonic fields} for more
details. Then, as explained in Section \ref{Coupled Systems}, all the
many-body Hamiltonians involved in this analysis act on the tensor product $%
\mathfrak{F}\doteq \mathfrak{H}\otimes \mathcal{F}_{+}$. These operators are
generally unbounded. In this section, we provide their rigorous construction
along with the essential (relative) bounds required to rigorously establish
a continuous flow of unbounded operators, as explained in Section \ref%
{Brockett-Wegner Flow}. Furthermore, to simplify the notation in what
follows, we write $\mathbf{1}\equiv \mathbf{1}_{\mathfrak{F}}$, unless this
leads to ambiguity.

\subsection{Extended Second-Quantization\label{extendedquantization}}

The main argument to obtain a proper definition of generalized spin-boson
models on the Hilbert space $\mathfrak{F}\doteq \mathfrak{H}\otimes \mathcal{%
F}_{+}$, like $\mathrm{H}_{0}$ (\ref{def_H0}) (see also (\ref{defHt})), is
the Kato-Rellich theorem \cite[Theorem 8.5]{Konrad}, which is used below
after showing that all individual terms in these models are relatively
bounded with respect to term of the form (\ref{Bosonic Hamilto}), i.e.,%
\begin{equation}
\mathrm{d}\Gamma ^{\otimes }\left( \mu \right) \doteq \sum_{k\in \mathbb{N}%
}\mu _{k}\otimes a_{k}^{\ast }a_{k}  \label{Bosonic Hamiltobis}
\end{equation}%
for some sequence $\mu \equiv \left( \mu _{k}\right) _{k\in \mathbb{N}}$ of
self-adjoint operators. Compare this definition with Equations (\ref{def_H0}%
) and (\ref{defHt}). As explained in Sections \ref{Bosonic fields}--\ref%
{Coupled Systems}, this expression is very similar to the well-known second
quantization (\ref{Model second quantized}) of some self-adjoint operator.
We thus extend this formalism to coefficients $\mu _{k}$ that are not
necessarily real numbers, that is $\mu _{k}\equiv \mu _{k}\mathbf{1}_{%
\mathfrak{H}}$ in (\ref{Bosonic Hamiltobis}), but general self-adjoint
operators 
\begin{equation*}
\mu _{k}\in \mathcal{B}\left( \mathfrak{H}\right) \supseteq \mathcal{B}%
\left( \mathfrak{h}_{k}\right) ^{\otimes }\equiv \mathcal{B}\left( \mathfrak{%
h}_{k}\right) .
\end{equation*}%
We explain in particular that this extended version of the second
quantization also leads to self-adjoint operators, acting now on the tensor
product $\mathfrak{F}\doteq \mathfrak{H}\otimes \mathcal{F}_{+}$.

For $n\in \mathbb{N}$ spaces $\mathcal{X}_{1},\ldots ,\mathcal{X}_{n}$, we
define $\mathcal{X}_{1}\vee \cdots \vee \mathcal{X}_{n}$ to be the subspace
of symmetric elements of the tensor product $\mathcal{X}_{1}\otimes \cdots
\otimes \mathcal{X}_{2}$. For $n\in \mathbb{N}$, the operator $A_{1}\vee
\cdots \vee A_{n}$ acting on $\mathcal{X}_{1}\vee \cdots \vee \mathcal{X}%
_{n} $ is then equal to the operator $A_{1}\otimes \cdots \otimes A_{n}$
restricted to $\mathcal{X}_{1}\vee \cdots \vee \mathcal{X}_{n}$ and with
range projected onto symmetric elements of $\mathcal{X}_{1}\otimes \cdots
\otimes \mathcal{X}_{2}$.

Now, let $\mathrm{S}\subseteq \mathbb{N}$ be a countable set and $(\mathcal{P%
}_{k})_{k\in \mathrm{S}}$ be a collection of mutually orthogonal subspaces
of the separable Hilbert space $\mathcal{H}$ such that 
\begin{equation}
\overline{\bigcup_{k\in \mathrm{S}}\mathcal{P}_{k}}=\mathcal{H}.
\label{coveringcondition}
\end{equation}%
For each natural number $k\in \mathrm{S}$, we denote the orthogonal
projection onto $\mathcal{P}_{k}$ by $P_{k}:\mathcal{H}\rightarrow \mathcal{P%
}_{k}$. For any family $\mu \equiv (\mu _{k})_{k\in \mathrm{S}}$ of
self-adjoint operators in $\mathcal{B}(\mathfrak{H})$, we use the formal
notation 
\begin{equation}
A_{\mu }\doteq \sum_{k\in \mathrm{S}}\mu _{k}\otimes P_{k}.  \label{Amu}
\end{equation}%
Then, for any $n\in \mathbb{N}$ we define the operator\footnote{%
Here we use the notation $\vee _{n}^{m}(\cdot )=\varnothing $ for natural
numbers $m>n$, meaning that this symbol is simply ignored. For instance, $%
\vee _{1}^{0}(\mathbf{1})\vee P_{k}\doteq P_{k}$ and $P_{k}\vee \vee
_{n+1}^{n}(\mathbf{1})\doteq P_{k}$.} 
\begin{equation*}
\Gamma _{n}^{\otimes }\left( A_{\mu }\right) \doteq \sum_{k\in \mathrm{S}%
}\mu _{k}\otimes \left( \sum_{m=1}^{n}\left( \bigvee_{j=1}^{m-1}\mathbf{1}_{%
\mathcal{H}}\right) \vee P_{k}\vee \left( \bigvee_{j=m+1}^{n}\mathbf{1}_{%
\mathcal{H}}\right) \right)
\end{equation*}%
acting on the domain 
\begin{equation}
\mathcal{D}_{n}\doteq \mathfrak{H}_{\mathrm{pre}}\otimes \bigvee^{n}\left(
\bigcup_{N\in \mathbb{N}}\left( \bigoplus_{k\in \mathrm{S}\cap \lbrack 0,N]}%
\mathcal{P}_{k}\right) \right) =\mathfrak{H}_{\mathrm{pre}}\otimes
\bigvee^{n}\left( \mathrm{span}\left( \bigcup_{k\in \mathbb{N}}\mathcal{P}%
_{k}\right) \right) \subseteq \mathfrak{H}\otimes \mathcal{F}_{+}\doteq 
\mathfrak{F},  \label{domain n}
\end{equation}%
where $\mathfrak{H}_{\mathrm{pre}}$ is the dense subset (\ref{pre-hilbert})
of $\mathfrak{H}$. For instance, if $\mathrm{S}$ is a finite set then $%
\mathcal{D}_{n}=\mathfrak{H}_{\mathrm{pre}}\otimes \vee ^{n}\mathcal{H}$.
Finally, using the additional definitions $\Gamma _{0}^{\otimes }\left(
A_{\mu }\right) \doteq 0$ and $\mathcal{D}_{0}\doteq \mathbb{C}$, we set 
\begin{equation}
\mathrm{d}\Gamma ^{\otimes }\left( A_{\mu }\right) _{0}\doteq
\bigoplus_{n\in \mathbb{N}_{0}}\Gamma _{n}^{\otimes }\left( A_{\mu }\right)
\qquad \text{and}\qquad \mathcal{D}_{\infty }\doteq \bigcup_{M\in \mathbb{N}%
}\left( \bigoplus_{n=0}^{M}\mathcal{D}_{n}\right) = \mathrm{span}\left(
\bigcup_{n\in \mathbb{N}}\mathcal{D}_{n}\right)  \label{second quantu}
\end{equation}%
for any family $\mu \equiv (\mu _{k})_{k\in \mathrm{S}}$ of self-adjoint
operators in $\mathcal{B}(\mathfrak{H})$. For any $n\in \mathbb{N}_{0}$, $%
\varphi \in \mathcal{D}_{n}$ is an analytic vector and $\mathrm{d}\Gamma
^{\otimes }(A_{\mu })_{0}$ is a symmetric operator. Consequently, by the
Nelson theorem \cite[Theorem 7.16]{Konrad}, since $\mathrm{d}\Gamma
^{\otimes }(A)_{0}$ possesses a dense set of analytic vectors in its domain,
it is essentially self-adjoint. Its unique self-adjoint extension is what we
call an \emph{extended }second quantization:

\begin{definition}[Extended second-quantization]
\label{extendsecondquanti}\mbox{}\newline
Let $\mathrm{S}\subseteq \mathbb{N}$ be a countable set and $(\mathcal{P}%
_{k})_{k\in \mathrm{S}}$ be a family of disjoint subspaces of the separable
Hilbert space $\mathcal{H}$ satisfying (\ref{coveringcondition}), with
associated orthogonal projector family $(P_{k})_{k\in \mathrm{S}}$. For any
family $\mu \equiv (\mu _{k})_{k\in \mathrm{S}}$ of self-adjoint operators
in $\mathcal{B}(\mathfrak{H})$, the self-adjoint extension of (\ref{second
quantu}) is called the extended second quantization of the operator $A_{\mu
} $ (\ref{Amu}) and is denoted by 
\begin{equation*}
\mathrm{d}\Gamma ^{\otimes }\left( \mu \right) \equiv \mathrm{d}\Gamma
^{\otimes }\left( A_{\mu }\right) \doteq \mathrm{d}\Gamma ^{\otimes }\left(
A_{\mu }\right) _{0}^{\ast \ast }.
\end{equation*}
\end{definition}

If $\mathrm{S}=\{1\}$, $\mu _{1}=\mathbf{1}_{\mathfrak{H}}$ and $\mathcal{P}%
_{1}=\mathcal{H}$, then $P_{1}=\mathbf{1}_{\mathcal{H}}$ and we obtain that 
\begin{equation}
\mathrm{d}\Gamma ^{\otimes }\left( \mu \right) =\mathbf{1}_{\mathfrak{H}%
}\otimes \mathrm{d}\Gamma (\mathbf{1}_{\mathcal{H}})=\mathbf{1}_{\mathfrak{H}%
}\otimes \mathrm{N},  \label{notation abuse0}
\end{equation}%
named here the extended particle number operator. When there is no
ambiguity, by a slight abuse of notation, we use the symbols $\mathbf{1}_{%
\mathfrak{H}}\otimes \mathrm{N}$ and\ $\mathrm{N\ }$(the standard notation
for the particle number operator) interchangeably, i.e., 
\begin{equation}
\mathrm{N}\equiv \mathbf{1}_{\mathfrak{H}}\otimes \mathrm{N}\mathbf{.}
\label{notation abuse}
\end{equation}%
This simplifies the equations that follow.

As is customary with the usual second quantization, the extended second
quantization defined above can be rewritten in terms of creation and
annihilation operators using a given orthonormal basis. Recall that, for any 
$\varphi \in \mathcal{H}$, $a(\varphi )$ ($a(\varphi )^{\ast }$) is the
corresponding annihilation (creation) operator acting on the bosonic Fock
space $\mathcal{F}_{+}$.

\begin{proposition}[Basis-dependent expression of extended second
quantizations]
\label{Proposition self-adjoint}\mbox{}\newline
The extended second quantization of Definition \ref{extendsecondquanti} is
equal on its core $\mathcal{D}_{\infty }$ to%
\begin{equation}
\mathrm{d}\Gamma ^{\otimes }\left( \mu \right) \equiv \mathrm{d}\Gamma
^{\otimes }\left( A_{\mu }\right) =\sum_{k\in \mathrm{S}}\mu _{k}\otimes
\sum_{j=1}^{\mathrm{dim}(\mathcal{P}_{k})}a(\psi _{k,j})^{\ast }a(\psi
_{k,j}),  \label{eqaution operators}
\end{equation}%
where $\{\psi _{k,j}\}_{j=1}^{\mathrm{dim}(\mathcal{P}_{k})}$ is an
orthonormal basis of the subspace $\mathcal{P}_{k}$ for each $k\in \mathrm{S}
$. In particular, the right-hand side of (\ref{eqaution operators}) does not
depend on the orthonormal basis used.
\end{proposition}

\begin{proof}
The proof is a standard computation using the fact that, for any $k\in 
\mathrm{S}$, the orthogonal projection $P_{k}$ on the subspace $\mathcal{P}%
_{k}$ is equal to%
\begin{equation*}
P_{k}=\sum_{j=1}^{\mathrm{dim}(\mathcal{P}_{k})}P_{k,j}
\end{equation*}%
for any orthonormal basis $\{\psi _{k,j}\}_{j=1}^{\mathrm{dim}(\mathcal{P}%
_{k})}$ of the subspace $\mathcal{P}_{k}$, where $P_{k,j}$ is the orthogonal
projection on the one-dimensional subspace spanned by each basis element $%
\psi _{k,j}$, with $k\in \mathrm{S}$ and $j\in \{1,\ldots ,\mathrm{dim}(%
\mathcal{P}_{k})\}$.
\end{proof}

\noindent Note that both sides of (\ref{eqaution operators}) are well
defined on the domain $\mathcal{D}_{\infty }$ as essentially self-adjoint
operators, by the Nelson theorem \cite[Theorem 7.16]{Konrad}. By uniqueness
of their closure, they define the same operator since they coincide on their
common core $\mathcal{D}_{\infty }$, by the above proposition. In
particular, we can use Equality (\ref{eqaution operators}) to write any
extended second quantization of some operator in terms of (bosonic) creation
and annihilation operators, as it is done in the usual second quantization.

\subsection{Spin-Boson Interactions of Order 1}

In this paper, we assume (without loss of generality) that the separable
Hilbert space $\mathcal{H}$ is infinite-dimensional; the finite-dimensional
case is obtained by even simpler arguments. From this point forward, we
focus exclusively on the case where $\mathcal{P}_{k}=\mathrm{span}\left(
\psi _{k}\right) $ for $k\in \mathbb{N}$, $\{\psi _{k}\}_{k\in \mathbb{N}}$
being some orthonormal basis of $\mathcal{H}$. In this situation, recall
that we write $a_{k}\doteq a(\psi _{k})$ and $a_{k}^{\ast }\doteq a(\psi
_{k})^{\ast }$ for $k\in \mathbb{N}$ and the set $\mathcal{D}_{n}$
introduced in the general setting through (\ref{domain n}) is simply equal to%
\begin{equation}
\mathcal{D}_{n}\doteq \mathfrak{H}_{\mathrm{pre}}\otimes \bigvee^{n}\left(
\bigcup_{N\in \mathbb{N}}\left( \bigoplus_{k\in \lbrack 0,N]}\mathrm{span}%
\left( \psi _{k}\right) \right) \right) \subseteq \mathfrak{H}\otimes 
\mathcal{F}_{+}\doteq \mathfrak{F}.  \label{domain nbis}
\end{equation}

Note that this expression reduces to 
\begin{equation*}
\mathcal{D}_{n}=\mathfrak{H}_{\mathrm{pre}}\otimes \mathrm{span}\left(
\bigvee^{n}\left( \mathrm{span}\{\psi _{k}\text{ }:\text{ }k\in \mathbb{N}%
\}\right) \right) ,
\end{equation*}%
which is consistent with (\ref{domain D}) and (\ref{second quantu}) for the
definition of the \emph{dense} domain $\mathcal{D}_{\infty }\subseteq 
\mathfrak{F}$.

Generalized spin-boson models (see, e.g., (\ref{def_H0})) include a sum of
interactions of the form $(\mathbb{X}+\mathbb{X}^{\dag })$, where 
\begin{equation}
\mathbb{X}\doteq \sum_{k\in \mathbb{N}}X_{k}\otimes a_{k}\qquad \text{and}%
\qquad \mathbb{X}^{\dag }\doteq \sum_{k\in \mathbb{N}}X_{k}^{\ast }\otimes
a_{k}^{\ast }  \label{interaction elementaire}
\end{equation}%
for any sequence $X\equiv (X_{k})_{k\in \mathbb{N}}$ of operators $X_{k}$ in 
\begin{equation}
\mathcal{B}\left( \mathfrak{h}_{k}\right) \equiv \mathcal{B}\left( \mathfrak{%
h}_{k}\right) ^{\otimes }\doteq \left\{ \bigotimes_{q\in \mathbb{N}%
}A_{q}:A_{k}\in \mathcal{B}\left( \mathfrak{h}_{k}\right) \text{ and }A_{q}=%
\mathbf{1}_{\mathfrak{h}_{q}}\text{ for all }q\in \mathbb{N}\backslash
\{k\}\right\} \subseteq \mathcal{B}\left( \mathfrak{H}\right)
\label{fgkljdflkgjdfklgjdfkgjbis}
\end{equation}%
(cf. (\ref{fgkljdflkgjdfklgjdfkgj})). Recall that the identification above
relies on \eqref{overloaded notation}. Clearly, $\mathbb{X}$ is well defined
on $\mathcal{D}_{\infty }$ and its adjoint of course should be equal to the
expression $\mathbb{X}^{\dag }$ given above, at least on $\mathcal{D}%
_{\infty }$. We thus begin with an elementary but useful lemma about such
spin-boson interaction terms. Recall from (\ref{lp-operator norm}) that 
\begin{equation*}
\left\Vert X\right\Vert _{\mathrm{op},2}\doteq \left( \sum_{k\in \mathbb{N}%
}\left\Vert X_{k}\right\Vert _{\mathrm{op}}^{2}\right) ^{1/2}
\end{equation*}%
for any sequence $X\equiv (X_{k})_{k\in \mathbb{N}}$ of operators in $%
\mathcal{B}(\mathfrak{h}_{k})\equiv \mathcal{B}(\mathfrak{h}_{k})^{\otimes }$%
.

\begin{lemma}[Relative boundedness of $\mathbb{X+X}^{\dag }$]
\label{Lemma-self-adjoint1}\mbox{}\newline
Let $\mu \equiv (\mu _{k})_{k\in \mathbb{N}}$ and $X\equiv (X_{k})_{k\in 
\mathbb{N}}$ be two sequences of operators in $\mathcal{B}(\mathfrak{h}%
_{k})\equiv \mathcal{B}(\mathfrak{h}_{k})^{\otimes }$. The first one
additionally satisfies $\mu _{k}=\mu _{k}^{\ast }$ and $\mu _{k}\geq c_{k}%
\mathbf{1}_{\mathfrak{h}_{k}}$ for some $c_{k}\in \mathbb{R}_{0}^{+}$ and
each $k\in \mathbb{N}$. Let $A\geq 0$ be any positive operator acting on $%
\mathfrak{F}$, whose domain contains $\mathcal{D}_{\infty }$ (\ref{second
quantu}). Then, for all vectors $\varphi \in \mathcal{D}_{\infty }$, 
\begin{equation*}
\left\Vert \mathbb{X}\varphi \right\Vert _{\mathfrak{F}}+\left\Vert \mathbb{X%
}^{\dag }\varphi \right\Vert _{\mathfrak{F}}\leq 2^{-1}\left\Vert \left( 
\mathrm{d}\Gamma ^{\otimes }\left( \mu \right) +A\right) \varphi \right\Vert
_{\mathfrak{F}}+\left( \left\Vert X\right\Vert _{\mathrm{op},2}+2\sum_{k\in 
\mathbb{N}}c_{k}^{-1}\left\Vert X_{k}\right\Vert _{\mathrm{op}}^{2}\right)
\left\Vert \varphi \right\Vert _{\mathfrak{F}},
\end{equation*}%
using the convention $0^{-1}\times 0\doteq 0$ and $0^{-1}\times x\doteq
\infty $ when $x\neq 0$, while%
\begin{equation*}
\left\Vert \mathbb{X}\varphi \right\Vert _{\mathfrak{F}}+\left\Vert \mathbb{X%
}^{\dag }\varphi \right\Vert _{\mathfrak{F}}\leq 2\left\Vert X\right\Vert _{%
\mathrm{op},2}\left( \left\Vert \mathrm{N}^{1/2}\varphi \right\Vert _{%
\mathfrak{F}}\ +\left\Vert \varphi \right\Vert _{\mathfrak{F}}\right) .
\end{equation*}
\end{lemma}

\begin{proof}
Fix all parameters of the lemma. First, for any $k\in \mathbb{N}$, $\mu
_{k}\geq c_{k}\mathbf{1}_{\mathfrak{h}_{k}}\geq 0$ implies that $\mu
_{k}^{1/2}\geq c_{k}^{1/2}\mathbf{1}_{\mathfrak{h}_{k}}$. By (\ref%
{fgkljdflkgjdfklgjdfkgjbis}) and since $X_{k}\in \mathcal{B}(\mathfrak{h}%
_{k})^{\otimes }$, for all $k,q\in \mathbb{N}$ with $k\neq q$, we have 
\begin{equation}
\left[ X_{k},X_{q}\right] =\left[ X_{k},X_{q}^{\ast }\right] =0\qquad \text{%
and}\qquad \left\Vert X_{q}^{\ast }X_{k}\right\Vert _{\mathrm{op}%
}=\left\Vert X_{k}\right\Vert _{\mathrm{op}}\left\Vert X_{q}\right\Vert _{%
\mathrm{op}},  \label{eq1eq12}
\end{equation}%
bearing in mind that the operator norm is a $C^{\ast }$-norm. Of course, $%
[X_{k},X_{k}]=0$ and $\Vert X_{k}^{\ast }X_{k}\Vert _{\mathrm{op}}=\Vert
X_{k}\Vert _{\mathrm{op}}^{2}$. Using these preliminary observations,
together with the Cauchy-Schwarz inequality, we obtain that, for all vectors 
$\varphi \in \mathcal{D}_{\infty }$,%
\begin{eqnarray}
\left\Vert \mathbb{X}\varphi \right\Vert _{\mathfrak{F}}^{2} &\leq
&\sum_{k\in \mathbb{N}}\Vert \mu _{k}^{1/2}\otimes a_{k}\varphi \Vert _{%
\mathfrak{F}}^{2}\ \ \sum_{k\in \mathbb{N}}c_{k}^{-1}\left\Vert
X_{k}\right\Vert _{\mathrm{op}}^{2}  \notag \\
&=&\left\langle \varphi ,\mathrm{d}\Gamma ^{\otimes }\left( \mu \right)
\varphi \right\rangle _{\mathfrak{F}}\ \sum_{k\in \mathbb{N}%
}c_{k}^{-1}\left\Vert X_{k}\right\Vert _{\mathrm{op}}^{2},
\label{sdsdfsdfsdfsdfsdfsdsdfsdfsdfsdfsdf}
\end{eqnarray}%
where the last identity follows from Proposition \ref{Proposition
self-adjoint}. We use above the convention $0^{-1}\times 0\doteq 0$ and $%
0^{-1}\times x\doteq \infty $ when $x\neq 0$. Note that the positive
operator $\mathrm{d}\Gamma ^{\otimes }(\mu )$ in the above inequality can
trivially be replaced with $\mathrm{d}\Gamma ^{\otimes }(\mu )+A$ for any
positive operator $A\geq 0$ acting on $\mathfrak{F}$, provided its domain
contains $\mathcal{D}_{\infty }$. Again by the Cauchy-Schwarz inequality, it
follows that, for any $\varphi \in \mathcal{D}_{\infty }$, 
\begin{equation}
\left\Vert \mathbb{X}\varphi \right\Vert _{\mathfrak{F}}\leq \left(
\sum_{k\in \mathbb{N}}c_{k}^{-1}\left\Vert X_{k}\right\Vert _{\mathrm{op}%
}^{2}\right) ^{1/2}\left\Vert \varphi \right\Vert _{\mathfrak{F}%
}^{1/2}\left\Vert \left( \mathrm{d}\Gamma ^{\otimes }\left( \mu \right)
+A\right) \varphi \right\Vert _{\mathfrak{F}}^{1/2}.  \label{sdsdsdssd1}
\end{equation}%
Meanwhile, observe from the CCR (\ref{CCR}) and the commutator identity of (%
\ref{eq1eq12}) together with the Cauchy-Schwarz inequality, that, for any $%
\varphi \in \mathcal{D}_{\infty }$, 
\begin{align}
\left\Vert \mathbb{X}^{\dag }\varphi \right\Vert _{\mathfrak{F}}^{2}& \leq
\sum_{k,q\in \mathbb{N}:k\neq q}\left\langle X_{k}\otimes a_{k}\varphi
,X_{q}\otimes a_{q}\varphi \right\rangle _{\mathfrak{F}}+\sum_{k\in \mathbb{N%
}}\left\langle X_{k}^{\ast }\otimes a_{k}\varphi ,X_{k}^{\ast }\otimes
a_{k}\varphi \right\rangle _{\mathfrak{F}}+\sum_{k\in \mathbb{N}}\left\Vert
X_{k}\right\Vert _{\mathrm{op}}^{2}\left\Vert \varphi \right\Vert _{%
\mathfrak{F}}^{2}  \notag \\
& \leq \sum_{k\in \mathbb{N}}\Vert \mu _{k}^{1/2}\otimes a_{k}\varphi \Vert
_{\mathfrak{F}}^{2}\ \ \sum_{k\in \mathbb{N}}c_{k}^{-1}\left\Vert
X_{k}\right\Vert _{\mathrm{op}}^{2}+\sum_{k\in \mathbb{N}}\left\Vert
X_{k}\right\Vert _{\mathrm{op}}^{2}\left\Vert \varphi \right\Vert _{%
\mathfrak{F}}^{2}.  \label{XstarboundllikeX}
\end{align}%
Proceeding as in the derivation of the upper bound (\ref{sdsdsdssd1}), and
additionally using the elementary inequality $\sqrt{x+y}\leq \sqrt{x}+\sqrt{y%
}$ for any $x,y\in \mathbb{R}^{+}$, we obtain that 
\begin{equation}
\left\Vert \mathbb{X}^{\dag }\varphi \right\Vert _{\mathfrak{F}}\leq \left(
\sum_{k\in \mathbb{N}}c_{k}^{-1}\left\Vert X_{k}\right\Vert _{\mathrm{op}%
}^{2}\right) ^{1/2}\left\Vert \varphi \right\Vert _{\mathfrak{F}%
}^{1/2}\left\Vert \left( \mathrm{d}\Gamma ^{\otimes }\left( \mu \right)
+A\right) \varphi \right\Vert _{\mathfrak{F}}^{1/2}+\left( \sum_{k\in 
\mathbb{N}}\left\Vert X_{k}\right\Vert _{\mathrm{op}}^{2}\right)
^{1/2}\left\Vert \varphi \right\Vert _{\mathfrak{F}}.  \label{sdsdsdssd2}
\end{equation}%
Employing again (\ref{sdsdsdssd1}), as well as the fact that $2xy\leq
x^{2}+y^{2}$ for any $x,y\in \mathbb{R}$ , used here with 
\begin{equation*}
x=2^{-1/2}\left\Vert \left( \mathrm{d}\Gamma ^{\otimes }\left( \mu \right)
+A\right) \varphi \right\Vert _{\mathfrak{F}}^{1/2}\qquad \text{and}\qquad
y=\left( 2\sum_{k\in \mathbb{N}}c_{k}^{-1}\left\Vert X_{k}\right\Vert _{%
\mathrm{op}}^{2}\right) ^{1/2}\left\Vert \varphi \right\Vert _{\mathfrak{F}%
}^{1/2},
\end{equation*}%
we arrive at the first inequality%
\begin{equation*}
\left\Vert \mathbb{X}\varphi \right\Vert _{\mathfrak{F}}+\left\Vert \mathbb{X%
}^{\dag }\varphi \right\Vert _{\mathfrak{F}}\leq 2^{-1}\left\Vert \left( 
\mathrm{d}\Gamma ^{\otimes }\left( \mu \right) +A\right) \varphi \right\Vert
_{\mathfrak{F}}+\left( 2\sum_{k\in \mathbb{N}}c_{k}^{-1}\left\Vert
X_{k}\right\Vert _{\mathrm{op}}^{2}+\left( \sum_{k\in \mathbb{N}}\left\Vert
X_{k}\right\Vert _{\mathrm{op}}^{2}\right) ^{1/2}\right) \left\Vert \varphi
\right\Vert _{\mathfrak{F}}.
\end{equation*}%
The second inequality follows directly from (\ref%
{sdsdfsdfsdfsdfsdfsdsdfsdfsdfsdfsdf}) and (\ref{XstarboundllikeX}) upon
choosing $\mu _{k}=\mathbf{1}_{\mathfrak{h}_{k}}$ and $c_{k}=1$ for all $%
k\in \mathbb{N}$.
\end{proof}

Lemma \ref{Lemma-self-adjoint1} is a key ingredient to rigorously define
generalized spin-boson Hamiltonians as well as the spin-boson interaction as
self-adjoint operators, as the next statement shows.

\begin{proposition}[Self-adjointness of $\mathbb{X+X}^{\dag }$]
\label{corollarire-selfadjoint2}\mbox{}\newline
For every sequence $X\equiv (X_{k})_{k\in \mathbb{N}}$ of operators in $%
\mathcal{B}(\mathfrak{h}_{k})\equiv \mathcal{B}(\mathfrak{h}_{k})^{\otimes }$
satisfying $\Vert X\Vert _{\mathrm{op},2}<\infty $, the expression $(\mathbb{%
X+X}^{\dag })$ given by (\ref{interaction elementaire}) defines an
essentially self-adjoint operator on $\mathcal{D}_{\infty }$ (\ref{second
quantu}). Furthermore, the domain of its self-adjoint extension contains the
domain\footnote{%
In particular, since it is essentially self-adjoint on $\mathcal{D}_{\infty
}\subseteq \mathcal{D}(\mathrm{N}^{1/2})$, it is also essentially
self-adjoint on $\mathcal{D}(\mathrm{N}^{1/2})$.} $\mathcal{D}(\mathrm{N}%
^{1/2})$ (cf. (\ref{notation abuse0})--(\ref{notation abuse})), since $(%
\mathbb{X+X}^{\dag })\left( \mathrm{N}+\mathbf{1}\right) ^{-1/2}$ defines a
bounded operator.
\end{proposition}

\begin{proof}
By Lemma \ref{Lemma-self-adjoint1}, the operator $\mathrm{X}\doteq (\mathbb{%
X+X}^{\dag })$ is well defined on the dense subset $\mathcal{D}_{\infty }$
and is symmetric. It is therefore closable, because any densely defined
symmetric operator is closable. Then, the proof that it is essentially
self-adjoint is inspired by the one of \cite[Theorem 5.3]%
{Bruneau-derezinski2007}, itself inspired by \cite[Theorem 6.1]{Berezin}. We
show that $\ker \left( \mathrm{X}^{\ast }-z\mathbf{1}\right) =\left\{
0\right\} $ for all $z\in \mathbb{\mathbb{C}}\backslash \mathbb{R}$ with $%
\mathbf{1}_{\mathfrak{F}}\equiv \mathbf{1}$. Applied to $z\in \{-i,i\}$, it
would yield the desired essential self-adjointness on $\mathcal{D}_{\infty }$%
, see, e.g., \cite[Proposition 3.8]{Konrad}. The orthogonal projection from $%
\mathfrak{F}$ onto $\mathcal{D}_{n}$ (\ref{domain nbis}) is denoted by $%
P_{n} $ for $n\in \mathbb{N}_{0}$. Then, for $n\in \mathbb{N}_{0}$, $z\in 
\mathbb{\mathbb{C}}\backslash \mathbb{R}$ and any vector $\varphi \in \ker
\left( \mathrm{X}^{\ast }-z\mathbf{1}\right) $, 
\begin{equation}
z\left\Vert P_{n}\varphi \right\Vert _{\mathfrak{F}}^{2}=z\langle
P_{n}\varphi ,\varphi \rangle _{\mathfrak{F}}=\langle P_{n}\varphi ,\mathrm{X%
}^{\ast }\varphi \rangle _{\mathfrak{F}}=\langle \mathrm{X}P_{n}\varphi
,\varphi \rangle _{\mathfrak{F}}  \label{eq 1 self}
\end{equation}%
as well as 
\begin{equation}
\overline{z}\left\Vert P_{n}\varphi \right\Vert _{\mathfrak{F}}^{2}=\langle
\varphi ,\mathrm{X}P_{n}\varphi \rangle _{\mathfrak{F}}.  \label{eq 2 self}
\end{equation}%
It follows that 
\begin{equation}
2i\mathrm{Im}\left( z\left\Vert P_{n}\varphi \right\Vert _{\mathfrak{F}%
}^{2}\right) =\langle \mathrm{X}P_{n}\varphi ,\varphi \rangle _{\mathfrak{F}%
}-\langle \varphi ,\mathrm{X}P_{n}\varphi \rangle _{\mathfrak{F}}
\label{eq 3 self}
\end{equation}%
for $n\in \mathbb{N}_{0}$, $z\in \mathbb{\mathbb{C}}\backslash \mathbb{R}$
and any vector $\varphi \in \ker \left( \mathrm{X}^{\ast }-z\mathbf{1}%
\right) $. By convention we set $P_{-1}\doteq 0$. It follows that 
\begin{equation}
2i\mathrm{Im}\left( z\left\Vert P_{n}\varphi \right\Vert _{\mathfrak{F}%
}^{2}\right) =\langle \mathbb{X}P_{n}\varphi ,P_{n-1}\varphi \rangle _{%
\mathfrak{F}}+\langle \mathbb{X}^{\dag }P_{n}\varphi ,P_{n+1}\varphi \rangle
_{\mathfrak{F}}-\langle \mathbb{X}P_{n+1}\varphi ,P_{n}\varphi \rangle _{%
\mathfrak{F}}-\langle \mathbb{X}^{\dag }P_{n-1}\varphi ,P_{n}\varphi \rangle
_{\mathfrak{F}}  \label{wet00}
\end{equation}%
for $n\in \mathbb{N}_{0}$, $z\in \mathbb{\mathbb{C}}\backslash \mathbb{R}$
and any vector $\varphi \in \ker \left( \mathrm{X}^{\ast }-z\mathbf{1}%
\right) $. By summing over $n\in \{0,\ldots ,N\}$ with $N\in \mathbb{N}$, we
obtain that 
\begin{equation}
2i\mathrm{Im}\left( z\right) \sum_{n=0}^{N}\left\Vert P_{n}\varphi
\right\Vert _{\mathfrak{F}}^{2}=\langle \mathbb{X}^{\dag }P_{N}\varphi
,P_{N+1}\varphi \rangle _{\mathfrak{F}}-\langle \mathbb{X}P_{N+1}\varphi
,P_{N}\varphi \rangle _{\mathfrak{F}}.  \label{wet0}
\end{equation}%
Then, using the Cauchy-Schwarz and triangle inequalities as well as Lemma %
\ref{Lemma-self-adjoint1}, we arrive at the inequalities 
\begin{eqnarray}
2\left\vert \mathrm{Im}\left( z\right) \right\vert \sum_{n=0}^{N}\left\Vert
P_{n}\varphi \right\Vert _{\mathfrak{F}}^{2} &\leq &2\left\Vert X\right\Vert
_{\mathrm{op},2}\left( N^{1/2}+\left( N+1\right) ^{1/2}+2\right) \left\Vert
P_{N}\varphi \right\Vert _{\mathfrak{F}}\left\Vert P_{N+1}\varphi
\right\Vert _{\mathfrak{F}}  \notag \\
&\leq &4\left\Vert X\right\Vert _{\mathrm{op},2}\sqrt{N+1}\left( \left\Vert
P_{N}\varphi \right\Vert _{\mathfrak{F}}^{2}+\left\Vert P_{N+1}\varphi
\right\Vert _{\mathfrak{F}}^{2}\right) .  \label{hjbis}
\end{eqnarray}%
(The last upper bound is obtained from the trivial inequality $2xy\leq
x^{2}+y^{2}$ for $x,y\in \mathbb{R}_{0}^{+}$.) By contradiction, suppose
that $\varphi \neq 0$. Necessarily there exist $N_{0}\in \mathbb{N}$ and a
constant $C\in \mathbb{R}^{+}$ such that for all natural numbers $N\geq
N_{0} $, 
\begin{equation}
\sum_{n=0}^{N}\left\Vert P_{n}\varphi \right\Vert _{\mathfrak{F}}^{2}\geq
\sum_{n=0}^{N_{0}}\left\Vert P_{n}\varphi \right\Vert _{\mathfrak{F}%
}^{2}=C>0.  \label{wet000}
\end{equation}%
As a consequence, for all natural numbers $N\geq N_{0}$, we infer from (\ref%
{hjbis}) that%
\begin{equation*}
\frac{2C\left\vert \mathrm{Im}\left( z\right) \right\vert }{\sqrt{N+1}}\leq
4\left\Vert X\right\Vert _{\mathrm{op},2}\left( \left\Vert P_{N}\varphi
\right\Vert _{\mathfrak{F}}^{2}+\left\Vert P_{N+1}\varphi \right\Vert _{%
\mathfrak{F}}^{2}\right) .
\end{equation*}%
If we sum up to $M\in \mathbb{N}$ with $M\geq N_{0}$ we find 
\begin{equation*}
\sum_{N=N_{0}}^{M}\frac{2C\left\vert \mathrm{Im}\left( z\right) \right\vert 
}{\sqrt{N+1}}\leq 4\left\Vert X\right\Vert _{\mathrm{op},2}%
\sum_{N=N_{0}}^{M}\left( \left\Vert P_{N}\varphi \right\Vert _{\mathfrak{F}%
}^{2}+\left\Vert P_{N+1}\varphi \right\Vert _{\mathfrak{F}}^{2}\right) \leq
8\left\Vert X\right\Vert _{\mathrm{op},2}\left\Vert \varphi \right\Vert _{%
\mathfrak{F}}^{2}<\infty .
\end{equation*}%
By letting $M\rightarrow \infty $, the left-hand side is a divergent series
and the above inequality cannot be true. Therefore $\varphi =0$ and, hence, $%
\ker \left( \mathrm{X}^{\ast }-z\mathbf{1}\right) =\left\{ 0\right\} $ for
all $z\in \mathbb{C}\backslash \mathbb{R}$. As a consequence, $\mathrm{X}$
is essentially self-adjoint on $\mathcal{D}_{\infty }$, by \cite[Proposition
3.8]{Konrad}. Finally, note from Lemma \ref{Lemma-self-adjoint1} and the
triangle inequality that the closure of the densely defined symmetric
operator $\mathrm{X}$ always includes the domain $\mathcal{D}(\mathrm{N}%
^{1/2})$ of the operator $\mathrm{N}^{1/2}$.\ 
\end{proof}

To implement the desired continuous flow of generalized spin-boson models,
we invoke the theory of non-autonomous (Kato-hyperbolic) evolution equations
in\ Section \ref{Unitary Flow}. This framework requires specific commutator
estimates involving the spin-boson interactions and the particle number
operator. These essential estimates are established in the two subsequent
lemmata.

\begin{lemma}[Commutator estimates -- I]
\label{Lemma-self-adjoint1 copy(2)}\mbox{}\newline
Let $X\equiv (X_{k})_{k\in \mathbb{N}}$ be any sequence of operators in $%
\mathcal{B}(\mathfrak{h}_{k})\equiv \mathcal{B}(\mathfrak{h}_{k})^{\otimes }$
satisfying $\Vert X\Vert _{\mathrm{op},2}<\infty $. Then, for any $n\in 
\mathbb{N}$, following (\ref{interaction elementaire}), 
\begin{equation*}
\left( \mathrm{N}+\mathbf{1}\right) ^{n}\left( \mathbb{X+X}^{\dag }\right)
\left( \mathrm{N}+\mathbf{1}\right) ^{-n}=\mathbb{X+X}^{\dag }+\mathbb{T}_{n}
\end{equation*}%
where 
\begin{equation*}
\left( \mathrm{N}+\mathbf{1}\right) ^{n}\left( \mathbb{X+X}^{\dag }\right)
\left( \mathrm{N}+\mathbf{1}\right) ^{-\left( n+n/2\right) }\quad \text{and}%
\quad \mathbb{T}_{n}\equiv \left[ (\mathrm{N}+\mathbf{1})^{n},\mathbb{X+X}%
^{\dag }\right] (\mathrm{N}+\mathbf{1})^{-n}
\end{equation*}%
are bounded operators acting on $\mathfrak{F}$. In particular, 
\begin{equation}
\left\Vert \mathbb{T}_{n}\right\Vert _{\mathrm{op}}\leq 4\left(
2^{n}-1\right) \left\Vert X\right\Vert _{\mathrm{op},2}.
\label{constant Cn0}
\end{equation}%
Note that the assertions also hold with $\overline{\mathbb{X+X}^{\dag }}$
replacing $\mathbb{X+X}^{\dag }$.
\end{lemma}

\begin{proof}
We divide the proof into several steps. In all the proof, via (\ref%
{interaction elementaire}), $\mathrm{X}\doteq (\mathbb{X+X}^{\dag })$ for
some fixed sequence $X\equiv (X_{k})_{k\in \mathbb{N}}$ of operators in $%
\mathcal{B}(\mathfrak{h}_{k})\equiv \mathcal{B}(\mathfrak{h}_{k})^{\otimes }$
satisfying $\Vert X\Vert _{\mathrm{op},2}<\infty $. \medskip

\noindent \underline{Step 1:} Let us formally define multi-commutators with
respect to the particle number operator $\mathrm{N}$ by induction as
follows: For any operator $A$ acting on $\mathfrak{F}$, 
\begin{equation*}
\left[ \mathrm{N},A\right] _{1}\doteq \left[ \mathrm{N},A\right] \doteq 
\mathrm{N}A-A\mathrm{N}
\end{equation*}%
while, for any natural number $n\geq 2$,%
\begin{equation*}
\left[ \mathrm{N},A\right] _{n}\doteq \left[ \mathrm{N},\left[ \mathrm{N},A%
\right] _{n-1}\right] .
\end{equation*}%
Since the operators $\mathrm{N}$ and $\mathrm{X}$ conserve the dense domain $%
\mathcal{D}_{\infty }$ (\ref{second quantu}) and 
\begin{equation*}
\left[ a_{q}^{\ast }a_{q},a_{k}\right] =-\delta _{k,q}a_{k}\qquad \text{and}%
\qquad \left[ a_{q}^{\ast }a_{q},a_{k}^{\ast }\right] =\delta
_{k,q}a_{k}^{\ast }
\end{equation*}%
according to the CCR (\ref{CCR}) ($\delta _{k,q}$ is the Kronecker delta),
we can compute the multi-commutator $\left[ \mathrm{N},\mathrm{X}\right]
_{n} $ using its inductive definition to arrive at the equality%
\begin{equation}
\left[ \mathrm{N},\mathrm{X}\right] _{n}=\sum_{k\in \mathbb{N}}\left( \left(
-1\right) ^{n}X_{k}\otimes a_{k}+X_{k}^{\ast }\otimes a_{k}^{\ast }\right) ,
\label{itercommutNG0}
\end{equation}%
which is well-defined at least on the domain $\mathcal{D}_{\infty }$.
\medskip

\noindent \underline{Step 2:} Now, observe that%
\begin{equation}
\left[ \left( \mathrm{N}+\mathbf{1}\right) ^{n},\mathrm{X}\right]
=\sum_{k=1}^{n}\left( 
\begin{array}{c}
n \\ 
k%
\end{array}%
\right) \left[ \mathrm{N},\mathrm{X}\right] _{k}\left( \mathrm{N}+\mathbf{1}%
\right) ^{n-k},  \label{itercommutNG}
\end{equation}%
at least on the domain $\mathcal{D}_{\infty }$. The proof of Equality (\ref%
{itercommutNG}) can be done by induction: The result clearly holds for $n=1$%
. If the equality (\ref{itercommutNG}) is true for some natural number $%
n\geq 1$, then%
\begin{eqnarray*}
\left[ \left( \mathrm{N}+\mathbf{1}\right) ^{n+1},\mathrm{X}\right] &=&\left[
\mathrm{N},\mathrm{X}\right] \left( \mathrm{N}+\mathbf{1}\right) ^{n}+\left( 
\mathrm{N}+\mathbf{1}\right) \left[ \left( \mathrm{N}+\mathbf{1}\right) ^{n},%
\mathrm{X}\right] \\
&=&\left[ \mathrm{N},\mathrm{X}\right] \left( \mathrm{N}+\mathbf{1}\right)
^{n}+\sum_{k=1}^{n}\left( 
\begin{array}{c}
n \\ 
k%
\end{array}%
\right) \left( \mathrm{N}+\mathbf{1}\right) \left[ \mathrm{N},\mathrm{X}%
\right] _{k}\left( \mathrm{N}+\mathbf{1}\right) ^{n-k} \\
&=&\left[ \mathrm{N},\mathrm{X}\right] \left( \mathrm{N}+\mathbf{1}\right)
^{n}+\sum_{k=1}^{n}\left( 
\begin{array}{c}
n \\ 
k%
\end{array}%
\right) \left[ \mathrm{N},\mathrm{X}\right] _{k}\left( \mathrm{N}+\mathbf{1}%
\right) ^{n+1-k} \\
&&\qquad \qquad \qquad \qquad \qquad +\sum_{k=1}^{n}\left( 
\begin{array}{c}
n \\ 
k%
\end{array}%
\right) \left[ \mathrm{N},\mathrm{X}\right] _{k+1}\left( \mathrm{N}+\mathbf{1%
}\right) ^{n-k}.
\end{eqnarray*}%
Re-indexing the last sum with $j=k+1$ and applying Pascal's identity%
\begin{equation*}
\left( 
\begin{array}{c}
n \\ 
k%
\end{array}%
\right) +\left( 
\begin{array}{c}
n \\ 
k-1%
\end{array}%
\right) =\left( 
\begin{array}{c}
n+1 \\ 
k%
\end{array}%
\right) ,
\end{equation*}%
we then obtain that%
\begin{eqnarray*}
&&\left[ \left( \mathrm{N}+\mathbf{1}\right) ^{n+1},\mathrm{X}\right] \\
&=&\left[ \mathrm{N},\mathrm{X}\right] \left( \mathrm{N}+\mathbf{1}\right)
^{n}+\sum_{k=1}^{n}\left( 
\begin{array}{c}
n \\ 
k%
\end{array}%
\right) \left[ \mathrm{N},\mathrm{X}\right] _{k}\left( \mathrm{N}+\mathbf{1}%
\right) ^{n+1-k}+\sum_{j=2}^{n+1}\left( 
\begin{array}{c}
n \\ 
j-1%
\end{array}%
\right) \left[ \mathrm{N},\mathrm{X}\right] _{j}\left( \mathrm{N}+\mathbf{1}%
\right) ^{n+1-j} \\
&=&\left( 
\begin{array}{c}
n+1 \\ 
1%
\end{array}%
\right) \left[ \mathrm{N},\mathrm{X}\right] \left( \mathrm{N}+\mathbf{1}%
\right) ^{n}+\sum_{k=2}^{n}\left( 
\begin{array}{c}
n+1 \\ 
k%
\end{array}%
\right) \left[ \mathrm{N},\mathrm{X}\right] _{k}\left( \mathrm{N}+\mathbf{1}%
\right) ^{n+1-k}+\left( 
\begin{array}{c}
n+1 \\ 
n+1%
\end{array}%
\right) \left[ \mathrm{N},\mathrm{X}\right] _{n+1},
\end{eqnarray*}%
which is nothing else than (\ref{itercommutNG}) for $\left( n+1\right) $.
\medskip

\noindent \underline{Step 3:} For any $n\in \mathbb{N}$, we define the
operator $\mathbb{T}_{n}$ on the dense domain $\mathcal{D}_{\infty }$ to be%
\begin{equation*}
\mathbb{T}_{n}\doteq \left[ (\mathrm{N}+\mathbf{1})^{n},\mathrm{X}\right] (%
\mathrm{N}+\mathbf{1})^{-n}.
\end{equation*}%
Observe in particular that 
\begin{equation}
\left( \mathrm{N}+\mathbf{1}\right) ^{n}\mathrm{X}\left( \mathrm{N}+\mathbf{1%
}\right) ^{-n}=\mathrm{X}+\mathbb{T}_{n}  \label{eqausdskldd}
\end{equation}%
on $\mathcal{D}_{\infty }$. Indeed, by (\ref{itercommutNG0})(\ref%
{itercommutNG}) combined with Lemma \ref{Lemma-self-adjoint1} and the
triangle inequality, for each $n\in \mathbb{N}$ and all $\varphi \in 
\mathcal{D}_{\infty }$, 
\begin{equation*}
\left\Vert \mathbb{T}_{n}\varphi \right\Vert _{\mathfrak{F}}\leq C_{n}\left(
\left\Vert \mathbb{X}\left( \mathrm{N}+\mathbf{1}\right) ^{-1}\varphi
\right\Vert _{\mathfrak{F}}+\left\Vert \mathbb{X}^{\dag }\left( \mathrm{N}+%
\mathbf{1}\right) ^{-1}\varphi \right\Vert _{\mathfrak{F}}\right) \leq
4C_{n}\left\Vert X\right\Vert _{\mathrm{op},2}\left\Vert \varphi \right\Vert
_{\mathfrak{F}},
\end{equation*}%
using the constant 
\begin{equation*}
C_{n}\doteq \sum_{k=1}^{n}\left( 
\begin{array}{c}
n \\ 
k%
\end{array}%
\right) =2^{n}-1.
\end{equation*}%
Therefore, because $\mathcal{D}_{\infty }\subseteq \mathfrak{F}$ is dense, $%
\mathbb{T}_{n}$ admits a unique bounded continuous extension to $\mathfrak{F}
$. By Proposition \ref{corollarire-selfadjoint2}, $\mathrm{X}(\mathrm{N}+%
\mathbf{1})^{-1/2}$ defines a bounded operator. So, according to (\ref%
{eqausdskldd}), 
\begin{equation*}
\left( \mathrm{N}+\mathbf{1}\right) ^{n}\mathrm{X}\left( \mathrm{N}+\mathbf{1%
}\right) ^{-\left( n+1/2\right) }
\end{equation*}%
is also a bounded operator.
\end{proof}

\begin{lemma}[Commutator estimates -- II]
\label{RelativeboundGlemma}\mbox{}\newline
Let $X\equiv (X_{k})_{k\in \mathbb{N}}$ be any sequence of operators in $%
\mathcal{B}(\mathfrak{h}_{k})\equiv \mathcal{B}(\mathfrak{h}_{k})^{\otimes }$
satisfying $\Vert X\Vert _{\mathrm{op},2}<\infty $. Then, for any $n\in 
\mathbb{N}_{0}$, following (\ref{interaction elementaire}), 
\begin{equation*}
\left( \mathrm{N}+\mathbf{1}\right) ^{n+1/2}\left( \mathbb{X+X}^{\dag
}\right) \left( \mathrm{N}+\mathbf{1}\right) ^{-\left( n+1/2\right) }=%
\mathbb{X+X}^{\dag }+\mathbb{T}_{n+1/2}
\end{equation*}%
where 
\begin{equation}
\left( \mathrm{N}+\mathbf{1}\right) ^{n+1/2}\left( \mathbb{X+X}^{\dag
}\right) \left( \mathrm{N}+\mathbf{1}\right) ^{-\left( n+1\right) }\quad 
\text{and}\quad \mathbb{T}_{n+1/2}\equiv \left[ (\mathrm{N}+\mathbf{1}%
)^{n+1/2},\mathbb{X+X}^{\dag }\right] (\mathrm{N}+\mathbf{1})^{-\left(
n+1/2\right) }  \label{fg00}
\end{equation}%
are bounded operators acting on $\mathfrak{F}$. In particular, 
\begin{equation*}
\left\Vert \mathbb{T}_{n+1/2}\right\Vert _{\mathrm{op}}\leq \left( 12\times
2^{n}-4\right) \left\Vert X\right\Vert _{\mathrm{op},2}.
\end{equation*}%
Note that the assertions also hold with $\overline{\mathbb{X+X}^{\dag }}%
\doteq \left( \mathbb{X+X}^{\dag }\right) ^{\ast \ast }$ replacing $\mathbb{%
X+X}^{\dag }$.
\end{lemma}

\begin{proof}
In all the proof, $\mathrm{X}\doteq (\mathbb{X+X}^{\dag })$ (see (\ref%
{interaction elementaire})) for some fixed sequence $X\equiv (X_{k})_{k\in 
\mathbb{N}}$ of operators in $\mathcal{B}(\mathfrak{h}_{k})\equiv \mathcal{B}%
(\mathfrak{h}_{k})^{\otimes }$ satisfying $\Vert X\Vert _{\mathrm{op}%
,2}<\infty $. We distinguish between the cases $n=0$ and $n\in \mathbb{N}$.
In fact, we start with the first case, which is used as a springboard for
the case with $n\in \mathbb{N}$. We do this in several steps:\medskip

\noindent \underline{Step 1:} We first make sense of the following
(Banach-valued) Riemann integral 
\begin{equation*}
\int_{0}^{\infty }\lambda ^{\alpha -1}\left[ \left( \mathrm{N}+\mathbf{1}%
\right) \left( \mathrm{N}+\left( \lambda +1\right) \mathbf{1}\right) ^{-1},%
\mathrm{X}\right] \left( \mathrm{N}+\mathbf{1}\right) ^{-1/2}\mathrm{d}%
\lambda ,\qquad \alpha \in (0,1).
\end{equation*}%
Observe that the operators $\mathrm{N}$ and $\mathrm{X}$ conserve the dense
domain $\mathcal{D}_{\infty }$ (\ref{second quantu}) and, for any $\lambda
\in \mathbb{R}_{0}^{+}$,%
\begin{eqnarray*}
\left[ \left( \mathrm{N}+\mathbf{1}\right) \left( \mathrm{N}+\left( \lambda
+1\right) \mathbf{1}\right) ^{-1},\mathrm{X}\right] &=&\left( \mathrm{N}+%
\mathbf{1}\right) \left( \mathrm{N}+\left( \lambda +1\right) \mathbf{1}%
\right) ^{-1}\left[ \mathrm{X},\mathrm{N}\right] \left( \mathrm{N}+\left(
\lambda +1\right) \mathbf{1}\right) ^{-1} \\
&&+\left[ \mathrm{N},\mathrm{X}\right] \left( \mathrm{N}+\left( \lambda
+1\right) \mathbf{1}\right) ^{-1}
\end{eqnarray*}%
at least on the domain $\mathcal{D}_{\infty }$, so that one can formally
write%
\begin{align}
& \int_{0}^{\infty }\lambda ^{\alpha -1}\left[ \left( \mathrm{N}+\mathbf{1}%
\right) \left( \mathrm{N}+\left( \lambda +1\right) \mathbf{1}\right) ^{-1},%
\mathrm{X}\right] \left( \mathrm{N}+\mathbf{1}\right) ^{-1/2}\mathrm{d}%
\lambda  \notag \\
=& \int_{0}^{\infty }\lambda ^{\alpha -1}\left( \mathrm{N}+\mathbf{1}\right)
\left( \mathrm{N}+\left( \lambda +1\right) \mathbf{1}\right) ^{-1}\left[ 
\mathrm{X},\mathrm{N}\right] \left( \mathrm{N}+\mathbf{1}\right)
^{-1/2}\left( \mathrm{N}+\left( \lambda +1\right) \mathbf{1}\right) ^{-1}%
\mathrm{d}\lambda  \notag \\
& +\int_{0}^{\infty }\lambda ^{\alpha -1}\left[ \mathrm{N},\mathrm{X}\right]
\left( \mathrm{N}+\mathbf{1}\right) ^{-1/2}\left( \mathrm{N}+\left( \lambda
+1\right) \mathbf{1}\right) ^{-1}\mathrm{d}\lambda  \label{eq:integraleent}
\end{align}%
for any $\alpha \in (0,1)$. For each $\alpha \in (0,1)$, the mapping 
\begin{equation*}
\mathbb{R}_{0}^{+}\ni \lambda \mapsto \lambda ^{\alpha -1}\left( \mathrm{N}%
+\left( \lambda +1\right) \mathbf{1}\right) ^{-1}\in \mathcal{B}\left( 
\mathcal{\mathfrak{F}}\right)
\end{equation*}%
is Riemann integrable on the entire non-negative real line $\mathbb{R}%
_{0}^{+}$ because it is clearly continuous and 
\begin{equation}
\int_{0}^{\infty }\lambda ^{\alpha -1}\left\Vert \left( \mathrm{N}+\left(
\lambda +1\right) \mathbf{1}\right) ^{-1}\right\Vert _{\mathrm{op}}\mathrm{d}%
\lambda \leq \int_{0}^{\infty }\frac{\lambda ^{\alpha -1}}{1+\lambda }%
\mathrm{d}\lambda =\frac{\pi }{\sin \left( \alpha \pi \right) }<\infty .
\label{eq sup integral}
\end{equation}%
Therefore, the last term on the right-hand side of (\ref{eq:integraleent})
is well-defined as a Riemann integral provided 
\begin{equation*}
\left\Vert \left[ \mathrm{N},\mathrm{X}\right] \left( \mathrm{N}+\mathbf{1}%
\right) ^{-1/2}\right\Vert _{\mathrm{op}}<\infty .
\end{equation*}%
Mutatis mutandis for the other term on the right-hand side of (\ref%
{eq:integraleent}). This last inequality is satisfied, as a consequence of
Equation (\ref{itercommutNG0}) and Lemma \ref{Lemma-self-adjoint1}. Indeed,
we obtain that 
\begin{equation}
\left\Vert \left[ \mathrm{N},\mathrm{X}\right] \left( \mathrm{N}+\mathbf{1}%
\right) ^{-1/2}\right\Vert _{\mathrm{op}}<4\left\Vert X\right\Vert _{\mathrm{%
op},2}.  \label{boundednessbis}
\end{equation}%
We deduce from (\ref{eq:integraleent}) that the mapping 
\begin{equation*}
\mathbb{R}_{0}^{+}\ni \lambda \mapsto \lambda ^{\alpha -1}\left[ \left( 
\mathrm{N}+\mathbf{1}\right) \left( \mathrm{N}+\left( \lambda +1\right) 
\mathbf{1}\right) ^{-1},\mathrm{X}\right] \left( \mathrm{N}+\mathbf{1}%
\right) ^{-1/2}
\end{equation*}%
is Riemann-integrable for any $\alpha \in (0,1)$. Together with (\ref%
{boundednessbis}), this justifies the equalities%
\begin{eqnarray*}
&&\left[ \left( \mathrm{N}+\mathbf{1}\right) ^{\alpha },\mathrm{X}\right]
\left( \mathrm{N}+\mathbf{1}\right) ^{-1/2} \\
&=&\frac{\sin \left( \alpha \pi \right) }{\pi }\int_{0}^{\infty }\lambda
^{\alpha -1}\left( \mathrm{N}+\mathbf{1}\right) \left( \mathrm{N}+\left(
\lambda +1\right) \mathbf{1}\right) ^{-1}\mathrm{X}\left( \mathrm{N}+\mathbf{%
1}\right) ^{-1/2}\mathrm{d}\lambda \\
&&-\frac{\sin \left( \alpha \pi \right) }{\pi }\mathrm{X}\left( \mathrm{N}+%
\mathbf{1}\right) ^{-1/2}\int_{0}^{\infty }\lambda ^{\alpha -1}\left( 
\mathrm{N}+\mathbf{1}\right) \left( \mathrm{N}+\left( \lambda +1\right) 
\mathbf{1}\right) ^{-1}\mathrm{d}\lambda \\
&=&\frac{\sin \left( \alpha \pi \right) }{\pi }\int_{0}^{\infty }\lambda
^{\alpha -1}\left[ \left( \mathrm{N}+\mathbf{1}\right) \left( \mathrm{N}%
+\left( \lambda +1\right) \mathbf{1}\right) ^{-1},\mathrm{X}\right] \left( 
\mathrm{N}+\mathbf{1}\right) ^{-1/2}\mathrm{d}\lambda ,
\end{eqnarray*}%
for any $\alpha \in (0,1)$, where we use the integral representation of $(%
\mathrm{N}+\mathbf{1})^{\alpha }$ given by \cite[1.4.7 (e)]{Caps}. Indeed, $(%
\mathrm{N}+\mathbf{1})$ is obviously a self-adjoint operator that is
semibounded from below by one and, thus, of positive type according to \cite[%
Definition 1.4.1]{Caps}. It means in particular that%
\begin{equation}
\left[ \left( \mathrm{N}+\mathbf{1}\right) ^{\alpha },\mathrm{X}\right] =%
\frac{\sin \left( \alpha \pi \right) }{\pi }\int_{0}^{\infty }\lambda
^{\alpha -1}\left[ \left( \mathrm{N}+\mathbf{1}\right) \left( \mathrm{N}%
+\left( \lambda +1\right) \mathbf{1}\right) ^{-1},\mathrm{X}\right] \mathrm{d%
}\lambda ,  \label{integral representation}
\end{equation}%
at least on the (dense) domain $\mathcal{D}_{\infty }\subseteq \mathcal{D}(%
\mathrm{N}^{1/2})$. \medskip

\noindent \underline{Step 2:} We define the operator $\mathbb{T}_{1/2}$ on
the dense domain $\mathcal{D}_{\infty }$ to be 
\begin{equation*}
\mathbb{T}_{1/2}\doteq \lbrack (\mathrm{N}+\mathbf{1)}^{1/2},\mathrm{X}](%
\mathrm{N}+\mathbf{1})^{-1/2}.
\end{equation*}%
Observe in particular that 
\begin{equation}
\left( \mathrm{N}+\mathbf{1}\right) ^{1/2}\mathrm{X}\left( \mathrm{N}+%
\mathbf{1}\right) ^{-1/2}=\mathrm{X}+\mathbb{T}_{1/2}  \label{eqausdsklddbis}
\end{equation}%
on $\mathcal{D}_{\infty }$. By Equations (\ref{eq:integraleent})--(\ref%
{integral representation}), it follows that 
\begin{equation}
\left\Vert \mathbb{T}_{1/2}\varphi \right\Vert _{\mathfrak{F}}\leq
8\left\Vert X\right\Vert _{\mathrm{op},2}\left\Vert \varphi \right\Vert _{%
\mathfrak{F}},\qquad \varphi \in \mathcal{D}_{\infty }.
\label{eq:comutNalphabound}
\end{equation}%
Therefore, because $\mathcal{D}_{\infty }\subseteq \mathfrak{F}$ is dense, $%
\mathbb{T}_{1/2}$ admits a unique bounded continuous extension to $\mathfrak{%
F}$. By Proposition \ref{corollarire-selfadjoint2}, $\mathrm{X}(\mathrm{N}+%
\mathbf{1})^{-1/2}$ defines a bounded operator. So, according to (\ref%
{eqausdsklddbis}), 
\begin{equation*}
\left( \mathrm{N}+\mathbf{1}\right) ^{1/2}\mathrm{X}\left( \mathrm{N}+%
\mathbf{1}\right) ^{-1}
\end{equation*}%
is also a bounded operator.\medskip

\noindent \underline{Step 3:} We are now in a position to prove the
assertion for $n\in \mathbb{N}$. Observe that, for any $n\in \mathbb{N}$,%
\begin{eqnarray}
\left[ (\mathrm{N}+\mathbf{1})^{n+1/2},\mathrm{X}\right] \left( \mathrm{N}+%
\mathbf{1}\right) ^{-\left( n+1/2\right) } &=&(\mathrm{N}+\mathbf{1})^{1/2}%
\left[ (\mathrm{N}+\mathbf{1})^{n},\mathrm{X}\right] (\mathrm{N}+\mathbf{1}%
)^{-\left( n+1/2\right) }  \notag \\
&&+\left[ (\mathrm{N}+\mathbf{1})^{1/2},\mathrm{X}\right] (\mathrm{N}+%
\mathbf{1})^{-1/2}  \notag \\
&=&\left[ (\mathrm{N}+\mathbf{1})^{1/2},\left[ (\mathrm{N}+\mathbf{1})^{n},%
\mathrm{X}\right] \right] (\mathrm{N}+\mathbf{1})^{-\left( n+1/2\right) }
\label{rpoyiptru} \\
&&+\left[ (\mathrm{N}+\mathbf{1})^{n},\mathrm{X}\right] (\mathrm{N}+\mathbf{1%
})^{-n}+\left[ (\mathrm{N}+\mathbf{1})^{1/2},\mathrm{X}\right] (\mathrm{N}+%
\mathbf{1})^{-1/2}  \notag
\end{eqnarray}%
on the dense domain $\mathcal{D}_{\infty }$. Because of (\ref{itercommutNG0}%
)--(\ref{itercommutNG}) and \cite[1.4.7 (e)]{Caps}, we get that, for any $%
n\in \mathbb{N}$ and all $\varphi \in \mathcal{D}_{\infty }$, 
\begin{equation}
\left\Vert \left[ (\mathrm{N}+\mathbf{1})^{1/2},\left[ (\mathrm{N}+\mathbf{1}%
)^{n},\mathrm{X}\right] \right] (\mathrm{N}+\mathbf{1})^{-\left(
n+1/2\right) }\varphi \right\Vert _{\mathfrak{F}}\leq 8\left( 2^{n}-1\right)
\left\Vert X\right\Vert _{\mathrm{op},2}\left\Vert \varphi \right\Vert _{%
\mathfrak{F}},  \label{eq:commutNnwithNalphabound}
\end{equation}%
exactly as is done to prove (\ref{eq:comutNalphabound}). Combining Equality (%
\ref{rpoyiptru}) with Lemma \ref{Lemma-self-adjoint1 copy(2)} and Inequality
(\ref{eq:comutNalphabound}), we deduce that 
\begin{equation*}
\left\Vert \left[ (\mathrm{N}+\mathbf{1})^{n+1/2},\mathrm{X}\right] (\mathrm{%
N}+\mathbf{1})^{-\left( n+1/2\right) }\varphi \right\Vert _{\mathfrak{F}%
}\leq 4\left( 3\left( 2^{n}-1\right) +2\right) \Vert X\Vert _{\mathrm{op}%
,2}\left\Vert \varphi \right\Vert _{\mathfrak{F}}
\end{equation*}%
for each $n\in \mathbb{N}$ and all $\varphi \in \mathcal{D}_{\infty }$. All
that remains now is to invoke the same arguments used in Step 2 or at the
end of the proof of Lemma \ref{Lemma-self-adjoint1 copy(2)} to obtain the
assertion. The proof can be carried out in an identical manner by replacing $%
\mathrm{X}$ with its closure $\mathrm{X}=\mathrm{X}^{\ast \ast }$; we omit
the details.
\end{proof}

\subsection{Generalized Spin-Boson Hamiltonians}

We now establish the rigorous definition of generalized spin-boson
Hamiltonians, formally introduced in Equation (\ref{def_H0}). This can be
approached in two distinct ways, each yielding slightly different results.
First, we use the Berezin-Bruneau-Derezinski (BBD) approach \cite%
{Bruneau-derezinski2007,Berezin}, as in the proof of Proposition \ref%
{corollarire-selfadjoint2}:

\begin{theorem}[Generalized spin-boson Hamiltonians -- BBD]
\label{HamilSA copy(2)}\mbox{}\newline
\emph{SB0.} Let $(\mathfrak{h}_{k})_{k\in \mathbb{N}}$ be a countable family
of Hilbert spaces, and $\mathcal{F}_{+}$ be the bosonic Fock space
constructed over an infinite-dimensional, separable Hilbert space $\mathcal{H%
}$ with an orthonormal basis $\{\psi _{k}\}_{k\in \mathbb{N}}$. \newline
\emph{SB1. }Let $\mu \equiv (\mu _{k})_{k\in \mathbb{N}}$ be a sequence of
bounded self-adjoint operators $\mu _{k}\in \mathcal{B}(\mathfrak{h}%
_{k})\equiv \mathcal{B}(\mathfrak{h}_{k})^{\otimes }$. \newline
\emph{SB2.} Let $X\equiv (X_{k})_{k\in \mathbb{N}}$ be a sequence of
operators $X_{k}\in \mathcal{B}(\mathfrak{h}_{k})\equiv \mathcal{B}(%
\mathfrak{h}_{k})^{\otimes }$ satisfying $\Vert X\Vert _{\mathrm{op}%
,2}<\infty $. \newline
\emph{SB3.} Let $H$ be an operator on $\mathfrak{F}\doteq \mathfrak{H}%
\otimes \mathcal{F}_{+}$ such that $H+\mathrm{d}\Gamma ^{\otimes }(\mu )$ is
self-adjoint with core $\mathcal{D}_{\infty }$ (\ref{second quantu}%
).\smallskip \newline
If $H\left( \mathcal{D}_{n}\right) \subseteq \mathcal{D}_{n}$ for all $n\in 
\mathbb{N}_{0}$ (cf. (\ref{domain nbis})) and $\tilde{H}$ is a bounded
self-adjoint operator on $\mathfrak{F}$, then the expression 
\begin{equation*}
\tilde{H}+H+\mathrm{d}\Gamma ^{\otimes }\left( \mu \right) +\mathbb{X+X}%
^{\dag }
\end{equation*}%
(cf. (\ref{interaction elementaire})) defines an essentially self-adjoint
operator on $\mathcal{D}_{\infty }$.
\end{theorem}

\begin{proof}
The proof directly extends that of Proposition \ref{corollarire-selfadjoint2}%
. We give it for completeness even if similar arguments can be found in \cite%
[Theorem 5.3]{Bruneau-derezinski2007}. The case $X=0$ is obvious from
Assumption SB3 and the boundedness of $\tilde{H}$ (see, e.g., \cite[%
Proposition 1.6 (vii)]{Konrad}). If $X\neq 0$ then the expression 
\begin{equation*}
H+\mathrm{d}\Gamma ^{\otimes }\left( \mu \right) +\mathbb{X+X}^{\dag }
\end{equation*}%
defines a symmetric operator $\mathrm{H}$ with domain given by the dense set 
$\mathcal{D}_{\infty }$. Thus we proceed as in Proposition \ref%
{corollarire-selfadjoint2} by proving that $\ker \left( \mathrm{H}^{\ast
}-z\right) =\left\{ 0\right\} $ for all $z\in \mathbb{\mathbb{C}}\backslash 
\mathbb{R}$. As compared to the proof of Proposition \ref%
{corollarire-selfadjoint2}, we do not directly use a vector $\varphi \in
\ker \left( \mathrm{H}^{\ast }-z\mathbf{1}\right) $, but rather use an
approximation of it given by 
\begin{equation*}
\varphi _{\pm \epsilon }\doteq \left( 1\pm i\epsilon \left( H+\mathrm{d}%
\Gamma ^{\otimes }\left( \mu \right) \right) \right) ^{-1}\varphi ,\qquad
\epsilon \in \mathbb{R}^{+},
\end{equation*}%
on which we eventually take the limit $\epsilon \rightarrow 0^{+}$, using
that $\lim_{\epsilon \rightarrow 0^{+}}\varphi _{\pm \epsilon }=\varphi $
and $P_{n}\varphi _{\pm \epsilon }=\left( P_{n}\varphi \right) _{\pm
\epsilon }$, provided that $H\left( \mathcal{D}_{n}\right) \subseteq 
\mathcal{D}_{n}$ for $n\in \mathbb{N}_{0}$. The resolvent is well-defined by
SB3. Recall from the proof of Proposition \ref{corollarire-selfadjoint2}
that $P_{n}$, $n\in \mathbb{N}_{0}$, are the orthogonal projections from $%
\mathfrak{F}$ onto $\mathcal{D}_{n}$. Then, one checks that, for each $n\in 
\mathbb{N}_{0}$, $z\in \mathbb{\mathbb{C}}\backslash \mathbb{R}$ and any
vector $\varphi \in \ker \left( \mathrm{H}^{\ast }-z\right) $, 
\begin{equation*}
z\left\Vert P_{n}\varphi \right\Vert _{\mathfrak{F}}^{2}=z\langle
P_{n}\varphi ,\varphi \rangle _{\mathfrak{F}}=\lim_{\epsilon \rightarrow
0^{+}}z\langle P_{n}\varphi _{\epsilon },\varphi \rangle _{\mathfrak{F}%
}=\lim_{\epsilon \rightarrow 0^{+}}\langle P_{n}\varphi _{\epsilon },\mathrm{%
H}^{\ast }\varphi \rangle _{\mathfrak{F}}=\lim_{\epsilon \rightarrow
0^{+}}\langle \mathrm{H}P_{n}\varphi _{\epsilon },\varphi \rangle _{%
\mathfrak{F}}
\end{equation*}%
and, in the same way, 
\begin{equation*}
\overline{z}\left\Vert P_{n}\varphi \right\Vert _{\mathfrak{F}%
}^{2}=\lim_{\epsilon \rightarrow 0^{+}}\langle \varphi ,\mathrm{H}%
P_{n}\varphi _{-\epsilon }\rangle _{\mathfrak{F}}.
\end{equation*}%
Compare these last equations with (\ref{eq 1 self})--(\ref{eq 2 self}). As
we get (\ref{eq 3 self}), we thus find that 
\begin{eqnarray*}
2i\mathrm{Im}\left( z\left\Vert P_{n}\varphi \right\Vert _{\mathfrak{F}%
}^{2}\right) &=&\lim_{\epsilon \rightarrow 0^{+}}\left\{ \langle \left( H+%
\mathrm{d}\Gamma ^{\otimes }\left( \mu \right) \right) P_{n}\varphi
_{\epsilon },\varphi \rangle _{\mathfrak{F}}-\langle \varphi ,\left( H+%
\mathrm{d}\Gamma ^{\otimes }\left( \mu \right) \right) P_{n}\varphi
_{-\epsilon }\rangle _{\mathfrak{F}}\right. \\
&&\left. +\langle \mathrm{X}P_{n}\varphi _{\epsilon },\varphi \rangle _{%
\mathfrak{F}}-\langle \varphi ,\mathrm{X}P_{n}\varphi _{-\epsilon }\rangle _{%
\mathfrak{F}}\right\}
\end{eqnarray*}%
for each $n\in \mathbb{N}_{0}$, $z\in \mathbb{\mathbb{C}}\backslash \mathbb{R%
}$ and any vector $\varphi \in \ker \left( \mathrm{H}^{\ast }-z\right) $,
with the operator $\mathrm{X}\doteq (\mathbb{X+X}^{\dag })$ defined as in
the proof of Proposition \ref{corollarire-selfadjoint2}. If $H\left( 
\mathcal{D}_{n}\right) \subseteq \mathcal{D}_{n}$ for each $n\in \mathbb{N}%
_{0}$, the first two terms of the right-hand side of the last equation
cancel out. Therefore, because $\mathrm{X}P_{n}$ is bounded (and thus
continuous), we can take the limit $\epsilon \rightarrow 0^{+}$ to arrive
exactly at the identity (\ref{eq 3 self}). All we have to do now is follow
the end of the proof of Proposition \ref{corollarire-selfadjoint2} to prove
the assertion when $\tilde{H}=0$ and $H\left( \mathcal{D}_{n}\right)
\subseteq \mathcal{D}_{n}$ for all $n\in \mathbb{N}_{0}$. Adding any bounded
self-adjoint operator $\tilde{H}$ to this last Hamiltonian does not change
the statement, thanks to \cite[Proposition 1.6 (vii)]{Konrad}.
\end{proof}

Note that the condition $H\left( \mathcal{D}_{n}\right) \subseteq \mathcal{D}%
_{n}$ for all $n\in \mathbb{N}_{0}$ in Theorem \ref{HamilSA copy(2)} for
unbounded $H$ can be eliminated by making additional assumptions, which in
any case are used to set up our flow. This is explained in a second approach
based on the Kato-Rellich (KR) theorem. Indeed, Lemma \ref%
{Lemma-self-adjoint1} together with the triangle inequality essentially
means that the (closure of the) operator $(\mathbb{X+X}^{\dag })$, as
defined by (\ref{interaction elementaire}), is relatively bounded with
respect to any self-adjoint operator of the form $B=\mathrm{d}\Gamma
^{\otimes }\left( \mu \right) +A$ with $A\geq 0$. The so-called
\textquotedblleft $B$-bound\textquotedblright\ of (the closure of) $(\mathbb{%
X+X}^{\dag })$ \cite[Definition 8.1]{Konrad} is in particular strictly less
than one, making the use of the Kato-Rellich theorem \cite[Theorem 8.5]%
{Konrad} possible. We provide further details in the proof of the following
theorem, which simultaneously establishes again the definition of
generalized spin-boson models:

\begin{theorem}[Generalized spin-boson Hamiltonians -- KR]
\label{HamilSA}\mbox{}\newline
Assume SB0--SB3 of Theorem \ref{HamilSA copy(2)} together with the following
ones: \newline
\emph{SB4. }Assume that $H$ is a lower semibounded self-adjoint operator on $%
\mathfrak{F}$, whose domain contains the dense subset $\mathcal{D}_{\infty }$%
. \newline
\emph{SB5. }Assume that $\mu _{k}\geq c_{k}\mathbf{1}_{\mathfrak{h}_{k}}$
for some $c_{k}\in \mathbb{R}_{0}^{+}$ and each $k\in \mathbb{N}$ such that 
\begin{equation*}
\sum_{k\in \mathbb{N}}c_{k}^{-1}\left\Vert X_{k}\right\Vert _{\mathrm{op}%
}^{2}<\infty ,
\end{equation*}%
with the convention that $0^{-1}\times 0\doteq 0$ if $c_{k}=\Vert X_{k}\Vert
_{\mathrm{op}}=0$.\smallskip \newline
Then, the expressions 
\begin{equation}
H+\mathrm{d}\Gamma ^{\otimes }\left( \mu \right) +\mathbb{X+X}^{\dag }\text{%
\qquad and\qquad }H+\mathrm{d}\Gamma ^{\otimes }\left( \mu \right) +%
\overline{\mathbb{X+X}^{\dag }}  \label{expression KR models}
\end{equation}%
define the same lower semibounded self-adjoint operator on the domain of $(H+%
\mathrm{d}\Gamma ^{\otimes }(\mu ))$. Additionally, if $(H+\mathrm{d}\Gamma
^{\otimes }(\mu ))$ is essentially self-adjoint\ on a core such as $\mathcal{%
D}_{\infty }$, then the above operators are also essentially self-adjoint on
that same domain.
\end{theorem}

\begin{proof}
Because it is symmetric and densely defined, $(\mathbb{X+X}^{\dag })$
defined on $\mathcal{D}_{\infty }$ by (\ref{interaction elementaire}) is
closable. By Proposition \ref{corollarire-selfadjoint2} and SB0--SB2, its
closure $\overline{\mathbb{X+X}^{\dag }}$ \ is a self-adjoint operator. By
Assumption SB4, there is a real number $b\in \mathbb{R}$ such that 
\begin{equation}
H_{b}=H-b\mathbf{1}_{\mathfrak{F}}\geq 0.  \label{sdssdsd}
\end{equation}%
Hence, using in particular Assumptions SB1 and SB3--SB5, we infer from Lemma %
\ref{Lemma-self-adjoint1} with $A=H_{b}$\ (whose domain contains $\mathcal{D}%
_{\infty }$) that the domain of $\overline{\mathbb{X+X}^{\dag }}$ must
always contain\footnote{$\overline{\left( \mathbb{X+X}^{\ast }\right) }$ is
defined on $\mathcal{D}(H_{b}+\mathrm{d}\Gamma ^{\otimes }(\mu ))$ as a
bounded operator acting on the Banach space $\mathcal{D}(H_{b}+\mathrm{d}%
\Gamma ^{\otimes }(\mu ))$ endowed with the graph norm associated with the
self-adjoint operator $(H_{b}+\mathrm{d}\Gamma ^{\otimes }\left( \mu \right)
)$. Indeed, it is the continuous extension of a densely defined operator
that is uniformly bounded (on the dense subset $\mathcal{D}_{\infty }$ of
this Banach space).} the domain $\mathcal{D}(H_{b}+\mathrm{d}\Gamma
^{\otimes }(\mu ))$. In particular, since $\mathcal{D}_{\infty }$ is a core
of $H_{b}+\mathrm{d}\Gamma ^{\otimes }(\mu )$, using the triangle inequality
one can extend the estimate of Lemma \ref{Lemma-self-adjoint1} so that, for
any $\varphi \in \mathcal{D}(H_{b}+\mathrm{d}\Gamma ^{\otimes }(\mu
))\supseteq \mathcal{D}_{\infty }$, 
\begin{equation}
\left\Vert \overline{\left( \mathbb{X+X}^{\dag }\right) }\varphi \right\Vert
_{\mathfrak{F}}\leq 2^{-1}\left\Vert \left( H_{b}+\mathrm{d}\Gamma ^{\otimes
}\left( \mu \right) \right) \varphi \right\Vert _{\mathfrak{F}}+\left\Vert
\varphi \right\Vert _{\mathfrak{F}}\left( \left\Vert X\right\Vert _{\mathrm{%
op},2}+2\sum_{k\in \mathbb{N}}c_{k}^{-1}\left\Vert X_{k}\right\Vert _{%
\mathrm{op}}^{2}\right)  \label{sfsfssdf}
\end{equation}%
with the convention that $0^{-1}\times 0\doteq 0$ if $c_{k}=\Vert X_{k}\Vert
_{\mathrm{op}}=0$. So, we can now invoke the Kato-Rellich theorem \cite[%
Theorem 8.5]{Konrad} to get the assertion, but with the operator $H_{b}$ (%
\ref{sdssdsd}) instead of $H$. This in turn implies the theorem (i.e., when $%
H=H_{b}-b\mathbf{1}_{\mathfrak{F}}$ now replaces $H_{b}$), because $b\mathbf{%
1}_{\mathfrak{F}}$ is obviously a bounded operator and so, $(T+b\mathbf{1}_{%
\mathfrak{F}})^{\ast }=T^{\ast }+b\mathbf{1}_{\mathfrak{F}}$ for any
possibly unbounded operator $T$ acting on $\mathfrak{F}$, see \cite[%
Proposition 1.6 (vii)]{Konrad}. Note from SB4--SB5 that $H_{b}+\mathrm{d}%
\Gamma ^{\otimes }(\mu )$ is a lower semibounded self-adjoint operator, and
so are the models defined by (\ref{expression KR models}). See \cite[Section
8.7, Exercice 8]{Konrad}.
\end{proof}

Theorems \ref{HamilSA copy(2)} and \ref{HamilSA} are pivotal because they
allows us to define generalized spin-boson models as (essentially)
self-adjoint operators acting on the tensor product $\mathfrak{F}\doteq 
\mathfrak{H}\otimes \mathcal{F}_{+}$. In Equation (\ref{def_H0}), we use an
operator $H$ of the form 
\begin{equation}
H\left( h\right) \doteq \sum_{k\in \mathbb{N}}h_{k}\otimes \mathbf{1}_{%
\mathcal{F}_{+}}  \label{sdsdsdsdsd}
\end{equation}%
(see also (\ref{HS})) for some sequence $h\equiv (h_{k})_{k\in \mathbb{N}}$
of bounded self-adjoint operators $h_{k}\in \mathcal{B}(\mathfrak{h}%
_{k})\equiv \mathcal{B}(\mathfrak{h}_{k})^{\otimes }$. We therefore give
below sufficient conditions for this type of operator to satisfy conditions
SB3 and SB4 of Theorems \ref{HamilSA copy(2)} and \ref{HamilSA}:

\begin{proposition}[Generalized spin-boson models without spin-boson
interactions]
\label{Proposition self-adjoint copy(1)}\mbox{}\newline
Let $u\equiv (u_{k})_{k\in \mathbb{N}}$, $v\equiv (v_{k})_{k\in \mathbb{N}}$
and $\mu \equiv (\mu _{k})_{k\in \mathbb{N}}$ be three sequences of
self-adjoint operators $u_{k},v_{k},\mu _{k}\in \mathcal{B}(\mathfrak{h}%
_{k}) $. Assume that $u$ satisfies Assumption \ref{Assumption S}, and $v$ is
operator-norm summable, meaning 
\begin{equation*}
\left\Vert v\right\Vert _{\mathrm{op},1}\doteq \sum_{k\in \mathbb{N}%
}\left\Vert v_{k}\right\Vert _{\mathrm{op}}<\infty .
\end{equation*}

\begin{itemize}
\item[\emph{(i)}] $H(u)$ is a lower semibounded self-adjoint operator with
core $\mathcal{D}_{\infty }$. Furthermore, $H(u)$ leaves the domains $%
\mathcal{D}_{n}$ invariant, i.e., $H(u)\left( \mathcal{D}_{n}\right)
\subseteq \mathcal{D}_{n}$ for all $n\in \mathbb{N}_{0}$.

\item[\emph{(ii)}] $H(v)$ is a bounded self-adjoint operator on $\mathfrak{F}
$.

\item[\emph{(iii)}] If $h=u+v$ then $H\left( h\right) $ is a lower
semibounded self-adjoint operator with core $\mathcal{D}_{\infty }$.
Furthermore, $(H(h)+\mathrm{d}\Gamma ^{\otimes }(\mu ))$ defines a
self-adjoint operator on $\mathfrak{F}$ with core $\mathcal{D}_{\infty }$.
\end{itemize}
\end{proposition}

\begin{proof}
Recall that the Hilbert space $\mathfrak{H}$ is constructed as the
restricted tensor product $\bigotimes_{k\in \mathbb{N}}\mathfrak{h}_{k}$
with respect to a fixed sequence of reference vectors $(\varsigma
_{k})_{k\in \mathbb{N}}$ with $\varsigma _{k}\in \mathfrak{h}_{k}$.
Specifically, $\mathfrak{H}$ is the completion of $\mathfrak{H}_{\mathrm{pre}%
}$ (\ref{pre-hilbert}), which is the linear span of elementary tensor
products of the form $\otimes _{k\in \mathbb{N}}\varphi _{k}$, where $%
\varphi _{k}\in \mathfrak{h}_{k}$ and $\varphi _{k}=\varsigma _{k}$ for all
but finitely many $k\in \mathbb{N}$. Given these considerations, we are
prepared to establish the proposition, beginning with the first
assertion:\medskip

\noindent \underline{(i):} Assumption \ref{Assumption S} for the sequence $%
(u_{k})_{k\in \mathbb{N}}$ means the existence of $k_{0}\in \mathbb{N}$ such
that $\varsigma _{k}\in \ker u_{k}$ for all natural numbers $k\geq k_{0}$,
and 
\begin{equation}
\sum_{k\in \mathbb{N}}\left\Vert u_{k}\mathbf{1}\left[ u_{k}\leq 0\right]
\right\Vert _{\mathrm{op}}<\infty ,  \label{sdfsdfsfsd}
\end{equation}%
where $\mathbf{1}\left[ A\leq 0\right] $ denotes the spectral projection of
a self-adjoint operator $A$ onto its negative spectrum. Therefore, under
Assumption \ref{Assumption S}, the expression (\ref{sdsdsdsdsd}) defines an
operator $H(u)$ on the dense domain $\mathcal{D}_{\infty }\subseteq 
\mathfrak{H}_{\mathrm{pre}}\otimes \mathcal{F}_{+}$ (see again Equation (\ref%
{pre-hilbert}) and (\ref{second quantu})). If $u$ satisfies Assumption \ref%
{Assumption S}, then (\ref{sdsdsdsdsd}) implies that for any subset subset $%
\mathcal{X\subseteq F}_{+}$, 
\begin{equation*}
H(u)\left( \mathfrak{H}_{\mathrm{pre}}\otimes \mathcal{X}\right) \subseteq 
\mathfrak{H}_{\mathrm{pre}}\otimes \mathcal{X}.
\end{equation*}%
Consequently, $H(u)$ leaves the domains $\mathcal{D}_{n}$ invariant for all $%
n\in \mathbb{N}_{0}$. Furthermore, $H(u)$ is symmetric because $u_{k}$ is
bounded and self-adjoint for all $k\in \mathbb{N}$. As all elements of $%
\mathcal{D}_{\infty }$ are analytic vectors for $H(u)$, by the Nelson
theorem \cite[Theorem 7.16]{Konrad}, it is essentially self-adjoint on $%
\mathcal{D}_{\infty }$. Finally, one checks from (\ref{sdfsdfsfsd}) that $%
H(u)$ is lower semibounded: 
\begin{equation*}
H(u)\geq -\sum_{k\in \mathbb{N}}\left\Vert u_{k}\mathbf{1}\left[ u_{k}\leq 0%
\right] \right\Vert _{\mathrm{op}}\mathbf{1}_{\mathfrak{F}}.
\end{equation*}%
With this, the proof of Assertion (i) is complete.\medskip

\noindent \underline{(ii):} If $\Vert v\Vert _{\mathrm{op},1}<\infty $ then $%
H(v)\in \mathcal{B}(\mathfrak{F})$. Indeed, the operator-norm summability of 
$v$, combined with the triangle inequality, ensures the absolute convergence
of the series (\ref{sdsdsdsdsd}) for $h=v$ in the operator norm of the
Banach space $\mathcal{B}(\mathfrak{F})$. The operator $H(v)$ is symmetric
because $v_{k}$ is bounded and self-adjoint for all $k\in \mathbb{N}$. Since 
$H(v)$ is bounded, it is automatically self-adjoint. Thus, Assertion (ii) is
proven.\medskip

\noindent \underline{(iii):} Applying \cite[Proposition 1.6 (vii)]{Konrad}
and Assertions (i)--(ii), we deduce that 
\begin{equation*}
H\left( h\right) =H\left( u\right) +H\left( v\right)
\end{equation*}%
is essentially self-adjoint on $\mathcal{D}_{\infty }$ and its extension is
a lower semibounded self-adjoint operator. In addition, one verifies that
all elements of $\mathcal{D}_{\infty }\subseteq \mathfrak{H}_{\mathrm{pre}%
}\otimes \mathcal{F}_{+}$\ are analytic vectors of the sum $(H(h)+\mathrm{d}%
\Gamma ^{\otimes }(\mu ))$, which is meanwhile symmetric and densely defined
and, hence, closable. By the Nelson theorem \cite[Theorem 7.16]{Konrad}, $%
(H(h)+\mathrm{d}\Gamma ^{\otimes }(\mu ))$ is essentially self-adjoint on $%
\mathcal{D}_{\infty }$, leading to Assertion (iii).
\end{proof}

Combined with Proposition \ref{Proposition self-adjoint copy(1)}, Theorems %
\ref{HamilSA copy(2)} or \ref{HamilSA} allow us to rigorously define
Hamiltonians of the form (\ref{def_H0}); see Theorem \ref{HamilSA copy(1)}.
While the formulation of the latter requires stronger structural assumptions
than Theorem \ref{HamilSA}, our ultimate goal of achieving an (approximate)
diagonalization of the generalized spin-boson model forces even more
restrictive conditions upon us anyway. These are explored in Section \ref%
{Fiberflowsection}. For our purposes, the scenario covered by Theorem \ref%
{HamilSA copy(1)} is the most mathematically relevant.

We conclude this section with two technical lemmata concerning the
invariance of the domain $\mathcal{D}(\mathrm{N})$ of the particle number
operator $\mathrm{N}$ (and certain of its powers) under the action of
resolvents of generalized spin-boson Hamiltonians. These results provide
powerful tools for constructing an approximate solution to the
Brockett-Wegner flow (\ref{brockettWegner}); see Section \ref{Approximated
Solution} for more details.

\begin{lemma}[Conservation of $\mathcal{D}(\mathrm{N}^{n/2})$ by generalized
spin-boson Hamiltonians]
\label{diversPreservedom}\mbox{}\newline
Suppose that hypotheses SB0--SB3 of Theorem \ref{HamilSA copy(2)} are
satisfied with $\left[ H,\mathrm{N}\right] =0$. Furthermore, assume that 
\begin{equation}
\mathrm{H}\doteq H+\mathrm{d}\Gamma ^{\otimes }\left( \mu \right) +\mathbb{%
X+X}^{\dag }  \label{notation H}
\end{equation}%
is self-adjoint with core $\mathcal{D}_{\infty }$ (cf. Theorems \ref{HamilSA
copy(2)} or \ref{HamilSA}). For any $n\in \mathbb{N}_{0}$,%
\begin{eqnarray}
\left\Vert \left( \mathrm{N}+\mathbf{1}\right) ^{\frac{n}{2}}\left( \mathrm{H%
}+i\lambda \mathbf{1}\right) ^{-1}\left( \mathrm{N}+\mathbf{1}\right) ^{-%
\frac{n}{2}}\right\Vert _{\mathrm{op}} &\leq &\frac{1}{\lambda -2^{3+\frac{n%
}{2}}\left\Vert X\right\Vert _{\mathrm{op},2}},\qquad \lambda >2^{3+\frac{n}{%
2}}\left\Vert X\right\Vert _{\mathrm{op},2},  \label{ResolventHsurN} \\
\left\Vert \left( \mathrm{N}+\mathbf{1}\right) \mathrm{H}\left( \mathrm{H}%
+i\lambda \mathbf{1}\right) ^{-1}\left( \mathrm{N}+\mathbf{1}\right)
^{-1}\right\Vert _{\mathrm{op}} &\leq &1+\frac{\lambda }{\lambda
-2^{4}\left\Vert X\right\Vert _{\mathrm{op},2}},\qquad \lambda
>2^{4}\left\Vert X\right\Vert _{\mathrm{op},2}.  \label{HresolveHperserve}
\end{eqnarray}
\end{lemma}

\begin{proof}
Observe that for any $n\in \mathbb{N}_{0}$, 
\begin{eqnarray*}
\left( \mathrm{N}+\mathbf{1}\right) ^{\frac{n}{2}}\left( \mathrm{H}+i\lambda 
\mathbf{1}\right) ^{-1}\left( \mathrm{N}+\mathbf{1}\right) ^{-\frac{n}{2}}
&=&\left( \left( \mathrm{N}+\mathbf{1}\right) ^{\frac{n}{2}}\mathrm{H}\left( 
\mathrm{N}+\mathbf{1}\right) ^{-\frac{n}{2}}+i\lambda \mathbf{1}\right) ^{-1}
\\
&=&\left( H+\mathrm{d}\Gamma ^{\otimes }\left( \mu \right) +\left( \mathrm{N}%
+\mathbf{1}\right) ^{\frac{n}{2}}\mathrm{X}\left( \mathrm{N}+\mathbf{1}%
\right) ^{-\frac{n}{2}}+i\lambda \mathbf{1}\right) ^{-1},
\end{eqnarray*}%
which, combined with Lemmata \ref{Lemma-self-adjoint1 copy(2)} and \ref%
{RelativeboundGlemma}, in turn implies that 
\begin{equation}
\left( \mathrm{N}+\mathbf{1}\right) ^{\frac{n}{2}}\left( \mathrm{H}+i\lambda 
\mathbf{1}\right) ^{-1}\left( \mathrm{N}+\mathbf{1}\right) ^{-\frac{n}{2}%
}=\left( \mathrm{H}+\mathbb{T}_{n/2}+i\lambda \mathbf{1}\right) ^{-1}.
\label{fg1}
\end{equation}%
Expanding in a Neumann series, we now obtain 
\begin{equation}
\left( \mathrm{H}+\mathbb{T}_{n/2}+i\lambda \mathbf{1}\right) ^{-1}=\left( 
\mathrm{H}+i\lambda \mathbf{1}\right) ^{-1}\sum_{k\in \mathbb{N}_{0}}\left(
-1\right) ^{k}\left\{ \left( \mathrm{H}+i\lambda \mathbf{1}\right) ^{-1}%
\mathbb{T}_{n/2}\right\} ^{k},  \label{condition lambdabis}
\end{equation}%
which is well-defined as an absolutely convergent series when $\lambda
>\Vert \mathbb{T}_{n/2}\Vert _{\mathrm{op}}$, since in this case, by the
triangle inequality, 
\begin{equation}
\left\Vert \left( \mathrm{H}+\mathbb{T}_{n/2}+i\lambda \mathbf{1}\right)
^{-1}\right\Vert _{\mathrm{op}}\leq \sum_{k\in \mathbb{N}_{0}}\lambda
^{-\left( k+1\right) }\left\Vert \mathbb{T}_{n/2}\right\Vert _{\mathrm{op}%
}^{k}=\frac{1}{\lambda -\left\Vert \mathbb{T}_{n/2}\right\Vert _{\mathrm{op}}%
}<\infty .  \label{fg2}
\end{equation}%
According to Lemmata \ref{Lemma-self-adjoint1 copy(2)} and \ref%
{RelativeboundGlemma}, for any $n\in \mathbb{N}_{0}$, we obtain the simple
(unified\footnote{%
Here we use that, for $n=2p$, $4\left( 2^{p}-1\right) \leq 8\times 2^{p}$
while for $n=2p+1$, $12\times 2^{p}-4\leq 8\times 2^{p+1/2}$.}) bound 
\begin{equation}
\Vert \mathbb{T}_{n/2}\Vert _{\mathrm{op}}\leq 2^{3+n/2}\left\Vert
X\right\Vert _{\mathrm{op},2}.  \label{fg3}
\end{equation}%
Equation (\ref{ResolventHsurN}) thus follows from (\ref{fg1}) and (\ref{fg2}%
)--(\ref{fg3}). To obtain (\ref{HresolveHperserve}), it suffices to combine
the equality 
\begin{equation*}
\mathrm{H}\left( \mathrm{H}+i\lambda \mathbf{1}\right) ^{-1}=\mathbf{1}%
-i\lambda \left( \mathrm{H}+i\lambda \mathbf{1}\right) ^{-1},\qquad \lambda
\in \mathbb{R},
\end{equation*}%
with (\ref{ResolventHsurN}) and the triangle inequality.
\end{proof}

\begin{lemma}[Particular case of gapped Hamiltonians]
\label{diversPreservedom copy(1)}\mbox{}\newline
Assume SB0--SB5 of Theorems \ref{HamilSA copy(2)} and \ref{HamilSA} with $%
c_{k}\geq c\in \mathbb{R}^{+}$ for all $k\in \mathbb{N}$. Then, $\mathrm{H}$
(\ref{notation H}) has domain $\mathcal{D}\left( \mathrm{H}\right) \subseteq 
\mathcal{D}\left( \mathrm{N}\right) $ and satisfies the following bounds: If 
$\lambda >2^{7/2}\Vert X\Vert _{\mathrm{op},2}$ and $\zeta \doteq \lambda
^{-1}\left\Vert X\right\Vert _{\mathrm{op},2}$, then 
\begin{eqnarray*}
\left\Vert \left( \mathrm{N}+\mathbf{1}\right) ^{\frac{3}{2}}\left( \mathrm{H%
}+i\lambda \mathbf{1}\right) ^{-1}\left( \mathrm{N}+\mathbf{1}\right)
^{-1}\right\Vert _{\mathrm{op}} &\leq &\frac{\mathrm{K}_{c,b}}{1-4\zeta }%
\left( 1+\frac{12\zeta }{1-2^{7/2}\zeta }\right) , \\
\left\Vert \left( \mathrm{N}+\mathbf{1}\right) \left( \mathrm{H}+i\lambda 
\mathbf{1}\right) ^{-1}\right\Vert _{\mathrm{op}} &\leq &\mathrm{K}%
_{c,b}\left( 1+12\mathrm{K}_{c,b}\left( 1+\frac{12\zeta }{1-2^{7/2}\zeta }%
\right) \frac{\zeta }{1-8\zeta }\right) ,
\end{eqnarray*}%
where the parameter $b\in \mathbb{R}$ is taken such that $H\geq b\mathbf{1}$
and%
\begin{equation}
\mathrm{K}_{c,b}\doteq \sup_{k\in \mathbb{N}_{0}}\sup_{x\in \left[ b,\infty
\right) }\frac{\left( k+1\right) }{\sqrt{\left( ck+x\right) ^{2}+\lambda ^{2}%
}}<\infty .  \label{def Kcb}
\end{equation}
\end{lemma}

\begin{proof}
In the same way as before (cf. (\ref{fg1})--(\ref{condition lambdabis})), we
start by noting that%
\begin{align}
\left( \mathrm{N}+\mathbf{1}\right) ^{\frac{3}{2}}\left( \mathrm{H}+i\lambda 
\mathbf{1}\right) ^{-1}\left( \mathrm{N}+\mathbf{1}\right) ^{-1}& =\left( 
\mathrm{N}+\mathbf{1}\right) ^{\frac{1}{2}}\left( \left( \mathrm{N}+\mathbf{1%
}\right) \mathrm{H}\left( \mathrm{N}+\mathbf{1}\right) ^{-1}+i\lambda 
\mathbf{1}\right) ^{-1}  \notag \\
& =\left( \mathrm{N}+\mathbf{1}\right) ^{\frac{1}{2}}\left( \mathrm{H}%
+i\lambda \mathbf{1}\right) ^{-1}\sum_{k\in \mathbb{N}_{0}}\left( -1\right)
^{k}\left\{ \left( \mathrm{H}+i\lambda \mathbf{1}\right) ^{-1}\mathbb{T}%
_{1}\right\} ^{k}  \label{sdfsgdfhfghjfhjjhk}
\end{align}%
for any real parameter $\lambda >4\Vert X\Vert _{\mathrm{op},2}$, using
Lemma \ref{Lemma-self-adjoint1 copy(2)}. Indeed, the above Neumann series
absolutely converges when $\lambda >4\Vert X\Vert _{\mathrm{op},2}$: 
\begin{equation}
\sum_{k\in \mathbb{N}_{0}}\left\Vert \left( \mathrm{H}+i\lambda \mathbf{1}%
\right) ^{-1}\mathbb{T}_{1}\right\Vert _{\mathrm{op}}^{k}\leq \sum_{k\in 
\mathbb{N}_{0}}\left( 4\lambda ^{-1}\Vert X\Vert _{\mathrm{op},2}\right)
^{k}=\frac{1}{1-4\lambda ^{-1}\Vert X\Vert _{\mathrm{op},2}}<\infty ,
\label{qweqweqe}
\end{equation}%
thanks to Lemma \ref{Lemma-self-adjoint1 copy(2)} for $n=1$. Thus, the
problem reduces to establishing the inequality 
\begin{equation}
\left\Vert \left( \mathrm{N}+\mathbf{1}\right) ^{\frac{1}{2}}\left( \mathrm{H%
}+i\lambda \mathbf{1}\right) ^{-1}\right\Vert _{\mathrm{op}}<\infty ,
\label{sfsdfsdfsfssfsdf}
\end{equation}%
at least for sufficiently large $\lambda \in \mathbb{R}^{+}$. In fact, for
any $\lambda \in \mathbb{R}\backslash \{0\}$, using the resolvent identity
and the triangle inequality combined with the fact that $\mathrm{N}\geq 0$, 
\begin{eqnarray}
&&\left\Vert \left( \mathrm{N}+\mathbf{1}\right) ^{\frac{1}{2}}\left( 
\mathrm{H}+i\lambda \mathbf{1}\right) ^{-1}\right\Vert _{\mathrm{op}}  \notag
\\
&\leq &\left\Vert \left( \mathrm{N}+\mathbf{1}\right) ^{\frac{1}{2}}\left( H+%
\mathrm{d}\Gamma ^{\otimes }\left( \mu \right) +i\lambda \mathbf{1}\right)
^{-1}\right\Vert _{\mathrm{op}}  \notag \\
&&+\left\Vert \left( \mathrm{N}+\mathbf{1}\right) ^{\frac{1}{2}}\left(
\left( \mathrm{H}+i\lambda \mathbf{1}\right) ^{-1}-\left( H+\mathrm{d}\Gamma
^{\otimes }\left( \mu \right) +i\lambda \mathbf{1}\right) ^{-1}\right)
\right\Vert _{\mathrm{op}}  \notag \\
&\leq &\left\Vert \left( \mathrm{N}+\mathbf{1}\right) \left( H+\mathrm{d}%
\Gamma ^{\otimes }\left( \mu \right) +i\lambda \mathbf{1}\right)
^{-1}\right\Vert _{\mathrm{op}}  \notag \\
&&\times \left( 1+\left\Vert \left( \mathrm{N}+\mathbf{1}\right) ^{\frac{1}{2%
}}\left( \mathrm{H}+i\lambda \mathbf{1}\right) ^{-1}\left( \mathrm{N}+%
\mathbf{1}\right) ^{-\frac{1}{2}}\right\Vert _{\mathrm{op}}\left\Vert \left( 
\mathrm{N}+\mathbf{1}\right) ^{\frac{1}{2}}\mathrm{X}\left( \mathrm{N}+%
\mathbf{1}\right) ^{-1}\right\Vert _{\mathrm{op}}\right) .
\label{gfhfghfghh}
\end{eqnarray}%
According to Lemma \ref{diversPreservedom}, if $\lambda >2^{7/2}\Vert X\Vert
_{\mathrm{op},2}$ then 
\begin{equation}
\left\Vert \left( \mathrm{N}+\mathbf{1}\right) ^{\frac{1}{2}}\left( \mathrm{H%
}+i\lambda \mathbf{1}\right) ^{-1}\left( \mathrm{N}+\mathbf{1}\right) ^{-%
\frac{1}{2}}\right\Vert _{\mathrm{op}}\leq \frac{1}{\lambda
-2^{7/2}\left\Vert X\right\Vert _{\mathrm{op},2}}<\infty ,
\label{gfhfghfghh1}
\end{equation}%
while, by Lemmata \ref{Lemma-self-adjoint1} and \ref{RelativeboundGlemma}
for $n=0$, 
\begin{equation}
\left\Vert \left( \mathrm{N}+\mathbf{1}\right) ^{\frac{1}{2}}\mathrm{X}%
\left( \mathrm{N}+\mathbf{1}\right) ^{-1}\right\Vert _{\mathrm{op}}\leq
12\left\Vert X\right\Vert _{\mathrm{op},2}<\infty .  \label{gfhfghfghh2}
\end{equation}%
Therefore, by (\ref{gfhfghfghh}) it suffices to show now that 
\begin{equation}
\left\Vert \left( \mathrm{N}+\mathbf{1}\right) \left( H+\mathrm{d}\Gamma
^{\otimes }\left( \mu \right) +i\lambda \mathbf{1}\right) ^{-1}\right\Vert _{%
\mathrm{op}}<\infty ,  \label{fgin}
\end{equation}%
at least for sufficiently large $\lambda \in \mathbb{R}^{+}$. In fact, using
that $\mu _{k}\geq c\mathbf{1}_{\mathfrak{h}_{k}}$ for all $k\in \mathbb{N}$%
, where $c\in \mathbb{R}^{+}$, one verifies that $\mathrm{d}\Gamma ^{\otimes
}(\mu )\geq c\mathrm{N}$. Clearly, 
\begin{equation*}
\left[ \mathrm{N},H\right] =\left[ \mathrm{N},\mathrm{d}\Gamma ^{\otimes
}\left( \mu \right) \right] =\left[ \mathrm{N},H+\mathrm{d}\Gamma ^{\otimes
}\left( \mu \right) \right] =0
\end{equation*}%
and, in fact, one checks that $\mathrm{N}$ and $H+\mathrm{d}\Gamma ^{\otimes
}\left( \mu \right) $ strongly commute in the sense of \cite[Definition 5.2]%
{Konrad}. By\ Assumption SB4 in Theorem \ref{HamilSA}, $H$ is a lower
semibounded self-adjoint operator: $H\geq b\mathbf{1}$ for some $b\in 
\mathbb{R}$. Then, by the functional calculus,\ one thus deduces that, for
any $\lambda \in \mathbb{R}\backslash \{0\}$, 
\begin{eqnarray}
\left\Vert \left( \mathrm{N}+\mathbf{1}\right) \left( H+\mathrm{d}\Gamma
^{\otimes }\left( \mu \right) +i\lambda \mathbf{1}\right) ^{-1}\right\Vert _{%
\mathrm{op}} &\leq &\left\Vert \left( \mathrm{N}+\mathbf{1}\right) \left( H+c%
\mathrm{N}+i\lambda \mathbf{1}\right) ^{-1}\right\Vert _{\mathrm{op}}  \notag
\\
&\leq &\sup_{k\in \mathbb{N}_{0}}\sup_{x\in \left[ b,\infty \right) }\frac{%
k+1}{\sqrt{\left( ck+x\right) ^{2}+\lambda ^{2}}}\doteq \mathrm{K}%
_{c,b}<\infty .  \label{inequality Q1+}
\end{eqnarray}%
Hence, by (\ref{gfhfghfghh})--(\ref{gfhfghfghh2}) and (\ref{inequality Q1+}%
), if $\lambda >2^{7/2}\Vert X\Vert _{\mathrm{op},2}$ then 
\begin{equation}
\left\Vert \left( \mathrm{N}+\mathbf{1}\right) ^{\frac{1}{2}}\left( \mathrm{H%
}+i\lambda \mathbf{1}\right) ^{-1}\right\Vert _{\mathrm{op}}\leq \mathrm{K}%
_{c,b}\left( 1+\frac{12\left\Vert X\right\Vert _{\mathrm{op},2}}{\lambda
-2^{7/2}\left\Vert X\right\Vert _{\mathrm{op},2}}\right) ,  \label{fdg1}
\end{equation}%
which, combined with (\ref{sdfsgdfhfghjfhjjhk})--(\ref{qweqweqe}) and (\ref%
{gfhfghfghh1})--(\ref{gfhfghfghh2}), implies the first inequality.

Observe from (\ref{inequality Q1+}) that $\mathcal{D}\left( H+\mathrm{d}%
\Gamma ^{\otimes }\left( \mu \right) \right) \subseteq \mathcal{D}\left( 
\mathrm{N}\right) $, which, combined with Theorem \ref{HamilSA}, in turn
implies that $\mathcal{D}\left( \mathrm{H}\right) \subseteq \mathcal{D}%
\left( \mathrm{N}\right) $. In fact, in the same way we deduce (\ref%
{sdfsgdfhfghjfhjjhk}) and (\ref{gfhfghfghh}), we have 
\begin{eqnarray}
&&\left\Vert \left( \mathrm{N}+\mathbf{1}\right) \left( \mathrm{H}+i\lambda 
\mathbf{1}\right) ^{-1}\right\Vert _{\mathrm{op}}  \notag \\
&\leq &\left\Vert \left( \mathrm{N}+\mathbf{1}\right) \left( H+\mathrm{d}%
\Gamma ^{\otimes }\left( \mu \right) +i\lambda \mathbf{1}\right)
^{-1}\right\Vert _{\mathrm{op}}  \notag \\
&&\times \left( 1+\left\Vert \left( \mathrm{N}+\mathbf{1}\right) \left( 
\mathrm{H}+i\lambda \mathbf{1}\right) ^{-1}\left( \mathrm{N}+\mathbf{1}%
\right) ^{-\frac{1}{2}}\right\Vert _{\mathrm{op}}\left\Vert \left( \mathrm{N}%
+\mathbf{1}\right) ^{\frac{1}{2}}\mathrm{X}\left( \mathrm{N}+\mathbf{1}%
\right) ^{-1}\right\Vert _{\mathrm{op}}\right) ,  \label{dfg1}
\end{eqnarray}%
where%
\begin{equation}
\left( \mathrm{N}+\mathbf{1}\right) \left( \mathrm{H}+i\lambda \mathbf{1}%
\right) ^{-1}\left( \mathrm{N}+\mathbf{1}\right) ^{-\frac{1}{2}}=\left( 
\mathrm{N}+\mathbf{1}\right) ^{\frac{1}{2}}\left( \mathrm{H}+i\lambda 
\mathbf{1}\right) ^{-1}\sum_{k\in \mathbb{N}_{0}}\left( -1\right)
^{k}\left\{ \left( \mathrm{H}+i\lambda \mathbf{1}\right) ^{-1}\mathbb{T}%
_{1/2}\right\} ^{k}  \label{dfg2}
\end{equation}%
for $\lambda >8\Vert X\Vert _{\mathrm{op},2}$. Indeed, the above Neumann
series absolutely converges when $\lambda >8\Vert X\Vert _{\mathrm{op},2}$,
thanks to Lemma \ref{RelativeboundGlemma} for $n=0$. Using (\ref{gfhfghfghh2}%
) and (\ref{inequality Q1+}) we infer from (\ref{dfg1}) that, for $\lambda
>2^{7/2}\Vert X\Vert _{\mathrm{op},2}$, 
\begin{equation}
\left\Vert \left( \mathrm{N}+\mathbf{1}\right) \left( \mathrm{H}+i\lambda 
\mathbf{1}\right) ^{-1}\right\Vert _{\mathrm{op}}\leq \mathrm{K}_{c,b}\left(
1+12\left\Vert X\right\Vert _{\mathrm{op},2}\left\Vert \left( \mathrm{N}+%
\mathbf{1}\right) \left( \mathrm{H}+i\lambda \mathbf{1}\right) ^{-1}\left( 
\mathrm{N}+\mathbf{1}\right) ^{-\frac{1}{2}}\right\Vert _{\mathrm{op}%
}\right) ,  \label{as1}
\end{equation}%
while, by (\ref{fdg1}), (\ref{dfg2}) and Lemma \ref{RelativeboundGlemma} for 
$n=0$, 
\begin{equation}
\left\Vert \left( \mathrm{N}+\mathbf{1}\right) \left( \mathrm{H}+i\lambda 
\mathbf{1}\right) ^{-1}\left( \mathrm{N}+\mathbf{1}\right) ^{-\frac{1}{2}%
}\right\Vert _{\mathrm{op}}\leq \mathrm{K}_{c,b}\left( 1+\frac{12\left\Vert
X\right\Vert _{\mathrm{op},2}}{\lambda -2^{7/2}\left\Vert X\right\Vert _{%
\mathrm{op},2}}\right) \frac{1}{1-8\lambda ^{-1}\left\Vert X\right\Vert _{%
\mathrm{op},2}}.  \label{as2}
\end{equation}%
The final upper bound of the lemma follows immediately by combining
Inequalities (\ref{as1})--(\ref{as2}).
\end{proof}

\subsection{Spin-Boson Interactions of Order 3}

We now turn our attention to an alternative class of spin-boson interactions
given by 
\begin{equation}
\mathbb{Y}\doteq \frac{1}{2}\sum_{k\in \mathbb{N}}Y_{k}\otimes \left(
a_{k}a_{k}^{\ast }a_{k}+a_{k}^{\ast }a_{k}a_{k}\right) \qquad \text{and}%
\qquad \mathbb{Y}^{\dag }\doteq \frac{1}{2}\sum_{k\in \mathbb{N}}Y_{k}^{\ast
}\otimes \left( a_{k}^{\ast }a_{k}a_{k}^{\ast }+a_{k}^{\ast }a_{k}^{\ast
}a_{k}\right)  \label{Ydef}
\end{equation}%
for any sequence $Y\equiv (Y_{k})_{k\in \mathbb{N}}$ of operators $Y_{k}\in 
\mathcal{B}(\mathfrak{h}_{k})\equiv \mathcal{B}(\mathfrak{h}_{k})^{\otimes }$%
. They do not appear in the expression of generalized spin-boson
Hamiltonians per se, but they are generated by our method of analysis; see
Equation (\ref{ldfj}). Clearly, $\mathbb{Y}$ is well defined on $\mathcal{D}%
_{\infty }$ and its adjoint is equal to the expression $\mathbb{Y}^{\dag }$
on $\mathcal{D}_{\infty }$. These correspond to spin-boson interactions of
order three in the creation and annihilation operators. We study here their
relative boundedness with respect to $\mathrm{N}^{3/2}$.

\begin{lemma}[Relative boundedness of $\mathbb{Y+Y}^{\dag }$ with respect to 
$\mathrm{N}^{3/2}$]
\label{Lemma-self-adjoint1 copy(1)}\mbox{}\newline
Let $Y\equiv (Y_{k})_{k\in \mathbb{N}}$ be a sequence of operators $Y_{k}\in 
\mathcal{B}(\mathfrak{h}_{k})\equiv \mathcal{B}(\mathfrak{h}_{k})^{\otimes }$%
. Then, for all vectors $\varphi \in \mathcal{D}_{\infty }$, 
\begin{equation*}
\left\Vert \mathbb{Y}\varphi \right\Vert _{\mathfrak{F}}+\left\Vert \mathbb{Y%
}^{\dag }\varphi \right\Vert _{\mathfrak{F}}\leq 10\left\Vert Y\right\Vert _{%
\mathrm{op},2}\left( \left\Vert \mathrm{N}^{3/2}\varphi \right\Vert _{%
\mathfrak{F}}+\left\Vert \varphi \right\Vert _{\mathfrak{F}}\right) .
\end{equation*}
\end{lemma}

\begin{proof}
Fix all parameters of the lemma. Note also from the CCR (\ref{CCR}) that%
\begin{equation}
\mathbb{Y}=\mathbb{A}-\frac{1}{2}\sum_{k\in \mathbb{N}}Y_{k}\otimes
a_{k}\qquad \text{and}\qquad \mathbb{Y}^{\dag }=\mathbb{A}^{\dag }-\frac{1}{2%
}\sum_{k\in \mathbb{N}}Y_{k}^{\ast }\otimes a_{k}^{\ast },  \label{Y}
\end{equation}%
where $\mathbb{A}$ and $\mathbb{A}^{\dag }$ are the operators defined on $%
\mathcal{D}_{\infty }$ by 
\begin{equation*}
\mathbb{A}\doteq \sum_{k\in \mathbb{N}}Y_{k}\otimes a_{k}a_{k}^{\ast
}a_{k}\qquad \text{and}\qquad \mathbb{A}^{\dag }\doteq \sum_{k\in \mathbb{N}%
}Y_{k}^{\ast }\otimes a_{k}^{\ast }a_{k}a_{k}^{\ast }.
\end{equation*}%
See also Lemma \ref{Lemma-self-adjoint1}. Then, for any vector $\varphi \in 
\mathcal{D}_{\infty }$, the Cauchy-Schwarz inequality yields 
\begin{eqnarray}
\left\Vert \mathbb{A}\varphi \right\Vert _{\mathfrak{F}}^{2} &\leq
&\sum_{k\in \mathbb{N}}\left\Vert \mathbf{1}_{\mathfrak{H}}\otimes
a_{k}a_{k}^{\ast }a_{k}\varphi \right\Vert _{\mathfrak{F}}^{2}\ \ \sum_{k\in 
\mathbb{N}}\left\Vert Y_{k}\right\Vert _{\mathrm{op}}^{2}  \notag \\
&=&\sum_{k\in \mathbb{N}}\left\langle \varphi ,\mathbf{1}_{\mathfrak{H}%
}\otimes \left( a_{k}^{\ast }a_{k}\right) ^{3}\varphi \right\rangle _{%
\mathfrak{F}}\ \ \sum_{k\in \mathbb{N}}\left\Vert Y_{k}\right\Vert _{\mathrm{%
op}}^{2}  \notag \\
&=&\left\Vert \mathrm{N}^{3/2}\varphi \right\Vert _{\mathfrak{F}}^{2}\
\sum_{k\in \mathbb{N}}\left\Vert Y_{k}\right\Vert _{\mathrm{op}}^{2}.
\label{wet1}
\end{eqnarray}%
Observe meanwhile that, for any $k\in \mathbb{N}$,%
\begin{equation*}
a_{k}a_{k}^{\ast }a_{k}a_{k}^{\ast }a_{k}a_{k}^{\ast }=a_{k}^{\ast
}a_{k}a_{k}^{\ast }a_{k}a_{k}^{\ast }a_{k}+3a_{k}^{\ast }a_{k}a_{k}^{\ast
}a_{k}+3a_{k}^{\ast }a_{k}+\mathbf{1}_{\mathcal{F}_{+}},
\end{equation*}%
using the CCR (\ref{CCR}). It follows that, for any $\varphi \in \mathcal{D}%
_{\infty }$,%
\begin{eqnarray*}
\left\Vert \mathbb{A}^{\dag }\varphi \right\Vert _{\mathfrak{F}}^{2}
&=&\sum_{k,q\in \mathbb{N}}\left\langle \varphi ,Y_{k}Y_{q}^{\ast }\otimes
a_{k}a_{k}^{\ast }a_{k}a_{q}^{\ast }a_{q}a_{q}^{\ast }\varphi \right\rangle
_{\mathfrak{F}} \\
&=&\sum_{k,q\in \mathbb{N}}\left\langle \varphi ,Y_{k}Y_{q}^{\ast }\otimes
a_{q}^{\ast }a_{q}a_{q}^{\ast }a_{k}a_{k}^{\ast }a_{k}\varphi \right\rangle
_{\mathfrak{F}}+3\sum_{k\in \mathbb{N}}\left\langle \varphi
,Y_{k}Y_{k}^{\ast }\otimes a_{k}^{\ast }a_{k}a_{k}^{\ast }a_{k}\varphi
\right\rangle _{\mathfrak{F}} \\
&&+3\sum_{k\in \mathbb{N}}\left\langle \varphi ,Y_{k}Y_{k}^{\ast }\otimes
a_{k}^{\ast }a_{k}\varphi \right\rangle _{\mathfrak{F}}+\sum_{k\in \mathbb{N}%
}\left\langle \varphi ,Y_{k}Y_{k}^{\ast }\otimes \mathbf{1}_{\mathcal{F}%
_{+}}\varphi \right\rangle _{\mathfrak{F}}.
\end{eqnarray*}%
Therefore, by the Cauchy-Schwarz inequality, we get that 
\begin{eqnarray*}
\left\Vert \mathbb{A}^{\dag }\varphi \right\Vert _{\mathfrak{F}}^{2} &\leq
&\sum_{k\in \mathbb{N}}\left\langle \varphi ,\mathbf{1}_{\mathfrak{H}%
}\otimes \left( a_{k}^{\ast }a_{k}\right) ^{3}\varphi \right\rangle _{%
\mathfrak{F}}\ \ \sum_{k\in \mathbb{N}}\left\Vert Y_{k}\right\Vert _{\mathrm{%
op}}^{2} \\
&&+3\sup_{k\in \mathbb{N}}\left\Vert Y_{k}\right\Vert _{\mathrm{op}%
}^{2}\sum_{k\in \mathbb{N}}\left\langle \varphi ,\mathbf{1}_{\mathfrak{H}%
}\otimes \left( a_{k}^{\ast }a_{k}\right) ^{2}\varphi \right\rangle \\
&&+3\sup_{k\in \mathbb{N}}\left\Vert Y_{k}\right\Vert _{\mathrm{op}%
}^{2}\sum_{k\in \mathbb{N}}\left\langle \varphi ,\mathbf{1}_{\mathfrak{H}%
}\otimes a_{k}^{\ast }a_{k}\varphi \right\rangle _{\mathfrak{F}}+\sum_{k\in 
\mathbb{N}}\left\Vert Y_{k}\right\Vert _{\mathrm{op}}^{2}\left\Vert \varphi
\right\Vert _{\mathfrak{F}}^{2}
\end{eqnarray*}%
for any $\varphi \in \mathcal{D}_{\infty }$. We thus deduce that 
\begin{equation}
\left\Vert \mathbb{A}^{\dag }\varphi \right\Vert _{\mathfrak{F}}^{2}\leq
\left\Vert \left( \mathrm{N}^{3}+\mathbf{1}\right) ^{1/2}\varphi \right\Vert
_{\mathfrak{F}}^{2}\ \ \sum_{k\in \mathbb{N}}\left\Vert Y_{k}\right\Vert _{%
\mathrm{op}}^{2}+3\left\Vert \left( \mathrm{N}\left( \mathrm{N}+\mathbf{1}%
\right) \right) ^{1/2}\varphi \right\Vert _{\mathfrak{F}}^{2}\sup_{k\in 
\mathbb{N}}\left\Vert Y_{k}\right\Vert _{\mathrm{op}}^{2}.  \label{wet2}
\end{equation}%
Combining this last inequality with (\ref{Y})--(\ref{wet1}), Lemma \ref%
{Lemma-self-adjoint1} and elementary estimates\footnote{%
E.g., $6n^{3}\geq 3n\left( n+1\right) $ and $n^{3}\geq n$ for all $n\in 
\mathbb{N}$.}, we arrive at the assertion.
\end{proof}

\section{Unitary Flow on Spin-Boson Models\label{Unitary Flow}}

In Section \ref{Fiberflowsection}, we study the non-linear operator-valued
differential equations (\ref{flowoneparticle}) for each $k\in \mathbb{N}$ on
the corresponding Hilbert space $\mathfrak{h}_{k}$. The solution to these
differential equations allows us to rigorously define and analyze the
time-dependent generalized spin-boson model (\ref{defHt}), as explained in
Section \ref{Small-Coupling Limit}.

\subsection{Generators of the Unitary Flow}

As explained in Section \ref{Brockett-Wegner Flow}, we setup a strongly
continuous two-parameter flow of unitary operators generated by the
time-dependent generator $\mathrm{G}_{t}$, $t\in \mathbb{R}_{0}^{+}$,
defined by (\ref{generator}), that is,%
\begin{equation}
\mathrm{G}_{t}\doteq i\sum_{k\in \mathbb{N}}\left( X_{k,t}^{\ast }\otimes
a_{k}^{\ast }-X_{k,t}\otimes a_{k}\right) ,\qquad t\in \mathbb{R}_{0}^{+},
\label{def_Gt}
\end{equation}%
where, for each $t\in \mathbb{R}_{0}^{+}$, $X_{t}\equiv (X_{k,t})_{k\in 
\mathbb{N}}$ is a sequence of operators $X_{k,t}\in \mathcal{B}(\mathfrak{h}%
_{k})\equiv \mathcal{B}(\mathfrak{h}_{k})^{\otimes }$ satisfying 
\begin{equation*}
\left\Vert X_{t}\right\Vert _{\mathrm{op},2}\doteq \left( \sum_{k\in \mathbb{%
N}}\left\Vert X_{k,t}\right\Vert _{\mathrm{op}}^{2}\right) ^{1/2}<\infty
,\qquad t\in \mathbb{R}_{0}^{+},
\end{equation*}%
(see (\ref{lp-operator norm})). By Proposition \ref{corollarire-selfadjoint2}%
, for each time $t\in \mathbb{R}_{0}^{+}$, this operator is rigorously
defined on the domain $\mathcal{D}(\mathrm{N}^{1/2})$, where it is
essentially self-adjoint. To simplify, we use the same notation $\mathrm{G}%
_{t}\equiv \mathrm{G}_{t}^{\ast \ast }$, $t\in \mathbb{R}_{0}^{+}$, for
their self-adjoint extension. Note also from Lemma \ref{Lemma-self-adjoint1}
that the (now self-adjoint) generators $\mathrm{G}_{t}$, $t\in \mathbb{R}%
_{0}^{+}$, are relatively bounded with respect to the particle number
operator $\mathrm{N}$ (cf. (\ref{notation abuse0})--(\ref{notation abuse})).

Indeed, according to Lemma \ref{Lemma-self-adjoint1}, 
\begin{equation}
\beta _{0}\left( t\right) \doteq \Vert \mathrm{G}_{t}\left( \mathrm{N}+%
\mathbf{1}\right) ^{-1/2}\Vert _{\mathrm{op}}\leq 4\left\Vert
X_{t}\right\Vert _{\mathrm{op},2},\qquad t\in \mathbb{R}_{0}^{+},
\label{relativeboundG}
\end{equation}%
and 
\begin{equation}
\left\vert \beta _{0}\left( t\right) -\beta _{0}\left( s\right) \right\vert
\leq \Vert \left( \mathrm{G}_{t}-\mathrm{G}_{s}\right) \left( \mathrm{N}+%
\mathbf{1}\right) ^{-1/2}\Vert _{\mathrm{op}}\leq 4\left\Vert
X_{t}-X_{s}\right\Vert _{\mathrm{op},2},\qquad s,t\in \mathbb{R}_{0}^{+},
\label{relativeboundG2}
\end{equation}%
As in the previous inequalities, recall that we adopt the shorthand notation 
$\mathbf{1}_{\mathfrak{F}}\equiv \mathbf{1}$ throughout, unless the context
requires explicit clarification. The final inequality implies that $\beta
_{0}$ is continuous provided that the mapping $t\mapsto X_{t}$ is continuous
with respect to the norm $\Vert \cdot \Vert _{\mathrm{op},2}$, i.e., $X\in C(%
\mathbb{R}_{0}^{+};\ell ^{2}(\mathcal{B}(\mathfrak{h})))$.

Note also from Lemmata \ref{Lemma-self-adjoint1 copy(2)} and \ref%
{RelativeboundGlemma} (see (\ref{fg3})) that for all $n\in \mathbb{N}$, the
expression 
\begin{equation*}
\mathbb{T}_{n/2}\left( t\right) \equiv \left[ (\mathrm{N}+\mathbf{1})^{n/2},%
\mathrm{G}_{t}\right] (\mathrm{N}+\mathbf{1})^{-n/2},\qquad t\in \mathbb{R}%
_{0}^{+},
\end{equation*}%
defines a family of bounded operators acting on $\mathfrak{F}$ with an
operator norm always bounded by 
\begin{equation}
\beta _{n/2}\left( t\right) \doteq \left\Vert \mathbb{T}_{n/2}\left(
t\right) \right\Vert _{\mathrm{op}}\leq 2^{3+n/2}\left\Vert X_{t}\right\Vert
_{\mathrm{op},2},\qquad t\in \mathbb{R}_{0}^{+},  \label{relativeboundG3}
\end{equation}%
and 
\begin{equation}
\left\vert \beta _{n/2}\left( t\right) -\beta _{n/2}\left( s\right)
\right\vert \leq \left\Vert \mathbb{T}_{n/2}\left( t\right) -\mathbb{T}%
_{n/2}\left( s\right) \right\Vert _{\mathrm{op}}\leq 2^{3+n/2}\left\Vert
X_{t}-X_{s}\right\Vert _{\mathrm{op},2},\qquad s,t\in \mathbb{R}_{0}^{+}.
\label{relativeboundG4}
\end{equation}%
The last equation shows that the function $\beta _{n/2}$ is also continuous
when the mapping $t\mapsto X_{t}$ is continuous in the norm $\Vert \cdot
\Vert _{\mathrm{op},2}$.

To construct the desired continuous flow, the self-adjointness of $\mathrm{G}%
_{t}$ for $t\in \mathbb{R}_{0}^{+}$ along with the above relative bounds
play a pivotal role. Indeed, in the next subsection, we invoke the theory of
non-autonomous (Kato-hyperbolic) evolution equations following the approach
explained in \cite[Section VII.1]{bach-bru-memo}.

\subsection{Unitary Propagators}

Given a closed set $I\subseteq \mathbb{R}^{2}$, an evolution system is a
family $(U_{t,s})_{(s,t)\in I}$ of bounded operators acting on a Banach
space that is jointly strongly continuous in $s$ and $t$ and satisfy $%
U_{s,s}\doteq \mathbf{1}$ as well as the cocycle (or Chapman--Kolmogorov)
property: $U_{t,s}=U_{t,x}U_{x,s}$ for any $(s,t),(t,x),(x,s)\in I$. If the
operators are always unitary, the system is called a unitary propagator.

In this section, we set up a unitary propagator generated by the
time-dependent operators $\mathrm{G}_{t}$ for $t\in \mathbb{R}_{0}^{+}$, as
defined by (\ref{def_Gt}). In other words, we establish a solution $(\mathrm{%
U}_{t,s})_{s,t\in \mathbb{R}_{0}^{+}}\subseteq \mathcal{B}\left( \mathfrak{F}%
\right) $ to the non-autonomous evolution equation%
\begin{equation}
\forall s,t\in \mathbb{R}_{0}^{+}:\qquad 
\begin{array}{llll}
\partial _{t}\mathrm{U}_{t,s}=-i\mathrm{G}_{t}\mathrm{U}_{t,s} & , & \mathrm{%
U}_{s,s}\doteq \mathbf{1} & ,%
\end{array}
\label{poui1}
\end{equation}%
on some dense subset of $\mathfrak{F}$. This can be done using Kato's
results \cite{Kato,Kato1973} on the well-posedness of non-autonomous
evolution equations. His idea was to discretize the evolution equation in
order to use the Hille-Yosida generation theorems and take the continuous
limit to obtain a well-defined solution. Another strategy is to use Dyson
series with approximate generators obtained with the Yosida approximation.
That is what Ishii did \cite{Ishii1,Ishii2}. This strategy was re-examined
in the monograph \cite[Section VII.1]{bach-bru-memo}, with arguments
technically different from those of Ishii. We use this more recent approach
here, as it has the advantage of being well adapted to our problem.

As shown in Inequality (\ref{relativeboundG2}), the mapping $t\mapsto X_{t}$
must be continuous with respect to the norm $\Vert \cdot \Vert _{\mathrm{op}%
,2}$, i.e., $X\in C(\mathbb{R}_{0}^{+};\ell ^{2}(\mathcal{B}(\mathfrak{h})))$
Recall that $\ell ^{2}(\mathcal{B}(\mathfrak{h}))$ denotes the Banach space (%
\ref{L2}) of sequences $X_{k}\in \mathcal{B}(\mathfrak{h}_{k})$, $k\in 
\mathbb{N}$, equipped with the $\ell ^{2}$--operator norm $\Vert \cdot \Vert
_{\mathrm{op},2}$. As for notations, we use $A\lesssim B$ to denote $A\leq
DB $ for some positive constant $D$. We write $A\lesssim _{n}B$ to indicate
that the implicit constant $D=D_{n}$ depends on the parameter $n$. If $%
s,t\in \mathbb{R}$ are real numbers, then $s\wedge t$\ and $s\vee t$ stand
respectively for the minimum and the maximum of $s,t$.

We are now in a position to show the existence of the unitary propagator as
a solution to the non-autonomous evolution equation (\ref{poui1}).

\begin{theorem}[Unitary propagators]
\label{propagators}\mbox{}\newline
Assume $X\in C(\mathbb{R}_{0}^{+};\ell ^{2}(\mathcal{B}(\mathfrak{h})))$.
Then, there is a family $(\mathrm{U}_{t,s})_{s,t\in \mathbb{R}%
_{0}^{+}}\subseteq \mathcal{B}\left( \mathfrak{F}\right) $ of bounded
operators satisfying the following properties:

\begin{enumerate}
\item[\emph{(i)}] For any $s,t\in \mathbb{R}_{0}^{+}$, $\mathrm{U}_{s,t}=%
\mathrm{U}_{t,s}^{\ast }$ is a unitary operator.

\item[\emph{(ii)}] It satisfies the cocycle property\ $\mathrm{U}_{t,x}%
\mathrm{U}_{x,s}=\mathrm{U}_{t,s}$ for any $s,x,t\in \mathbb{R}_{0}^{+}$.

\item[\emph{(iii)}] It is strongly continuous on $(\mathbb{R}_{0}^{+})^{2}$.

\item[\emph{(iv)}] For any $s,t\in \mathbb{R}_{0}^{+}$, and all $n\in 
\mathbb{N}_{0}$, 
\begin{equation*}
\ln \Vert (\mathrm{N}+\mathbf{1})^{\frac{n}{2}}\mathrm{U}_{t,s}(\mathrm{N}+%
\mathbf{1})^{-\frac{n}{2}}\Vert _{\mathrm{op}}\lesssim _{n}\int_{s\wedge
t}^{s\vee t}\left\Vert X_{\tau }\right\Vert _{\mathrm{op},2}\mathrm{d}\tau
\end{equation*}%
and the mapping $\left( s,t\right) \mapsto (\mathrm{N}+\mathbf{1})^{n+1/2}%
\mathrm{U}_{t,s}(\mathrm{N}+\mathbf{1})^{-n-1/2}$ is jointly strongly
continuous on $(\mathbb{R}_{0}^{+})^{2}$.

\item[\emph{(v)}] It is the unique\footnote{%
Note that $\partial _{t}\mathrm{U}_{t,s}=-i\mathrm{G}_{t}\mathrm{U}_{t,s}$
with $\mathrm{U}_{s,s}\doteq \mathbf{1}$ for $s,t\in \mathbb{R}_{0}^{+}$ has
already a unique solution.} evolution system satisfying on the domain $%
\mathcal{D}(\mathrm{N}^{1/2})$ the non-autonomous evolution equations 
\begin{equation*}
\forall s,t\in \mathbb{R}_{0}^{+}:\qquad \left\{ 
\begin{array}{llll}
\partial _{t}\mathrm{U}_{t,s}=-i\mathrm{G}_{t}\mathrm{U}_{t,s} & , & \mathrm{%
U}_{s,s}\doteq \mathbf{1} & . \\ 
\partial _{s}\mathrm{U}_{t,s}=i\mathrm{U}_{t,s}\mathrm{G}_{s} & , & \mathrm{U%
}_{t,t}\doteq \mathbf{1} & .%
\end{array}%
\right.
\end{equation*}
\end{enumerate}
\end{theorem}

\begin{proof}
The proof is done in several steps: \medskip

\noindent \underline{Step 1:} The existence of a unique evolution system $%
\left( \mathrm{U}_{t,s}\right) _{t\geq s\geq 0}$ satisfying on the domain $%
\mathcal{D}(\mathrm{N}^{1/2})$ the non-autonomous evolution equations%
\footnote{\label{try}The derivatives $\partial _{t}$ and $\partial _{s}$ on
the borderline $t=s$ or $s=0$ have to be understood as either right or left
derivatives.} 
\begin{equation}
\forall t\geq s\geq 0:\qquad \left\{ 
\begin{array}{llll}
\partial _{t}\mathrm{U}_{t,s}=-i\mathrm{G}_{t}\mathrm{U}_{t,s} & , & \mathrm{%
U}_{s,s}\doteq \mathbf{1} & , \\ 
\partial _{s}\mathrm{U}_{t,s}=i\mathrm{U}_{t,s}\mathrm{G}_{s} & , & \mathrm{U%
}_{t,t}\doteq \mathbf{1} & ,%
\end{array}%
\right.  \label{flow equationbis-newbis}
\end{equation}%
is a direct consequence of \cite[Theorem 88]{bach-bru-memo}: The
infinitesimal generator $\mathrm{G}_{t}$ is self-adjoint for all times $t\in 
\mathbb{R}_{0}^{+}$ and the assumption B1 of \cite[Section VII.1]%
{bach-bru-memo} is immediate. Using the closed auxiliary operator $\Theta
\doteq (\mathrm{N}+\mathbf{1})^{1/2}$ and the space 
\begin{equation*}
\mathcal{Y}\doteq \mathcal{D}(\Theta )=\left\{ \varphi \in \mathfrak{F}%
:\left\Vert \varphi \right\Vert _{\mathcal{Y}}\doteq \Vert (\mathrm{N}+%
\mathbf{1})^{1/2}\varphi \Vert _{\mathfrak{F}}<\infty \right\} ,
\end{equation*}%
we deduce the assumptions B2 and B3 of \cite[Section VII.1]{bach-bru-memo}
from Inequalities (\ref{relativeboundG})--(\ref{relativeboundG3}) and $X\in
C(\mathbb{R}_{0}^{+};\ell ^{2}(\mathcal{B}(\mathfrak{h})))$. So we can
invoke \cite[Theorem 88]{bach-bru-memo} to get the unique solution $\left( 
\mathrm{U}_{t,s}\right) _{t\geq s\geq 0}$ to (\ref{flow equationbis-newbis}%
). Note that $\mathrm{U}_{t,x}\mathrm{U}_{x,s}=\mathrm{U}_{t,s}$ for any $%
s,x,t\in \mathbb{R}_{0}^{+}$ satisfying $s\leq x\leq t$. Furthermore,
because of Inequalities (\ref{relativeboundG3})--(\ref{relativeboundG4}) and 
\cite[Lemma 92]{bach-bru-memo}, for all $s,t\in \mathbb{R}_{0}^{+}$, $t\geq
s $, and each $n\in \mathbb{N}_{0}$, there exists a family $(D_{n})_{n\in 
\mathbb{N}_{0}}\subseteq \mathbb{R}^{+}$ of strictly positive constants --
depending solely on $n\in \mathbb{N}_{0}$ -- such that 
\begin{equation}
\Vert (\mathrm{N}+\mathbf{1})^{\frac{n}{2}}\mathrm{U}_{t,s}(\mathrm{N}+%
\mathbf{1})^{-\frac{n}{2}}\Vert _{\mathrm{op}}\leq \mathrm{\exp }\left\{
D_{n}\int_{s}^{t}\Vert X_{\tau }\Vert _{\mathrm{op},2}\mathrm{d}\tau \right\}
\label{BounddefDn_ordre}
\end{equation}%
and the mapping $\left( s,t\right) \mapsto (\mathrm{N}+\mathbf{1})^{n/2}%
\mathrm{U}_{t,s}(\mathrm{N}+\mathbf{1})^{-n/2}$ is jointly strongly
continuous in $s$ and $t$. \medskip

\noindent \underline{Step 2:} Step 1 neither implies that the adjoint $%
\mathrm{U}_{t,s}^{\ast }$ of the evolution operator $\mathrm{U}_{t,s}$
conserves the dense domain $\mathcal{D}(\mathrm{N}^{1/2})$, nor that it is
jointly strongly continuous in $s$ and $t$ for all $t\geq s\geq 0$. However,
in the same way we prove Step 1, one checks the existence of a unique
evolution system $(\mathrm{V}_{t,s})_{t\geq s\geq 0}$ satisfying on the
domain $\mathcal{D}(\mathrm{N}^{1/2})$ the non-autonomous evolution
equations $^{\ref{try}}$ 
\begin{equation*}
\forall t\geq s\geq 0:\qquad \left\{ 
\begin{array}{llll}
\partial _{s}\mathrm{V}_{t,s}=-i\mathrm{G}_{s}\mathrm{V}_{t,s} & , & \mathrm{%
V}_{t,t}\doteq \mathbf{1} & . \\ 
\partial _{t}\mathrm{V}_{t,s}=i\mathrm{V}_{t,s}\mathrm{G}_{t} & , & \mathrm{V%
}_{s,s}\doteq \mathbf{1} & .%
\end{array}%
\right.
\end{equation*}%
Moreover, because of Inequalities (\ref{relativeboundG3})--(\ref%
{relativeboundG4}) and \cite[Lemma 92]{bach-bru-memo}, $\mathrm{V}_{t,s}$
conserves the domain $\mathcal{D}(\mathrm{N}^{1/2})$ for all $s,t\in \mathbb{%
R}_{0}^{+}$, $t\geq s$, and for each each $n\in \mathbb{N}_{0}$, there
exists a family $(D_{n}^{\prime })_{n\in \mathbb{N}_{0}}\subseteq \mathbb{R}%
^{+}$ of strictly positive constants -- depending solely on $n\in \mathbb{N}%
_{0}$ -- such that 
\begin{equation*}
\Vert (\mathrm{N}+\mathbf{1})^{\frac{n}{2}}\mathrm{V}_{t,s}(\mathrm{N}+%
\mathbf{1})^{-\frac{n}{2}}\Vert _{\mathrm{op}}\leq \mathrm{\exp }\left\{
D_{n}^{\prime }\int_{s}^{t}\Vert X_{\tau }\Vert _{\mathrm{op},2}\mathrm{d}%
\tau \right\} .
\end{equation*}%
Furthermore, the mapping $\left( s,t\right) \mapsto (\mathrm{N}+\mathbf{1}%
)^{n/2}\mathrm{V}_{t,s}(\mathrm{N}+\mathbf{1})^{-n/2}$ is jointly strongly
continuous in $s$ and $t$.\medskip

\noindent \underline{Step 3:} We prove here the unitarity of the operator $%
\mathrm{U}_{t,s}$. When $t\geq s\geq 0$, note that \cite[Theorem 88]%
{bach-bru-memo} gives both the operators $\mathrm{V}_{t,s}$ and $\mathrm{U}%
_{t,s}$ as limit $\lambda \rightarrow \infty $ of operators $\mathrm{V}%
_{t,s,\lambda }=\mathrm{U}_{t,s,\lambda }^{\ast }$ and $\mathrm{U}%
_{t,s,\lambda }$ defined as in \cite[Equation (VII.6)]{bach-bru-memo}, both
having a norm convergent Dyson series representation. From this, one deduces
that $\mathrm{V}_{t,s}=\mathrm{U}_{t,s}^{\ast }$ for all times satisfying $%
t\geq s\geq 0$. For more details, see the arguments proving \cite[Lemma 68]%
{bach-bru-memo}. Using now Steps 1 and 2, we deduce that, when $t\geq s\geq
0 $, 
\begin{equation*}
(\mathbf{1}-\mathrm{U}_{t,s}\mathrm{U}_{t,s}^{\ast })(\mathrm{N}+\mathbf{1}%
)^{-1/2}=\int_{s}^{t}\partial _{\tau }\left\{ \mathrm{U}_{t,\tau }\mathrm{U}%
_{t,\tau }^{\ast }\right\} (\mathrm{N}+\mathbf{1})^{-1/2}\mathrm{d}\tau =0
\end{equation*}%
and 
\begin{equation*}
(\mathbf{1}-\mathrm{U}_{t,s}^{\ast }\mathrm{U}_{t,s})(\mathrm{N}+\mathbf{1}%
)^{-1/2}=-\int_{s}^{t}\partial _{\tau }\left\{ \mathrm{U}_{\tau ,s}^{\ast }%
\mathrm{U}_{\tau ,s}\right\} (\mathrm{N}+\mathbf{1})^{-1/2}\mathrm{d}\tau
=0\ .
\end{equation*}%
Thus, for all times such that $t\geq s\geq 0$, $\mathrm{U}_{t,s}^{\ast }%
\mathrm{U}_{t,s}=\mathrm{U}_{t,s}\mathrm{U}_{t,s}^{\ast }=\mathbf{1}$ on the
dense domain $\mathcal{D}(\mathrm{N}^{1/2})\subseteq \mathfrak{F}$, which
immediately implies the unitarity of $\mathrm{U}_{t,s}\in \mathcal{B}(%
\mathfrak{F})$. \medskip

\noindent \underline{Step 4:} We finally use the definition $\mathrm{U}%
_{t,s}\doteq \mathrm{U}_{s,t}^{\ast }$ when $s\geq t\geq 0$. Then, by
combining Steps 1-3, one deduces all the assertions of the theorem. In
particular, (\ref{BounddefDn_ordre}) now reads as 
\begin{equation}
\Vert (\mathrm{N}+\mathbf{1})^{\frac{n}{2}}\mathrm{U}_{t,s}(\mathrm{N}+%
\mathbf{1})^{-\frac{n}{2}}\Vert _{\mathrm{op}}\leq \mathrm{\exp }\left\{
D_{n}\int_{s\wedge t}^{s\vee t}\Vert X_{\tau }\Vert _{\mathrm{op},2}\mathrm{d%
}\tau \right\}  \label{BounddefDn}
\end{equation}%
for all $s,t\in \mathbb{R}_{0}^{+}$ and some family $(D_{n})_{n\in \mathbb{N}%
_{0}}\subseteq \mathbb{R}^{+}$ of strictly positive constants that depends
solely on $n\in \mathbb{N}_{0}$.
\end{proof}

\subsection{The Differential Flow of Spin-Boson Hamiltonians\label%
{Approximated Solution}}

Let $(\mathfrak{h}_{k})_{k\in \mathbb{N}}$ be a countable family of
separable Hilbert spaces. Let $\mathcal{F}_{+}$ denote the bosonic Fock
space built over an infinite-dimensional separable Hilbert space $\mathcal{H}
$, with orthonormal basis $\{\psi _{k}\}_{k\in \mathbb{N}}$. This
mathematical framework corresponds to Assumption BW0 in Condition \ref%
{Assumption BWMmainresult}. Henceforth, we assume this hypothesis holds
implicitly throughout the paper, even when not explicitly stated.

For each $k\in \mathbb{N}$, one takes a solution $(h_{k},\mu _{k},X_{k})\in
C(\mathbb{R}_{0}^{+},\mathcal{B}(\mathfrak{h}_{k}))^{3}$ to the system (\ref%
{goodflow}) of non-linear operator-valued differential equations for some
fixed initial data. Then, for any time $t\in \mathbb{R}_{0}^{+}$, using the
notation 
\begin{equation*}
h_{t}\equiv \left( h_{k,t}\right) _{k\in \mathbb{N}},\quad \mu _{t}\equiv
\left( \mu _{k,t}\right) _{k\in \mathbb{N}},\quad X_{t}X_{t}^{\ast }\equiv
\left( X_{k,t}X_{k,t}^{\ast }\right) _{k\in \mathbb{N}}\quad \text{and}\quad %
\left[ A,B\right] \equiv \left( \left[ A_{k},B_{k}\right] \right) _{k\in 
\mathbb{N}}
\end{equation*}%
for some given $A\equiv \left( A_{k}\right) _{k\in \mathbb{N}}$, $B\equiv
\left( B_{k}\right) _{k\in \mathbb{N}}$ with $A_{k},B_{k}\in \mathcal{B}(%
\mathfrak{h}_{k})\equiv \mathcal{B}(\mathfrak{h}_{k})^{\otimes }$ for $k\in 
\mathbb{N}$, as well as 
\begin{equation}
L_{k,t}\doteq h_{k,t}-\mu _{k,t}/2,\qquad R_{k,t}\doteq -h_{k,t}-\mu
_{k,t}/2,\qquad k\in \mathbb{N},  \label{relation bis03}
\end{equation}%
(cf. (\ref{relation bis0}) or (\ref{relation bis02})), we study the
expression 
\begin{equation}
\mathrm{H}_{t}\doteq H\left( h_{t}\right) +\mathrm{d}\Gamma ^{\otimes
}\left( \mu _{t}\right) +\sum_{k\in \mathbb{N}}\left( X_{k,t}\otimes
a_{k}+X_{k,t}^{\ast }\otimes a_{k}^{\ast }\right) ,\qquad t\in \mathbb{R}%
_{0}^{+},  \label{defHtbis}
\end{equation}%
as well as its derivative which, according to (\ref{goodflow}), should be
formally given by the second expression 
\begin{eqnarray}
\partial _{t}\mathrm{H}_{t} &\doteq &-2H\left( X_{t}X_{t}^{\ast }\right) +2%
\mathrm{d}\Gamma ^{\otimes }\left( \left[ X_{t}^{\ast },X_{t}\right] \right)
\label{defHtbis-derivative} \\
&&+\sum_{k\in \mathbb{N}}\left( \left( L_{k,t}X_{k,t}+X_{k,t}R_{k,t}\right)
\otimes a_{k}+\left( X_{k,t}^{\ast }L_{k,t}+R_{k,t}X_{k,t}^{\ast }\right)
\otimes a_{k}^{\ast }\right) .  \notag
\end{eqnarray}%
Here, for any sequence $A\equiv (A_{k})_{k\in \mathbb{N}}$ of self-adjoint
operators $A_{k}\in \mathcal{B}(\mathfrak{h}_{k})\equiv \mathcal{B}(%
\mathfrak{h}_{k})^{\otimes }$, $\mathrm{d}\Gamma ^{\otimes }\left( A\right) $
is the (extended) second quantization given by Definition \ref%
{extendsecondquanti}, while%
\begin{equation}
H\left( A\right) \doteq \sum_{k\in \mathbb{N}}A_{k}\otimes \mathbf{1}_{%
\mathcal{F}_{+}},  \label{expression H}
\end{equation}%
provided this expression make sense on some domain. We also use the operator 
$\mathrm{W}_{t}$ formally given by the third expression%
\begin{equation}
\mathrm{W}_{t}\doteq \frac{1}{2}\sum_{k\in \mathbb{N}}\left( \left[ \mu
_{k,t},X_{k,t}\right] \otimes \left( a_{k}a_{k}^{\ast }a_{k}+a_{k}^{\ast
}a_{k}a_{k}\right) +\left[ X_{k,t}^{\ast },\mu _{k,t}\right] \otimes \left(
a_{k}^{\ast }a_{k}a_{k}^{\ast }+a_{k}^{\ast }a_{k}^{\ast }a_{k}\right)
\right) .  \label{expression Wt}
\end{equation}

Certain conditions on the operator-valued coefficients are necessary for a
rigorous definition of the three expressions $\mathrm{H}_{t}$, $\partial _{t}%
\mathrm{H}_{t}$ and $\mathrm{W}_{t}$ as densely defined symmetric operators
acting on $\mathfrak{F}\doteq \mathfrak{H}\otimes \mathcal{F}_{+}$. In order
to define and investigate the properties of the Hamiltonians above, one
needs some assumptions, given by Condition \ref{AbstractAssumptions}, which
allow us first of all to define all the Hamiltonians mentioned above, at
least for all strictly positive times $t\in \mathbb{R}^{+}$:

\begin{proposition}[Flow of generalized spin-boson Hamiltonians]
\label{Lemma Htbis}\mbox{}\newline
Assuming Condition \ref{AbstractAssumptions} (i)--(iv), the following
Hamiltonians are well-defined:

\begin{enumerate}
\item[\emph{(i)}] For each $t\in \mathbb{R}_{0}^{+}$, the expression (\ref%
{defHtbis}) defines a lower semibounded self-adjoint operator $\mathrm{H}%
_{t} $ on the domain of the Hamiltonian 
\begin{equation*}
\mathrm{\tilde{H}}_{t}\doteq H\left( h_{t}\right) +\mathrm{d}\Gamma
^{\otimes }\left( \mu _{t}\right) =\mathrm{\tilde{H}}_{t}^{\ast }.
\end{equation*}%
If $\mathrm{\tilde{H}}_{t}$ is essentially self-adjoint\ on a core such as $%
\mathcal{D}_{\infty }$, then $\mathrm{H}_{t}$ is also essentially
self-adjoint on that same domain.

\item[\emph{(ii)}] For each $t\in \mathbb{R}^{+}$, the expression (\ref%
{defHtbis-derivative}) defines a self-adjoint operator $\partial _{t}\mathrm{%
H}_{t}$ with cores given by $\mathcal{D}_{\infty }$ and the domain $\mathcal{%
D}(\mathrm{N})\supseteq \mathcal{D}_{\infty }$ of the particle number
operator $\mathrm{N}$.

\item[\emph{(iii)}] For each $t\in \mathbb{R}_{0}^{+}$, the expression (\ref%
{expression Wt}) with domain $\mathcal{D}_{\infty }$ defines a closable
symmetric operator $\mathrm{W}_{t}$ and $\mathcal{D}(\mathrm{N}%
^{3/2})\supseteq \mathcal{D}_{\infty }$ is a core of it.
\end{enumerate}
\end{proposition}

\begin{proof}
By Proposition \ref{Proposition self-adjoint copy(1)} (i)--(ii) and
Conditions \ref{AbstractAssumptions} (ii)--(iii), for each $t\in \mathbb{R}%
_{0}^{+}$, the expression (\ref{expression H}) for $A=h_{t}$ and $%
A=X_{t}X_{t}^{\ast }$ respectively defines a lower semibounded self-adjoint
operator $H\left( h_{t}\right) $ and a bounded self-adjoint operator $%
H(X_{t}X_{t}^{\ast })$, both acting on $\mathfrak{F}$ with domain including
(at least) the dense subset $\mathcal{D}_{\infty }$ (\ref{second quantu}).
Furthermore, by Proposition \ref{Proposition self-adjoint copy(1)} (iii),
for each $t\in \mathbb{R}_{0}^{+}$, the operators $\mathrm{\tilde{H}}_{t}$
and 
\begin{equation*}
\partial _{t}\mathrm{\tilde{H}}_{t}\doteq -2H\left( X_{t}X_{t}^{\ast
}\right) +2\mathrm{d}\Gamma ^{\otimes }\left( \left[ X_{t}^{\ast },X_{t}%
\right] \right)
\end{equation*}%
are two self-adjoint operators on $\mathfrak{F}$, with core $\mathcal{D}%
_{\infty }$. These results are related to the conditions SB3 and SB4 of
Theorems \ref{HamilSA copy(2)} and \ref{HamilSA}. Additionally, the
condition SB0 of Theorem \ref{HamilSA copy(2)} refers here to the condition
BW0 -- assumed in this section and the following ones -- which adds the
separability of Hilbert spaces $\mathfrak{h}_{k}$ for $k\in \mathbb{N}$. The
conditions SB1--SB2 and SB5 of Theorems \ref{HamilSA copy(2)} and \ref%
{HamilSA} follow directly from Conditions \ref{AbstractAssumptions} (i),
(iii) and (iv). Therefore, we can apply Theorem \ref{HamilSA} to deduce
Assertion (i) about the expression (\ref{defHtbis}). Concerning $\partial
_{t}\mathrm{H}_{t}$ defined via the expression (\ref{defHtbis-derivative}),
we note that SB1--SB2 of Theorem \ref{HamilSA copy(2)} are satisfied for any
strictly positive $t\in \mathbb{R}^{+}$ because of Conditions \ref%
{AbstractAssumptions} (i) and (iii). So, in this case, we can apply Theorem %
\ref{HamilSA copy(2)} to show that the expression (\ref{defHtbis-derivative}%
) defines a self-adjoint operator $\partial _{t}\mathrm{H}_{t}$ with core $%
\mathcal{D}_{\infty }$, provided $t\in \mathbb{R}^{+}$. Note that $t=0$ is
not included here because we need to ensure the operator-norm summability of
the families $L_{t}X_{t}\equiv (L_{k,t}X_{k,t})_{k\in \mathbb{N}}$ and $%
X_{t}R_{t}\equiv \left( X_{k,t}R_{k,t}\right) _{k\in \mathbb{N}}$: 
\begin{equation}
\left\Vert L_{t}X_{t}\right\Vert _{\mathrm{op},2}^{2}+\left\Vert
X_{t}R_{t}\right\Vert _{\mathrm{op},2}^{2}<\infty .  \label{dfg0-bis}
\end{equation}%
By Inequality (\ref{dfg0}), it is only proven for strictly positive times $%
t\in \mathbb{R}^{+}$ when $\Vert X_{0}\Vert _{\mathrm{op},2}<\infty $.
Moreover, we use the Cauchy-Schwarz inequality two times to estimate that,
for any $\varphi \in \mathcal{D(}\mathrm{N})$ and each sequence $Z\equiv
(Z_{k})_{k\in \mathbb{N}}$ of self-adjoint operators $Z_{k}=Z_{k}^{\ast }\in 
\mathcal{B}(\mathfrak{h}_{k})$,%
\begin{eqnarray}
\left\Vert \sum_{k\in \mathbb{N}}Z_{k}\otimes a_{k}^{\ast }a_{k}\varphi
\right\Vert _{\mathfrak{F}}^{2} &=&\sum_{k,p\in \mathbb{N}}\left\langle
\varphi ,Z_{k}Z_{q}\otimes a_{k}^{\ast }a_{k}a_{p}^{\ast }a_{p}\varphi
\right\rangle _{\mathfrak{F}}  \notag \\
&\leq &\left\Vert Z_{k}\right\Vert _{\mathrm{op},2}^{2}\sum_{k\in \mathbb{N}%
}\left\Vert \mathbf{1}_{\mathfrak{H}}\otimes a_{k}^{\ast }a_{k}\varphi
\right\Vert _{\mathfrak{F}}^{2}  \notag \\
&\leq &\left\Vert Z_{k}\right\Vert _{\mathrm{op},2}^{2}\left\Vert \mathrm{N}%
\varphi \right\Vert _{\mathfrak{F}}^{2}.  \label{erret1}
\end{eqnarray}%
Using additionally Inequality (\ref{dfg0-bis}) for $t>0$ (cf. (\ref{dfg0})),
Lemma \ref{Lemma-self-adjoint1} and $H(X_{t}X_{t}^{\ast })\in \mathcal{B}(%
\mathfrak{F})$, we conclude that $\mathcal{D}(\mathrm{N})\supseteq \mathcal{D%
}_{\infty }$ is also on the domain of $\partial _{t}\mathrm{H}_{t}$ for any $%
t\in \mathbb{R}^{+}$. This completes the proof of Assertion (ii). Finally,
Assertion (iii) is a consequence of Lemma \ref{Lemma-self-adjoint1 copy(1)}
and Conditions \ref{AbstractAssumptions} (i) (on the self-adjointness of
operator-valued coefficients) and (iii). Here, we use the fact that any
densely defined symmetric operator, such as $\mathrm{W}_{t}$, is closable.
Additionally by Lemma \ref{Lemma-self-adjoint1 copy(1)} and Conditions \ref%
{AbstractAssumptions} (iii), the domain of $\mathrm{W}_{t}^{\ast \ast }$
contains $\mathcal{D}(\mathrm{N}^{3/2})\supseteq \mathcal{D}_{\infty }$,
which is thus a core.
\end{proof}

We now study the flow $\mathrm{H}_{(\cdot )}$ of spin-boson Hamiltonians,
which is well-defined as shown above. While our ultimate goal is to
differentiate the time evolution of this operator family, we must first
investigate the continuity of both the family $\mathrm{H}_{(\cdot )}$ and
its resolvent. These preliminary results serve as key technical tools in the
sequel:

\begin{lemma}[Continuity of time-dependent spin-boson Hamiltonians]
\label{lemmabounddiffH}\mbox{}\newline
Under Condition \ref{AbstractAssumptions} (i)--(iv), for all $s,t\in \mathbb{%
R}_{0}^{+}$, 
\begin{equation}
\left\Vert \left( \mathrm{H}_{t}-\mathrm{H}_{s}\right) \left( \mathrm{N}+%
\mathbf{1}\right) ^{-1}\right\Vert _{\mathrm{op}}\leq 3\left\Vert
X_{t}-X_{s}\right\Vert _{\mathrm{op},2}+6\int_{s\wedge t}^{s\vee
t}\left\Vert X_{\tau }\right\Vert _{\mathrm{op},2}^{2}\mathrm{d}\tau
\label{differenceHonN}
\end{equation}%
and 
\begin{equation}
\lim_{t\rightarrow s}\left\Vert \left( \mathrm{H}_{t}-\mathrm{H}_{s}\right)
\left( \mathrm{N}+\mathbf{1}\right) ^{-1}\right\Vert _{\mathrm{op}}=0.
\label{limXtmoinsXs}
\end{equation}
\end{lemma}

\begin{proof}
We first observe from the system (\ref{goodflow}) of non-linear
operator-valued differential equations and Propositions \ref{Proposition
self-adjoint} and \ref{Lemma Htbis} (in particular (\ref{defHtbis})) that,
for any $s,t\in \mathbb{R}_{0}^{+}$, 
\begin{eqnarray}
\mathrm{H}_{t}-\mathrm{H}_{s} &=&-2\sum_{k\in \mathbb{N}}\left( \int_{s}^{t}%
\left[ X_{k,\tau },X_{k,\tau }^{\ast }\right] \mathrm{d}\tau \otimes
a_{k}^{\ast }a_{k}+\int_{s}^{t}X_{k,\tau }X_{k,\tau }^{\ast }\mathrm{d}\tau
\otimes \mathbf{1}_{\mathcal{F}_{+}}\right)  \notag \\
&&+\sum_{k\in \mathbb{N}}\left( \left( X_{k,t}-X_{k,s}\right) \otimes
a_{k}+\left( X_{k,t}^{\ast }-X_{k,s}^{\ast }\right) \otimes a_{k}^{\ast
}\right) .  \label{equality1}
\end{eqnarray}%
The right-hand side of this equality defines a self-adjoint operator with
core $\mathcal{D}_{\infty }$; see, e.g., Condition \ref{AbstractAssumptions}
(iii) and Theorem \ref{HamilSA copy(2)}. Moreover, using Proposition \ref%
{Proposition self-adjoint copy(1)}, the triangle inequality and Tonelli's
theorem, we get that 
\begin{equation}
\left\Vert \sum_{k\in \mathbb{N}}\int_{s}^{t}X_{k,\tau }X_{k,\tau }^{\ast }%
\mathrm{d}\tau \otimes \mathbf{1}_{\mathcal{F}_{+}}\right\Vert _{\mathrm{op}%
}\leq \sum_{k\in \mathbb{N}}\left\Vert \int_{s}^{t}X_{k,\tau }X_{k,\tau
}^{\ast }\mathrm{d}\tau \right\Vert _{\mathrm{op}}\leq \int_{s\wedge
t}^{s\vee t}\left\Vert X_{\tau }\right\Vert _{\mathrm{op},2}^{2}\mathrm{d}%
\tau .  \label{erret2}
\end{equation}%
Therefore, the first assertion is a consequence of Inequalities (\ref{erret1}%
) and (\ref{erret2}), Equation (\ref{equality1}) and Lemma \ref%
{Lemma-self-adjoint1} together with the triangle inequality and the
elementary inequality $\sqrt{x}/(x+1)\leq 1/2$ for $x\in \mathbb{R}_{0}^{+}$%
. Equation (\ref{differenceHonN}) and Proposition \ref{prop useful} yield (%
\ref{limXtmoinsXs}).
\end{proof}

\begin{lemma}[Continuity of time-dependent resolvents]
\label{coro-Ht copy(1)}\mbox{}\newline
Under Condition \ref{AbstractAssumptions} (i)--(iv), for any $T\in \mathbb{R}%
^{+}$ and $\lambda \gtrsim _{T}\Vert X_{0}\Vert _{\mathrm{op},2}$, the
mappings 
\begin{equation*}
t\mapsto \left( \mathrm{H}_{t}+i\lambda \mathbf{1}\right) ^{-1}\qquad \text{%
and}\qquad t\mapsto \left( \mathrm{N}+\mathbf{1}\right) ^{\frac{3}{2}}\left( 
\mathrm{H}_{t}+i\lambda \mathbf{1}\right) ^{-1}\left( \mathrm{N}+\mathbf{1}%
\right) ^{-\frac{3}{2}}
\end{equation*}%
are both strongly continuous on $[0,T]$. If Condition \ref%
{AbstractAssumptions} (v) additionally holds, the assertions remain true on $%
\mathbb{R}_{0}^{+}$ (i.e., for $T=\infty $) with $\lambda
>2^{5+1/2}\left\Vert X_{0}\right\Vert _{\mathrm{op},2}$ and the mapping 
\begin{equation*}
t\mapsto \left( \mathrm{N}+\mathbf{1}\right) ^{\frac{3}{2}}\left( \mathrm{H}%
_{t}+i\lambda \mathbf{1}\right) ^{-1}\left( \mathrm{N}+\mathbf{1}\right)
^{-1}
\end{equation*}%
is also strongly continuous.
\end{lemma}

\begin{proof}
Fix $T\in \mathbb{R}^{+}$. Note from Proposition \ref{Lemma Htbis} that $%
\mathrm{H}_{t}$ is a well-defined Hamiltonian. So, using the resolvent
identity\footnote{%
The right-hand side alone is a bounded operator, already by Lemmata \ref%
{diversPreservedom copy(1)} and \ref{lemmabounddiffH} (e.g., $\mathcal{D}(%
\mathrm{H}_{s})\subseteq \mathcal{D}(\mathrm{N})$).} 
\begin{equation*}
\left( \mathrm{H}_{t}+i\lambda \mathbf{1}\right) ^{-1}-\left( \mathrm{H}%
_{s}+i\lambda \mathbf{1}\right) ^{-1}=-\left( \mathrm{H}_{t}+i\lambda 
\mathbf{1}\right) ^{-1}\left( \mathrm{H}_{t}-\mathrm{H}_{s}\right) \left( 
\mathrm{H}_{s}+i\lambda \mathbf{1}\right) ^{-1},
\end{equation*}%
and Lemmata \ref{diversPreservedom} and \ref{lemmabounddiffH} together with
Condition \ref{AbstractAssumptions} (iii), we deduce that 
\begin{equation*}
t\mapsto \left( \mathrm{H}_{t}+i\lambda \mathbf{1}\right) ^{-1}\in C\left( %
\left[ 0,T\right] ;\mathcal{B}(\mathcal{D}\left( \mathrm{N}\right) ,%
\mathfrak{F})\right)
\end{equation*}%
for 
\begin{equation*}
\lambda >2^{4}\sup_{t\in \left[ 0,T\right] }\left\Vert X_{t}\right\Vert _{%
\mathrm{op},2}.
\end{equation*}%
Since $\mathcal{D}\left( \mathrm{N}\right) $ is a dense subset of $\mathfrak{%
F}$ and the time-dependent resolvents are uniformly bounded by $1/\lambda $ (%
$\lambda >0$), the mapping $t\mapsto \left( \mathrm{H}_{t}+i\lambda \mathbf{1%
}\right) ^{-1}$ is strongly continuous on $\mathbb{R}_{0}^{+}$. Note that
the strong continuity is established by exploiting the density of the
subspace $\mathcal{D}\left( \mathrm{N}\right) \subseteq \mathfrak{F}$.
Moreover, using Equations (\ref{erret1}), (\ref{equality1}) and (\ref{erret2}%
), Lemmata \ref{RelativeboundGlemma}, \ref{diversPreservedom} as well as
Proposition \ref{prop useful} together with the triangle inequality and the
resolvent identity, we get that 
\begin{equation*}
t\mapsto \left( \mathrm{N}+\mathbf{1}\right) ^{\frac{3}{2}}\left( \mathrm{H}%
_{t}+i\lambda \mathbf{1}\right) ^{-1}\left( \mathrm{N}+\mathbf{1}\right) ^{-%
\frac{3}{2}}\in C\left( \left[ 0,T\right] ;\mathcal{B}(\mathcal{D}\left( 
\mathrm{N}\right) ,\mathfrak{F})\right)
\end{equation*}%
for 
\begin{equation*}
\lambda >2^{5+1/2}\sup_{t\in \left[ 0,T\right] }\left\Vert X_{t}\right\Vert
_{\mathrm{op},2},
\end{equation*}%
which, again by Lemma \ref{diversPreservedom} and density of $\mathcal{D}%
\left( \mathrm{N}\right) \subseteq \mathfrak{F}$, in turn implies the strong
continuity of this mapping on $\mathbb{R}_{0}^{+}$. Note from Condition \ref%
{AbstractAssumptions} (i) that 
\begin{equation*}
\sup_{t\in \left[ 0,T\right] }\left\Vert X_{t}\right\Vert _{\mathrm{op}%
,2}\leq D_{T}^{1/2}\left\Vert X_{0}\right\Vert _{\mathrm{op},2}
\end{equation*}%
and it suffices to take%
\begin{equation}
\lambda >2^{5+1/2}D_{T}^{1/2}\Vert X_{0}\Vert _{\mathrm{op},2}
\label{dfsdfsgsf}
\end{equation}%
to get all the asserted continuities under Condition \ref%
{AbstractAssumptions} (i)--(iv). Now, if Condition \ref{AbstractAssumptions}
(v) additionally holds, then for the same reasons, the mapping%
\begin{equation*}
\left( \mathrm{N}+\mathbf{1}\right) ^{\frac{3}{2}}\left( \mathrm{H}%
_{t}+i\lambda \mathbf{1}\right) ^{-1}\left( \mathrm{N}+\mathbf{1}\right)
^{-1}
\end{equation*}%
becomes strongly continuous on $[0,T]$ for 
\begin{equation*}
\lambda >2^{5}\sup_{t\in \left[ 0,T\right] }\left\Vert X_{t}\right\Vert _{%
\mathrm{op},2},
\end{equation*}%
in particular because of Lemmata \ref{diversPreservedom} and \ref%
{diversPreservedom copy(1)}. Furthermore, Corollary \ref{pseudoHS} (with $T_{%
\mathrm{max}}=\infty $) can be used. In particular, 
\begin{equation}
\sup_{t\in \left[ 0,T\right] }\left\Vert X_{t}\right\Vert _{\mathrm{op}%
,2}=\left\Vert X_{0}\right\Vert _{\mathrm{op},2}  \label{sdadasdasdasdas}
\end{equation}%
and the constant $D_{T}$ can be taken as $D_{T}=1$ for all $T\in \mathbb{R}%
^{+}$.
\end{proof}

We are now in a position to study the flow $\mathrm{H}_{(\cdot )}$ of
spin-boson Hamiltonians. Due to the unboundedness of the Hamiltonians, it is
more convenient to analyze the flow of their corresponding resolvents. This
approach is pursued in the following proposition.

\begin{proposition}[Time-derivative of the resolvent flow]
\label{Resolventeq}\mbox{}\newline
Under Condition \ref{AbstractAssumptions} (i)--(iv), for any $T\in \mathbb{R}%
^{+}$, $\lambda \gtrsim _{T}\Vert X_{0}\Vert _{\mathrm{op},2}$ and $t\in
(0,T]$, the resolvent $\left( \mathrm{H}_{t}+i\lambda \mathbf{1}\right)
^{-1} $ satisfies the differential equation 
\begin{equation}
\left\langle \psi ,\left( \partial _{t}\mathrm{R}_{t}\right) \varphi
\right\rangle _{\mathfrak{F}}=\left\langle \psi ,\left( i\left[ \mathrm{R}%
_{t},\mathrm{G}_{t}\right] -\mathrm{R}_{t}\mathrm{W}_{t}\mathrm{R}%
_{t}\right) \varphi \right\rangle _{\mathfrak{F}},\qquad \mathrm{R}%
_{t}\doteq \left( \mathrm{H}_{t}+i\lambda \mathbf{1}\right) ^{-1},
\label{DeriveWeaksenseResolv}
\end{equation}%
for any $\varphi ,\psi \in \mathcal{D}(\mathrm{N}^{3/2})$, where $\mathrm{W}%
_{t}$ is the symmetric operator defined by (\ref{expression Wt}) via
Proposition \ref{Lemma Htbis} (iii). If we further assume Condition \ref%
{AbstractAssumptions} (v), then (\ref{DeriveWeaksenseResolv}) holds for any $%
\psi ,\varphi \in \mathcal{D}(\mathrm{N})$ and $t\in \mathbb{R}^{+}$,
provided $\lambda >2^{5+1/2}\left\Vert X_{0}\right\Vert _{\mathrm{op},2}$.
\end{proposition}

\begin{proof}
Fix $T\in \mathbb{R}^{+}$. The proof is done in several steps: \medskip

\noindent \underline{Step 1:} By Lemma \ref{lemmabounddiffH} and $\Vert
\cdot \Vert _{\mathrm{op},2}$--Lipschitz continuity of $t\mapsto X_{t}$
(Condition \ref{AbstractAssumptions} (iii)) together with the Lebesgue
differentiation theorem, for any $t\in (0,T]$, 
\begin{equation}
\lim_{\varepsilon \rightarrow 0}\varepsilon ^{-1}\left\Vert \left( \mathrm{H}%
_{(t+\varepsilon )\vee 0}-\mathrm{H}_{t}\right) \left( \mathrm{N}+\mathbf{1}%
\right) ^{-1}\right\Vert _{\mathrm{op}}<\infty .  \label{deriveHsurN}
\end{equation}%
As a consequence, using the resolvent identity and the Cauchy-Schwarz
inequality, one shows that, for any $\varphi ,\psi \in \mathcal{D}\left( 
\mathrm{N}\right) $ and $\varepsilon \in \mathbb{R}^{+}$, 
\begin{multline*}
\left\vert \left\langle \varphi ,\left( \left( \mathrm{H}_{t+\varepsilon
}+i\lambda \mathbf{1}\right) ^{-1}-\left( \mathrm{H}_{t}+i\lambda \mathbf{1}%
\right) ^{-1}-\left( \mathrm{H}_{t}+i\lambda \mathbf{1}\right) ^{-1}\left( 
\mathrm{H}_{t}-\mathrm{H}_{t+\varepsilon }\right) \left( \mathrm{H}%
_{t}+i\lambda \mathbf{1}\right) ^{-1}\right) \psi \right\rangle _{\mathfrak{F%
}}\right\vert \\
\leq \left\Vert \left( \left( \mathrm{H}_{t+\varepsilon }+i\lambda \mathbf{1}%
\right) ^{-1}-\left( \mathrm{H}_{t}+i\lambda \mathbf{1}\right) ^{-1}\right)
\varphi \right\Vert _{\mathfrak{F}}\left\Vert \left( \mathrm{H}_{t}-\mathrm{H%
}_{t+\varepsilon }\right) \left( \mathrm{N}+\mathbf{1}\right)
^{-1}\right\Vert _{\mathrm{op}} \\
\left\Vert \left( \mathrm{N}+\mathbf{1}\right) \left( \mathrm{H}%
_{t}+i\lambda \mathbf{1}\right) ^{-1}\left( \mathrm{N}+\mathbf{1}\right)
^{-1}\right\Vert _{\mathrm{op}}\left\Vert \left( \mathrm{N}+\mathbf{1}%
\right) \psi \right\Vert _{\mathfrak{F}}.
\end{multline*}%
By combining this last inequality with (\ref{deriveHsurN}), Lemmata \ref%
{diversPreservedom} and \ref{lemmabounddiffH}--\ref{coro-Ht copy(1)}, we
arrive at the equality 
\begin{equation}
\lim_{\varepsilon \rightarrow 0}\varepsilon ^{-1}\left\vert \left\langle
\varphi ,\left( \left( \mathrm{H}_{t+\varepsilon }+i\lambda \mathbf{1}%
\right) ^{-1}-\left( \mathrm{H}_{t}+i\lambda \mathbf{1}\right) ^{-1}-\left( 
\mathrm{H}_{t}+i\lambda \mathbf{1}\right) ^{-1}\left( \mathrm{H}_{t}-\mathrm{%
H}_{t+\varepsilon }\right) \left( \mathrm{H}_{t}+i\lambda \mathbf{1}\right)
^{-1}\right) \psi \right\rangle _{\mathfrak{F}}\right\vert =0
\label{proofweakderiveH}
\end{equation}%
for any $\varphi ,\psi \in \mathcal{D}\left( \mathrm{N}\right) $ and
sufficiently large $\lambda \in \mathbb{R}^{+}$ (see, e.g., (\ref{dfsdfsgsf}%
)).\medskip

\noindent \underline{Step 2:} By adapting the proof of Lemma \ref%
{lemmabounddiffH} and using Proposition \ref{prop useful}, we can explicitly
evaluate (\ref{deriveHsurN}) by showing that, for any strictly positive $%
t\in (0,T]$, 
\begin{equation}
\lim_{\varepsilon \rightarrow 0}\left\Vert \left( \varepsilon ^{-1}\left( 
\mathrm{H}_{(t+\varepsilon )\vee 0}-\mathrm{H}_{t}\right) -\partial _{t}%
\mathrm{H}_{t}\right) \left( \mathrm{N}+\mathbf{1}\right) ^{-1}\right\Vert _{%
\mathrm{op}}=0,  \label{deriveHsurN2}
\end{equation}%
where $\partial _{t}\mathrm{H}_{t}$ is the Hamiltonian defined from (\ref%
{defHtbis-derivative}) via Proposition \ref{Lemma Htbis} (ii) for strictly
positive times $t\in (0,T]$. Using now (\ref{proofweakderiveH}) and Lemma %
\ref{diversPreservedom}, we thus deduce that%
\begin{equation*}
\lim_{\varepsilon \rightarrow 0}\left\vert \langle \varphi ,(\varepsilon
^{-1}((\mathrm{H}_{(t+\varepsilon )\vee 0}+i\lambda \mathbf{1})^{-1}-(%
\mathrm{H}_{t}+i\lambda \mathbf{1})^{-1})+(\mathrm{H}_{t}+i\lambda \mathbf{1}%
)^{-1}\partial _{t}\mathrm{H}_{t}(\mathrm{H}_{t}+i\lambda \mathbf{1}%
)^{-1})\psi \rangle _{\mathfrak{F}}\right\vert =0
\end{equation*}%
for any $\varphi ,\psi \in \mathcal{D}\left( \mathrm{N}\right) $, $t\in 
\mathbb{R}^{+}$ and sufficiently large $\lambda \in \mathbb{R}^{+}$ (see,
e.g., (\ref{dfsdfsgsf})). In other words,%
\begin{equation}
\partial _{t}\left( \mathrm{H}_{t}+i\lambda \mathbf{1}\right) ^{-1}=-\left( 
\mathrm{H}_{t}+i\lambda \mathbf{1}\right) ^{-1}\left( \partial _{t}\mathrm{H}%
_{t}\right) \left( \mathrm{H}_{t}+i\lambda \mathbf{1}\right) ^{-1},\qquad
t\in \mathbb{R}^{+},  \label{weakderiveH}
\end{equation}%
in the weak sense on the domain $\mathcal{D}(\mathrm{N})\times \mathcal{D}(%
\mathrm{N})$, where $\lambda \in \mathbb{R}^{+}$ is chosen sufficiently
large, for instance, such that (\ref{dfsdfsgsf}) holds. \medskip

\noindent \underline{Step 3:} For strictly positive parameters $t,\lambda
\in \mathbb{R}^{+}$ ($\lambda $ chosen sufficiently large) we can perform
explicit computations using the flow (\ref{goodflow}) and the CCR as is done
in (\ref{generator})--(\ref{ldfj}) to arrive at the equalities%
\begin{eqnarray}
&&\left( \mathrm{H}_{t}+i\lambda \mathbf{1}\right) ^{-1}\left( \partial _{t}%
\mathrm{H}_{t}\right) \left( \mathrm{H}_{t}+i\lambda \mathbf{1}\right)
^{-1}\left( \mathrm{N}+\mathbf{1}\right) ^{-3/2}  \notag \\
&=&\left( \mathrm{H}_{t}+i\lambda \mathbf{1}\right) ^{-1}\left( i\left[ 
\mathrm{H}_{t},\mathrm{G}_{t}\right] +\mathrm{W}_{t}\right) \left( \mathrm{H}%
_{t}+i\lambda \mathbf{1}\right) ^{-1}\left( \mathrm{N}+\mathbf{1}\right)
^{-3/2}  \notag \\
&=&-i\left[ \left( \mathrm{H}_{t}+i\lambda \mathbf{1}\right) ^{-1},\mathrm{G}%
_{t}\right] \left( \mathrm{N}+\mathbf{1}\right) ^{-3/2}+\left( \mathrm{H}%
_{t}+i\lambda \mathbf{1}\right) ^{-1}\mathrm{W}_{t}\left( \mathrm{H}%
_{t}+i\lambda \mathbf{1}\right) ^{-1}\left( \mathrm{N}+\mathbf{1}\right)
^{-3/2}  \label{inequality 1bis}
\end{eqnarray}%
with $\mathrm{W}_{t}$ being the operator defined by (\ref{expression Wt})
via Proposition \ref{Lemma Htbis} (iii). Note that both sides of the
equality are in fact bounded operators for any sufficiently large $\lambda
\in \mathbb{R}^{+}$ (see, e.g., (\ref{dfsdfsgsf})) and $t\in (0,T]$. Indeed,
using Lemmata \ref{Lemma-self-adjoint1}, \ref{diversPreservedom} and \ref%
{Lemma-self-adjoint1 copy(1)} (extended to all $\varphi \in \mathcal{D}(%
\mathrm{N}^{3/2})$, see Proposition \ref{Lemma Htbis} (iii)), for any $t\in
(0,T]$\ and $\lambda \in \mathbb{R}^{+}$ satisfying (\ref{dfsdfsgsf}), the
following bounds hold:

\begin{itemize}
\item[(a)] 
\begin{multline*}
\left\Vert \left( \mathrm{H}_{t}+i\lambda \mathbf{1}\right) ^{-1}\left(
\partial _{t}\mathrm{H}_{t}\right) \left( \mathrm{H}_{t}+i\lambda \mathbf{1}%
\right) ^{-1}\left( \mathrm{N}+\mathbf{1}\right) ^{-1}\right\Vert _{\mathrm{%
op}} \\
\leq \lambda ^{-1}\left\Vert \left( \partial _{t}\mathrm{H}_{t}\right)
\left( \mathrm{N}+\mathbf{1}\right) ^{-1}\right\Vert _{\mathrm{op}%
}\left\Vert \left( \mathrm{N}+\mathbf{1}\right) \left( \mathrm{H}%
_{t}+i\lambda \mathbf{1}\right) ^{-1}\left( \mathrm{N}+\mathbf{1}\right)
^{-1}\right\Vert _{\mathrm{op}}<\infty ;
\end{multline*}

\item[(b)] 
\begin{multline*}
\left\Vert \left( \mathrm{H}_{t}+i\lambda \mathbf{1}\right) ^{-1}\mathrm{G}%
_{t}\mathrm{H}_{t}\left( \mathrm{H}_{t}+i\lambda \mathbf{1}\right)
^{-1}\left( \mathrm{N}+\mathbf{1}\right) ^{-1}\right\Vert _{\mathrm{op}} \\
\leq \lambda ^{-1}\left\Vert \mathrm{G}_{t}\left( \mathrm{N}+\mathbf{1}%
\right) ^{-1}\right\Vert _{\mathrm{op}}\left\Vert \left( \mathrm{N}+\mathbf{1%
}\right) \mathrm{H}_{t}\left( \mathrm{H}_{t}+i\lambda \mathbf{1}\right)
^{-1}\left( \mathrm{N}+\mathbf{1}\right) ^{-1}\right\Vert _{\mathrm{op}%
}<\infty ;
\end{multline*}

\item[(c)] 
\begin{multline*}
\left\Vert \left( \mathrm{H}_{t}+i\lambda \mathbf{1}\right) ^{-1}\mathrm{W}%
_{t}\left( \mathrm{H}_{t}+i\lambda \mathbf{1}\right) ^{-1}\left( \mathrm{N}+%
\mathbf{1}\right) ^{-3/2}\right\Vert _{\mathrm{op}} \\
\leq \lambda ^{-1}\left\Vert \mathrm{W}_{t}\left( \mathrm{N}+\mathbf{1}%
\right) ^{-3/2}\right\Vert _{\mathrm{op}}\left\Vert \left( \mathrm{N}+%
\mathbf{1}\right) ^{3/2}\left( \mathrm{H}_{t}+i\lambda \mathbf{1}\right)
^{-1}\left( \mathrm{N}+\mathbf{1}\right) ^{-3/2}\right\Vert _{\mathrm{op}%
}<\infty .
\end{multline*}
\end{itemize}

\noindent According to (\ref{inequality 1bis}), (a)--(c) necessarily imply
that 
\begin{equation*}
\left\Vert \left( \mathrm{H}_{t}+i\lambda \mathbf{1}\right) ^{-1}\mathrm{H}%
_{t}\mathrm{G}_{t}\left( \mathrm{H}_{t}+i\lambda \mathbf{1}\right)
^{-1}\left( \mathrm{N}+\mathbf{1}\right) ^{-1}\right\Vert _{\mathrm{op}%
}<\infty .
\end{equation*}%
Therefore, the first assertion follows by combining (\ref{weakderiveH}) and (%
\ref{inequality 1bis}). Finally, if $c_{k}\geq c\in \mathbb{R}^{+}$ for all $%
k\in \mathbb{N}$ (Condition \ref{AbstractAssumptions} (v)), then (\ref%
{inequality 1bis}) can be replaced by 
\begin{eqnarray*}
&&\left( \mathrm{H}_{t}+i\lambda \mathbf{1}\right) ^{-1}\left( \partial _{t}%
\mathrm{H}_{t}\right) \left( \mathrm{H}_{t}+i\lambda \mathbf{1}\right)
^{-1}\left( \mathrm{N}+\mathbf{1}\right) ^{-1} \\
&=&-i\left[ \left( \mathrm{H}_{t}+i\lambda \mathbf{1}\right) ^{-1},\mathrm{G}%
_{t}\right] \left( \mathrm{N}+\mathbf{1}\right) ^{-1}+\left( \mathrm{H}%
_{t}+i\lambda \mathbf{1}\right) ^{-1}\mathrm{W}_{t}\left( \mathrm{H}%
_{t}+i\lambda \mathbf{1}\right) ^{-1}\left( \mathrm{N}+\mathbf{1}\right)
^{-1}
\end{eqnarray*}%
and, thanks to Lemma \ref{diversPreservedom copy(1)}, (c) can be substituted
with 
\begin{multline*}
\left\Vert \left( \mathrm{H}_{t}+i\lambda \mathbf{1}\right) ^{-1}\mathrm{W}%
_{t}\left( \mathrm{H}_{t}+i\lambda \mathbf{1}\right) ^{-1}\left( \mathrm{N}+%
\mathbf{1}\right) ^{-1}\right\Vert _{\mathrm{op}} \\
\leq \lambda ^{-1}\left\Vert \mathrm{W}_{t}\left( \mathrm{N}+\mathbf{1}%
\right) ^{-3/2}\right\Vert _{\mathrm{op}}\left\Vert \left( \mathrm{N}+%
\mathbf{1}\right) ^{3/2}\left( \mathrm{H}_{t}+i\lambda \mathbf{1}\right)
^{-1}\left( \mathrm{N}+\mathbf{1}\right) ^{-1}\right\Vert _{\mathrm{op}%
}<\infty
\end{multline*}%
in order to obtain again (\ref{DeriveWeaksenseResolv}), but now for all
vectors $\psi ,\varphi \in \mathcal{D}(\mathrm{N})$ , $t\in \mathbb{R}^{+}$
and $\lambda \in \mathbb{R}^{+}$ chosen sufficiently large, for instance,
such that $\lambda >2^{5+1/2}\left\Vert X_{0}\right\Vert _{\mathrm{op},2}$
(cf. (\ref{dfsdfsgsf})--(\ref{sdadasdasdasdas})).
\end{proof}

\begin{corollary}[The resolvent flow]
\label{Resolvent expression}\mbox{}\newline
Under Condition \ref{AbstractAssumptions} (i)--(iv), for any $T\in \mathbb{R}%
^{+}$ and $\lambda \gtrsim _{T}\Vert X_{0}\Vert _{\mathrm{op},2}$, $s,t\in %
\left[ 0,T\right] $ and $\varphi ,\psi \in \mathcal{D}\left( \mathrm{N}%
\right) $, the resolvent $\mathrm{R}_{t}\doteq \left( \mathrm{H}%
_{t}+i\lambda \mathbf{1}\right) ^{-1}$ satisfies the equation 
\begin{equation*}
\left\langle \psi ,\mathrm{R}_{t}\varphi \right\rangle _{\mathfrak{F}%
}=\left\langle \psi ,\left( \mathrm{U}_{t,s}\mathrm{R}_{s}\mathrm{U}%
_{s,t}-\int_{s}^{t}\mathrm{U}_{t,\tau }\mathrm{R}_{\tau }\mathrm{W}_{\tau }%
\mathrm{R}_{\tau }\mathrm{U}_{\tau ,t}\mathrm{d}\tau \right) \varphi
\right\rangle _{\mathfrak{F}},
\end{equation*}%
for any $\varphi ,\psi \in \mathcal{D}\left( \mathrm{N}^{3/2}\right) $,
where $\mathrm{W}_{\tau }\equiv \mathrm{W}_{\tau }^{\ast \ast }$ is the
closure of the densely defined symmetric operator defined by (\ref%
{expression Wt}) via Proposition \ref{Lemma Htbis} (iii). If we further
assume Condition \ref{AbstractAssumptions} (v), then the assertion holds on $%
\mathcal{D}(\mathrm{N})$ for all $s,t\in \mathbb{R}_{0}^{+}$ and $\lambda
>2^{5+1/2}\left\Vert X_{0}\right\Vert _{\mathrm{op},2}$.
\end{corollary}

\begin{proof}
By using Lemma \ref{TechnicalIntegralWandRWR} below, note that we can make
sense of the expression 
\begin{equation*}
\int_{s}^{t}\mathrm{U}_{t,\tau }\mathrm{R}_{\tau }\mathrm{W}_{\tau }\mathrm{R%
}_{\tau }\mathrm{U}_{\tau ,t}\mathrm{d}\tau
\end{equation*}%
as a densely defined symmetric operator with domain $\mathcal{D}(\mathrm{N}%
^{3/2})$. We combine Theorem \ref{propagators} and Proposition \ref%
{Resolventeq} with explicit computations to obtain that, for any $\varphi
,\psi \in \mathcal{D}(\mathrm{N}^{3/2})$ and every $\epsilon \in \mathbb{R}%
^{+}$, 
\begin{multline*}
\left\langle \psi ,\left( \mathrm{U}_{t,s}\mathrm{R}_{s}\mathrm{U}%
_{s,t}-\int_{s}^{t}\mathrm{U}_{t,\tau }\mathrm{R}_{\tau }\mathrm{W}_{\tau }%
\mathrm{R}_{\tau }\mathrm{U}_{\tau ,t}\mathrm{d}\tau \right) \right. \\
\left. -\left( \mathrm{U}_{t,t+\epsilon }\mathrm{R}_{t+\epsilon }\mathrm{U}%
_{t+\epsilon ,t}-\int_{t+\epsilon }^{t}\mathrm{U}_{t,\tau }\mathrm{R}_{\tau }%
\mathrm{W}_{\tau }\mathrm{R}_{\tau }\mathrm{U}_{\tau ,t}\mathrm{d}\tau
\right) \varphi \right\rangle _{\mathfrak{F}}=0.
\end{multline*}%
We now take the limit $\epsilon \rightarrow 0^{+}$ using Theorem \ref%
{propagators} (iii)--(iv) as well as Lemmata \ref{diversPreservedom} and \ref%
{Lemma-self-adjoint1 copy(1)} in order to arrive at 
\begin{equation}
\left\langle \psi ,\mathrm{R}_{t}-\left( \mathrm{U}_{t,s}\mathrm{R}_{s}%
\mathrm{U}_{s,t}-\int_{s}^{t}\mathrm{U}_{t,\tau }\mathrm{R}_{\tau }\mathrm{W}%
_{\tau }\mathrm{R}_{\tau }\mathrm{U}_{\tau ,t}\mathrm{d}\tau \right) \varphi
\right\rangle _{\mathfrak{F}}=0  \label{eq:eqtempweakfini}
\end{equation}%
for any $\varphi ,\psi \in \mathcal{D}(\mathrm{N}^{3/2})$. Under Condition %
\ref{AbstractAssumptions} (v), a straightforward adaptation of the proof
extends the result on $\mathcal{D}(\mathrm{N})\supseteq \mathcal{D}(\mathrm{N%
}^{3/2})$ and for all times $s,t\in \mathbb{R}_{0}^{+}$ and $\lambda
>2^{5+1/2}\left\Vert X_{0}\right\Vert _{\mathrm{op},2}$ (cf. (\ref{dfsdfsgsf}%
)--(\ref{sdadasdasdasdas})).
\end{proof}

\begin{remark}[From weak to strong topology]
\label{Remarkresolveq}\mbox{}\newline
In Corollary \ref{Resolvent expression}, under Condition \ref%
{AbstractAssumptions} (v), the vector 
\begin{equation*}
\left( \mathrm{R}_{t}-\mathrm{U}_{t,s}\mathrm{R}_{s}\mathrm{U}%
_{s,t}+\int_{s}^{t}\mathrm{U}_{t,\tau }\mathrm{R}_{\tau }\mathrm{W}_{\tau }%
\mathrm{R}_{\tau }\mathrm{U}_{\tau ,t}\mathrm{d}\tau \right) \varphi
\end{equation*}%
itself in (\ref{eq:eqtempweakfini}) belongs to the domain $\mathcal{D}(%
\mathrm{N})$ for any $\varphi \in \mathcal{D}(\mathrm{N})$. Consequently,
this result even holds with respect to the strong topology. We refrain from
providing a rigorous proof here, however, since we establish a significantly
stronger result in the next section using Condition \ref{AbstractAssumptions}
(v).
\end{remark}

\subsection{Yosida Regularization of the Differential Flow\label%
{sectionYosida}}

Based on the previous section, and specifically Corollary \ref{Resolvent
expression}, it is natural to expect that the unitary transformation $%
\mathrm{U}_{t,s}\mathrm{H}_{s}\mathrm{U}_{s,t}$ (for $s,t\in \mathbb{R}^{+}$%
) is equal to $\mathrm{H}_{t}$ plus some remainder term $\mathbb{A}_{t,s}$.
Recall that $\mathrm{U}_{s,t}=\mathrm{U}_{t,s}^{\ast }$ for any $s,t\in 
\mathbb{R}_{0}^{+}$. However, the operator $\mathbb{A}_{t,s}$ represents a
major obstacle, because we lack information regarding its domain, as well as
how the unitary propagator transforms $\mathcal{D}\left( \mathrm{H}%
_{t}\right) $.

Indeed, for a closed flow for which the remainder vanishes (i.e., $\mathbb{A}%
_{t,s}=0$), such as in \cite{bach-bru-memo,QuadraHamilFermio}, the strategy
is to show 
\begin{equation*}
\mathrm{U}_{t,s}\left( \mathrm{H}_{s}+i\lambda \mathbf{1}\right) ^{-1}%
\mathrm{U}_{s,t}=\left( \mathrm{H}_{t}+i\lambda \mathbf{1}\right)
^{-1},\qquad s,t\in \mathbb{R}^{+},
\end{equation*}%
for a sufficiently large parameter $\lambda \in \mathbb{R}^{+}$. This
approach exploits our explicit knowledge of the differential equation
governing the evolution of the resolvent, which in our context would
correspond to setting the $\mathrm{W}$-dependent terms to zero in Corollary %
\ref{Resolventeq}. However, because our flow does not close\footnote{%
Specifically, the term involving $\mathrm{W}$ in Corollary \ref{Resolventeq}
is non-zero.},\emph{\ a direct adaptation of this method fails}: Such an
approach would require analyzing the differential equation satisfied by the
mapping 
\begin{equation*}
t\mapsto \left( \mathrm{H}_{t}+\mathbb{A}_{t,s}+i\lambda \mathbf{1}\right)
^{-1}
\end{equation*}%
in the hope of deducing\footnote{%
If $\mathrm{U}_{t,s}\mathrm{H}_{s}\mathrm{U}_{t,s}^{\ast }=\mathrm{H}_{t}+%
\mathbb{A}_{t,s}$, then we must have $\mathbb{A}_{s,s}=0$.} that 
\begin{equation*}
\mathrm{U}_{t,s}\left( \mathrm{H}_{s}+i\lambda \mathbf{1}\right) ^{-1}%
\mathrm{U}_{s,t}=\mathrm{U}_{t,s}\left( \mathrm{H}_{s}+\mathbb{A}%
_{s,s}+i\lambda \mathbf{1}\right) ^{-1}\mathrm{U}_{s,t}=\left( \mathrm{H}%
_{t}+\mathbb{A}_{t,s}+i\lambda \mathbf{1}\right) ^{-1},\qquad s,t\in \mathbb{%
R}^{+}.
\end{equation*}%
Unfortunately, without further control over $\mathbb{A}_{t,s}$, it is
unclear whether this resolvent is even well-defined.

To circumvent this difficulty, we introduce a regularized version of the
flow that mitigates the unboundedness of the operators. We then analyze the
dynamics entirely within this regularized framework, subsequently lifting
the resulting properties to the full Hamiltonian in a static manner, i.e.,
for each point in time.

The standard strategy for mitigating unbounded operators is the celebrated
Yosida approximation (or regularization). Applied to the Hamiltonian $%
\mathrm{H}_{t}$ it corresponds to the bounded operator 
\begin{equation}
\mathrm{H}_{t}^{(\lambda )}\doteq \frac{i\lambda \mathrm{H}_{t}}{\mathrm{H}%
_{t}+i\lambda \mathbf{1}}\in \mathcal{B}\left( \mathfrak{F}\right) ,\qquad
t\in \mathbb{R}_{0}^{+},\ \lambda \in \mathbb{R}^{+}.  \label{YosidaApproc}
\end{equation}%
Note from Lemma \ref{diversPreservedom} that $(\mathrm{N}+\mathbf{1})\mathrm{%
H}_{t}^{(\lambda )}(\mathrm{N}+\mathbf{1})^{-1}\in \mathcal{B}(\mathfrak{F})$
for sufficiently large $\lambda \in \mathbb{R}^{+}$. Using the Yosida
approximation, we define the following operators on the dense domain $%
\mathcal{D}\left( \mathrm{N}\right) $ of the particle number operator $%
\mathrm{N}$:%
\begin{eqnarray}
\mathrm{Bwd}_{t,s}^{(\lambda )} &\doteq &\mathrm{U}_{s,t}\mathrm{H}%
_{t}^{(\lambda )}\mathrm{U}_{t,s}+\lambda ^{2}\int_{s}^{t}\mathrm{U}_{s,\tau
}\left( \mathrm{H}_{\tau }+i\lambda \mathbf{1}\right) ^{-1}\mathrm{W}_{\tau
}\left( \mathrm{H}_{\tau }+i\lambda \mathbf{1}\right) ^{-1}\mathrm{U}_{\tau
,s}\mathrm{d}\tau ,  \label{DefRbb} \\
\mathrm{Fwd}_{t,s}^{(\lambda )} &\doteq &\mathrm{U}_{t,s}\mathrm{H}%
_{s}^{(\lambda )}\mathrm{U}_{s,t}-\lambda ^{2}\int_{s}^{t}\mathrm{U}_{t,\tau
}\left( \mathrm{H}_{\tau }+i\lambda \mathbf{1}\right) ^{-1}\mathrm{W}_{\tau
}\left( \mathrm{H}_{\tau }+i\lambda \mathbf{1}\right) ^{-1}\mathrm{U}_{\tau
,t}\mathrm{d}\tau ,  \label{DefRbbforward}
\end{eqnarray}%
where $\mathrm{W}_{\tau }\equiv \mathrm{W}_{\tau }^{\ast \ast }$ is the
closure of the densely defined symmetric operator defined by (\ref%
{expression Wt}) via Proposition \ref{Lemma Htbis} (iii). These operators
are well-defined on $\mathcal{D}(\mathrm{N})$ by virtue of Lemma \ref%
{diversPreservedom copy(1)} and Theorem \ref{propagators} (under sufficient
conditions, in particular SB0-SB5 of Theorems \ref{HamilSA copy(2)} and the
gap condition). Being densely defined and symmetric, these operators are
automatically closable and we tacitly identify them with their closures.
Although some unboundedness stemming from the second quantization framework
is removed, note that these expressions do not \emph{a priori} define
bounded operators. Furthermore, we must give rigorous meaning to the
integral terms in (\ref{DefRbb})--(\ref{DefRbbforward}). To this end, we
establish the following technical result:

\begin{lemma}[Riemann integration of densely defined operators]
\label{TechnicalIntegralWandRWR}\mbox{}\newline
Assume Condition \ref{AbstractAssumptions} (i)--(iv). Take $T\in \mathbb{R}%
^{+}$ and $\lambda \gtrsim _{T}\Vert X_{0}\Vert _{\mathrm{op},2}$ (see,
e.g., (\ref{dfsdfsgsf})). Let $\left( A_{t}\right) _{t\geq 0}$ and $\left(
B_{t}\right) _{t\geq 0}$ be two families of operators on $\mathfrak{F}$ that
satisfy the following conditions for $\kappa \in \{0,1,3/2\}$:

\begin{enumerate}
\item[(i)] For any $t\in \mathbb{R}_{0}^{+}$, 
\begin{equation*}
\left( \mathrm{N}+\mathbf{1}\right) ^{\kappa }A_{t}\left( \mathrm{N}+\mathbf{%
1}\right) ^{-\kappa },\left( \mathrm{N}+\mathbf{1}\right) ^{\kappa
}B_{t}\left( \mathrm{N}+\mathbf{1}\right) ^{-\kappa }\in \mathcal{B}(%
\mathfrak{F}).
\end{equation*}

\item[(ii)] The mappings 
\begin{equation*}
t\mapsto \left( \mathrm{N}+\mathbf{1}\right) ^{\kappa }A_{t}\left( \mathrm{N}%
+\mathbf{1}\right) ^{-\kappa }\qquad \text{and}\qquad t\mapsto \left( 
\mathrm{N}+\mathbf{1}\right) ^{\kappa }B_{t}\left( \mathrm{N}+\mathbf{1}%
\right) ^{-\kappa }
\end{equation*}%
are strongly continuous on $\mathbb{R}_{0}^{+}$.
\end{enumerate}

\noindent Then, for each $s,t\in \lbrack 0,T]$, the expressions%
\begin{equation*}
\mathbb{W}_{t}\doteq A_{t}\mathrm{W}_{t}B_{t}\qquad \text{and}\qquad \mathbb{%
W}_{t}^{R}\doteq A_{t}\left( \mathrm{H}_{t}+i\lambda \mathbf{1}\right) ^{-1}%
\mathrm{W}_{t}\left( \mathrm{H}_{t}+i\lambda \mathbf{1}\right) ^{-1}B_{t}
\end{equation*}%
and their (Riemann) integration over $[s,t]$%
\begin{equation}
\int_{s}^{t}\mathbb{W}_{\tau }\mathrm{d}\tau \qquad \text{and}\qquad
\int_{s}^{t}\mathbb{W}_{\tau }^{R}\mathrm{d}\tau  \label{integraleURWRU}
\end{equation}%
define operators on the dense domain $\mathcal{D}(\mathrm{N}^{3/2})$. If
Condition \ref{AbstractAssumptions} (v) also holds, then $\mathbb{W}_{t}^{R}$
and its integration over $[s,t]$ can even be defined on $\mathcal{D}(\mathrm{%
N})$. If $B_{t}=A_{t}^{\ast }$ for all $t\in \lbrack 0,T]$, then all of the
above operators are also symmetric and therefore closable.
\end{lemma}

\begin{proof}
Whether the two expressions in (\ref{integraleURWRU}) are well-defined is
not a priori clear, as they involve integrals of unbounded operators. We
therefore give a rigorous mathematical meaning to these formal expressions,
beginning with a study of their integrands in Step 1. Fix once and for all $%
T\in \mathbb{R}^{+}$. \medskip

\noindent \underline{Step 1:} First, according to Proposition \ref{Lemma
Htbis} (iii) and Lemmata \ref{diversPreservedom} and \ref%
{Lemma-self-adjoint1 copy(1)}, as well as Condition \ref{AbstractAssumptions}
(iii), 
\begin{equation}
\max \left\{ \sup_{s\in \mathbb{R}_{0}^{+}}\left\Vert \mathrm{W}_{s}\left( 
\mathrm{H}_{\tau }+i\lambda \mathbf{1}\right) ^{-1}\left( \mathrm{N}+\mathbf{%
1}\right) ^{-\kappa }\right\Vert _{\mathrm{op}},\sup_{s\in \mathbb{R}%
_{0}^{+}}\left\Vert \mathrm{W}_{s}(\mathrm{N}+\mathbf{1})^{-\frac{3}{2}%
}\right\Vert _{\mathrm{op}}\right\} <\infty ,  \label{rytry}
\end{equation}%
with $\kappa =3/2$, while (\ref{rytry}) holds also for $\kappa =1$ under
Condition \ref{AbstractAssumptions} (v), via Lemma \ref{diversPreservedom
copy(1)}. Then, using now Assumption (i), we thus deduce that the operators
under the integrals satisfy the respective operator bounds for any $t\in 
\mathbb{R}_{0}^{+}$: 
\begin{equation*}
\left\Vert \mathbb{W}_{t}(\mathrm{N}+\mathbf{1})^{-\frac{3}{2}}\right\Vert _{%
\mathrm{op}}\leq \left\Vert A_{t}\right\Vert _{\mathrm{op}}\left\Vert 
\mathrm{W}_{t}(\mathrm{N}+\mathbf{1})^{-\frac{3}{2}}\right\Vert _{\mathrm{op}%
}\left\Vert (\mathrm{N}+\mathbf{1})^{\frac{3}{2}}B_{t}(\mathrm{N}+\mathbf{1}%
)^{-\frac{3}{2}}\right\Vert _{\mathrm{op}}<\infty ,
\end{equation*}%
as well as 
\begin{multline*}
\left\Vert \mathbb{W}_{t}^{R}(\mathrm{N}+\mathbf{1})^{-\kappa }\right\Vert _{%
\mathrm{op}}\leq \lambda ^{-1}\left\Vert A_{t}\right\Vert _{\mathrm{op}%
}\left\Vert \mathrm{W}_{t}(\mathrm{N}+\mathbf{1})^{-\frac{3}{2}}\right\Vert
_{\mathrm{op}}\left\Vert (\mathrm{N}+\mathbf{1})^{\frac{3}{2}}\left( \mathrm{%
H}_{t}+i\lambda \mathbf{1}\right) ^{-1}(\mathrm{N}+\mathbf{1})^{-\kappa
}\right\Vert _{\mathrm{op}} \\
\left\Vert (\mathrm{N}+\mathbf{1})^{\kappa }B_{t}(\mathrm{N}+\mathbf{1}%
)^{-\kappa }\right\Vert _{\mathrm{op}}<\infty ,
\end{multline*}%
with $\kappa =3/2$, while the last inequality holds also for $\kappa =1$
under Condition \ref{AbstractAssumptions} (v), thanks again to Lemma \ref%
{diversPreservedom copy(1)}. We now focus on the formal expressions of (\ref%
{integraleURWRU}). Below, we establish the continuity of their integrands,
which immediately ensures their Riemann integrabilty on compact intervals.
We analyze the first integral of (\ref{integraleURWRU}) in the next step,
while the second is the subject of Step 3.\medskip

\noindent \underline{Step 2:} For any $s\in \mathbb{R}_{0}^{+}$, $%
t_{1},t_{2}\in \lbrack s,T]$ and $\varphi \in \mathcal{D}(\mathrm{N}^{3/2})$%
, observe from the triangle inequality that 
\begin{eqnarray}
\left\Vert \left( \mathbb{W}_{t_{2}}-\mathbb{W}_{t_{1}}\right) \varphi
\right\Vert _{\mathfrak{F}} &\leq &\left\Vert \left(
A_{t_{2}}-A_{t_{1}}\right) \mathrm{W}_{t_{2}}B_{t_{2}}\varphi \right\Vert _{%
\mathfrak{F}}+\left\Vert A_{t_{1}}\right\Vert _{\mathrm{op}}\left\Vert
\left( \mathrm{W}_{t_{2}}-\mathrm{W}_{t_{1}}\right) B_{t_{2}}\varphi
\right\Vert _{\mathfrak{F}}  \notag \\
&&+\left\Vert A_{t_{1}}\right\Vert _{\mathrm{op}}\left\Vert \mathrm{W}%
_{t_{1}}(\mathrm{N}+\mathbf{1})^{-\frac{3}{2}}\right\Vert _{\mathrm{op}%
}\left\Vert (\mathrm{N}+\mathbf{1})^{\frac{3}{2}}\left(
B_{t_{2}}-B_{t_{1}}\right) \varphi \right\Vert _{\mathfrak{F}}.
\label{qweqwe}
\end{eqnarray}%
By Assumptions (i)--(ii) and Inequality (\ref{rytry}), this upper bound
vanishes in the limit $t_{1}\rightarrow t_{2}$ for any $s\in \mathbb{R}%
_{0}^{+}$, $t_{2}\in (s,T]$ and $\varphi \in \mathcal{D}(\mathrm{N}^{3/2})$.
Indeed, in this situation, $B_{t_{2}}\varphi \in \mathcal{D}(\mathrm{N}%
^{3/2})$ and we deduce from Lemma \ref{Lemma-self-adjoint1 copy(1)}
(extended to all vectors of $\mathcal{D}(\mathrm{N}^{3/2})$) that 
\begin{equation}
\left\Vert \left( \mathrm{W}_{t_{2}}-\mathrm{W}_{t_{1}}\right)
B_{t_{2}}\varphi \right\Vert _{\mathfrak{F}}^{2}\leq 10\left\Vert \left[ \mu
_{t_{2}},X_{t_{2}}\right] -\left[ \mu _{t_{1}},X_{t_{1}}\right] \right\Vert
_{\mathrm{op},2}\left( \left\Vert \mathrm{N}^{3/2}B_{t_{2}}\varphi
\right\Vert _{\mathfrak{F}}+\left\Vert B_{t_{2}}\varphi \right\Vert _{%
\mathfrak{F}}\right) ,  \label{dfdfhdhdgfh}
\end{equation}%
where $[X_{t},\mu _{t}]\equiv ([X_{k,t},\mu _{k,t}])_{k\in \mathbb{N}}$. By
Condition \ref{AbstractAssumptions} (iii), the right-hand side of this last
inequality vanishes in the limit $t_{1}\rightarrow t_{2}$ (with $t_{2}\neq 0$%
). We thus conclude that 
\begin{equation*}
\lim_{t_{1}\rightarrow t_{2}}\left\Vert \left( \mathrm{W}_{t_{2}}-\mathrm{W}%
_{t_{1}}\right) B_{t_{2}}\varphi \right\Vert _{\mathfrak{F}}^{2}=0,
\end{equation*}%
and, by Assumptions (i)--(ii) and Inequalities (\ref{rytry})--(\ref{qweqwe}%
), it follows that 
\begin{equation}
\lim_{t_{1}\rightarrow t_{2}}\left\Vert \left( \mathbb{W}_{t_{2}}-\mathbb{W}%
_{t_{1}}\right) \varphi \right\Vert _{\mathfrak{F}}=0  \label{strongcontWbb}
\end{equation}%
for any $t_{2}\in (0,T]$ and $\varphi \in \mathcal{D}(\mathrm{N}^{3/2})$. We
can then interpret the formal expression 
\begin{equation*}
\int_{s}^{t}\mathbb{W}_{\tau }\mathrm{d}\tau ,\qquad s,t\in \left( 0,T\right]
,
\end{equation*}%
as the operator defined on $\mathcal{D}(\mathrm{N}^{3/2})$ via the following
(Banach-valued) Riemann integral on the Hilbert space $\mathfrak{F}$: 
\begin{equation*}
\left( \int_{s}^{t}\mathbb{W}_{\tau }\mathrm{d}\tau \right) (\mathrm{N}+%
\mathbf{1})^{-\frac{3}{2}}\varphi \doteq \int_{s}^{t}\mathbb{W}_{\tau }(%
\mathrm{N}+\mathbf{1})^{-\frac{3}{2}}\varphi \mathrm{d}\tau ,\qquad \varphi
\in \mathfrak{F}.
\end{equation*}%
Note that the continuity of the integrand only holds on $(0,T]$, but since
the integrand is bounded near $0$, it remains Riemann-integrable on $[0,T]$.
Indeed, using Condition \ref{AbstractAssumptions} (iii) and Lemma \ref%
{Lemma-self-adjoint1 copy(1)}, one verifies that%
\begin{equation*}
\lim_{s\rightarrow 0}\int_{s}^{t}\mathbb{W}_{\tau }(\mathrm{N}+\mathbf{1})^{-%
\frac{3}{2}}\varphi \mathrm{d}\tau =\int_{0}^{t}\mathbb{W}_{\tau }(\mathrm{N}%
+\mathbf{1})^{-\frac{3}{2}}\varphi \mathrm{d}\tau
\end{equation*}%
since 
\begin{equation*}
\sup_{\tau \in \lbrack 0,t]}\left\Vert \mathbb{W}_{\tau }(\mathrm{N}+\mathbf{%
1})^{-\frac{3}{2}}\varphi \right\Vert _{\mathfrak{F}}<\infty .
\end{equation*}%
The first expression of (\ref{integraleURWRU}) therefore defines an operator
on the (dense) domain $\mathcal{D}(\mathrm{N}^{3/2})$. \medskip

\noindent \underline{Step 3:} Now concerning the second integral of (\ref%
{integraleURWRU}), we first use the following upper bound for any $s\in 
\mathbb{R}_{0}^{+}$, $t_{1},t_{2}\in \lbrack s,T]$, $\kappa \in \{1,3/2\}$
and $\varphi \in \mathcal{D}(\mathrm{N}^{\kappa })$: 
\begin{eqnarray}
&&\left\Vert \left( \mathbb{W}_{t_{2}}^{R}-\mathbb{W}_{t_{1}}^{R}\right)
\varphi \right\Vert _{\mathfrak{F}}  \notag \\
&\leq &\left\Vert \left( A_{t_{2}}-A_{t_{1}}\right) \left( \mathrm{H}%
_{t_{2}}+i\lambda \mathbf{1}\right) ^{-1}\mathrm{W}_{t_{2}}\left( \mathrm{H}%
_{t_{2}}+i\lambda \mathbf{1}\right) ^{-1}B_{t_{2}}\varphi \right\Vert _{%
\mathfrak{F}}  \notag \\
&&+\left\Vert A_{t_{1}}\right\Vert _{\mathrm{op}}\left\Vert \left( \left( 
\mathrm{H}_{t_{2}}+i\lambda \mathbf{1}\right) ^{-1}-\left( \mathrm{H}%
_{t_{1}}+i\lambda \mathbf{1}\right) ^{-1}\right) \mathrm{W}_{t_{2}}\left( 
\mathrm{H}_{t_{2}}+i\lambda \mathbf{1}\right) ^{-1}B_{t_{2}}\varphi
\right\Vert _{\mathfrak{F}}  \notag \\
&&+\lambda ^{-1}\left\Vert A_{t_{1}}\right\Vert _{\mathrm{op}}\left\Vert
\left( \mathrm{W}_{t_{2}}-\mathrm{W}_{t_{1}}\right) \left( \mathrm{H}%
_{t_{2}}+i\lambda \mathbf{1}\right) ^{-1}B_{t_{2}}\varphi \right\Vert _{%
\mathfrak{F}}  \notag \\
&&+\lambda ^{-1}\left\Vert A_{t_{1}}\right\Vert _{\mathrm{op}}\left\Vert 
\mathrm{W}_{t_{1}}(\mathrm{N}+\mathbf{1})^{-\frac{3}{2}}\right\Vert _{%
\mathrm{op}}\left\Vert (\mathrm{N}+\mathbf{1})^{\frac{3}{2}}\left( \left( 
\mathrm{H}_{t_{2}}+i\lambda \mathbf{1}\right) ^{-1}-\left( \mathrm{H}%
_{t_{1}}+i\lambda \mathbf{1}\right) ^{-1}\right) B_{t_{2}}\varphi
\right\Vert _{\mathfrak{F}}  \notag \\
&&+\lambda ^{-1}\left\Vert A_{t_{1}}\right\Vert _{\mathrm{op}}\left\Vert 
\mathrm{W}_{t_{1}}(\mathrm{N}+\mathbf{1})^{-\frac{3}{2}}\right\Vert _{%
\mathrm{op}}\left\Vert (\mathrm{N}+\mathbf{1})^{\frac{3}{2}}\left( \mathrm{H}%
_{t_{1}}+i\lambda \mathbf{1}\right) ^{-1}(\mathrm{N}+\mathbf{1})^{-\kappa
}\right\Vert _{\mathrm{op}}  \notag \\
&&\qquad \qquad \qquad \qquad \qquad \qquad \qquad \qquad \qquad \qquad
\times \left\Vert (\mathrm{N}+\mathbf{1})^{\kappa }\left(
B_{t_{2}}-B_{t_{1}}\right) \varphi \right\Vert _{\mathfrak{F}}.
\label{qweqwe2}
\end{eqnarray}%
By Assumption (i), $\varphi \in \mathcal{D}(\mathrm{N}^{\kappa })$ yields $%
B_{t_{2}}\varphi \in \mathcal{D}(\mathrm{N}^{\kappa })$ for any $\kappa \in
\{1,3/2\}$. Furthermore, given $\kappa \in \{1,3/2\}$, we get from Lemma \ref%
{diversPreservedom} ($\kappa =3/2$) or Lemma \ref{diversPreservedom copy(1)}
under Condition \ref{AbstractAssumptions} (v) ($\kappa =1$), together with
Lemma \ref{Lemma-self-adjoint1 copy(1)}, that, for any $s\in \mathbb{R}%
_{0}^{+}$, $t_{1},t_{2}\in \lbrack s,T]$, and $\varphi \in \mathcal{D}(%
\mathrm{N}^{\kappa })$, 
\begin{equation}
\left\Vert \left( \mathrm{W}_{t_{2}}-\mathrm{W}_{t_{1}}\right) \left( 
\mathrm{H}_{t_{2}}+i\lambda \mathbf{1}\right) ^{-1}B_{t_{2}}\varphi
\right\Vert _{\mathfrak{F}}^{2}\leq \theta _{\kappa }\left\Vert \left[ \mu
_{t_{2}},X_{t_{2}}\right] -\left[ \mu _{t_{1}},X_{t_{1}}\right] \right\Vert
_{\mathrm{op},2}\left\Vert (\mathrm{N}+\mathbf{1})^{\kappa }B_{t_{2}}\varphi
\right\Vert _{\mathfrak{F}},  \label{sfsdsd1}
\end{equation}%
where the parameter $\lambda \in \mathbb{R}^{+}$ is chosen sufficiently large%
\footnote{%
Compare with (\ref{dfsdfsgsf}).} such that 
\begin{equation*}
\lambda >2^{4+\frac{1}{2}}D_{T}^{1/2}\left\Vert X_{0}\right\Vert _{\mathrm{op%
},2}
\end{equation*}%
(see Condition \ref{AbstractAssumptions} (i)) and the strictly positive
constants $\theta _{3/2}$ and $\theta _{1}$ are then respectively defined by%
\begin{equation*}
\theta _{3/2}\doteq \frac{20}{\lambda -2^{4+\frac{1}{2}}D_{T}^{1/2}\left%
\Vert X_{0}\right\Vert _{\mathrm{op},2}}\qquad \text{and}\qquad \theta
_{1}\doteq \frac{20\mathrm{K}_{c,b}}{1-4\zeta }\left( 1+\frac{12\zeta }{%
1-2^{7/2}\zeta }\right)
\end{equation*}%
with $\zeta \doteq \lambda ^{-1}D_{T}^{1/2}\Vert X_{0}\Vert _{\mathrm{op},2}$%
. Here, $\mathrm{K}_{c,b}\in \mathbb{R}^{+}$ is defined by (\ref{def Kcb})
for the constant $c\in \mathbb{R}^{+}$ of Condition \ref{AbstractAssumptions}
(v) and any parameter $b\in \mathbb{R}$ such that $H(h_{t})\geq b\mathbf{1}$
for all times $t\in \lbrack 0,T]$. The existence of such a $b\in \mathbb{R}$
is guaranteed by the following observations:

\begin{itemize}
\item By Condition \ref{AbstractAssumptions} (ii) and Proposition \ref%
{Proposition self-adjoint copy(1)}, $H(h_{t})$ is a lower semibounded
self-adjoint operator for all $t\in \mathbb{R}_{0}^{+}$.

\item By the triangle inequality and Condition \ref{AbstractAssumptions}
(i), for any $t\in \lbrack 0,T]$, 
\begin{equation*}
\left\Vert \int_{0}^{t}X_{\tau }X_{\tau }^{\ast }\mathrm{d}\tau \right\Vert
_{\mathrm{op},1}\leq \int_{0}^{T}\left\Vert X_{\tau }\right\Vert _{\mathrm{op%
},2}^{2}\mathrm{d}\tau \leq D_{T}T\left\Vert X_{0}\right\Vert _{\mathrm{op}%
,2}^{2}.
\end{equation*}%
By Propositions \ref{prop useful} and \ref{Proposition self-adjoint copy(1)}%
, it follows that 
\begin{equation*}
H(h_{t})-H(h_{0})=2H\left( \int_{0}^{t}X_{\tau }X_{\tau }^{\ast }\mathrm{d}%
\tau \right) \in \mathcal{B}(\mathfrak{F}).
\end{equation*}
\end{itemize}

\noindent Therefore, using Inequalities (\ref{qweqwe2})--(\ref{sfsdsd1})
together with Assumptions (i)--(ii) as well as Lemmata \ref%
{diversPreservedom} and \ref{coro-Ht copy(1)}, we obtain the following
limit: 
\begin{equation}
\lim_{t_{1}\rightarrow t_{2}}\left\Vert \left( \mathbb{W}_{t_{2}}^{R}-%
\mathbb{W}_{t_{1}}^{R}\right) \varphi \right\Vert _{\mathfrak{F}}=0
\label{StrongcontinuityRWR}
\end{equation}%
for any $s\in \mathbb{R}_{0}^{+}$, $t_{2}\in (s,T]$ and $\varphi \in 
\mathcal{D}(\mathrm{N}^{3/2})$. If Condition \ref{AbstractAssumptions} (v)
also holds, then Equation (\ref{StrongcontinuityRWR}) is satisfied for all $%
\varphi \in \mathcal{D}(\mathrm{N})$, by using additionally Lemma \ref%
{diversPreservedom copy(1)}. We can then interpret the formal expression 
\begin{equation*}
\int_{s}^{t}\mathbb{W}_{\tau }^{R}\mathrm{d}\tau ,\qquad s,t\in \left( 0,T%
\right] ,
\end{equation*}%
as the operator defined on $\mathcal{D}(\mathrm{N}^{\kappa })$ via the
following Riemann integral on the Hilbert space $\mathfrak{F}$:%
\begin{equation*}
\left( \int_{s}^{t}\mathbb{W}_{\tau }^{R}\mathrm{d}\tau \right) (\mathrm{N}+%
\mathbf{1})^{-\kappa }\varphi \doteq \int_{s}^{t}\mathbb{W}_{\tau }^{R}(%
\mathrm{N}+\mathbf{1})^{-\kappa }\varphi \mathrm{d}\tau ,\qquad \varphi \in 
\mathfrak{F},
\end{equation*}%
where $\kappa =3/2$, or $\kappa =1$ provided that Condition \ref%
{AbstractAssumptions} (v) additionally holds. Even though the integrand may
be continuous only on $(0,T]$, it remains measurable on every compact subset
of $[0,T]$. The details are omitted here, as the arguments are closely
parallel those presented in Step 2. The second expression in (\ref%
{integraleURWRU}) therefore defines an operator on the domain $\mathcal{D}(%
\mathrm{N}^{3/2})$, as well as on $\mathcal{D}(\mathrm{N})$ whenever
Condition \ref{AbstractAssumptions} (v) is satisfied.
\end{proof}

We must also consider the asymptotic behavior of the (Banach-valued) Riemann
integrals in Equation (\ref{integraleURWRU}) as $t\rightarrow \infty $. This
is addressed in the following lemma.

\begin{lemma}[Improper Riemann integration of operators]
\label{TechnicalIntegralWandRWR copy(1)}\mbox{}\newline
Assume Condition \ref{AbstractAssumptions} (i)--(iv) and let $A\equiv \left(
A_{t}\right) _{t\geq 0}$ and $B\equiv \left( B_{t}\right) _{t\geq 0}$ be two
families of operators on $\mathfrak{F}$ satisfying assumptions (i)--(ii) of
Lemma \ref{TechnicalIntegralWandRWR} for $\kappa \in \{0,1,3/2\}$.
Furthermore, suppose that:

\begin{enumerate}
\item[(i)] The constant $D_{T}$ in Condition \ref{AbstractAssumptions} (i)
is uniformly bounded\footnote{%
This is satisfied under Condition \ref{AbstractAssumptions} (v). Indeed, in
this case, $D_{T}=1$. See (\ref{sdadasdasdasdas}).} on $\mathbb{R}_{0}^{+}$.

\item[(ii)] The mapping $t\mapsto \Vert \lbrack X_{t},\mu _{t}]\Vert _{%
\mathrm{op,2}}$ is integrable on $\mathbb{R}_{0}^{+}$.

\item[(iii)] As $t\rightarrow \infty $, the operators $A_{t}$, $B_{t}$ and $(%
\mathrm{H}_{t}+i\lambda \mathbf{1})^{-1}$ converge strongly to three bounded
operators $A_{\infty }$, $B_{\infty }$ and $(\mathrm{H}_{\infty }+i\lambda 
\mathbf{1})^{-1}$, respectively, where $\lambda \gtrsim \Vert X_{0}\Vert _{%
\mathrm{op},2}$ is sufficiently large\footnote{%
The existence of such a $\lambda $ is guaranteed because the constant $D_{T}$
in Condition \ref{AbstractAssumptions} (i) is, by assumption, uniformly
bounded on $\mathbb{R}_{0}^{+}$. See (\ref{dfsdfsgsf}).}.

\item[(iv)] The following uniform bound holds:%
\begin{equation*}
\sup_{t\in \mathbb{R}_{0}^{+}}\left\{ \left\Vert A_{t}\right\Vert _{\mathrm{%
op}}+\left\Vert B_{t}\right\Vert _{\mathrm{op}}+\left\Vert \left( \mathrm{N}+%
\mathbf{1}\right) A_{t}\left( \mathrm{N}+\mathbf{1}\right) ^{-1}\right\Vert
_{\mathrm{op}}+\left\Vert \left( \mathrm{N}+\mathbf{1}\right) B_{t}\left( 
\mathrm{N}+\mathbf{1}\right) ^{-1}\right\Vert _{\mathrm{op}}\right\} <\infty
.
\end{equation*}
\end{enumerate}

\noindent Then the assertions of Lemma \ref{TechnicalIntegralWandRWR} remain
valid for $t=\infty $.
\end{lemma}

\begin{proof}
These additional hypotheses allow us to reproduce the same arguments (as in
Lemma \ref{TechnicalIntegralWandRWR}), thereby extending (\ref{strongcontWbb}%
) and (\ref{StrongcontinuityRWR}) to the case where $T\in \mathbb{R}%
_{0}^{+}\cup \{\infty \}$. However, we must still justify the finiteness of
the corresponding integrals as $t\rightarrow \infty $ as, obviously, $%
[0,+\infty )$ does not have finite Lebesgue measure. In fact, using
Condition \ref{AbstractAssumptions} (i) and (iii) together with together
Lemmata \ref{diversPreservedom} and \ref{Lemma-self-adjoint1 copy(1)} we
verify that for any $t\in \mathbb{R}_{0}^{+}$ and $\psi ,\varphi \in 
\mathcal{D}(\mathrm{N}^{3/2})$, 
\begin{multline*}
\max \left\{ \left\Vert \mathbb{W}_{t}\psi \right\Vert _{\mathfrak{F}%
},\left\Vert \mathbb{W}_{t}^{R}\varphi \right\Vert _{\mathfrak{F}}\right\}
=\max \left\{ \left\Vert A_{t}\mathrm{W}_{t}B_{t}\psi \right\Vert _{%
\mathfrak{F}},\left\Vert A_{t}\left( \mathrm{H}_{t}+i\lambda \mathbf{1}%
\right) ^{-1}\mathrm{W}_{t}\left( \mathrm{H}_{t}+i\lambda \mathbf{1}\right)
^{-1}B_{t}\varphi \right\Vert _{\mathfrak{F}}\right\} \\
\lesssim \left\Vert \left[ X_{t},\mu _{t}\right] \right\Vert _{\mathrm{op,2}%
},
\end{multline*}%
where the upper-bound is integrable by assumption. If Condition \ref%
{AbstractAssumptions} (v) is satisfied then the last inequality is true for $%
\varphi \in \mathcal{D}(\mathrm{N})$, thanks to Lemma \ref{diversPreservedom
copy(1)}. As a result, by the triangle inequality, for any $s\in \mathbb{R}%
_{0}^{+}$ we can then interpret the formal expression 
\begin{equation*}
\int_{s}^{\infty }\mathbb{W}_{\tau }\mathrm{d}\tau \qquad \text{and}\qquad
\int_{s}^{\infty }\mathbb{W}_{\tau }^{R}\mathrm{d}\tau
\end{equation*}%
as the operators defined on $\mathcal{D}(\mathrm{N}^{3/2})$ via the
following improper Riemann integrals on the Hilbert space $\mathfrak{F}$:%
\begin{equation*}
\left( \int_{s}^{\infty }\mathbb{W}_{\tau }\mathrm{d}\tau \right) \psi
\doteq \lim_{t\rightarrow \infty }\int_{s}^{t}\mathbb{W}_{\tau }\psi \mathrm{%
d}\tau \qquad \text{and}\qquad \left( \int_{s}^{\infty }\mathbb{W}_{\tau
}^{R}\mathrm{d}\tau \right) \varphi \doteq \lim_{t\rightarrow \infty
}\int_{s}^{t}\mathbb{W}_{\tau }^{R}\varphi \mathrm{d}\tau
\end{equation*}%
for any $\psi ,\varphi \in \mathcal{D}(\mathrm{N}^{3/2})$. If Condition \ref%
{AbstractAssumptions} (v) is satisfied, then the last improper Riemann
integral remains well-defined for $\varphi \in \mathcal{D}(\mathrm{N}%
)\varsupsetneq \mathcal{D}(\mathrm{N}^{3/2})$.
\end{proof}

\begin{remark}[Bochner integrals]
\label{remark utiles copy(3)}\mbox{}\newline
In our setting, all Riemann integrals (proper or improper) can be viewed as
Bochner integrals (even if they take values in a non-separable\footnote{%
Recall that the space of bounded operators on an infinite-dimensional
Hilbert space is always non-separable with respect to the operator norm.}
Banach space, as is the case here), because all integrands under
consideration are continuous and absolutely integrable over the domain. See,
e.g., \cite[Chapter III, Theorems 1.1 and 1.2]{pettis}.
\end{remark}

Lemmata \ref{TechnicalIntegralWandRWR} and \ref{TechnicalIntegralWandRWR
copy(1)} can be used to define Hamiltonians (\ref{DefRbb}) and (\ref%
{DefRbbforward}). Indeed, under Conditions \ref{AbstractAssumptions} (i) and
(iii), the family $(\mathrm{U}_{t,s})_{s,t\in \mathbb{R}_{0}^{+}}$ of
unitary operators satisfies Assumptions (i)--(ii) of Lemma \ref%
{TechnicalIntegralWandRWR}, thanks to Proposition \ref{prop useful} and
Theorem \ref{propagators}. The case $t=\infty $ is studied in Section \ref%
{Asymptotics}, which additionally involves Condition \ref{Assumptionsasympt}.

With the regularized Hamiltonians (\ref{DefRbb})--(\ref{DefRbbforward})
well-defined, we are now in a position to study their properties. In
particular, they generate a tractable time evolution. The following lemma
provides the explicit expression for the Yosida approximation of $\mathrm{H}%
_{t}$ for any time $t\in \mathbb{R}_{0}^{+}$. Our goal is then to leverage
this information to establish a parallel result for the unregularized
Hamiltonian $\mathrm{H}_{t}$ itself. Notably, this is the only instance in
this section where we exploit the dynamic, i.e., time-dependent properties.
In the sequel, the information established here is indeed transferred in a
\textquotedblleft static manner\textquotedblright , that is, at each fixed
time. This strategy allows us to bypass the highly non-trivial task of
rigorously computing the derivative of $s\mapsto \mathrm{U}_{t,s}\mathrm{H}%
_{s}\mathrm{U}_{s,t}$ at fixed $t\in \mathbb{R}_{0}^{+}$, which is severely
complicated by the unboundedness of the Hamiltonians.

\begin{proposition}[Yosida regularization of the differential flow]
\label{EvolRegu}\mbox{}\newline
Under Condition \ref{AbstractAssumptions} (i)--(v), for $\lambda
>2^{5+1/2}\Vert X_{0}\Vert _{\mathrm{op},2}$, $s,t\in \mathbb{R}_{0}^{+}$
and $\varphi \in \mathcal{D}(\mathrm{N})$, 
\begin{equation*}
\mathrm{Bwd}_{t,s}^{(\lambda )}\varphi =\mathrm{H}_{s}^{(\lambda )}\varphi
\qquad \text{and}\qquad \mathrm{Fwd}_{t,s}^{(\lambda )}\varphi =\mathrm{H}%
_{t}^{(\lambda )}\varphi ,
\end{equation*}%
where $\mathrm{Bwd}_{t,s}^{(\lambda )}$ and $\mathrm{Fwd}_{t,s}^{(\lambda )}$
are the densely defined operators respectively defined by (\ref{DefRbb})--(%
\ref{DefRbbforward}) via Lemma \ref{TechnicalIntegralWandRWR}, while $%
\mathrm{H}_{s}^{(\lambda )}$ is the Yosida approximation (\ref{YosidaApproc}%
) of $\mathrm{H}_{s}$.
\end{proposition}

\begin{proof}
Fix in all the proof the parameter $\lambda \in \mathbb{R}^{+}$ such that $%
\lambda >2^{5+1/2}\Vert X_{0}\Vert _{\mathrm{op},2}$ (cf. (\ref{dfsdfsgsf})
with $D_{T}=1$). Take $\epsilon \in \mathbb{R}^{+}$ and $\varphi ,\psi \in 
\mathcal{D}\left( \mathrm{N}\right) $. Recall that, by Lemma \ref%
{diversPreservedom copy(1)} (for which Condition \ref{AbstractAssumptions}
(v) is crucial) together with Proposition \ref{Lemma Htbis} (iii) and Lemma %
\ref{Lemma-self-adjoint1 copy(1)}, relying on Condition \ref%
{AbstractAssumptions} (iii), the operators $\left( \mathrm{H}_{t}+i\lambda 
\mathbf{1}\right) ^{-1}\mathrm{W}_{t}\left( \mathrm{H}_{t}+i\lambda \mathbf{1%
}\right) ^{-1}$, $t\in \mathbb{R}_{0}^{+}$, are always well-defined on the
dense domain $\mathcal{D}(\mathrm{N})$. Additionally, using Theorem \ref%
{propagators}, we ensure that everything is well-defined in the following
computation done for any $\varphi ,\psi \in \mathcal{D}\left( \mathrm{N}%
\right) $ and $t\in \mathbb{R}^{+}$:%
\begin{eqnarray*}
\mathbf{X}_{\epsilon } &\doteq &\left\langle \varphi ,\left( \epsilon
^{-1}\left( \mathrm{U}_{s,t+\epsilon }\mathrm{H}_{t+\epsilon }^{(\lambda )}%
\mathrm{U}_{t+\epsilon ,s}-\mathrm{U}_{s,t}\mathrm{H}_{t}^{(\lambda )}%
\mathrm{U}_{t,s}\right) +\lambda ^{2}\mathrm{U}_{s,t}\mathrm{R}_{t}\mathrm{W}%
_{t}\mathrm{R}_{t}\mathrm{U}_{t,s}\right) \psi \right\rangle _{\mathfrak{F}}
\\
&=&\left\langle \varphi ,\epsilon ^{-1}\left( \mathrm{U}_{s,t+\epsilon }-%
\mathrm{U}_{s,t}\right) \left( i\lambda +\lambda ^{2}\mathrm{R}_{t}\right) 
\mathrm{U}_{t,s}\psi \right\rangle _{\mathfrak{F}}+\left\langle \varphi
,\lambda ^{2}\mathrm{U}_{s,t}\mathrm{R}_{t}\mathrm{W}_{t}\mathrm{R}_{t}%
\mathrm{U}_{t,s}\psi \right\rangle _{\mathfrak{F}} \\
&&+\left\langle \mathrm{U}_{t+\epsilon ,s}\varphi ,\lambda ^{2}\epsilon
^{-1}\left( \mathrm{R}_{t+\epsilon }-\mathrm{R}_{t}\right) \mathrm{U}%
_{t,s}\psi \right\rangle _{\mathfrak{F}}+\left\langle \varphi ,\epsilon ^{-1}%
\mathrm{U}_{s,t+\epsilon }\left( i\lambda +\lambda ^{2}\mathrm{R}%
_{t+\epsilon }\right) \left( \mathrm{U}_{t+\epsilon ,s}-\mathrm{U}%
_{t,s}\right) \psi \right\rangle _{\mathfrak{F}},
\end{eqnarray*}%
with the notation $\mathrm{R}_{t}\doteq \left( \mathrm{H}_{t}+i\lambda 
\mathbf{1}\right) ^{-1}$ and 
\begin{equation}
\mathrm{H}_{t}^{(\lambda )}\doteq \frac{i\lambda \mathrm{H}_{t}}{\mathrm{H}%
_{t}+i\lambda \mathbf{1}}=i\lambda +\frac{\lambda ^{2}}{\mathrm{H}%
_{t}+i\lambda \mathbf{1}}=i\lambda +\lambda ^{2}\mathrm{R}_{t}.
\label{dfsdfsfsdfsf}
\end{equation}%
By Theorem \ref{propagators} (iv) we know that $\mathrm{U}_{t,s}\psi ,%
\mathrm{U}_{t+\epsilon ,s}\varphi \in \mathcal{D}\left( \mathrm{N}\right) $
and we can therefore use Proposition \ref{Resolventeq} (with Condition \ref%
{AbstractAssumptions} (v)). As $\mathcal{D}(\mathrm{N})\subseteq \mathcal{D}(%
\mathrm{N}^{1/2})$, Theorem \ref{propagators} (v) also applies. By combining
these observations and going further into the computation we get that%
\begin{eqnarray*}
\mathbf{X}_{\epsilon } &=&\left\langle \left( \epsilon ^{-1}\left( \mathrm{U}%
_{t+\epsilon ,s}-\mathrm{U}_{t,s}\right) +i\mathrm{G}_{t}\mathrm{U}%
_{t,s}\right) \varphi ,\left( i\lambda +\lambda ^{2}\mathrm{R}_{t}\right) 
\mathrm{U}_{t,s}\psi \right\rangle _{\mathfrak{F}} \\
&&+\left\langle \mathrm{U}_{t+\epsilon ,s}\varphi ,\lambda ^{2}\left(
\epsilon ^{-1}\left( \mathrm{R}_{t+\epsilon }-\mathrm{R}_{t}\right) -i\left[ 
\mathrm{R}_{t},\mathrm{G}_{t}\right] +\mathrm{R}_{t}\mathrm{W}_{t}\mathrm{R}%
_{t}\right) \mathrm{U}_{t,s}\psi \right\rangle _{\mathfrak{F}} \\
&&+\left\langle \varphi ,\mathrm{U}_{s,t+\epsilon }\left( i\lambda +\lambda
^{2}\mathrm{R}_{t+\epsilon }\right) \left( \epsilon ^{-1}\left( \mathrm{U}%
_{t+\epsilon ,s}-\mathrm{U}_{t,s}\right) +i\mathrm{G}_{t}\mathrm{U}%
_{t,s}\right) \psi \right\rangle _{\mathfrak{F}} \\
&&+\left\langle \left( \mathrm{U}_{t,s}-\mathrm{U}_{t+\epsilon ,s}\right)
\varphi ,i\mathrm{G}_{t}\left( i\lambda +\lambda ^{2}\mathrm{R}_{t}\right) 
\mathrm{U}_{t,s}\psi \right\rangle _{\mathfrak{F}} \\
&&+\left\langle \mathrm{U}_{t+\epsilon ,s}\varphi ,\lambda ^{2}\left( 
\mathrm{R}_{t}-\mathrm{R}_{t+\epsilon }\right) i\mathrm{G}_{t}\mathrm{U}%
_{t,s}\psi \right\rangle \\
&&+\left\langle \left( \mathrm{U}_{t,s}-\mathrm{U}_{t+\epsilon ,s}\right)
\varphi ,\lambda ^{2}\mathrm{R}_{t}\mathrm{W}_{t}\mathrm{R}_{t}\mathrm{U}%
_{t,s}\psi \right\rangle _{\mathfrak{F}},
\end{eqnarray*}%
where all the terms cancel each other in the limit $\epsilon \rightarrow 0$
at any $t\in \mathbb{R}^{+}$ and $s\in \mathbb{R}_{0}^{+}$ for the reasons
stated above, together with the strong continuity of $(\mathrm{R}_{t})_{t\in 
\mathbb{R}_{0}^{+}}$ (Lemma \ref{coro-Ht copy(1)}). To compute this limit
note that we apply the equality 
\begin{eqnarray*}
&&\left\langle \mathrm{U}_{t+\epsilon ,s}\varphi ,\lambda ^{2}\left(
\epsilon ^{-1}\left( \mathrm{R}_{t+\epsilon }-\mathrm{R}_{t}\right) -i\left[ 
\mathrm{R}_{t},\mathrm{G}_{t}\right] +\mathrm{R}_{t}\mathrm{W}_{t}\mathrm{R}%
_{t}\right) \mathrm{U}_{t,s}\psi \right\rangle _{\mathfrak{F}} \\
&=&\left\langle \left( \epsilon ^{-1}\left( \mathrm{U}_{t+\epsilon ,s}-%
\mathrm{U}_{t,s}\right) +i\mathrm{G}_{t}\mathrm{U}_{t,s}\right) \varphi
,\lambda ^{2}\left( \mathrm{R}_{t+\epsilon }-\mathrm{R}_{t}\right) \mathrm{U}%
_{t,s}\psi \right\rangle _{\mathfrak{F}}-\left\langle i\mathrm{G}_{t}\mathrm{%
U}_{t,s}\varphi ,\lambda ^{2}\left( \mathrm{R}_{t+\epsilon }-\mathrm{R}%
_{t}\right) \mathrm{U}_{t,s}\psi \right\rangle _{\mathfrak{F}} \\
&&+\left\langle \left( \mathrm{U}_{t+\epsilon ,s}-\mathrm{U}_{t,s}\right)
\varphi ,\lambda ^{2}\left( -i\left[ \mathrm{R}_{t},\mathrm{G}_{t}\right] +%
\mathrm{R}_{t}\mathrm{W}_{t}\mathrm{R}_{t}\right) \mathrm{U}_{t,s}\psi
\right\rangle _{\mathfrak{F}} \\
&&+\left\langle \mathrm{U}_{t,s}\varphi ,\lambda ^{2}\left( \epsilon
^{-1}\left( \mathrm{R}_{t+\epsilon }-\mathrm{R}_{t}\right) -i\left[ \mathrm{R%
}_{t},\mathrm{G}_{t}\right] +\mathrm{R}_{t}\mathrm{W}_{t}\mathrm{R}%
_{t}\right) \mathrm{U}_{t,s}\psi \right\rangle _{\mathfrak{F}},
\end{eqnarray*}%
which yields 
\begin{equation*}
\lim_{\epsilon \rightarrow 0}\left\langle \mathrm{U}_{t+\epsilon ,s}\varphi
,\lambda ^{2}\left( \epsilon ^{-1}\left( \mathrm{R}_{t+\epsilon }-\mathrm{R}%
_{t}\right) -i\left[ \mathrm{R}_{t},\mathrm{G}_{t}\right] +\mathrm{R}_{t}%
\mathrm{W}_{t}\mathrm{R}_{t}\right) \mathrm{U}_{t,s}\psi \right\rangle _{%
\mathfrak{F}}=0,
\end{equation*}%
by Lemma \ref{coro-Ht copy(1)}, Theorem \ref{propagators} and Proposition %
\ref{Resolventeq}. In other words, in the weak sense on the domain $\mathcal{%
D}(\mathrm{N})$ and for any $t\in \mathbb{R}^{+}$ and $s\in \mathbb{R}%
_{0}^{+}$, we have 
\begin{equation}
\partial _{t}\left\{ \mathrm{U}_{s,t}\mathrm{H}_{t}^{(\lambda )}\mathrm{U}%
_{t,s}\right\} =-\lambda ^{2}\mathrm{U}_{s,t}\mathrm{R}_{t}\mathrm{W}_{t}%
\mathrm{R}_{t}\mathrm{U}_{t,s}.  \label{eq:deriveTransfoYosiApprox}
\end{equation}%
By Theorem \ref{propagators} combined with Lemma \ref%
{TechnicalIntegralWandRWR} (under Condition \ref{AbstractAssumptions}
(i)--(v)), for any $s,t\in \mathbb{R}_{0}^{+}$, 
\begin{equation*}
\mathbb{I}_{t,s}\left( \mathrm{N}+\mathbf{1}\right) ^{-1}\doteq \int_{s}^{t}%
\mathrm{U}_{s,\tau }\mathrm{R}_{\tau }\mathrm{W}_{\tau }\mathrm{R}_{\tau }%
\mathrm{U}_{\tau ,s}\left( \mathrm{N}+\mathbf{1}\right) ^{-1}\mathrm{d}\tau
\end{equation*}%
defines a bounded operator. The strong continuity of the mapping 
\begin{equation*}
t\mapsto \mathrm{U}_{s,t}\mathrm{R}_{t}\mathrm{W}_{t}\mathrm{R}_{t}\mathrm{U}%
_{t,s}\left( \mathrm{N}+\mathbf{1}\right) ^{-1}
\end{equation*}%
for any $s\in \mathbb{R}_{0}^{+}$ (see (\ref{StrongcontinuityRWR})) together
with the Lebesgue differentiation theorem yield 
\begin{equation}
\partial _{t}\mathbb{I}_{t,s}=\mathrm{U}_{s,t}\mathrm{R}_{t}\mathrm{W}_{t}%
\mathrm{R}_{t}\mathrm{U}_{t,s}  \label{eq:deriveIbb}
\end{equation}%
in the strong sense on the domain $\mathcal{D}(\mathrm{N})$ (and thus in the
weak one on the same domain). Now use Proposition \ref{Resolventeq} together
with (\ref{eq:deriveTransfoYosiApprox})--(\ref{eq:deriveIbb}) (both in the
weak sense on $\mathcal{D}(\mathrm{N})$) to obtain from (\ref{DefRbb}) that 
\begin{equation*}
\partial _{t}\mathrm{Bwd}_{t,s}^{(\lambda )}=0,
\end{equation*}%
again in the weak sense on $\mathcal{D}(\mathrm{N})$. I.e., $\langle \varphi
,\mathrm{Bwd}_{\cdot ,s}^{(\lambda )}\psi \rangle _{\mathfrak{F}}$ is
constant on $\mathbb{R}^{+}$ for any $\varphi ,\psi \in \mathcal{D}(\mathrm{N%
})$. By observing that $\mathbb{I}_{s,s}=0$ and that $\mathrm{U}_{s,s}=%
\mathbf{1}$, for any $s\in \mathbb{R}_{0}^{+}$, we deduce from (\ref{DefRbb}%
) the following identity: 
\begin{equation}
\left\langle \varphi ,\mathrm{Bwd}_{t,s}^{(\lambda )}\psi \right\rangle _{%
\mathfrak{F}}=\left\langle \varphi ,\mathrm{H}_{s}^{(\lambda )}\psi
\right\rangle _{\mathfrak{F}},\qquad \psi ,\varphi \in \mathcal{D}\left( 
\mathrm{N}\right) ,\ t,s\in \mathbb{R}^{+}.  \label{TemporaryWeaaksense}
\end{equation}%
Finally, observe from Lemma \ref{diversPreservedom} and Theorem \ref%
{propagators} (iv) (see also (\ref{dfsdfsfsdfsf})) that 
\begin{equation*}
\mathrm{U}_{s,t}\mathrm{H}_{t}^{(\lambda )}\mathrm{U}_{t,s}\mathcal{D}\left( 
\mathrm{N}\right) \subseteq \mathcal{D}\left( \mathrm{N}\right) .
\end{equation*}%
Furthermore, by Lemma \ref{diversPreservedom copy(1)} and Theorem \ref%
{propagators} (iv), the expression 
\begin{equation*}
\mathrm{W}_{t}\left( \mathrm{H}_{t}+i\lambda \mathbf{1}\right) ^{-1}\mathrm{U%
}_{t,s}\psi \in \mathfrak{F}
\end{equation*}%
is always a well-defined vector for any $t\in \mathbb{R}_{0}^{+}$ and $\psi
\in \mathcal{D}\left( \mathrm{N}\right) $, and 
\begin{equation*}
\left( \mathrm{H}_{t}+i\lambda \mathbf{1}\right) ^{-1}\mathrm{W}_{t}\left( 
\mathrm{H}_{t}+i\lambda \mathbf{1}\right) ^{-1}\mathrm{U}_{t,s}\mathcal{D}%
\left( \mathrm{N}\right) \subseteq \mathcal{D}\left( \mathrm{H}_{t}\right)
\subseteq \mathcal{D}\left( \mathrm{N}\right) .
\end{equation*}%
Note that $\mathcal{D}\left( \mathrm{H}_{t}\right) \subseteq \mathcal{D}%
\left( \mathrm{N}\right) $ for all times $t\in \mathbb{R}_{0}^{+}$, because
of Lemma \ref{diversPreservedom copy(1)} (under Condition \ref%
{AbstractAssumptions} (v)). In fact, by Lemma \ref{diversPreservedom copy(1)}%
, 
\begin{equation*}
\left\Vert \left( \mathrm{N}+\mathbf{1}\right) \left( \mathrm{H}%
_{t}+i\lambda \mathbf{1}\right) ^{-1}\mathrm{W}_{t}\left( \mathrm{H}%
_{t}+i\lambda \mathbf{1}\right) ^{-1}\mathrm{U}_{t,s}\left( \mathrm{N}+%
\mathbf{1}\right) ^{-1}\right\Vert _{\mathrm{op}}<\infty ,
\end{equation*}%
because $\lambda >2^{5+1/2}\Vert X_{0}\Vert _{\mathrm{op},2}>2^{7/2}\Vert
X_{0}\Vert _{\mathrm{op},2}$ and $\Vert X_{t}\Vert _{\mathrm{op},2}\leq
\Vert X_{0}\Vert _{\mathrm{op},2}$ for all $t\in \mathbb{R}_{0}^{+}$
(Corollary \ref{pseudoHS} with $T_{\mathrm{max}}=\infty $). So, since $%
\mathrm{N}$ is a closed operator, applying again Theorem \ref{propagators}
and ensuring that the Riemann integral is a well-defined object by virtue of
Lemma \ref{TechnicalIntegralWandRWR}, we conclude that 
\begin{equation*}
\left( \int_{s}^{t}\mathrm{U}_{s,\tau }\left( \mathrm{H}_{\tau }+i\lambda 
\mathbf{1}\right) ^{-1}\mathrm{W}_{\tau }\left( \mathrm{H}_{\tau }+i\lambda 
\mathbf{1}\right) ^{-1}\mathrm{U}_{\tau ,s}\mathrm{d}\tau \right) \mathcal{D}%
\left( \mathrm{N}\right) \subseteq \mathcal{D}(\mathrm{N})
\end{equation*}%
and, consequently, 
\begin{equation*}
\left( \mathrm{U}_{s,t}\mathrm{H}_{t}^{(\lambda )}\mathrm{U}_{t,s}+\lambda
^{2}\int_{s}^{t}\mathrm{U}_{s,\tau }\left( \mathrm{H}_{\tau }+i\lambda 
\mathbf{1}\right) ^{-1}\mathrm{W}_{\tau }\left( \mathrm{H}_{\tau }+i\lambda 
\mathbf{1}\right) ^{-1}\mathrm{U}_{\tau ,s}\mathrm{d}\tau \right) \mathcal{D}%
\left( \mathrm{N}\right) \subseteq \mathcal{D}(\mathrm{N}).
\end{equation*}%
Therefore, by taking 
\begin{equation*}
\varphi \doteq \left( \mathrm{Bwd}_{t,s}^{(\lambda )}-\mathrm{H}%
_{s}^{(\lambda )}\right) \psi \in \mathcal{D}(\mathrm{N})
\end{equation*}%
in (\ref{TemporaryWeaaksense}) we arrive at the equality 
\begin{equation*}
\mathrm{Bwd}_{t,s}^{(\lambda )}=\mathrm{H}_{s}^{(\lambda )},\qquad t,s\in 
\mathbb{R}^{+},
\end{equation*}%
in the strong sense on the domain $\mathcal{D}(\mathrm{N})$. The study of $%
\mathrm{Fwd}_{t,s}^{(\lambda )}$ is very much similar: One proves in the
same way the Leibniz rule that allows use to compute that, for any $s,t\in 
\mathbb{R}^{+}$, 
\begin{equation*}
\partial _{s}\mathrm{Fwd}_{t,s}^{(\lambda )}=0
\end{equation*}%
on the weak sense on the domain $\mathcal{D}(\mathrm{N})$. From this, we
deduce that 
\begin{equation*}
\mathrm{Fwd}_{t,s}^{(\lambda )}=\mathrm{Fwd}_{t,t}^{(\lambda )}=\mathrm{H}%
_{t}^{(\lambda )},\qquad t,s\in \mathbb{R}^{+},
\end{equation*}%
still in the weak sense on the domain $\mathcal{D}(\mathrm{N})$. The
arguments used above then allow this last equation to be strengthened as an
operator equality over the entire domain $\mathcal{D}(\mathrm{N})$.

At this stage, the statement is only established when $t,s>0$ are strictly
positive numbers. We extend it by continuity to all $s,t\in \mathbb{R}%
_{0}^{+}$: Recall that $\lambda >2^{5+1/2}\Vert X_{0}\Vert _{\mathrm{op},2}$%
. Take $s,t\in \mathbb{R}^{+}$. Combine (\ref{dfsdfsfsdfsf}) with Lemma \ref%
{coro-Ht copy(1)} to obtain 
\begin{equation}
\lim_{s\rightarrow 0}\left\Vert \left( \mathrm{H}_{s}^{(\lambda )}-\mathrm{H}%
_{0}^{(\lambda )}\right) \varphi \right\Vert _{\mathfrak{F}%
}=\lim_{s\rightarrow 0}\lambda ^{2}\left\Vert \left( \mathrm{R}_{s}-\mathrm{R%
}_{0}\right) \varphi \right\Vert _{\mathfrak{F}}=0,\qquad \varphi \in 
\mathfrak{F}.  \label{ApproxCont}
\end{equation}%
Additionally, using Lemma \ref{TechnicalIntegralWandRWR} and the triangle
inequality we get that, for any $\varphi \in \mathcal{D}(\mathrm{N})$, 
\begin{eqnarray*}
\left\Vert \left( \mathrm{Bwd}_{t,s}^{(\lambda )}-\mathrm{Bwd}%
_{t,0}^{(\lambda )}\right) \varphi \right\Vert _{\mathfrak{F}} &\leq
&\lambda ^{2}\int_{s}^{t}\left\Vert \mathrm{U}_{s,\tau }\left( \mathrm{H}%
_{\tau }+i\lambda \mathbf{1}\right) ^{-1}\mathrm{W}_{\tau }\left( \mathrm{H}%
_{\tau }+i\lambda \mathbf{1}\right) ^{-1}\left( \mathrm{U}_{\tau ,s}-\mathrm{%
U}_{\tau ,0}\right) \varphi \right\Vert _{\mathfrak{F}}\mathrm{d}\tau \\
&&+\lambda ^{2}\int_{s}^{t}\left\Vert \left( \mathrm{U}_{s,\tau }-\mathrm{U}%
_{0,\tau }\right) \left( \mathrm{H}_{\tau }+i\lambda \mathbf{1}\right) ^{-1}%
\mathrm{W}_{\tau }\left( \mathrm{H}_{\tau }+i\lambda \mathbf{1}\right) ^{-1}%
\mathrm{U}_{\tau ,0}\varphi \right\Vert _{\mathfrak{F}}\mathrm{d}\tau \\
&&+\lambda ^{2}\int_{0}^{s}\left\Vert \mathrm{U}_{0,\tau }\left( \mathrm{H}%
_{\tau }+i\lambda \mathbf{1}\right) ^{-1}\mathrm{W}_{\tau }\left( \mathrm{H}%
_{\tau }+i\lambda \mathbf{1}\right) ^{-1}\mathrm{U}_{\tau ,0}\varphi
\right\Vert _{\mathfrak{F}}\mathrm{d}\tau \\
&&+\left\Vert \left( \mathrm{U}_{s,t}-\mathrm{U}_{0,t}\right) \mathrm{H}%
_{t}^{(\lambda )}\mathrm{U}_{t,0}\varphi \right\Vert _{\mathfrak{F}%
}+\left\Vert \mathrm{U}_{s,t}\mathrm{H}_{t}^{(\lambda )}\left( \mathrm{U}%
_{t,s}-\mathrm{U}_{t,0}\right) \varphi \right\Vert _{\mathfrak{F}}.
\end{eqnarray*}%
(The integrands appearing in this upper bound are continuous functions, and
are therefore integrable; see also Lemma \ref{TechnicalIntegralWandRWR} and 
\cite[Chapter III, Theorem 1.2]{pettis}.) By Theorem \ref{propagators} as
well as Lemmata \ref{diversPreservedom copy(1)} and \ref{Lemma-self-adjoint1
copy(1)} (under Condition \ref{AbstractAssumptions} (v)), the last four
terms on the right-hand side of this inequality go to zero in the limit $%
s\rightarrow 0$. As a consequence, for any $\varphi \in \mathcal{D}(\mathrm{N%
})$, 
\begin{eqnarray}
&&\lim_{s\rightarrow 0}\left\Vert \left( \mathrm{Bwd}_{t,s}^{(\lambda )}-%
\mathrm{Bwd}_{t,0}^{(\lambda )}\right) \varphi \right\Vert _{\mathfrak{F}}
\label{decompodecompo} \\
&\leq &\lambda ^{2}\lim_{s\rightarrow 0}\int_{0}^{t}\left\Vert \left( 
\mathrm{H}_{\tau }+i\lambda \mathbf{1}\right) ^{-1}\mathrm{W}_{\tau }\left( 
\mathrm{H}_{\tau }+i\lambda \mathbf{1}\right) ^{-1}\left( \mathrm{U}_{\tau
,s}-\mathrm{U}_{\tau ,0}\right) \varphi \right\Vert _{\mathfrak{F}}\mathrm{d}%
\tau ,  \notag
\end{eqnarray}%
using the additional fact that $\mathrm{U}_{s,\tau }$ is always a unitary
operator. Note that for all times $\tau \in \mathbb{R}_{0}^{+}$, $\mathrm{W}%
_{\tau }$ is not only well-defined on $\mathcal{D}(\mathrm{N}^{3/2})$ (cf.
Lemma \ref{Lemma-self-adjoint1 copy(1)}), but also symmetric since $\mathcal{%
D}(\mathrm{N}^{3/2})$ is a core of its (symmetric) closure by Proposition %
\ref{Lemma Htbis} (iii). Using this and the Cauchy-Schwarz inequality
together with Lemma \ref{diversPreservedom copy(1)} and Theorem \ref%
{propagators}, we deduce that, for any $\tau \in \mathbb{R}_{0}^{+}$,%
\begin{equation*}
\lim_{s\rightarrow 0}\left\langle \mathrm{W}_{\tau }\left( \mathrm{H}_{\tau
}+i\lambda \mathbf{1}\right) ^{-1}\psi ,\left( \mathrm{H}_{\tau }+i\lambda 
\mathbf{1}\right) ^{-1}\left( \mathrm{U}_{\tau ,s}-\mathrm{U}_{\tau
,0}\right) \varphi \right\rangle _{\mathfrak{F}}=0,\qquad \varphi ,\psi \in 
\mathcal{D}(\mathrm{N}),
\end{equation*}%
while, for any $T\in \mathbb{R}_{0}^{+}$, 
\begin{eqnarray}
&&\sup_{t,s\in \lbrack 0,T]}\left\Vert \left( \mathrm{H}_{t}+i\lambda 
\mathbf{1}\right) ^{-1}\mathrm{W}_{t}\left( \mathrm{H}_{t}+i\lambda \mathbf{1%
}\right) ^{-1}\left( \mathrm{U}_{t,s}-\mathrm{U}_{t,0}\right) \left( \mathrm{%
N}+\mathbf{1}\right) ^{-1}\right\Vert _{\mathrm{op}}  \notag \\
&\leq &2\sup_{t,s\in \lbrack 0,T]}\left\Vert \left( \mathrm{H}_{t}+i\lambda 
\mathbf{1}\right) ^{-1}\mathrm{W}_{t}\left( \mathrm{H}_{t}+i\lambda \mathbf{1%
}\right) ^{-1}\mathrm{U}_{t,s}\left( \mathrm{N}+\mathbf{1}\right)
^{-1}\right\Vert _{\mathrm{op}}<\infty ,  \label{ddddd}
\end{eqnarray}%
thanks to Lemmata \ref{diversPreservedom copy(1)}--\ref{Lemma-self-adjoint1
copy(1)} and Theorem \ref{propagators}. It follows from (\ref{decompodecompo}%
) and the last observations, combined with the density of $\mathcal{D}(%
\mathrm{N})$ and Lebesgue's dominated convergence theorem, that for any $%
\varphi \in \mathcal{D}(\mathrm{N})$, 
\begin{equation*}
\lim_{s\rightarrow 0}\left\Vert \left( \mathrm{Bwd}_{t,s}^{(\lambda )}-%
\mathrm{Bwd}_{t,0}^{(\lambda )}\right) \varphi \right\Vert _{\mathfrak{F}}=0,
\end{equation*}%
which, combined with (\ref{ApproxCont}), yields 
\begin{equation}
\mathrm{Bwd}_{t,s}^{(\lambda )}\varphi =\mathrm{H}_{s}^{(\lambda )}\varphi
,\qquad s\in \mathbb{R}_{0}^{+},t\in \mathbb{R}^{+}.  \label{extenSequal0}
\end{equation}%
We perform similar arguments with the other variable: For any $(s,t)\in 
\mathbb{R}_{0}^{+}\times \mathbb{R}^{+}$ and $\varphi \in \mathcal{D}(%
\mathrm{N})$,%
\begin{eqnarray*}
\left\Vert \left( \mathrm{Bwd}_{t,s}^{(\lambda )}-\mathrm{Bwd}%
_{0,s}^{(\lambda )}\right) \varphi \right\Vert _{\mathfrak{F}} &\leq
&\left\Vert \left( \mathrm{U}_{s,t}-\mathrm{U}_{s,0}\right) \mathrm{H}%
_{0}^{(\lambda )}\mathrm{U}_{0,s}\varphi \right\Vert _{\mathfrak{F}%
}+\left\Vert \mathrm{U}_{s,t}\left( \mathrm{H}_{t}^{(\lambda )}-\mathrm{H}%
_{0}^{(\lambda )}\right) \mathrm{U}_{0,s}\varphi \right\Vert _{\mathfrak{F}}
\\
&&+\left\Vert \mathrm{U}_{s,t}\mathrm{H}_{t}^{(\lambda )}\left( \mathrm{U}%
_{t,s}-\mathrm{U}_{0,s}\right) \varphi \right\Vert _{\mathfrak{F}} \\
&&+\int_{0}^{t}\left\Vert \mathrm{U}_{s,\tau }\left( \mathrm{H}_{\tau
}+i\lambda \mathbf{1}\right) ^{-1}\mathrm{W}_{\tau }\left( \mathrm{H}_{\tau
}+i\lambda \mathbf{1}\right) ^{-1}\mathrm{U}_{\tau ,s}\varphi \right\Vert _{%
\mathfrak{F}}\mathrm{d}\tau .
\end{eqnarray*}%
So, by combining this inequality with Theorem \ref{propagators} and
Equations (\ref{ApproxCont}) and (\ref{ddddd}), we can extend (\ref%
{extenSequal0}) to $t=0$. We analyze the second identity $\mathrm{Fwd}%
_{t,s}^{(\lambda )}=\mathrm{H}_{t}^{(\lambda )}$ on $\mathcal{D}(\mathrm{N})$
in an identical manner. We omit the details.
\end{proof}

By Theorem \ref{HamilSA copy(2)}, we already know that $\mathrm{H}_{s}$ is a
self-adjoint operator for any $s\in \mathbb{R}_{0}^{+}$ and that $\mathcal{D}%
_{\infty }$ is a core for $\mathrm{H}_{s}$. Now let us introduce the new
linear mapping 
\begin{equation}
\mathbb{B}_{t,s}\varphi \doteq \left( \mathrm{U}_{s,t}\mathrm{H}_{t}\mathrm{U%
}_{t,s}-\int_{s}^{t}\mathrm{U}_{s,\tau }\mathrm{W}_{\tau }\mathrm{U}_{\tau
,s}\mathrm{d}\tau \right) \varphi ,\qquad \varphi \in \mathbb{D}_{t,s},
\label{B blackbb}
\end{equation}%
where%
\begin{equation*}
\mathbb{D}_{t,s}\doteq \mathrm{U}_{s,t}\mathcal{D}\left( \mathrm{H}%
_{t}\right) \cap \mathcal{D}\left( \mathrm{N}^{3/2}\right) .
\end{equation*}%
It is not immediately clear whether the integral term is properly defined.
In fact, it is a Riemann integral on the Hilbert space $\mathfrak{F}$,
thanks to Theorem \ref{propagators} and Lemma \ref{TechnicalIntegralWandRWR}%
. Observe that $\mathcal{D}_{\infty }\subseteq \mathcal{D}(\mathrm{N}%
^{3/2})\cap \mathcal{D}(\mathrm{H}_{t})$, which is dense, and $\mathrm{U}%
_{s,t}\mathcal{D}_{\infty }\subseteq \mathbb{D}_{t,s}$, thanks to Theorem %
\ref{propagators}. As $\mathrm{U}_{s,t}$ is a unitary operator, $\mathrm{U}%
_{s,t}\mathcal{D}_{\infty }$ is also a dense subset and $\mathbb{B}_{t,s}$
is thus a densely defined operator. One checks that $\mathbb{B}_{t,s}$ is
additionally symmetric, and hence closable. Henceforth, we identify $\mathbb{%
B}_{t,s}$ with its closure $\mathbb{B}_{t,s}\equiv \mathbb{B}_{t,s}^{\ast
\ast }$ and denote its domain as $\mathcal{D}(\mathbb{B}_{t,s})\supseteq 
\mathbb{D}_{t,s}$. Consequently, we have a closed symmetric operator at our
disposal, though it remains to be checked whether $\mathbb{B}_{t,s}$ is
self-adjoint.

\begin{proposition}[Convergence of the Yosida regularization -- I]
\label{ConvergenceYosida}\mbox{}\newline
Assume Condition \ref{AbstractAssumptions} (i)--(v) and take $s,t\in \mathbb{%
R}_{0}^{+}$. Then, 
\begin{eqnarray*}
\lim_{\lambda \rightarrow \infty }\mathrm{Bwd}_{t,s}^{(\lambda )}\varphi &=&%
\mathrm{H}_{s}\varphi ,\qquad \varphi \in \mathcal{D}\left( \mathrm{H}%
_{s}\right) , \\
\lim_{\lambda \rightarrow \infty }\mathrm{Bwd}_{t,s}^{(\lambda )}\varphi &=&%
\mathbb{B}_{t,s}\varphi ,\qquad \varphi \in \mathrm{U}_{s,t}\mathcal{D}%
_{\infty }\subseteq \mathcal{D}(\mathrm{H}_{s}).
\end{eqnarray*}
\end{proposition}

\begin{proof}
Fix $s,t\in \mathbb{R}_{0}^{+}$ and $\lambda >\lambda _{0}\doteq
2^{5+1/2}\Vert X_{0}\Vert _{\mathrm{op},2}$; see (\ref{dfsdfsgsf}) with $%
D_{T}=1$. If $(\mathrm{Bwd}_{t,s}^{(\lambda )}\varphi )_{\lambda >\lambda
_{0}}\subseteq \mathfrak{F}$ is a Cauchy net for some $\varphi \in \mathcal{D%
}(\mathrm{N})$, then it converges, because $\mathfrak{F}$ is complete. We
define the operator $\mathrm{E}_{t,s}$ to be the limit%
\begin{equation}
\mathrm{E}_{t,s}\varphi \doteq \lim_{\lambda \rightarrow \infty }\mathrm{Bwd}%
_{t,s}^{(\lambda )}\varphi ,\qquad \varphi \in \mathcal{D}\left( \mathrm{E}%
_{t,s}\right) ,  \label{DefEts}
\end{equation}%
where 
\begin{equation*}
\mathcal{D}\left( \mathrm{E}_{t,s}\right) \doteq \left\{ \varphi \in 
\mathcal{D}(\mathrm{N}):(\mathrm{Bwd}_{t,s}^{(\lambda )}\varphi )_{\lambda
>\lambda _{0}}\text{ is a Cauchy net}\right\} .
\end{equation*}%
This set is non-empty and, in fact, dense, because it contains $\mathcal{D}%
\left( \mathrm{H}_{s}\right) $. To see this, let $\varphi \in \mathcal{D}%
\left( \mathrm{H}_{s}\right) $. By Lemma \ref{diversPreservedom copy(1)}, $%
\mathcal{D}(\mathrm{H}_{s})\subseteq \mathcal{D}(\mathrm{N})$. Therefore,
one can apply Proposition \ref{EvolRegu} to obtain that, for any $\varphi
\in \mathcal{D}(\mathrm{H}_{s})$, 
\begin{equation}
\mathrm{Bwd}_{t,s}^{(\lambda )}\varphi -\mathrm{H}_{s}\varphi =\left( 
\mathrm{H}_{s}^{(\lambda )}-\mathrm{H}_{s}\right) \varphi =\left( \frac{%
i\lambda }{\left( \mathrm{H}_{s}+i\lambda \right) }-\mathbf{1}\right) 
\mathrm{H}_{s}\varphi .  \label{sdsdsdsdsdsd}
\end{equation}%
In the limit $\lambda \rightarrow \infty $, it is well known that the Yosida
approximation $\mathrm{H}_{s}^{(\lambda )}$ converges strongly to $\mathrm{H}%
_{s}$ on the domain $\mathcal{D}(\mathrm{H}_{s})$: For any $\varphi \in 
\mathcal{D}(\mathrm{H}_{s})$, 
\begin{equation}
\left\Vert \left( \frac{i\lambda \mathbf{1}}{\mathrm{H}_{s}+i\lambda \mathbf{%
1}}-\mathbf{1}\right) \varphi \right\Vert _{\mathfrak{F}}\leq \lambda
^{-1}\left\Vert \mathrm{H}_{s}\varphi \right\Vert _{\mathfrak{F}}
\label{stronglimitpieceYosida0}
\end{equation}%
and, as a consequence, 
\begin{equation*}
\lim_{\lambda \rightarrow \infty }\frac{i\lambda \mathbf{1}}{\mathrm{H}%
_{s}+i\lambda \mathbf{1}}\varphi =\varphi ,\qquad \varphi \in \mathcal{D}%
\left( \mathrm{H}_{s}\right) .
\end{equation*}%
The same limit can be extended to all $\varphi \in \mathfrak{F}$, by density
of $\mathcal{D}(\mathrm{H}_{s})\subseteq \mathfrak{F}$ and the uniform
boundedness\footnote{%
Clearly, $\Vert \lambda \mathbf{1}\left( \mathrm{H}_{s}+i\lambda \mathbf{1}%
\right) ^{-1}\Vert _{\mathrm{op}}\leq 1$.} of operators $\lambda \mathbf{1}%
\left( \mathrm{H}_{s}+i\lambda \mathbf{1}\right) ^{-1}$. In other words,%
\begin{equation}
\lim_{\lambda \rightarrow \infty }\frac{i\lambda \mathbf{1}}{\mathrm{H}%
_{s}+i\lambda \mathbf{1}}\varphi =\varphi ,\qquad \varphi \in \mathfrak{F}.
\label{stronglimitpieceYosida}
\end{equation}%
Combining (\ref{stronglimitpieceYosida}) with (\ref{sdsdsdsdsdsd}) we deduce
that $\mathcal{D}(\mathrm{H}_{s})\subseteq \mathcal{D}\left( \mathrm{E}%
_{t,s}\right) $ and 
\begin{equation}
\lim_{\lambda \rightarrow \infty }\mathrm{Bwd}_{t,s}^{(\lambda )}\varphi
=\lim_{\lambda \rightarrow \infty }\mathrm{H}_{s}^{(\lambda )}\varphi =%
\mathrm{H}_{s}\varphi ,\qquad \varphi \in \mathcal{D}\left( \mathrm{H}%
_{s}\right) .  \label{Esymmetric0}
\end{equation}%
Hence, $\mathrm{E}_{t,s}$ is a proper densely defined operator.
Additionally, for any $\psi ,\phi \in \mathcal{D}(\mathrm{E}_{t,s})$, the
linear mappings $\langle \cdot ,\psi \rangle _{\mathfrak{F}}$ and $\overline{%
\langle \phi ,\cdot \rangle _{\mathfrak{F}}}$ are bounded, and hence
continuous. It follows that, for any $\psi ,\phi \in \mathcal{D}(\mathrm{E}%
_{t,s})$, 
\begin{equation}
\left\langle \phi ,\mathrm{E}_{t,s}\psi \right\rangle _{\mathfrak{F}%
}=\lim_{\lambda \rightarrow \infty }\left\langle \phi ,\mathrm{Bwd}%
_{t,s}^{(\lambda )}\psi \right\rangle _{\mathfrak{F}}=\lim_{\lambda
\rightarrow \infty }\left\langle \mathrm{Bwd}_{t,s}^{(\lambda )}\phi ,\psi
\right\rangle _{\mathfrak{F}}=\left\langle \mathrm{E}_{t,s}\phi ,\psi
\right\rangle _{\mathfrak{F}}.  \label{Esymmetric}
\end{equation}%
In other words, $\mathrm{E}_{t,s}$ is also symmetric. Hence, $\mathrm{E}%
_{t,s}$ is closable. We identify $\mathrm{E}_{t,s}$ with its closure: $%
\mathrm{E}_{t,s}\equiv \mathrm{E}_{t,s}^{\ast \ast }$. By (\ref{Esymmetric0}%
), this closed symmetric operator extends the Hamiltonian $\mathrm{H}_{s}$,
meaning that $\mathrm{H}_{s}\subseteq \mathrm{E}_{t,s}$. Taking adjoints
reverses this inclusion, yielding $\mathrm{E}_{t,s}^{\ast }\subseteq \mathrm{%
H}_{s}^{\ast }$. Since $\mathrm{H}_{s}$ is self-adjoint and $\mathrm{E}%
_{t,s} $ is symmetric ($\mathrm{E}_{t,s}\subseteq \mathrm{E}_{t,s}^{\ast }$%
), we obtain that 
\begin{equation}
\mathrm{H}_{s}\subseteq \mathrm{E}_{t,s}\subseteq \mathrm{E}_{t,s}^{\ast
}\subseteq \mathrm{H}_{s}^{\ast }=\mathrm{H}_{s}.  \label{inclusions}
\end{equation}%
This implies that all inclusions are equalities, and we deduce that $\mathrm{%
E}_{t,s}=\mathrm{H}_{s}$ is self-adjoint. Now, take $\varphi \in \mathbb{D}%
_{t,s}$ and, using the triangle inequality and the unitarity of the
operators $\mathrm{U}_{s,t}$ for $s,t\in \mathbb{R}_{0}^{+}$ (Theorem \ref%
{propagators}), we get from (\ref{DefRbb}) and (\ref{B blackbb}) that 
\begin{eqnarray}
&&\left\Vert \left( \mathrm{Bwd}_{t,s}^{(\lambda )}-\mathbb{B}_{t,s}\right)
\varphi \right\Vert _{\mathfrak{F}}  \notag \\
&\leq &\left\Vert \left( \mathrm{H}_{t}^{(\lambda )}-\mathrm{H}_{t}\right) 
\mathrm{U}_{t,s}\varphi \right\Vert _{\mathfrak{F}}  \label{IN1IN1} \\
&&+\int_{s\wedge t}^{s\vee t}\left\Vert \left( i\lambda \left( \mathrm{H}%
_{\tau }+i\lambda \mathbf{1}\right) ^{-1}-\mathbf{1}\right) \mathrm{W}_{\tau
}\mathrm{U}_{\tau ,s}\varphi \right\Vert _{\mathfrak{F}}\mathrm{d}\tau 
\notag \\
&&+\int_{s\wedge t}^{s\vee t}\left\Vert i\lambda \left( \mathrm{H}_{\tau
}+i\lambda \mathbf{1}\right) ^{-1}\mathrm{W}_{\tau }\left( i\lambda \left( 
\mathrm{H}_{\tau }+i\lambda \mathbf{1}\right) ^{-1}-\mathbf{1}\right) 
\mathrm{U}_{\tau ,s}\varphi \right\Vert _{\mathfrak{F}}\mathrm{d}\tau .
\label{Thirdtermilfautlebesgue}
\end{eqnarray}%
The strong convergence of $\mathrm{H}_{s}^{(\lambda )}$ towards $\mathrm{H}%
_{s}$ (see (\ref{Esymmetric0})) on the domain $\mathcal{D}(\mathrm{H}_{s})$
implies that 
\begin{equation}
\lim_{\lambda \rightarrow \infty }\left\Vert \left( \mathrm{H}_{t}^{(\lambda
)}-\mathrm{H}_{t}\right) \mathrm{U}_{t,s}\varphi \right\Vert _{\mathfrak{F}%
}=0,\qquad \varphi \in \mathbb{D}_{t,s}\subseteq \mathrm{U}_{s,t}\mathcal{D}%
\left( \mathrm{H}_{t}\right) .  \label{IN1IN10}
\end{equation}%
Using also (\ref{stronglimitpieceYosida}) together with Lemma \ref%
{Lemma-self-adjoint1 copy(1)}, Theorem \ref{propagators}, and Lebesgue's
dominated convergence theorem, we also show the convergence to zero of the
second term: 
\begin{equation}
\lim_{\lambda \rightarrow \infty }\int_{s\wedge t}^{s\vee t}\left\Vert
\left( i\lambda \left( \mathrm{H}_{\tau }+i\lambda \mathbf{1}\right) ^{-1}-%
\mathbf{1}\right) \mathrm{W}_{\tau }\mathrm{U}_{\tau ,s}\varphi \right\Vert
_{\mathfrak{F}}\mathrm{d}\tau =0,\qquad \varphi \in \mathbb{D}_{t,s}.
\label{IN1IN11}
\end{equation}%
Concerning the third term, observe from (\ref{YosidaApproc}) that, for any $%
\varphi \in \mathcal{D}(\mathrm{N}^{3/2})$ and $\tau \in \lbrack s,t]$,%
\begin{eqnarray}
&&\left\Vert i\lambda \left( \mathrm{H}_{\tau }+i\lambda \mathbf{1}\right)
^{-1}\mathrm{W}_{\tau }\left( i\lambda \left( \mathrm{H}_{\tau }+i\lambda 
\mathbf{1}\right) ^{-1}-\mathbf{1}\right) \mathrm{U}_{\tau ,s}\varphi
\right\Vert _{\mathfrak{F}}  \label{sdsdsd0} \\
&\leq &\left\Vert \mathrm{W}_{\tau }(\mathrm{N}+\mathbf{1})^{-\frac{3}{2}%
}\right\Vert _{\mathrm{op}}\left\Vert \left( i\lambda (\mathrm{N}+\mathbf{1}%
)^{\frac{3}{2}}\left( \mathrm{H}_{\tau }+i\lambda \mathbf{1}\right) ^{-1}(%
\mathrm{N}+\mathbf{1})^{-\frac{3}{2}}-\mathbf{1}\right) \mathrm{U}_{\tau
,s}\varphi \right\Vert _{\mathfrak{F}}.  \notag
\end{eqnarray}%
Since $\lambda >2^{5+1/2}\Vert X_{0}\Vert _{\mathrm{op},2}$, we can apply (%
\ref{fg1}) to obtain that 
\begin{equation}
(\mathrm{N}+\mathbf{1})^{\frac{3}{2}}\left( \mathrm{H}_{\tau }+i\lambda 
\mathbf{1}\right) ^{-1}(\mathrm{N}+\mathbf{1})^{-\frac{3}{2}}=\left( \mathrm{%
H}_{\tau }+\mathbb{T}_{3/2}+i\lambda \mathbf{1}\right) ^{-1},\qquad \tau \in 
\mathbb{R}_{0}^{+},  \label{ssdsd1}
\end{equation}%
where $\mathbb{T}_{3/2}$ is a $\tau $-dependent bounded operator with
operator norm uniformly bounded with respect to $\tau \in \mathbb{R}_{0}^{+}$
by 
\begin{equation}
\Vert \mathbb{T}_{3/2}\Vert _{\mathrm{op}}\leq 2^{3+3/2}\left\Vert X_{\tau
}\right\Vert _{\mathrm{op},2}\leq 2^{4+1/2}\left\Vert X_{0}\right\Vert _{%
\mathrm{op},2}<2^{5+1/2}\Vert X_{0}\Vert _{\mathrm{op},2}<\lambda .
\label{dfdfdff}
\end{equation}%
See Inequality (\ref{fg3}) and Corollary \ref{pseudoHS}. By Inequality (\ref%
{fg2}), the operator (\ref{ssdsd1}) is thus well defined: 
\begin{equation}
\sup_{\tau \in \mathbb{R}_{0}^{+}}\left\Vert \left( \mathrm{H}_{\tau }+%
\mathbb{T}_{3/2}+i\lambda \mathbf{1}\right) ^{-1}\right\Vert _{\mathrm{op}%
}\leq \frac{1}{\lambda -2^{5+1/2}\Vert X_{0}\Vert _{\mathrm{op},2}}<\infty .
\label{dfdfdff2}
\end{equation}%
Furthermore, by (\ref{dfdfdff})--(\ref{dfdfdff2}), for any $\tau \in \mathbb{%
R}_{0}^{+}$ and $\varphi \in \mathcal{D}(\mathrm{H}_{\tau })$, 
\begin{equation*}
\left\Vert \left( \frac{i\lambda }{\mathrm{H}_{\tau }+\mathbb{T}%
_{3/2}+i\lambda \mathbf{1}}-\mathbf{1}\right) \varphi \right\Vert _{%
\mathfrak{F}}\leq \frac{1}{\lambda -2^{5+1/2}\Vert X_{0}\Vert _{\mathrm{op}%
,2}}\left( \left\Vert \mathrm{H}_{\tau }\varphi \right\Vert _{\mathfrak{F}%
}+2^{5+1/2}\Vert X_{0}\Vert _{\mathrm{op},2}\left\Vert \varphi \right\Vert _{%
\mathfrak{F}}\right) .
\end{equation*}%
In the same manner that (\ref{stronglimitpieceYosida0}) implies (\ref%
{stronglimitpieceYosida}), we can deduce from this last inequality and the
density of $\mathcal{D}(\mathrm{H}_{\tau })\subseteq \mathfrak{F}$ that, for
any $\tau \in \mathbb{R}_{0}^{+}$, 
\begin{equation*}
\lim_{\lambda \rightarrow \infty }\frac{i\lambda \mathbf{1}}{\mathrm{H}%
_{\tau }+\mathbb{T}_{3/2}+i\lambda \mathbf{1}}\varphi =\varphi ,\qquad
\varphi \in \mathfrak{F}.
\end{equation*}%
Combining now this last limit with Inequality (\ref{sdsdsd0}), Lemma \ref%
{Lemma-self-adjoint1 copy(1)} and Theorem \ref{propagators} (iv), we get
that, for any $\varphi \in \mathcal{D}(\mathrm{N}^{3/2})$ and $\tau \in
\lbrack s,t]$, 
\begin{equation}
\lim_{\lambda \rightarrow \infty }\left\Vert i\lambda \left( \mathrm{H}%
_{\tau }+i\lambda \mathbf{1}\right) ^{-1}\mathrm{W}_{\tau }\left( i\lambda
\left( \mathrm{H}_{\tau }+i\lambda \mathbf{1}\right) ^{-1}-\mathbf{1}\right) 
\mathrm{U}_{\tau ,s}\varphi \right\Vert _{\mathfrak{F}}=0,\qquad \varphi \in 
\mathcal{D}(\mathrm{N}^{3/2})\supseteq \mathbb{D}_{t,s}.  \label{lasteq}
\end{equation}%
Applying Lemmata \ref{diversPreservedom} and \ref{Lemma-self-adjoint1
copy(1)} as well as Theorem \ref{propagators}, we obtain easily an
integrable bound of the integrand of (\ref{Thirdtermilfautlebesgue}), which
is independent of $\lambda $. Consequently, Lebesgue's dominated convergence
theorem and the limit (\ref{lasteq}) yields 
\begin{equation}
\lim_{\lambda \rightarrow \infty }\int_{s\wedge t}^{s\vee t}\left\Vert
i\lambda \left( \mathrm{H}_{\tau }+i\lambda \mathbf{1}\right) ^{-1}\mathrm{W}%
_{\tau }\left( i\lambda \left( \mathrm{H}_{\tau }+i\lambda \mathbf{1}\right)
^{-1}-\mathbf{1}\right) \mathrm{U}_{\tau ,s}\varphi \right\Vert _{\mathfrak{F%
}}\mathrm{d}\tau =0,\qquad \varphi \in \mathbb{D}_{t,s}.  \label{IN1IN12}
\end{equation}%
By combining (\ref{IN1IN1})--(\ref{IN1IN11}) with (\ref{IN1IN12}) we arrive
at the following limit:%
\begin{equation}
\lim_{\lambda \rightarrow \infty }\mathrm{Bwd}_{t,s}^{(\lambda )}\varphi =%
\mathbb{B}_{t,s}\varphi ,\qquad \varphi \in \mathbb{D}_{t,s}.
\label{RlambdaconvergeKbb}
\end{equation}%
As already explained before the proposition, note that $\mathrm{U}_{s,t}%
\mathcal{D}_{\infty }\subseteq \mathbb{D}_{t,s}$ and (\ref%
{RlambdaconvergeKbb}) means that $\mathbb{D}_{t,s}\subseteq \mathcal{D}(%
\mathrm{E}_{t,s})$. Therefore, $\mathrm{U}_{s,t}\mathcal{D}_{\infty
}\subseteq \mathcal{D}(\mathrm{E}_{t,s})=\mathcal{D}(\mathrm{H}_{s})$.
\end{proof}

Since Proposition \ref{ConvergenceYosida} shows that $\mathrm{U}_{s,t}%
\mathcal{D}_{\infty }\subseteq \mathcal{D}(\mathrm{H}_{s})$ for any $s,t\in 
\mathbb{R}_{0}^{+}$, we can take advantage of this information to define a
new operator: 
\begin{equation}
\mathbb{F}_{t,s}\varphi \doteq \left( \mathrm{U}_{t,s}\mathrm{H}_{s}\mathrm{U%
}_{s,t}+\int_{s}^{t}\mathrm{U}_{t,\tau }\mathrm{W}_{\tau }\mathrm{U}_{\tau
,t}\mathrm{d}\tau \right) \varphi ,\qquad \varphi \in \mathcal{D}_{\infty }.
\label{defFbb}
\end{equation}%
Indeed, $\mathcal{D}_{\infty }\subseteq \mathcal{D}(\mathrm{N}^{3/2})$ and
using Theorem \ref{propagators} and Lemma \ref{TechnicalIntegralWandRWR}, we
can make sense of the integral part. $\mathbb{F}_{t,s}$ is therefore a
symmetric, densely defined operator. In particular, it is closable. We
identify $\mathbb{F}_{t,s}$ with its closure $\mathbb{F}_{t,s}\equiv \mathbb{%
F}_{t,s}^{\ast \ast }$ and denote its domain by $\mathcal{D}(\mathbb{F}%
_{t,s})\supseteq \mathcal{D}_{\infty }$.

\begin{proposition}[Convergence of the Yosida regularization -- II]
\label{ConvergenceFwdToFbb}\mbox{}\newline
Assume Condition \ref{AbstractAssumptions} (i)--(v). Then, 
\begin{equation*}
\lim_{\lambda \rightarrow \infty }\mathrm{Fwd}_{t,s}^{(\lambda )}\varphi =%
\mathbb{F}_{t,s}\varphi ,\qquad \varphi \in \mathcal{D}_{\infty },\ s,t\in 
\mathbb{R}_{0}^{+}.
\end{equation*}
\end{proposition}

\begin{proof}
Fix once and for all $\varphi \in \mathcal{D}_{\infty }$. Take $s,t\in 
\mathbb{R}_{0}^{+}$ and recall that $\mathrm{U}_{s,t}\varphi \in \mathcal{D}(%
\mathrm{H}_{s})$, thanks to Proposition \ref{ConvergenceYosida}. Using the
triangle inequality and the unitarity of the operators $\mathrm{U}_{s,t}$
for $s,t\in \mathbb{R}_{0}^{+}$ (Theorem \ref{propagators}), we compute from
(\ref{DefRbbforward}) and (\ref{defFbb}) that 
\begin{eqnarray}
&&\left\Vert \mathrm{Fwd}_{t,s}^{(\lambda )}\varphi -\mathbb{F}_{t,s}\varphi
\right\Vert _{\mathfrak{F}}  \notag \\
&\leq &\left\Vert \left( \mathrm{H}_{s}^{(\lambda )}-\mathrm{H}_{s}\right) 
\mathrm{U}_{s,t}\varphi \right\Vert _{\mathfrak{F}}  \label{sdsdsfgghghjjhkj}
\\
&&+\int_{s\wedge t}^{s\vee t}\left\Vert \left( \frac{i\lambda \mathbf{1}}{%
\mathrm{H}_{\tau }+i\lambda \mathbf{1}}-\mathbf{1}\right) \mathrm{W}_{\tau }%
\mathrm{U}_{\tau ,t}\varphi \right\Vert _{\mathfrak{F}}\mathrm{d}\tau  \notag
\\
&&+\int_{s\wedge t}^{s\vee t}\left\Vert \left( \frac{i\lambda \mathbf{1}}{%
\mathrm{H}_{\tau }+i\lambda \mathbf{1}}\right) \mathrm{W}_{\tau }\left( 
\frac{i\lambda \mathbf{1}}{\mathrm{H}_{\tau }+i\lambda \mathbf{1}}-\mathbf{1}%
\right) \mathrm{U}_{\tau ,t}\varphi \right\Vert _{\mathfrak{F}}\mathrm{d}%
\tau .  \notag
\end{eqnarray}%
The first term in the right-hand side of the above inequality goes to zero
as $\lambda \rightarrow \infty $, because $\mathrm{U}_{s,t}\varphi \in 
\mathcal{D}(\mathrm{H}_{s})$ (Proposition \ref{ConvergenceYosida}) and the
Yosida approximation $\mathrm{H}_{s}^{(\lambda )}$ strongly converges to $%
\mathrm{H}_{s}$ on the domain $\mathcal{D}(\mathrm{H}_{s})$ (see the second
equality of (\ref{Esymmetric0})). Combining Equation (\ref%
{stronglimitpieceYosida}) with Lebesgue's dominated convergence theorem, we
find that the second term on the right-hand side of the above inequality
converges to zero in the same way we proceeded in the proof of Proposition %
\ref{ConvergenceYosida}. Again mimicking the proof of Proposition \ref%
{ConvergenceYosida}, by applying (\ref{lasteq}) together with Lemmata \ref%
{diversPreservedom} and \ref{Lemma-self-adjoint1 copy(1)}, Theorem \ref%
{propagators} as well as Lebesgue's dominated convergence theorem, we
conclude that the third term in the right-hand side of Inequality (\ref%
{sdsdsfgghghjjhkj}) also vanishes in the limit $\lambda \rightarrow \infty $%
, like the other two terms. This leads to the asserted statement.
\end{proof}

Similar to $\mathbb{B}_{t,s}$\ (Proposition \ref{ConvergenceYosida}), the
operator $\mathbb{F}_{t,s}$ reproduces the Hamiltonian $\mathrm{H}_{t}$ at
any time $t\in \mathbb{R}^{+}$. This is proven in the next statement:

\begin{theorem}[Limit of the forward Yosida regularization]
\label{Gluing}\mbox{}\newline
Assume Condition \ref{AbstractAssumptions} (i)--(v). For any $s,t\in \mathbb{%
R}_{0}^{+}$, the operator defined by (\ref{defFbb}) on $\mathcal{D}_{\infty
} $ is essentially self-adjoint and its closure equals $\mathbb{F}_{t,s}=%
\mathrm{H}_{t}=\mathbb{F}_{t,s}^{\ast }$.
\end{theorem}

\begin{proof}
Fix once and for all $t\in \mathbb{R}_{0}^{+}$ and $\varphi \in \mathcal{D}(%
\mathrm{H}_{t})$. Recall that $\mathcal{D}_{\infty }$ is a core for $\mathrm{%
H}_{t}$, i.e., it is dense in $\mathcal{D}(\mathrm{H}_{t})$ with respect of
the graph-norm topology of $\mathrm{H}_{t}$. Therefore, for any $\varphi \in 
\mathcal{D}(\mathrm{H}_{t})$, there exists a sequence\footnote{%
For any metric space -- such as $\mathcal{D}(\mathrm{H}_{t})$ endowed with
the graph-norm topology -- density and sequential density are well-known to
be equivalent.} $(\varphi _{n})_{n\in \mathbb{N}}\subseteq \mathcal{D}%
_{\infty }$ converging to $\varphi $ in the graph-norm topology of $\mathrm{H%
}_{t}$. By Proposition \ref{ConvergenceFwdToFbb}, we know that%
\begin{equation}
\mathbb{F}_{t,s}\varphi _{n}=\lim_{\lambda \rightarrow \infty }\mathrm{Fwd}%
_{t,s}^{(\lambda )}\varphi _{n},\qquad n\in \mathbb{N}.  \label{sdsdsddsd}
\end{equation}%
We aim at taking the limit $n\rightarrow \infty $\ and then reversing the
order of the limits. To this end we show the uniform convergence of $\mathrm{%
Fwd}_{t,s}^{(\lambda )}\varphi _{n}$ as $n\rightarrow \infty $. By virtue of
Lemma \ref{diversPreservedom copy(1)} and Proposition \ref{EvolRegu}, for
any $\lambda >2^{5+1/2}\Vert X_{0}\Vert _{\mathrm{op},2}$, 
\begin{equation*}
\mathrm{Fwd}_{t,s}^{(\lambda )}\phi =\mathrm{H}_{t}^{(\lambda )}\phi ,\qquad
\phi \in \mathcal{D}\left( \mathrm{H}_{t}\right) \subseteq \mathcal{D}\left( 
\mathrm{N}\right) ,
\end{equation*}%
which, together with standard estimates on the resolvent in $\mathrm{H}%
_{t}^{(\lambda )}$ (see (\ref{YosidaApproc})), implies that%
\begin{equation*}
\left\Vert \mathrm{Fwd}_{t,s}^{(\lambda )}\phi \right\Vert _{\mathfrak{F}%
}\leq \left\Vert \mathrm{H}_{t}\phi \right\Vert _{\mathfrak{F}},\qquad \phi
\in \mathcal{D}\left( \mathrm{H}_{t}\right) .
\end{equation*}%
Applying this upper bound to the vectors $(\varphi _{n}-\varphi )\in 
\mathcal{D}\left( \mathrm{H}_{t}\right) $, $n\in \mathbb{N}$, we obtain that 
\begin{equation*}
\sup_{\lambda >2^{5+1/2}\Vert X_{0}\Vert _{\mathrm{op},2}}\left\{ \left\Vert 
\mathrm{Fwd}_{t,s}^{(\lambda )}\left( \varphi _{n}-\varphi \right)
\right\Vert _{\mathfrak{F}}\right\} \leq \left\Vert \mathrm{H}_{t}\varphi
_{n}-\mathrm{H}_{t}\varphi \right\Vert _{\mathfrak{F}},\qquad n\in \mathbb{N}%
.
\end{equation*}%
Our choice of the sequence $(\varphi _{n})_{n\in \mathbb{N}}\subseteq 
\mathcal{D}_{\infty }$ ensures that 
\begin{equation*}
\lim_{n\rightarrow \infty }\sup_{\lambda >2^{5+1/2}\Vert X_{0}\Vert _{%
\mathrm{op},2}}\left\{ \left\Vert \mathrm{Fwd}_{t,s}^{(\lambda )}\left(
\varphi _{n}-\varphi \right) \right\Vert _{\mathfrak{F}}\right\} =0.
\end{equation*}%
In other words, as $n\rightarrow \infty $, the sequence of vector $\mathrm{%
Fwd}_{t,s}^{(\lambda )}\varphi _{n}$, $n\in \mathbb{N}$, converges to $%
\mathrm{Fwd}_{t,s}^{(\lambda )}\varphi $, uniformly for $\lambda
>2^{5+1/2}\Vert X_{0}\Vert _{\mathrm{op},2}$. As a consequence, we can apply
the Iterated Limit Theorem \cite[Chapter 2, Theorem 4]{topology} with the
directed sets $\mathbb{N}$ and $[2^{5+1/2}\Vert X_{0}\Vert _{\mathrm{op}%
,2},\infty )$ to deduce that%
\begin{equation}
\lim_{n\rightarrow \infty }\mathbb{F}_{t,s}\varphi _{n}=\lim_{n\rightarrow
\infty }\lim_{\lambda \rightarrow \infty }\mathrm{Fwd}_{t,s}^{(\lambda
)}\varphi _{n}=\lim_{\lambda \rightarrow \infty }\lim_{n\rightarrow \infty }%
\mathrm{Fwd}_{t,s}^{(\lambda )}\varphi _{n},  \label{weinverselimit}
\end{equation}%
whenever this last iterated limit exists. See also (\ref{sdsdsddsd}). We
first verify that%
\begin{equation}
\lim_{n\rightarrow \infty }\mathrm{Fwd}_{t,s}^{(\lambda )}\varphi _{n}=%
\mathrm{Fwd}_{t,s}^{(\lambda )}\varphi ,  \label{limit Fwd}
\end{equation}%
since subsequently taking the limit as $\lambda \rightarrow \infty $ yields $%
\mathrm{H}_{t}\varphi $. This last observation about the limit $\lambda
\rightarrow \infty $ indeed relies on Lemma \ref{diversPreservedom copy(1)},
Proposition \ref{EvolRegu} and Equation (\ref{Esymmetric0}), and is again
explained at the end of this proof. To prove Equation (\ref{limit Fwd}), for
any $\tau \in \lbrack s,t]$ and $n\in \mathbb{N}$, we use the following
upper bound: 
\begin{eqnarray}
&&\left\Vert \mathrm{U}_{t,\tau }\left( \mathrm{H}_{\tau }+i\lambda \mathbf{1%
}\right) ^{-1}\mathrm{W}_{\tau }\left( \mathrm{H}_{\tau }+i\lambda \mathbf{1}%
\right) ^{-1}\mathrm{U}_{\tau ,t}\left( \varphi _{n}-\varphi \right)
\right\Vert _{\mathfrak{F}}  \notag \\
&\leq &\lambda ^{-1}\left\Vert \mathrm{W}_{\tau }(\mathrm{N}+\mathbf{1})^{-%
\frac{3}{2}}\right\Vert _{\mathrm{op}}\left\Vert (\mathrm{N}+\mathbf{1})^{%
\frac{3}{2}}\left( \mathrm{H}_{\tau }+i\lambda \mathbf{1}\right) ^{-1}(%
\mathrm{N}+\mathbf{1})^{-1}\right\Vert _{\mathrm{op}}  \label{adfsdfgsdfgsfg}
\\
&&\left\Vert (\mathrm{N}+\mathbf{1})\mathrm{U}_{\tau ,t}(\mathrm{N}+\mathbf{1%
})^{-1}\right\Vert _{\mathrm{op}}\left\Vert (\mathrm{N}+\mathbf{1})\left( 
\mathrm{H}_{t}+i\lambda \mathbf{1}\right) ^{-1}\right\Vert _{\mathfrak{F}%
}\left\Vert \left( \mathrm{H}_{t}+i\lambda \mathbf{1}\right) \left( \varphi
_{n}-\varphi \right) \right\Vert _{\mathfrak{F}}.  \notag
\end{eqnarray}%
So, applying Lemmata \ref{diversPreservedom copy(1)}--\ref%
{Lemma-self-adjoint1 copy(1)} and Theorem \ref{propagators}, and keeping in
mind that $(\varphi _{n})_{n\in \mathbb{N}}\subseteq \mathcal{D}_{\infty }$
converges to $\varphi $ in the graph-norm topology of $\mathrm{H}_{t}$, we
infer from (\ref{adfsdfgsdfgsfg}) that 
\begin{equation*}
\lim_{n\rightarrow \infty }\int_{s}^{t}\left\Vert \mathrm{U}_{t,\tau }\left( 
\mathrm{H}_{\tau }+i\lambda \mathbf{1}\right) ^{-1}\mathrm{W}_{\tau }\left( 
\mathrm{H}_{\tau }+i\lambda \mathbf{1}\right) ^{-1}\mathrm{U}_{\tau
,t}\left( \varphi _{n}-\varphi \right) \right\Vert _{\mathfrak{F}}\mathrm{d}%
\tau =0.
\end{equation*}%
Combining this limit with the boundedness of the operator $\mathrm{U}_{t,s}%
\mathrm{H}_{s}^{(\lambda )}\mathrm{U}_{s,t}$ we deduce from (\ref%
{DefRbbforward}) the limit (\ref{limit Fwd}), which, combined with (\ref%
{weinverselimit}), in turn implies that 
\begin{equation*}
\lim_{n\rightarrow \infty }\mathbb{F}_{t,s}\varphi _{n}=\lim_{\lambda
\rightarrow \infty }\mathrm{Fwd}_{t,s}^{(\lambda )}\varphi ,
\end{equation*}%
whenever this last limit exists. As already mentioned, it suffices now to
invoke Proposition \ref{EvolRegu} and the strong convergence of $\mathrm{H}%
_{t}^{(\lambda )}$ to $\mathrm{H}_{t}$ on $\mathcal{D}(\mathrm{H}%
_{t})\subseteq \mathcal{D}(\mathrm{N})$ (see Lemma \ref{diversPreservedom
copy(1)} and (\ref{Esymmetric0})) to establish that 
\begin{equation*}
\lim_{n\rightarrow \infty }\mathbb{F}_{t,s}\varphi _{n}=\lim_{\lambda
\rightarrow \infty }\mathrm{Fwd}_{t,s}^{(\lambda )}\varphi =\mathrm{H}%
_{t}\varphi .
\end{equation*}%
Because $\mathbb{F}_{t,s}$ is closed, this implies that $\mathrm{H}%
_{t}\subseteq \mathbb{F}_{t,s}$. It is easy to check that $\mathbb{F}_{t,s}$
is also symmetric and we know from Theorem \ref{HamilSA} that $\mathrm{H}%
_{t} $ is self-adjoint with core $\mathcal{D}_{\infty }$. Much like in (\ref%
{inclusions}), we can readily deduce that 
\begin{equation*}
\mathrm{H}_{t}\subseteq \mathbb{F}_{t,s}\subseteq \mathbb{F}_{t,s}^{\ast
}\subseteq \mathrm{H}_{t}^{\ast }=\mathrm{H}_{t}
\end{equation*}%
for any $s,t\in \mathbb{R}_{0}^{+}$. From this, the desired assertion
follows.
\end{proof}

\subsection{Asymptotics of the Differential Flow\label{Asymptotics}}

We study the infinite-time limit of the differential flow. Given a triplet $%
(h_{k,t},\mu _{k,t},X_{k,t})\in \mathcal{B}(\mathfrak{h}_{k})^{3}$ that
converges in the operator norm as $t\rightarrow \infty $ to a limit $%
(h_{k,\infty },\mu _{k,\infty },0)\in \mathcal{B}(\mathfrak{h}_{k})^{3}$ for
any $k\in \mathbb{N}$, we can define the Hamiltonian 
\begin{equation}
\mathrm{H}_{\infty }\doteq \sum_{k\in \mathbb{N}}\left( h_{k,\infty }\otimes 
\mathbf{1}_{\mathcal{F}_{+}}+\mu _{k,\infty }\otimes a_{k}^{\ast
}a_{k}\right) .  \label{HinftyDef}
\end{equation}%
Compare this definition with (\ref{def_H0}). As long as the hypotheses of
Theorem \ref{HamilSA copy(1)} are fulfilled, the expression (\ref{HinftyDef}%
) defines an essentially self-adjoint operator. See also Proposition \ref%
{Proposition self-adjoint copy(1)}. As before, we identify $\mathrm{H}%
_{\infty }$ with its self-adjoint extension on a domain $\mathcal{D}(\mathrm{%
H}_{\infty })$ that contains $\mathcal{D}_{\infty }$ as a core. Because $%
\mathrm{H}_{\infty }$ is $\mathrm{N}$-diagonal, it preserves the number of
particles. However, $\mu _{k,\infty }$ is not necessarily a scalar multiple
of the identity operator for every $k\in \mathbb{N}$. Consequently, the bath
is generally not fully decoupled from the system but is instead \emph{%
pseudo-decoupled}, as announced in Section \ref{SectionGSBMain}. In other
words, the two quantum systems may remain strongly correlated, yet they
undergo no further direct energy exchange via the creation or absorption of
bosonic fields.

In this section, we establish our main result by extending Theorem \ref%
{Gluing} and Corollary \ref{Resolvent expression} to the limit $t\rightarrow
\infty $. To achieve this, we extend Condition \ref{AbstractAssumptions} to
the asymptotic regime $t\rightarrow \infty $, by requiring the integrability
of the non-negative function $t\mapsto \Vert X_{t}\Vert _{\mathrm{op},2}$.

Observe from Condition \ref{Assumptionsasympt} that $\mathrm{H}_{\infty }$
is a properly defined self-adjoint operator which admits $\mathcal{D}%
_{\infty }$ as a core, thanks to Proposition \ref{Proposition self-adjoint
copy(1)}. By Lemma \ref{lemmabounddiffH}, this Hamiltonian is the strong
limit on the domain $\mathcal{D}(\mathrm{N})$ of the family $(\mathrm{H}%
_{t})_{t\in \mathbb{R}_{0}^{+}}$, defined by (\ref{defHtbis}) via Theorem %
\ref{HamilSA} and Proposition \ref{Proposition self-adjoint copy(1)}:

\begin{lemma}[Convergence of generalized spin-boson Hamiltonians on $%
\mathcal{D}(\mathrm{N})$]
\label{strongConv}\mbox{}\newline
Under Conditions \ref{AbstractAssumptions}--\ref{Assumptionsasympt}, 
\begin{equation*}
\lim_{t\rightarrow \infty }\left\Vert \left( \mathrm{H}_{t}-\mathrm{H}%
_{\infty }\right) \left( \mathrm{N}+\mathbf{1}\right) ^{-1}\right\Vert _{%
\mathrm{op}}=0,
\end{equation*}%
where $\mathrm{H}_{\infty }$ is the Hamiltonian (\ref{HinftyDef}).
\end{lemma}

\begin{proof}
One checks that Lemma \ref{lemmabounddiffH} holds with $s=\infty $ (by
changing the integral from $s\wedge t$ to $s\vee t$ by an integral from $t$
to $\infty $). Then the assertion follows from Condition \ref%
{Assumptionsasympt} (vi).
\end{proof}

Observe that 
\begin{equation*}
\left\Vert \left( \mathrm{N}+\mathbf{1}\right) ^{-1}\left( \mathrm{H}_{t}-%
\mathrm{H}_{\infty }\right) \right\Vert _{\mathrm{op}}\leq \left\Vert \left( 
\mathrm{H}_{t}-\mathrm{H}_{\infty }\right) \left( \mathrm{N}+\mathbf{1}%
\right) ^{-1}\right\Vert _{\mathrm{op}}.
\end{equation*}%
See the discussion following Theorem \ref{MainDiago} on Sobolev-type
estimates.

By Theorem \ref{HamilSA} and Proposition \ref{Proposition self-adjoint
copy(1)}, recall that $\mathcal{D}_{\infty }$ is a core for $\mathrm{H}_{t}$%
, $t\in \left[ 0,\infty \right] $. From Lemma \ref{strongConv} we obtain in
particular that 
\begin{equation*}
\lim_{t\rightarrow \infty }\left( \mathrm{H}_{t}-\mathrm{H}_{\infty }\right)
\varphi =0,\qquad \varphi \in \mathcal{D}_{\infty }\subseteq \mathcal{D}(%
\mathrm{N}).
\end{equation*}%
It suggests that the domains $\mathcal{D}(\mathrm{H}_{t})$, $t\in \left[
0,\infty \right] $, are all the same. This property is proven in the next
lemma.

\begin{lemma}[Conservation of domains]
\label{DomHconst}\mbox{}\newline
Keeping in mind (\ref{sdsdsdsdsd}) and Definition \ref{extendsecondquanti},
under Conditions \ref{AbstractAssumptions}--\ref{Assumptionsasympt} we have 
\begin{equation*}
\mathcal{D}\left( \mathrm{H}_{t}\right) =\mathcal{D}\left( \mathrm{H}%
_{0}\right) =\mathcal{D}\left( H\left( h_{0}\right) +\mathrm{d}\Gamma
^{\otimes }\left( \mu _{0}\right) \right) ,\qquad t\in \left[ 0,\infty %
\right] .
\end{equation*}
\end{lemma}

\begin{proof}
Assume Conditions \ref{AbstractAssumptions}--\ref{Assumptionsasympt}. Fix $%
t\in \lbrack 0,\infty ]$. By Theorem \ref{HamilSA}, note that 
\begin{equation}
\mathcal{D}\left( H\left( h_{t}\right) +\mathrm{d}\Gamma^\otimes \left( \mu
_{t}\right) \right) =\mathcal{D}\left( \mathrm{H}_{t}\right) ,
\label{sdasdadasdasd}
\end{equation}%
where $H(h_{t})$ is defined by (\ref{sdsdsdsdsd}) as a lower semibounded
operator. There is in particular a real number $b\in \mathbb{R}$ such that 
\begin{equation*}
H_{b}=H\left( h_{t}\right) -b\mathbf{1}_{\mathfrak{F}}\geq 0.
\end{equation*}%
(See, for instance, Step 3 of the proof of Lemma \ref%
{TechnicalIntegralWandRWR}.) Keeping this information in mind, we give the
proof in two steps:\medskip

\noindent \underline{Step 1:} We prove here the following inequality:%
\begin{equation*}
\left\Vert \mathrm{N}\psi \right\Vert _{\mathfrak{F}}\lesssim \left\Vert
\left( H_{b}+\mathrm{d}\Gamma ^{\otimes }\left( \mu _{t}\right) \right) \psi
\right\Vert _{\mathfrak{F}}.
\end{equation*}%
To this end, using Conditions \ref{AbstractAssumptions} (v) (if $t\in 
\mathbb{R}_{0}^{+}$) or \ref{Assumptionsasympt} (if $t=\infty $), we find
that, for any $\varphi \in \mathcal{D}_{\infty }$, 
\begin{align}
\left\langle \varphi ,\mathrm{N}\varphi \right\rangle _{\mathfrak{F}}& \leq
\sum_{k\in \mathbb{N}}\left\langle \mathbf{1}\otimes a_{k}\varphi ,\mathbf{1}%
\otimes a_{k}\varphi \right\rangle _{\mathfrak{F}}\leq c^{-1}\sum_{k\in 
\mathbb{N}}\left\langle \varphi ,\mu _{k,t}\otimes a_{k}^{\ast }a_{k}\varphi
\right\rangle _{\mathfrak{F}}  \label{jkl} \\
& \leq c^{-1}\left\langle \varphi ,\left( H_{b}+\mathrm{d}\Gamma ^{\otimes
}\left( \mu _{t}\right) \right) \varphi \right\rangle _{\mathfrak{F}}. 
\notag
\end{align}%
For any $n\in \mathbb{N}$, $\mathfrak{H}\otimes \vee ^{n}\mathcal{H}$ is the
eigenspace of $\mathrm{N}$ associated with the eigenvalue $n$. If $\varphi $
belongs to both $\mathfrak{H}\otimes \vee ^{n}\mathcal{H}$ and the domain of 
$(H_{b}+\mathrm{d}\Gamma ^{\otimes }(\mu _{t}))$, then its image $(H_{b}+%
\mathrm{d}\Gamma ^{\otimes }(\mu _{t}))\varphi $ also lies in $\mathfrak{H}%
\otimes \vee ^{n}\mathcal{H}$. In particular, by using the spectral theorem%
\footnote{%
In fact, $(H_{b}+\mathrm{d}\Gamma ^{\otimes }(\mu _{t}))$ and $\mathrm{N}$
are strongly commuting in the sense of \cite[Definition 5.2]{Konrad} and one
could even use \cite[Theorem 5.23]{Konrad}.} \cite[Theorem 5.7]{Konrad} in
relation to the restriction of positive operators $(H_{b}+\mathrm{d}\Gamma
^{\otimes }(\mu _{t}))$ to each eigensubspace $\mathfrak{H}\otimes \vee ^{n}%
\mathcal{H}$ of $\mathrm{N}$ together with the monotonicity of the function $%
x\mapsto x^{2}$ on $\mathbb{R}_{0}^{+}$, one verifies from (\ref{jkl}) that 
\begin{equation}
\left\Vert \mathrm{N}\varphi \right\Vert _{\mathfrak{F}}^{2}\leq
c^{-2}\left\Vert \left( H_{b}+\mathrm{d}\Gamma ^{\otimes }\left( \mu
_{t}\right) \right) \varphi \right\Vert _{\mathfrak{F}}^{2}  \notag
\end{equation}%
for any $\varphi \in \mathcal{D}_{\infty }$. The space $\mathcal{D}_{\infty
} $ is a core for the (positive) self-adjoint operator $(H_{b}+\mathrm{d}%
\Gamma ^{\otimes }(\mu _{t}))$ and $\mathrm{N}$ is closed. We thus deduce
from the last inequality that 
\begin{equation}
\left\Vert \mathrm{N}\varphi \right\Vert _{\mathfrak{F}}\leq
c^{-1}\left\Vert \left( H_{b}+\mathrm{d}\Gamma ^{\otimes }\left( \mu
_{t}\right) \right) \varphi \right\Vert _{\mathfrak{F}}
\label{eq:borneNbydGamma}
\end{equation}%
for all vectors $\varphi \in \mathcal{D}(H_{b}+\mathrm{d}\Gamma ^{\otimes
}(\mu _{t}))$ and $t\in \lbrack 0,\infty ]$.\medskip

\noindent \underline{Step 2:} Take $\psi \in \mathcal{D}(\mathrm{H}_{t})$.
Let $(\varphi _{n})_{n\in \mathbb{N}}\subseteq \mathcal{D}_{\infty }$ be a
sequence such that 
\begin{equation*}
\varphi _{n}\rightarrow \psi \qquad \text{and}\qquad \left( H\left(
h_{t}\right) +\mathrm{d}\Gamma ^{\otimes }\left( \mu _{t}\right) \right)
\varphi _{n}\rightarrow \left( H\left( h_{t}\right) +\mathrm{d}\Gamma
^{\otimes }\left( \mu _{t}\right) \right) \psi .
\end{equation*}%
Such a sequence exists because $\mathcal{D}_{\infty }$ is a core of the
closed operator $(H\left( h_{t}\right) +\mathrm{d}\Gamma ^{\otimes }(\mu
_{t}))$ and $\psi $ is in its domain, by (\ref{sdasdadasdasd}). Therefore, 
\begin{equation*}
\left( H_{b}+\mathrm{d}\Gamma ^{\otimes }\left( \mu _{t}\right) \right)
\varphi _{n}\rightarrow \left( H_{b}+\mathrm{d}\Gamma ^{\otimes }\left( \mu
_{t}\right) \right) \psi
\end{equation*}%
and Inequality (\ref{eq:borneNbydGamma}) then yields $\mathrm{N}\varphi
_{n}\rightarrow \mathrm{N}\psi $ with $\psi \in \mathcal{D}(\mathrm{N})$, as 
$\mathrm{N}$ is a closed operator. For any $s\in \mathbb{R}_{0}^{+}$ and $%
n\in \mathbb{N}$, let us consider the equality 
\begin{eqnarray}
&&\left( H\left( h_{s}\right) +\mathrm{d}\Gamma ^{\otimes }\left( \mu
_{s}\right) \right) \varphi _{n}  \label{diffGamma} \\
&=&\left( H\left( h_{s}\right) +\mathrm{d}\Gamma ^{\otimes }\left( \mu
_{s}\right) -H\left( h_{t}\right) -\mathrm{d}\Gamma ^{\otimes }\left( \mu
_{t}\right) \right) \left( \mathrm{N}+\mathbf{1}\right) ^{-1}\left( \mathrm{N%
}+\mathbf{1}\right) \varphi _{n}+\left( H\left( h_{t}\right) +\mathrm{d}%
\Gamma ^{\otimes }\left( \mu _{t}\right) \right) \varphi _{n}.  \notag
\end{eqnarray}%
Inspired by the proof of Lemma \ref{lemmabounddiffH} -- specifically, by
applying (\ref{equality1}) without the final term on its right-hand side
while also invoking (\ref{equality1}) itself -- one establishes in the same
manner that 
\begin{equation}
\left( H\left( h_{s}\right) +\mathrm{d}\Gamma ^{\otimes }\left( \mu
_{s}\right) -H\left( h_{t}\right) -\mathrm{d}\Gamma ^{\otimes }\left( \mu
_{t}\right) \right) \left( \mathrm{N}+\mathbf{1}\right) ^{-1}\in \mathcal{B}%
\left( \mathfrak{F}\right) .  \label{diifborne}
\end{equation}%
Hence, in the limit $n\rightarrow \infty $, we deduce from (\ref{diffGamma})
that 
\begin{equation*}
\left( \left( H\left( h_{s}\right) +\mathrm{d}\Gamma ^{\otimes }\left( \mu
_{s}\right) \right) \varphi _{n}\right) _{n\in \mathbb{N}}
\end{equation*}%
is a Cauchy sequence (with $\varphi _{n}\rightarrow \psi \in \mathcal{D}(%
\mathrm{H}_{t})$, by assumption). So, this sequence converges, and the limit
must be 
\begin{equation*}
\left( H\left( h_{s}\right) +\mathrm{d}\Gamma ^{\otimes }\left( \mu
_{s}\right) \right) \psi ,\qquad \psi \in \mathcal{D}\left( H\left(
h_{s}\right) +\mathrm{d}\Gamma ^{\otimes }\left( \mu _{s}\right) \right) ,
\end{equation*}%
because $\left( H\left( h_{s}\right) +\mathrm{d}\Gamma ^{\otimes }\left( \mu
_{s}\right) \right) $ is a closed operator (see Proposition \ref{Proposition
self-adjoint copy(1)} (iii)). Applying Equation (\ref{sdasdadasdasd}) (i.e.,
Theorem \ref{HamilSA}) we conclude that, for any $t\in \lbrack 0,\infty ]$
and $s\in \mathbb{R}_{0}^{+}$, 
\begin{equation*}
\mathcal{D}\left( \mathrm{H}_{t}\right) \subseteq \mathcal{D}\left( H\left(
h_{s}\right) +\mathrm{d}\Gamma ^{\otimes }\left( \mu _{s}\right) \right) =%
\mathcal{D}\left( \mathrm{H}_{s}\right) .
\end{equation*}%
Interchanging the roles of $s$ and $t$ yields the reverse inclusion, which
completes the proof.
\end{proof}

The conservation of the domain given in Lemma \ref{DomHconst} is an
important property allowing us to compare the Hamiltonians $\mathrm{H}_{t}$
for all times $t\in \mathbb{R}^{+}\cup \{\infty \}$. We show next the
convergence of Hamiltonian family $(\mathrm{H}_{t})_{t\geq 0}$ to the
Hamiltonian $\mathrm{H}_{\infty }$ as $t\rightarrow \infty $, in the strong
resolvent sense.

\begin{lemma}[Convergence in the strong resolvent sense]
\label{strongResConv}\mbox{}\newline
Under Conditions \ref{AbstractAssumptions}--\ref{Assumptionsasympt}, for any
positive number $\lambda >2^{5+1/2}\Vert X_{0}\Vert _{\mathrm{op},2}$, in
the strong topology, 
\begin{equation*}
\lim_{t\rightarrow \infty }\left( \mathrm{H}_{t}+i\lambda \mathbf{1}\right)
^{-1}=\left( \mathrm{H}_{\infty }+i\lambda \mathbf{1}\right) ^{-1}.
\end{equation*}
\end{lemma}

\begin{proof}
Using the resolvent inequality with $\lambda >2^{5+1/2}\Vert X_{0}\Vert _{%
\mathrm{op},2}$, one obtains the upper bound 
\begin{eqnarray*}
&&\left\Vert \left( \left( \mathrm{H}_{t}+i\lambda \mathbf{1}\right)
^{-1}-\left( \mathrm{H}_{\infty }+i\lambda \mathbf{1}\right) ^{-1}\right)
\left( \mathrm{N}+\mathbf{1}\right) ^{-1}\right\Vert _{\mathrm{op}} \\
&\leq &\left\Vert \left( \mathrm{H}_{\infty }+i\lambda \mathbf{1}\right)
^{-1}\right\Vert _{\mathrm{op}}\left\Vert \left( \mathrm{H}_{t}-\mathrm{H}%
_{\infty }\right) \left( \mathrm{N}+\mathbf{1}\right) ^{-1}\right\Vert _{%
\mathrm{op}}\left\Vert \left( \left( \mathrm{N}+\mathbf{1}\right) \left( 
\mathrm{H}_{t}+i\lambda \mathbf{1}\right) ^{-1}\right) \left( \mathrm{N}+%
\mathbf{1}\right) ^{-1}\right\Vert _{\mathrm{op}},
\end{eqnarray*}%
which, combined with Lemmata \ref{diversPreservedom} and \ref{strongConv},
in turn implies that 
\begin{equation*}
\lim_{t\rightarrow \infty }\left\Vert \left( \left( \mathrm{H}_{t}+i\lambda 
\mathbf{1}\right) ^{-1}-\left( \mathrm{H}_{\infty }+i\lambda \mathbf{1}%
\right) ^{-1}\right) \left( \mathrm{N}+\mathbf{1}\right) ^{-1}\right\Vert _{%
\mathrm{op}}=0.
\end{equation*}%
The result then follows from the boundedness of the two resolvents combined
with the fact that $\mathcal{D}(\mathrm{N})$ is a dense subset of $\mathfrak{%
F}$.
\end{proof}

As we said above, $\mathrm{H}_{\infty }$ is the Hamiltonian of a
pseudo-decoupled system. We compare it with $\mathrm{U}_{t,s}\mathrm{H}_{s}%
\mathrm{U}_{s,t}$ as $t\rightarrow \infty $, in the light of Theorem \ref%
{Gluing}. The first step is to study the asymptotics of the propagators
introduced in Theorem \ref{propagators}.

\begin{theorem}[Unitary propagators at infinite times]
\label{Uinfinitime}\mbox{}\newline
Under Conditions \ref{AbstractAssumptions}--\ref{Assumptionsasympt}
(i)--(iv) and (vi), the family of unitary propagators of Theorem \ref%
{propagators} can be extended to a family $(\mathrm{U}_{t,s})_{s,t\in
\lbrack 0,\infty ]}$ of unitary operators satisfying, mutatis mutandis on $%
[0,\infty ]$, all properties (i)--(v) of Theorem \ref{propagators}.
\end{theorem}

\begin{proof}
We use Theorem \ref{propagators} and (\ref{relativeboundG}) together with
Conditions \ref{Assumptionsasympt} (vi) to obtain that, for any $%
s_{1},t_{1},s_{2},t_{2}\in \mathbb{R}_{0}^{+}$,%
\begin{eqnarray}
&&\left\Vert \left( \mathrm{U}_{t_{2},s_{2}}-\mathrm{U}_{t_{1},s_{1}}\right)
(\mathrm{N}+\mathbf{1})^{-\frac{1}{2}}\right\Vert _{\mathrm{op}}  \notag \\
&\leq &4\int_{s_{1}\wedge s_{2}}^{s_{1}\vee s_{2}}\left\Vert X_{\tau
}\right\Vert _{\mathrm{op},2}\mathrm{d}\tau +4\int_{t_{1}\wedge
t_{2}}^{t_{1}\vee t_{2}}\left\Vert X_{\tau _{1}}\right\Vert _{\mathrm{op},2}%
\mathrm{e}^{D\int_{s\wedge \tau _{1}}^{s\vee \tau _{1}}\left\Vert X_{\tau
_{2}}\right\Vert _{\mathrm{op},2}\mathrm{d}\tau _{2}}\mathrm{d}\tau _{1} 
\notag \\
&\leq &4\int_{s_{1}\wedge s_{2}}^{s_{1}\vee s_{2}}\left\Vert X_{\tau
}\right\Vert _{\mathrm{op},2}\mathrm{d}\tau +4\mathrm{e}^{D\int_{0}^{\infty
}\left\Vert X_{\tau }\right\Vert _{\mathrm{op},2}\mathrm{d}\tau
}\int_{t_{1}\wedge t_{2}}^{t_{1}\vee t_{2}}\left\Vert X_{\tau }\right\Vert _{%
\mathrm{op},2}\mathrm{d}\tau ,  \label{flow equation convergence unitary 1}
\end{eqnarray}%
where $D\in \mathbb{R}^{+}$ is a fixed, time-independent, constant coming
from (\ref{BounddefDn}) for $n=1$. By Conditions \ref{Assumptionsasympt}
(vi) and Inequality (\ref{flow equation convergence unitary 1}), we deduce
that, for any $s_{1},t_{1}\in \mathbb{R}_{0}^{+}$ and $\varphi \in \mathcal{D%
}(\mathrm{N}^{1/2})$,%
\begin{equation}
\left( \mathrm{U}_{t,s_{1}}\varphi \right) _{t\in \mathbb{R}_{0}^{+}}\qquad 
\text{and}\qquad \left( \mathrm{U}_{t_{1},s}\varphi \right) _{s\in \mathbb{R}%
_{0}^{+}}  \label{cauchy1}
\end{equation}%
are Cauchy nets and thus converge to two vectors respectively denoted by 
\begin{equation}
\mathrm{U}_{\infty ,s_{1}}\varphi \in \mathfrak{F}\qquad \text{and}\qquad 
\mathrm{U}_{t_{1},\infty }\varphi \in \mathfrak{F}.  \label{cauchy2}
\end{equation}%
In this case, Inequality (\ref{flow equation convergence unitary 1}) can
thus be extended to $s_{1},t_{1},s_{2},t_{2}\in \lbrack 0,\infty ]$. Then,
the nets (\ref{cauchy1}) are also Cauchy for any $s_{1},t_{1}\in \lbrack
0,\infty ]$ and $\varphi \in \mathcal{D}(\mathrm{N}^{1/2})$ and we extend
the notation (\ref{cauchy2}) to all $s_{1},t_{1}\in \lbrack 0,\infty ]$.
Since $\mathcal{D}(\mathrm{N}^{1/2})$ is dense in $\mathfrak{F}$, we can
thus infer the existence of two families $(\mathrm{U}_{\infty ,s})_{s\in
\lbrack 0,\infty ]}$ and $(\mathrm{U}_{t,\infty })_{t\in \lbrack 0,\infty ]}$
of bounded operators respectively defined for any $\varphi \in \mathfrak{F}$
and $s,t\in \mathbb{R}_{0}^{+}$, by the strong limits 
\begin{equation}
\mathrm{U}_{\infty ,s}\varphi \doteq \lim_{t\rightarrow \infty }\mathrm{U}%
_{t,s}\varphi \qquad \text{and}\qquad \mathrm{U}_{t,\infty }\varphi \doteq
\lim_{s\rightarrow \infty }\mathrm{U}_{t,s}\varphi ,
\label{Uinftystronglimit}
\end{equation}%
as well as 
\begin{equation}
\mathrm{U}_{\infty ,\infty }\varphi \doteq \lim_{t\rightarrow \infty
}\lim_{s\rightarrow \infty }\mathrm{U}_{t,s}\varphi =\lim_{s\rightarrow
\infty }\lim_{t\rightarrow \infty }\mathrm{U}_{t,s}\varphi =\varphi ,
\label{flow equation convergence unitary 1bis}
\end{equation}%
using again (\ref{flow equation convergence unitary 1}). In particular, $%
\mathrm{U}_{\infty ,\infty }=\mathbf{1}$. Additionally, by the unitarity of
the operator family $(\mathrm{U}_{t,s})_{s,t\in \mathbb{R}_{0}^{+}}$, we
must have the bound 
\begin{equation}
\sup_{s,t\in \left[ 0,\infty \right] }\max \left\{ \left\Vert \mathrm{U}%
_{\infty ,s}\right\Vert _{\mathrm{op}},\left\Vert \mathrm{U}_{t,\infty
}\right\Vert _{\mathrm{op}}\right\} \leq 1.
\label{bound for limit unitarity}
\end{equation}%
Recall that $\mathrm{U}_{s,t}=\mathrm{U}_{t,s}^{\ast }\ $for any $s,t\in 
\mathbb{R}_{0}^{+}$ (Theorem \ref{propagators} (i)). As a consequence,
without giving the details, we repeat the same process with the family $(%
\mathrm{U}_{t,s}^{\ast })_{s,t\in \mathbb{R}_{0}^{+}}$, and we obtain that $%
\mathrm{U}_{\infty ,s}^{\ast }=\mathrm{U}_{s,\infty }$ and $\mathrm{U}%
_{t,\infty }^{\ast }=\mathrm{U}_{\infty ,t}$ for any $s,t\in \mathbb{R}%
_{0}^{+}$. In particular, 
\begin{equation}
\mathrm{U}_{\infty ,s}^{\ast }\varphi \doteq \lim_{t\rightarrow \infty }%
\mathrm{U}_{t,s}^{\ast }\varphi \qquad \text{and}\qquad \mathrm{U}_{t,\infty
}^{\ast }\varphi \doteq \lim_{s\rightarrow \infty }\mathrm{U}_{t,s}^{\ast
}\varphi .  \label{Uinftystronglimitbis}
\end{equation}%
Then, for any $s,t\in \mathbb{R}_{0}^{+}$ and $\varphi \in \mathfrak{F}$,
the unitarity of $\mathrm{U}_{t,s}$ (Theorem \ref{propagators} (i)) can be
used to get the inequalities 
\begin{align}
\left\Vert \left( \mathrm{U}_{\infty ,s}\mathrm{U}_{\infty ,s}^{\ast }-%
\mathbf{1}\right) \varphi \right\Vert _{\mathfrak{F}}& =\left\Vert \left( 
\mathrm{U}_{\infty ,s}\mathrm{U}_{\infty ,s}^{\ast }-\mathrm{U}_{t,s}\mathrm{%
U}_{t,s}^{\ast }\right) \varphi \right\Vert _{\mathfrak{F}}  \notag \\
& \leq \left\Vert \left( \mathrm{U}_{\infty ,s}-\mathrm{U}_{t,s}\right) 
\mathrm{U}_{\infty ,s}^{\ast }\varphi \right\Vert _{\mathfrak{F}}+\left\Vert
\left( \mathrm{U}_{\infty ,s}^{\ast }-\mathrm{U}_{t,s}^{\ast }\right)
\varphi \right\Vert _{\mathfrak{F}}  \label{unitarityinfty1}
\end{align}%
and 
\begin{align}
\left\Vert \left( \mathrm{U}_{\infty ,s}^{\ast }\mathrm{U}_{\infty ,s}-%
\mathbf{1}\right) \varphi \right\Vert _{\mathfrak{F}}& =\left\Vert \left( 
\mathrm{U}_{\infty ,s}^{\ast }\mathrm{U}_{\infty ,s}-\mathrm{U}_{t,s}^{\ast }%
\mathrm{U}_{t,s}\right) \varphi \right\Vert _{\mathfrak{F}}  \notag \\
& \leq \left\Vert \left( \mathrm{U}_{\infty ,s}^{\ast }-\mathrm{U}%
_{t,s}^{\ast }\right) \mathrm{U}_{\infty ,s}\varphi \right\Vert _{\mathfrak{F%
}}+\left\Vert \left( \mathrm{U}_{\infty ,s}-\mathrm{U}_{t,s}\right) \varphi
\right\Vert _{\mathfrak{F}}.  \label{unitarityinfty2}
\end{align}%
So taking the limit $t\rightarrow \infty $ in (\ref{unitarityinfty1})--(\ref%
{unitarityinfty2}) together with the use of (\ref{Uinftystronglimit}) and (%
\ref{Uinftystronglimitbis}), we deduce that $\mathrm{U}_{\infty ,s}$ is a
unitary operator for any $s\in \mathbb{R}_{0}^{+}$. Mutatis mutandis for $%
\mathrm{U}_{t,\infty }$, $t\in \mathbb{R}_{0}^{+}$. This proves Assertion
(i) of Theorem \ref{propagators}, extended to $[0,\infty ]$.

Observe that, for any $s,x\in \mathbb{R}_{0}^{+}$ and $\varphi \in \mathfrak{%
F}$, 
\begin{equation*}
\left( \mathrm{U}_{\infty ,s}-\mathrm{U}_{\infty ,x}\mathrm{U}_{x,s}\right)
\varphi =\lim_{t\rightarrow \infty }\left( \mathrm{U}_{t,s}-\mathrm{U}_{t,x}%
\mathrm{U}_{x,s}\right) \varphi =0.
\end{equation*}%
Mutatis mutandis for the identity $\mathrm{U}_{t,\infty }=\mathrm{U}_{t,x}%
\mathrm{U}_{x,\infty }$ for $t,x\in \mathbb{R}_{0}^{+}$. This proves
Assertion (ii) of Theorem \ref{propagators}, extended to $[0,\infty ]$. It
is also straightforward to verify from (\ref{flow equation convergence
unitary 1}) that $(\mathrm{U}_{\infty ,s})_{s\in \mathbb{R}_{0}^{+}}$ and $(%
\mathrm{U}_{t,\infty })_{t\in \mathbb{R}_{0}^{+}}$ are strongly continuous
in $s$ and $t$, respectively. This corresponds to Assertion (iii) of Theorem %
\ref{propagators}, extended to $[0,\infty ]$.

Similarly to (\ref{flow equation convergence unitary 1}), using Theorem \ref%
{propagators} and Inequality (\ref{relativeboundG3}), we also obtain that,
for any $s_{1},t_{1},s_{2},t_{2}\in \mathbb{R}_{0}^{+}$ and $n\in \mathbb{N}$%
,%
\begin{equation}
\Vert \left( \mathrm{N}+\mathbf{1}\right) ^{\frac{n}{2}}\left( \mathrm{U}%
_{t_{2},s_{2}}-\mathrm{U}_{t_{1},s_{1}}\right) \left( \mathrm{N}+\mathbf{1}%
\right) ^{-\frac{1+n}{2}}\Vert _{\mathrm{op}}\lesssim _{n}\int_{s_{1}\wedge
s_{2}}^{s_{1}\vee s_{2}}\left\Vert X_{\tau }\right\Vert _{\mathrm{op},2}%
\mathrm{d}\tau +\int_{t_{1}\wedge t_{2}}^{t_{1}\vee t_{2}}\left\Vert X_{\tau
}\right\Vert _{\mathrm{op},2}\mathrm{d}\tau .  \label{diffUpreversedomainN}
\end{equation}%
Here, $\lesssim _{n}$ indicates that the implicit constant $D=D_{n}$ depends
only on $n\in \mathbb{N}$ (and not on $s_{1},t_{1},s_{2},t_{2}\in \mathbb{R}%
_{0}^{+}$). Using Conditions \ref{Assumptionsasympt} (vi) and the fact that
for all $n\in \mathbb{N}$, $(\mathrm{N}+\mathbf{1})^{n/2}$ is a closed
operator together with (\ref{Uinftystronglimit}), (\ref{diffUpreversedomainN}%
) and Theorem \ref{propagators} (iv), we deduce that, for any $\varphi \in 
\mathfrak{F}$, $s,t\in \mathbb{R}_{0}^{+}$ and $n\in \mathbb{N}$, 
\begin{eqnarray}
\left( \mathrm{N}+\mathbf{1}\right) ^{\frac{n}{2}}\mathrm{U}_{\infty
,s}\left( \mathrm{N}+\mathbf{1}\right) ^{-\frac{n}{2}}\varphi
&=&\lim_{t\rightarrow \infty }\left( \mathrm{N}+\mathbf{1}\right) ^{\frac{n}{%
2}}\mathrm{U}_{t,s}\left( \mathrm{N}+\mathbf{1}\right) ^{-\frac{n}{2}%
}\varphi \ ,  \label{limitUpreservedomN} \\
\left( \mathrm{N}+\mathbf{1}\right) ^{\frac{n}{2}}\mathrm{U}_{t,\infty
}\left( \mathrm{N}+\mathbf{1}\right) ^{-\frac{n}{2}}\varphi
&=&\lim_{s\rightarrow \infty }\left( \mathrm{N}+\mathbf{1}\right) ^{\frac{n}{%
2}}\mathrm{U}_{t,s}\left( \mathrm{N}+\mathbf{1}\right) ^{-\frac{n}{2}%
}\varphi \ ,  \label{limitUpreservedomNbis}
\end{eqnarray}%
where, for all $s\in \mathbb{R}_{0}^{+}$ and $n\in \mathbb{N}$, 
\begin{equation}
\max \left\{ \ln \left\Vert \left( \mathrm{N}+\mathbf{1}\right) ^{\frac{n}{2}%
}\mathrm{U}_{\infty ,s}\left( \mathrm{N}+\mathbf{1}\right) ^{-\frac{n}{2}%
}\right\Vert _{\mathrm{op}},\ln \left\Vert \left( \mathrm{N}+\mathbf{1}%
\right) ^{\frac{n}{2}}\mathrm{U}_{s,\infty }\left( \mathrm{N}+\mathbf{1}%
\right) ^{-\frac{n}{2}}\right\Vert _{\mathrm{op}}\right\} \lesssim
_{n}\int_{s}^{\infty }\left\Vert X_{\tau }\right\Vert _{\mathrm{op},2}%
\mathrm{d}\tau .  \label{assertion ivassertion iv}
\end{equation}%
Similar to (\ref{flow equation convergence unitary 1bis}), from (\ref%
{diffUpreversedomainN}) and the fact that $(\mathrm{N}+\mathbf{1})^{n/2}$ is
a closed operator for all $n\in \mathbb{N}$, one verifies that, for any $%
n\in \mathbb{N}$, 
\begin{align*}
\lim_{t\rightarrow \infty }\lim_{s\rightarrow \infty }\left( \mathrm{N}+%
\mathbf{1}\right) ^{\frac{n}{2}}\mathrm{U}_{t,s}\left( \mathrm{N}+\mathbf{1}%
\right) ^{-\frac{n}{2}}\varphi & =\lim_{s\rightarrow \infty
}\lim_{t\rightarrow \infty }\left( \mathrm{N}+\mathbf{1}\right) ^{\frac{n}{2}%
}\mathrm{U}_{t,s}\left( \mathrm{N}+\mathbf{1}\right) ^{-\frac{n}{2}}\varphi
\\
& =\varphi =\left( \mathrm{N}+\mathbf{1}\right) ^{\frac{n}{2}}\mathrm{U}%
_{\infty ,\infty }\left( \mathrm{N}+\mathbf{1}\right) ^{-\frac{n}{2}}
\end{align*}%
and%
\begin{equation*}
\left( \mathrm{N}+\mathbf{1}\right) ^{\frac{n}{2}}\mathrm{U}_{\infty
,s}\left( \mathrm{N}+\mathbf{1}\right) ^{-\frac{n}{2}}\qquad \text{and}%
\qquad \left( \mathrm{N}+\mathbf{1}\right) ^{\frac{n}{2}}\mathrm{U}%
_{t,\infty }\left( \mathrm{N}+\mathbf{1}\right) ^{-\frac{n}{2}}
\end{equation*}%
are strongly continuous in $s$ and $t$, respectively. Together with (\ref%
{assertion ivassertion iv}), this proves Assertion (iv) of Theorem \ref%
{propagators}, extended to $[0,\infty ]$. Now we extend Assertion (v) of
Theorem \ref{propagators}.

Using Theorem \ref{propagators}, Lemma \ref{RelativeboundGlemma}, the
triangle inequality and the positivity of $\mathrm{N}$, we find that, for
all $s,t\in \mathbb{R}_{0}^{+}$, any vector $\varphi \in \mathcal{D}(\mathrm{%
N}^{1/2})\subseteq \mathcal{D}(\mathrm{G}_{s})$ and $\epsilon \in \mathbb{R}$
such that $s+\epsilon \geq 0$, 
\begin{eqnarray}
&&\left\Vert \left\{ \left\vert \epsilon \right\vert ^{-1}\left( \mathrm{U}%
_{\infty ,s+\epsilon }-\mathrm{U}_{\infty ,s}\right) -i\mathrm{U}_{\infty ,s}%
\mathrm{G}_{s}\right\} \varphi \right\Vert _{\mathfrak{F}}  \notag \\
&\leq &\left\Vert \left( \mathrm{U}_{\infty ,s}-\mathrm{U}_{t,s}\right) 
\mathrm{G}_{s}\varphi \right\Vert _{\mathfrak{F}}+\left\vert \epsilon
\right\vert ^{-1}\left( \left\Vert \left( \mathrm{U}_{\infty ,s+\epsilon }-%
\mathrm{U}_{t,s+\epsilon }\right) \varphi \right\Vert _{\mathfrak{F}%
}+\left\Vert \left( \mathrm{U}_{\infty ,s}-\mathrm{U}_{t,s}\right) \varphi
\right\Vert _{\mathfrak{F}}\right)  \notag \\
&&+\Vert \left( \mathrm{N}+\mathbf{1}\right) ^{\frac{1}{2}}\varphi \Vert _{%
\mathfrak{F}}\left( \left\vert \epsilon \right\vert ^{-1}\int_{\left(
s+\epsilon \right) \wedge s}^{\left( s+\epsilon \right) \vee s}\Vert \left( 
\mathrm{U}_{t,\tau }-\mathrm{U}_{t,s}\right) \left( \mathrm{N}+\mathbf{1}%
\right) ^{-\frac{1}{2}}\Vert _{\mathrm{op}}\mathrm{d}\tau \right)  \notag \\
&&\times \left( \Vert \lbrack \left( \mathrm{N}+\mathbf{1}\right) ^{\frac{1}{%
2}},\mathrm{G}_{s}]\left( \mathrm{N}+\mathbf{1}\right) ^{-\frac{1}{2}}\Vert
_{\mathrm{op}}+\Vert \mathrm{G}_{s}\left( \mathrm{N}+\mathbf{1}\right) ^{-%
\frac{1}{2}}\Vert _{\mathrm{op}}\right)  \notag \\
&&+\Vert \left( \mathrm{N}+\mathbf{1}\right) ^{\frac{1}{2}}\varphi \Vert _{%
\mathfrak{F}}\ \left\vert \epsilon \right\vert ^{-1}\int_{\left( s+\epsilon
\right) \wedge s}^{\left( s+\epsilon \right) \vee s}\Vert \left( \mathrm{G}%
_{\tau }-\mathrm{G}_{s}\right) \left( \mathrm{N}+\mathbf{1}\right) ^{-\frac{1%
}{2}}\Vert _{\mathrm{op}}\mathrm{d}\tau ,  \label{sdfsdf}
\end{eqnarray}%
which, for some fixed $x\in \mathbb{R}_{0}^{+}$, yields%
\begin{eqnarray}
&&\left\Vert \left\{ \left\vert \epsilon \right\vert ^{-1}\left( \mathrm{U}%
_{\infty ,s+\epsilon }-\mathrm{U}_{\infty ,s}\right) -i\mathrm{U}_{\infty ,s}%
\mathrm{G}_{s}\right\} \varphi \right\Vert _{\mathfrak{F}}  \notag \\
&\leq &\left\Vert \left( \mathrm{U}_{\infty ,s}-\mathrm{U}_{t,s}\right) 
\mathrm{G}_{s}\varphi \right\Vert _{\mathfrak{F}}+\left\vert \epsilon
\right\vert ^{-1}\left( \left\Vert \left( \mathrm{U}_{\infty ,s+\epsilon }-%
\mathrm{U}_{t,s+\epsilon }\right) \varphi \right\Vert _{\mathfrak{F}%
}+\left\Vert \left( \mathrm{U}_{\infty ,s}-\mathrm{U}_{t,s}\right) \varphi
\right\Vert _{\mathfrak{F}}\right)  \notag \\
&&+\Vert \left( \mathrm{N}+\mathbf{1}\right) ^{\frac{1}{2}}\varphi \Vert _{%
\mathfrak{F}}\sup_{\tau \in \left[ \left( s+\epsilon \right) \wedge s,\left(
s+\epsilon \right) \vee s\right] }\Vert \left( \mathrm{U}_{x,\tau }-\mathrm{U%
}_{x,s}\right) \left( \mathrm{N}+\mathbf{1}\right) ^{-\frac{1}{2}}\Vert _{%
\mathrm{op}}  \notag \\
&&\times \left( \Vert \lbrack \left( \mathrm{N}+\mathbf{1}\right) ^{\frac{1}{%
2}},\mathrm{G}_{s}]\left( \mathrm{N}+\mathbf{1}\right) ^{-\frac{1}{2}}\Vert
_{\mathrm{op}}+\Vert \mathrm{G}_{s}\left( \mathrm{N}+\mathbf{1}\right) ^{-%
\frac{1}{2}}\Vert _{\mathrm{op}}\right)  \notag \\
&&+\Vert \left( \mathrm{N}+\mathbf{1}\right) ^{\frac{1}{2}}\varphi \Vert _{%
\mathfrak{F}}\sup_{\tau \in \left[ \left( s+\epsilon \right) \wedge s,\left(
s+\epsilon \right) \vee s\right] }\Vert \left( \mathrm{G}_{\tau }-\mathrm{G}%
_{s}\right) \left( \mathrm{N}+\mathbf{1}\right) ^{-\frac{1}{2}}\Vert _{%
\mathrm{op}},  \label{boundderiveUinftyDomainNcarre}
\end{eqnarray}%
using that, for any $t,\tau ,s,x\in \mathbb{R}_{0}^{+}$, 
\begin{equation*}
\left\Vert \left( \mathrm{U}_{t,\tau }-\mathrm{U}_{t,s}\right) \left( 
\mathrm{N}+\mathbf{1}\right) ^{-\frac{1}{2}}\right\Vert _{\mathrm{op}}\leq
\left\Vert \left( \mathrm{U}_{x,\tau }-\mathrm{U}_{x,s}\right) \left( 
\mathrm{N}+\mathbf{1}\right) ^{-\frac{1}{2}}\right\Vert _{\mathrm{op}},
\end{equation*}%
thanks to Theorem \ref{propagators} (i)--(ii). Therefore, using Equations (%
\ref{relativeboundG})--(\ref{relativeboundG3}), (\ref{flow equation
convergence unitary 1}) and (\ref{Uinftystronglimit}) together with Theorem %
\ref{propagators}, $X\in C(\mathbb{R}_{0}^{+};\ell ^{2}(\mathcal{B}(%
\mathfrak{h})))$ (Proposition \ref{prop useful}) and Conditions \ref%
{Assumptionsasympt} (vi), we can take first the limit $t\rightarrow \infty $
and then $\epsilon \rightarrow 0$ in Inequality (\ref%
{boundderiveUinftyDomainNcarre}) to deduce that 
\begin{equation}
\partial _{s}\mathrm{U}_{\infty ,s}\varphi =i\mathrm{U}_{\infty ,s}\mathrm{G}%
_{s}\varphi ,\qquad \varphi \in \mathcal{D}(\mathrm{N}^{1/2}).  \label{ds}
\end{equation}%
We now verify the other differential equation, i.e., the derivative with
respect to $t$. For all $s,t\in \mathbb{R}_{0}^{+}$, any vector $\varphi \in 
\mathcal{D}(\mathrm{N}^{1/2})$ and $\epsilon \in \mathbb{R}$ such that $%
t+\epsilon \geq 0$, 
\begin{eqnarray}
&&\left\Vert \left\{ \left\vert \epsilon \right\vert ^{-1}\left( \mathrm{U}%
_{t+\epsilon ,\infty }-\mathrm{U}_{t,\infty }\right) +i\mathrm{G}_{t}\mathrm{%
U}_{t,\infty }\right\} \varphi \right\Vert _{\mathfrak{F}}  \notag \\
&\leq &\Vert \mathrm{G}_{t}\left( \mathrm{N}+\mathbf{1}\right) ^{-\frac{1}{2}%
}\Vert _{\mathrm{op}}\Vert \left( \mathrm{N}+\mathbf{1}\right) ^{\frac{1}{2}%
}\left( \mathrm{U}_{t,\infty }-\mathrm{U}_{t,s}\right) \varphi \Vert _{%
\mathfrak{F}}  \notag \\
&&+\left\vert \epsilon \right\vert ^{-1}\left( \left\Vert \left( \mathrm{U}%
_{t+\epsilon ,\infty }-\mathrm{U}_{t+\epsilon ,s}\right) \varphi \right\Vert
_{\mathfrak{F}}+\left\Vert \left( \mathrm{U}_{t,\infty }-\mathrm{U}%
_{t,s}\right) \varphi \right\Vert _{\mathfrak{F}}\right)  \notag \\
&&+\Vert \left( \mathrm{N}+\mathbf{1}\right) ^{\frac{1}{2}}\varphi \Vert _{%
\mathfrak{F}}\ \left\vert \epsilon \right\vert ^{-1}\int_{\left( t+\epsilon
\right) \wedge t}^{\left( t+\epsilon \right) \vee t}\Vert \left( \mathrm{G}%
_{\tau }-\mathrm{G}_{t}\right) \left( \mathrm{N}+\mathbf{1}\right) ^{-\frac{1%
}{2}}\Vert _{\mathrm{op}}\Vert \left( \mathrm{N}+\mathbf{1}\right) ^{\frac{1%
}{2}}\mathrm{U}_{\tau ,s}\left( \mathrm{N}+\mathbf{1}\right) ^{-\frac{1}{2}%
}\Vert _{\mathrm{op}}\ \mathrm{d}\tau  \notag \\
&&+\left\vert \epsilon \right\vert ^{-1}\int_{\left( t+\epsilon \right)
\wedge t}^{\left( t+\epsilon \right) \vee t}\Vert \mathrm{G}_{t}\left( 
\mathrm{N}+\mathbf{1}\right) ^{-\frac{1}{2}}\Vert _{\mathrm{op}}\Vert \left( 
\mathrm{N}+\mathbf{1}\right) ^{\frac{1}{2}}\left( \mathrm{U}_{\tau ,s}-%
\mathrm{U}_{t,s}\right) \left( \mathrm{N}+\mathbf{1}\right) ^{-\frac{1}{2}%
}\varphi \Vert _{\mathfrak{F}}\ \mathrm{d}\tau .
\label{flow equation convergence unitary 2bis}
\end{eqnarray}%
Compare this upper bound with (\ref{sdfsdf}). So, in the same manner that we
prove (\ref{ds}), by taking the limit $s\rightarrow \infty $ and then $%
\epsilon \rightarrow 0$ in (\ref{flow equation convergence unitary 2bis})
and using additionally (\ref{limitUpreservedomNbis}), we deduce that 
\begin{equation*}
\partial _{t}\mathrm{U}_{t,\infty }\varphi =-i\mathrm{G}_{t}\mathrm{U}%
_{t,\infty }\varphi ,\qquad \varphi \in \mathcal{D}(\mathrm{N}^{1/2}),
\end{equation*}%
which, together with (\ref{ds}), leads to Assertion (v) of Theorem \ref%
{propagators}, extended to $[0,\infty ]$.
\end{proof}

In Section \ref{Approximated Solution}, we study the differential flow on
the corresponding resolvents (Proposition \ref{Resolventeq}) and establish
in Corollary \ref{Resolvent expression} its connection to the unitary flow
generated by the family $(\mathrm{U}_{t,s})_{s,t\in \mathbb{R}_{0}^{+}}$. We
extend the latter result for infinite times in the following assertion:

\begin{theorem}[Asymptotics of the resolvent flow]
\label{Resolvent expressioninfty}\mbox{}\newline
Under Conditions \ref{AbstractAssumptions}--\ref{Assumptionsasympt}
(i)--(iv) and (vi), for any positive number $\lambda >2^{5+1/2}\Vert
X_{0}\Vert _{\mathrm{op},2}$ and $s\in \mathbb{R}_{0}^{+}$, the resolvent $%
\mathrm{R}_{\infty }\doteq \left( \mathrm{H}_{\infty }+i\lambda \mathbf{1}%
\right) ^{-1}$ satisfies the equation 
\begin{equation*}
\left\langle \psi ,\mathrm{R}_{\infty }\varphi \right\rangle _{\mathfrak{F}%
}=\left\langle \psi ,\left( \mathrm{U}_{\infty ,s}\mathrm{R}_{s}\mathrm{U}%
_{s,\infty }-\int_{s}^{\infty }\mathrm{U}_{\infty ,\tau }\mathrm{R}_{\tau }%
\mathrm{W}_{\tau }\mathrm{R}_{\tau }\mathrm{U}_{\tau ,\infty }\mathrm{d}\tau
\right) \varphi \right\rangle _{\mathfrak{F}},
\end{equation*}%
for any $\varphi ,\psi \in \mathcal{D}(\mathrm{N}^{3/2})$, where $\mathrm{W}%
_{\tau }$ is the Hamiltonian defined by (\ref{expression Wt}) via
Proposition \ref{Lemma Htbis} (iii).
\end{theorem}

\begin{proof}
Fix once and for all $s\in \mathbb{R}$. Define the operators 
\begin{equation*}
\mathbf{X}_{t}\doteq \mathrm{U}_{t,s}\mathrm{R}_{s}\mathrm{U}%
_{s,t}-\int_{s}^{t}\mathrm{U}_{t,\tau }\mathrm{R}_{\tau }\mathrm{W}_{\tau }%
\mathrm{R}_{\tau }\mathrm{U}_{\tau ,t}\mathrm{d}\tau ,\qquad t\in \left[
s,\infty \right] ,
\end{equation*}%
on the domain $\mathcal{D}(\mathrm{N}^{3/2})$. This operator is well
defined, thanks to Lemmata \ref{TechnicalIntegralWandRWR}--\ref%
{TechnicalIntegralWandRWR copy(1)}, whose hypotheses are fulfilled by virtue
of Theorem \ref{Uinfinitime}. In particular, for any $\varphi \in \mathcal{D}%
(\mathrm{N}^{3/2})$, 
\begin{equation}
\lim_{t\rightarrow \infty }\left\Vert \int_{t}^{\infty }\mathrm{U}_{\infty
,\tau }\mathrm{R}_{\tau }\mathrm{W}_{\tau }\mathrm{R}_{\tau }\mathrm{U}%
_{\tau ,\infty }\varphi \mathrm{d}\tau \right\Vert _{\mathfrak{F}}=0.
\label{limit 1}
\end{equation}%
Using Theorem \ref{Uinfinitime} and the triangle inequality, we find that,
for any $\varphi \in \mathcal{D}(\mathrm{N}^{3/2})$,%
\begin{eqnarray}
\left\Vert \left( \mathbf{X}_{\infty }-\mathbf{X}_{t}\right) \varphi
\right\Vert _{\mathfrak{F}} &\leq &\left\Vert \mathrm{R}_{s}\left( \mathrm{U}%
_{s,t}-\mathrm{U}_{s,\infty }\right) \varphi \right\Vert _{\mathfrak{F}%
}+\left\Vert \left( \mathrm{U}_{\infty ,s}-\mathrm{U}_{t,s}\right) \mathrm{R}%
_{s}\mathrm{U}_{s,\infty }\varphi \right\Vert _{\mathfrak{F}}  \notag \\
&&+\left\Vert \int_{t}^{\infty }\mathrm{U}_{\infty ,\tau }\mathrm{R}_{\tau }%
\mathrm{W}_{\tau }\mathrm{R}_{\tau }\mathrm{U}_{\tau ,\infty }\varphi 
\mathrm{d}\tau \right\Vert _{\mathfrak{F}}  \notag \\
&&+\left\Vert \int_{s}^{\infty }\left( \mathrm{U}_{\infty ,\tau }-\mathrm{U}%
_{t,\tau }\right) \mathrm{R}_{\tau }\mathrm{W}_{\tau }\mathrm{R}_{\tau }%
\mathrm{U}_{\tau ,\infty }\varphi \mathrm{d}\tau \right\Vert _{\mathfrak{F}}
\notag \\
&&+\left\Vert \int_{s}^{\infty }\mathrm{U}_{t,\tau }\mathrm{R}_{\tau }%
\mathrm{W}_{\tau }\mathrm{R}_{\tau }\left( \mathrm{U}_{\tau ,t}-\mathrm{U}%
_{\tau ,\infty }\right) \varphi \mathrm{d}\tau \right\Vert _{\mathfrak{F}}.
\label{limit 1bis}
\end{eqnarray}%
By Equation (\ref{Uinftystronglimit}) and (\ref{Uinftystronglimitbis}), for
any $\varphi \in \mathcal{D}(\mathrm{N}^{3/2})$, we have%
\begin{equation}
\lim_{t\rightarrow \infty }\left\Vert \mathrm{R}_{s}\left( \mathrm{U}_{s,t}-%
\mathrm{U}_{s,\infty }\right) \varphi \right\Vert _{\mathfrak{F}%
}=\lim_{t\rightarrow \infty }\left\Vert \left( \mathrm{U}_{\infty ,s}-%
\mathrm{U}_{t,s}\right) \mathrm{R}_{s}\mathrm{U}_{s,\infty }\varphi
\right\Vert _{\mathfrak{F}}=0.  \label{limit 2}
\end{equation}%
The Riemann integral 
\begin{equation*}
\int_{s}^{\infty }\left( \mathrm{U}_{\infty ,\tau }-\mathrm{U}_{t,\tau
}\right) \mathrm{R}_{\tau }\mathrm{W}_{\tau }\mathrm{R}_{\tau }\mathrm{U}%
_{\tau ,\infty }\varphi \mathrm{d}\tau ,\qquad \varphi \in \mathcal{D}(%
\mathrm{N}^{3/2}),
\end{equation*}%
on the Hilbert space $\mathfrak{F}$ is well-defined, again because of
Theorem \ref{Uinfinitime} and Lemma \ref{TechnicalIntegralWandRWR copy(1)}
for the families $(\mathrm{U}_{\infty ,\tau }-\mathrm{U}_{t,\tau })_{\tau
\in \mathbb{R}_{0}^{+}}$ and $(\mathrm{U}_{\tau ,\infty })_{\tau \in \mathbb{%
R}_{0}^{+}}$. Note meanwhile that, for any $\varphi \in \mathcal{D}(\mathrm{N%
}^{3/2})$ and $\tau ,t\in \mathbb{R}_{0}^{+}$,%
\begin{equation*}
\left\Vert \left( \mathrm{U}_{\infty ,\tau }-\mathrm{U}_{t,\tau }\right) 
\mathrm{R}_{\tau }\mathrm{W}_{\tau }\mathrm{R}_{\tau }\mathrm{U}_{\tau
,\infty }\varphi \right\Vert _{\mathfrak{F}}\leq 2\left\Vert \mathrm{R}%
_{\tau }\mathrm{W}_{\tau }\mathrm{R}_{\tau }\mathrm{U}_{\tau ,\infty
}\varphi \right\Vert _{\mathfrak{F}}
\end{equation*}%
with%
\begin{equation*}
\int_{s}^{\infty }2\left\Vert \mathrm{R}_{\tau }\mathrm{W}_{\tau }\mathrm{R}%
_{\tau }\mathrm{U}_{\tau ,\infty }\varphi \right\Vert _{\mathfrak{F}}\mathrm{%
d}\tau <\infty ,
\end{equation*}%
thanks to Theorems \ref{Uinfinitime}, the triangle inequality and Lemmata %
\ref{diversPreservedom} and \ref{Lemma-self-adjoint1 copy(1)} together with
Condition \ref{Assumptionsasympt} (vi). By Lebesgue's dominated convergence
theorem and (\ref{Uinftystronglimit}), we thus deduce that, for any $\varphi
\in \mathcal{D}(\mathrm{N}^{3/2})$, 
\begin{equation}
\lim_{t\rightarrow \infty }\left\Vert \int_{s}^{\infty }\left( \mathrm{U}%
_{\infty ,\tau }-\mathrm{U}_{t,\tau }\right) \mathrm{R}_{\tau }\mathrm{W}%
_{\tau }\mathrm{R}_{\tau }\mathrm{U}_{\tau ,\infty }\varphi \mathrm{d}\tau
\right\Vert _{\mathfrak{F}}=0.  \label{limit 3}
\end{equation}%
Next, via Lemma \ref{TechnicalIntegralWandRWR copy(1)} and Theorem \ref%
{Uinfinitime} combined with the triangle inequality, we get that, for each $%
\varphi \in \mathcal{D}(\mathrm{N}^{3/2})$, 
\begin{eqnarray*}
&&\left\Vert \int_{s}^{\infty }\mathrm{U}_{t,\tau }\mathrm{R}_{\tau }\mathrm{%
W}_{\tau }\mathrm{R}_{\tau }\left( \mathrm{U}_{\tau ,t}-\mathrm{U}_{\tau
,\infty }\right) \varphi \mathrm{d}\tau \right\Vert _{\mathfrak{F}} \\
&\leq &\lambda ^{-1}\int_{s}^{\infty }\Vert \mathrm{W}_{\tau }(\mathrm{N}%
+1)^{-\frac{1}{2}}\Vert _{\mathrm{op}}\Vert (\mathrm{N}+1)^{\frac{1}{2}}%
\mathrm{R}_{\tau }(\mathrm{N}+1)^{-\frac{3}{2}}\Vert _{\mathrm{op}}\Vert (%
\mathrm{N}+1)^{\frac{3}{2}}\left( \mathrm{U}_{\tau ,t}-\mathrm{U}_{\tau
,\infty }\right) \varphi \Vert _{\mathfrak{F}}\ \mathrm{d}\tau .
\end{eqnarray*}%
Hence, by Lemmata \ref{diversPreservedom} and \ref{Lemma-self-adjoint1
copy(1)}, Theorem \ref{Uinfinitime} and Equation (\ref{limitUpreservedomNbis}%
) combined with Condition \ref{Assumptionsasympt} (vi) and Lebesgue's
dominated convergence theorem, for any $\varphi \in \mathcal{D}(\mathrm{N}%
^{3/2})$, 
\begin{equation}
\lim_{t\rightarrow \infty }\left\Vert \int_{s}^{\infty }\mathrm{U}_{t,\tau }%
\mathrm{R}_{\tau }\mathrm{W}_{\tau }\mathrm{R}_{\tau }\left( \mathrm{U}%
_{\tau ,t}-\mathrm{U}_{\tau ,\infty }\right) \varphi \mathrm{d}\tau
\right\Vert _{\mathfrak{F}}=0.  \label{limit 4}
\end{equation}%
Therefore, by combining (\ref{limit 1})--(\ref{limit 4}), we arrive at the
limit%
\begin{equation}
\lim_{t\rightarrow \infty }\mathbf{X}_{t}\varphi =\mathbf{X}_{\infty
}\varphi ,\qquad \varphi \in \mathcal{D}(\mathrm{N}^{3/2}).
\label{eq:lmiiteauxy}
\end{equation}%
To get the assertion, it suffices now to combine this limit with Corollary %
\ref{Resolvent expression} and Lemma \ref{strongResConv}.
\end{proof}

Note that Remark \ref{Remarkresolveq} still holds in the infinite-time
limit. Namely, if we further assume Condition \ref{AbstractAssumptions} (v),
then we could extend the result first in the weak sense on the domain $%
\mathcal{D}(\mathrm{N})$ and afterwards in the strong sense on the same
domain. However, we refrain from establishing this here, as a stronger
result is proved below. Indeed, we now have the necessary ingredients to
provide an alternative representation of $\mathrm{H}_{\infty }$: Similar to (%
\ref{defFbb}), given $s\in \mathbb{R}_{0}^{+}$ we propose the formal
expression: 
\begin{equation}
\mathbb{F}_{\infty ,s}\doteq \mathrm{U}_{\infty ,s}\mathrm{H}_{s}\mathrm{U}%
_{s,\infty }+\int_{s}^{\infty }\mathrm{U}_{\infty ,\tau }\mathrm{W}_{\tau }%
\mathrm{U}_{\tau ,\infty }\mathrm{d}\tau ,  \label{Defforwardinfty}
\end{equation}%
where $\mathrm{W}_{\tau }$ is the operator defined by (\ref{expression Wt})
via Proposition \ref{Lemma Htbis} (iii). In fact, we can make sense of this
expression at least on the ad-hoc domain 
\begin{equation}
\left( \mathrm{U}_{\infty ,s}\mathcal{D}(\mathrm{H}_{s})\right) \cap 
\mathcal{D}(\mathrm{N}^{3/2}).  \label{domain}
\end{equation}%
Indeed, under Conditions \ref{AbstractAssumptions}--\ref{Assumptionsasympt}
(i)--(iv) and (vi), the first term in the right-hand side of (\ref%
{Defforwardinfty}) clearly makes sense on this domain (cf. Theorems \ref%
{HamilSA} and \ref{propagators}), while the Riemann integral in (\ref%
{Defforwardinfty}) is defined on $\mathcal{D}(\mathrm{N}^{3/2})\subseteq 
\mathfrak{F}$ via Theorem \ref{propagators} and Lemma \ref%
{TechnicalIntegralWandRWR copy(1)}. Indeed, thanks to Condition \ref%
{Assumptionsasympt} (vi) together with Theorem \ref{Uinfinitime}, we can
give the following upper-bound for any $s,t\in \lbrack 0,\infty ]$:%
\begin{equation}
\ln \Vert (\mathrm{N}+\mathbf{1})^{\frac{3}{2}}\mathrm{U}_{t,s}(\mathrm{N}+%
\mathbf{1})^{-\frac{3}{2}}\Vert _{\mathrm{op}}\lesssim _{n}\int_{s\wedge
t}^{s\vee t}\Vert X_{\tau }\Vert _{\mathrm{op},2}\mathrm{d}\tau \leq
f_{1}(\Vert X_{0}\Vert _{\mathrm{op,2}})<\infty .
\label{verifNUNbornunifN32}
\end{equation}%
Moreover, Condition \ref{Assumptionsasympt} (vi) and Lemma \ref%
{Lemma-self-adjoint1 copy(1)} ensure that the real-valued function 
\begin{equation*}
t\mapsto \Vert \mathrm{W}_{t}(\mathrm{N}+\mathbf{1})^{-\frac{3}{2}}\Vert _{%
\mathrm{op}},
\end{equation*}%
decays sufficiently fast as $t\rightarrow \infty $.

However, it is not immediately clear whether the subspace defined in (\ref%
{domain}) is dense, or even non-trivial. Ideally, we want the operator $%
\mathbb{F}_{\infty ,s}$ to be symmetric and (densely) defined on $\mathcal{D}%
_{\infty }$ for all $s\in \mathbb{R}_{0}^{+}$, which would in particular
ensure its closability. Indeed, our goal is to extend Theorem \ref{Gluing}
to the infinite-time case $t=\infty $, thereby establishing that $\mathbb{F}%
_{\infty ,s}=\mathrm{H}_{\infty }$, which is a self-adjoint operator with
core $\mathcal{D}_{\infty }$.

To achieve this, rather than utilizing the operators $\mathrm{Bwd}%
_{t,s}^{(\lambda )}$ and $\mathbb{B}_{t,s}$ (see, e.g., Proposition \ref%
{ConvergenceYosida}), we directly study the limit $t\rightarrow \infty $ of $%
\mathrm{U}_{s,t}\varphi $ for elements $\varphi \in \mathcal{D}_{\infty }$,
which are known to belong to $\mathcal{D}(\mathrm{H}_{s})$ for all $s,t\in 
\mathbb{R}_{0}^{+}$. Proving that $\mathrm{U}_{s,t}\varphi $ does not leave $%
\mathcal{D}(\mathrm{H}_{s})$ as $t\rightarrow \infty $ enables us to extend
finite-time domain inclusions -- specifically $\mathrm{U}_{s,t}\mathcal{D}%
_{\infty }\subseteq \mathcal{D}(\mathrm{H}_{s})$ -- to infinite time $%
t=\infty $. In doing so, we establish that (\ref{Defforwardinfty}) yields a
well-defined symmetric operator $\mathbb{F}_{\infty ,s}$ on the core $%
\mathcal{D}_{\infty }$ of $\mathrm{H}_{\infty }$.

\begin{lemma}[Asymptotic graph norm convergence of $\mathrm{U}_{s,t}\protect%
\varphi $ under $\mathrm{H}_{s}$]
\label{UinftydomH}\mbox{}\newline
Under Conditions \ref{AbstractAssumptions}--\ref{Assumptionsasympt}, for any 
$s\in \mathbb{R}_{0}^{+}$, 
\begin{equation*}
\mathrm{U}_{s,\infty }\mathcal{D}_{\infty }\subseteq \mathcal{D}\left( 
\mathrm{H}_{s}\right) =\mathcal{D}(\mathrm{H}_{0})\qquad \text{and}\qquad
\lim_{t\rightarrow \infty }\mathrm{H}_{s}\mathrm{U}_{s,t}\varphi =\mathrm{H}%
_{s}\mathrm{U}_{s,\infty }\varphi ,\qquad \varphi \in \mathcal{D}_{\infty }.
\end{equation*}
\end{lemma}

\begin{proof}
For any $s,t\in \mathbb{R}_{0}^{+}$, note that $\mathrm{U}_{s,t}\mathcal{D}%
_{\infty }\subseteq \mathcal{D}\left( \mathrm{H}_{s}\right) =\mathcal{D}(%
\mathrm{H}_{0})$, by Proposition \ref{ConvergenceYosida} and Lemma \ref%
{DomHconst}. In particular, $\mathrm{H}_{s}\mathrm{U}_{s,t}\varphi $ is well
defined for any $s,t\in \mathbb{R}_{0}^{+}$ and $\varphi \in \mathcal{D}%
_{\infty }$. From now on, set $\varphi \in \mathcal{D}_{\infty }$ and $s\in 
\mathbb{R}_{0}^{+}$. By Theorem \ref{Uinfinitime} and its proof (cf. (\ref%
{Uinftystronglimit})), $\left( \mathrm{U}_{s,t}\varphi \right) _{t\in 
\mathbb{R}_{0}^{+}}$ is a Cauchy net in $\mathfrak{F}$. Using the unitarity
of $\mathrm{U}_{t,s}$ and the cocycle property, as given by Theorem \ref%
{propagators}, we observe that, for any $t_{1},t_{2}\in \mathbb{R}_{0}^{+}$
satisfying $t_{2}\geq t_{1}$, 
\begin{equation}
\left\Vert \left( \mathrm{H}_{s}\mathrm{U}_{s,t_{1}}-\mathrm{H}_{s}\mathrm{U}%
_{s,t_{2}}\right) \varphi \right\Vert _{\mathfrak{F}}=\left\Vert \left( 
\mathrm{U}_{t_{2},t_{1}}\mathrm{U}_{t_{1},s}\mathrm{H}_{s}\mathrm{U}%
_{s,t_{1}}-\mathrm{U}_{t_{2},s}\mathrm{H}_{s}\mathrm{U}_{s,t_{2}}\right)
\varphi \right\Vert _{\mathfrak{F}},  \label{estimcauchyinit}
\end{equation}%
which, combined with Theorem \ref{Gluing}, leads to the upper bound 
\begin{eqnarray}
\left\Vert \left( \mathrm{H}_{s}\mathrm{U}_{s,t_{1}}-\mathrm{H}_{s}\mathrm{U}%
_{s,t_{2}}\right) \varphi \right\Vert _{\mathfrak{F}} &\leq &\left\Vert
\left( \mathrm{U}_{t_{2},t_{1}}-\mathbf{1}\right) \mathrm{H}_{t_{2}}\varphi
\right\Vert _{\mathfrak{F}}+\left\Vert \mathrm{U}_{t_{2},t_{1}}\left( 
\mathrm{H}_{t_{1}}-\mathrm{H}_{t_{2}}\right) \varphi \right\Vert _{\mathfrak{%
F}}  \label{estimecauchybreak1} \\
&&+\left\Vert \mathrm{U}_{t_{2},t_{1}}\int_{s}^{t_{1}}\mathrm{U}_{t_{1},\tau
}\mathrm{W}_{\tau }\mathrm{U}_{\tau ,t_{1}}\varphi \mathrm{d}\tau
-\int_{s}^{t_{2}}\mathrm{U}_{t_{2},\tau }\mathrm{W}_{\tau }\mathrm{U}_{\tau
,t_{2}}\varphi \mathrm{d}\tau \right\Vert _{\mathfrak{F}}.  \notag
\end{eqnarray}%
By Lemma \ref{strongConv} and Theorem \ref{Uinfinitime} (see (\ref{flow
equation convergence unitary 1bis})), 
\begin{equation}
\lim_{t_{1}\rightarrow \infty }\lim_{t_{2}\rightarrow \infty }\left\Vert
\left( \mathrm{U}_{t_{2},t_{1}}-\mathbf{1}\right) \mathrm{H}_{t_{2}}\varphi
\right\Vert _{\mathfrak{F}}=\lim_{t_{1}\rightarrow \infty
}\lim_{t_{2}\rightarrow \infty }\left\Vert \mathrm{U}_{t_{2},t_{1}}\left( 
\mathrm{H}_{t_{1}}-\mathrm{H}_{t_{2}}\right) \varphi \right\Vert _{\mathfrak{%
F}}=0.  \label{limit F1}
\end{equation}%
Now observe that, using Theorem \ref{propagators} and Lemmata \ref%
{TechnicalIntegralWandRWR}--\ref{TechnicalIntegralWandRWR copy(1)}, the
Riemann integrals in the right-hand side of (\ref{estimecauchybreak1}) are
well-defined on $\mathcal{D}(\mathrm{N}^{3/2})\supseteq \mathcal{D}_{\infty
} $ for all intervals $[s,t_{1}]$ and $[s,t_{2}]$, even if one takes $%
t_{1}=t_{2}=\infty $, because of Condition \ref{Assumptionsasympt} (vi). In
particular, by the triangle inequality, the unitarity of $\mathrm{U}_{t,s}$
and the cocycle property, for any $t_{1},t_{2}\in \mathbb{R}_{0}^{+}$
satisfying $t_{2}\geq t_{1}$, 
\begin{eqnarray}
&&\left\Vert \mathrm{U}_{t_{2},t_{1}}\int_{s}^{t_{1}}\mathrm{U}_{t_{1},\tau }%
\mathrm{W}_{\tau }\mathrm{U}_{\tau ,t_{1}}\varphi \mathrm{d}\tau
-\int_{s}^{t_{2}}\mathrm{U}_{t_{2},\tau }\mathrm{W}_{\tau }\mathrm{U}_{\tau
,t_{2}}\varphi \mathrm{d}\tau \right\Vert _{\mathfrak{F}}  \notag \\
&\leq &\int_{s}^{t_{1}}\left\Vert \mathrm{W}_{\tau }\left( \mathrm{U}_{\tau
,t_{1}}-\mathrm{U}_{\tau ,t_{2}}\right) \varphi \right\Vert _{\mathfrak{F}}%
\mathrm{d}\tau +\int_{t_{1}}^{\infty }\left\Vert \mathrm{W}_{\tau }\mathrm{U}%
_{\tau ,t_{2}}\varphi \right\Vert _{\mathfrak{F}}\mathrm{d}\tau .
\label{eq:Grosseestimate}
\end{eqnarray}%
Combining Theorem \ref{Uinfinitime} with Lemma \ref{Lemma-self-adjoint1
copy(1)} and Condition \ref{Assumptionsasympt} (vi), we get that, for any $%
\tau \in \mathbb{R}_{0}^{+}$,%
\begin{equation}
\sup_{t_{1},t_{2}\in \lbrack 0,\infty ]}\max \left\{ \Vert \mathrm{W}_{\tau
}\left( \mathrm{U}_{\tau ,t_{2}}-\mathrm{U}_{\tau ,t_{1}}\right) \varphi
\Vert _{\mathfrak{F}},\Vert \mathrm{W}_{\tau }\mathrm{U}_{\tau
,t_{2}}\varphi \Vert _{\mathfrak{F}}\right\} \lesssim \Vert \left[ X_{\tau
},\mu _{\tau }\right] \Vert _{\mathrm{op,2}}\Vert (\mathrm{N}+\mathbf{1})^{%
\frac{3}{2}}\varphi \Vert _{\mathfrak{F}}.  \label{EstimDomconv4}
\end{equation}%
As a consequence, by Condition \ref{Assumptionsasympt} (vi), 
\begin{equation}
\lim_{t_{1}\rightarrow \infty }\sup_{t_{2}\in \lbrack 0,\infty
]}\int_{t_{1}}^{\infty }\left\Vert \mathrm{W}_{\tau }\mathrm{U}_{\tau
,t_{2}}\varphi \right\Vert _{\mathfrak{F}}\mathrm{d}\tau =0.
\label{limit F2}
\end{equation}%
Furthermore, using Equations (\ref{limitUpreservedomNbis})--(\ref{assertion
ivassertion iv}) and (\ref{EstimDomconv4}) together with Lemma \ref%
{Lemma-self-adjoint1 copy(1)}, Condition \ref{Assumptionsasympt} (vi) and
Lebesgue's dominated convergence theorem, we find that 
\begin{equation}
\lim_{t_{1}\rightarrow \infty }\lim_{t_{2}\rightarrow \infty
}\int_{s}^{t_{1}}\left\Vert \mathrm{W}_{\tau }\left( \mathrm{U}_{\tau
,t_{2}}-\mathrm{U}_{\tau ,t_{1}}\right) \varphi \right\Vert _{\mathfrak{F}}%
\mathrm{d}\tau =\lim_{t_{1}\rightarrow \infty }\int_{s}^{t_{1}}\left\Vert 
\mathrm{W}_{\tau }\left( \mathrm{U}_{\tau ,\infty }-\mathrm{U}_{\tau
,t_{1}}\right) \varphi \right\Vert _{\mathfrak{F}}\mathrm{d}\tau =0.
\label{limit F5}
\end{equation}%
Finally, combining (\ref{estimecauchybreak1})--(\ref{eq:Grosseestimate}) and
(\ref{limit F2})--(\ref{limit F5}), we arrive at the identity 
\begin{equation}
\lim_{t_{1}\rightarrow \infty }\lim_{t_{2}\rightarrow \infty }\left\Vert
\left( \mathrm{H}_{s}\mathrm{U}_{s,t_{1}}-\mathrm{H}_{s}\mathrm{U}%
_{s,t_{2}}\right) \varphi \right\Vert _{\mathfrak{F}}=0.  \label{limit F6}
\end{equation}%
Therefore, for any $s\in \mathbb{R}_{0}^{+}$, the net $\left( \mathrm{U}%
_{t,s}\varphi \right) _{t\in \mathbb{R}_{0}^{+}}$ is Cauchy with respect to
the graph norm of the closed operator $\mathrm{H}_{s}$. Since the graph of $%
\mathrm{H}_{s}$ is closed in $\mathfrak{F}\times \mathfrak{F}$, this net
converges within the graph. In particular, by (\ref{Uinftystronglimit}), it
follows that $\mathrm{U}_{s,\infty }\varphi \in \mathcal{D}\left( \mathrm{H}%
_{s}\right) =\mathcal{D}\left( \mathrm{H}_{0}\right) $ (see Lemma \ref%
{DomHconst}).
\end{proof}

We are now in a position to establish the main result of this subsection.
Indeed, having the operator $\mathbb{F}_{\infty ,s}$ now well-defined by (%
\ref{Defforwardinfty}) on the domain $\mathcal{D}_{\infty }$ for any $s\in 
\mathbb{R}_{0}^{+}$ (Lemma \ref{UinftydomH}), and noting that $\mathcal{D}%
_{\infty }$ is a core for $\mathrm{H}_{\infty }$, it follows that $\mathbb{F}%
_{\infty ,s}$ coincides with $\mathrm{H}_{\infty }$ for all $s\in \mathbb{R}%
_{0}^{+}$:

\begin{theorem}[Unitary Transformation at Infinite times]
\label{Hinftyunitequiv}\mbox{}\newline
Under Conditions \ref{AbstractAssumptions}--\ref{Assumptionsasympt}, 
\begin{equation*}
\mathrm{H}_{\infty }=\mathrm{U}_{\infty ,s}\mathrm{H}_{s}\mathrm{U}%
_{s,\infty }+\int_{s}^{\infty }\mathrm{U}_{\infty ,\tau }\mathrm{W}_{\tau }%
\mathrm{U}_{\tau ,\infty }\mathrm{d}\tau ,\qquad s\in \mathbb{R}_{0}^{+}.
\end{equation*}%
If $x=\Vert X_{0}\Vert _{\mathrm{op},2}$ and $[\mu _{k,0},X_{k,t}]=[\mu
_{k,0},X_{k,t}^{\ast }]=0$ for any $t\in \mathbb{R}$ and $k\in \mathbb{N}$
such that $X_{k,0}\neq 0$, then 
\begin{equation*}
\left\Vert \left( \mathrm{H}_{\infty }-\mathrm{U}_{\infty ,0}\mathrm{H}_{0}%
\mathrm{U}_{0,\infty }\right) \left( \mathrm{N}+\mathbf{1}\right) ^{-\frac{3%
}{2}}\right\Vert _{\mathrm{op}}\lesssim f_{1}\left( x\right) f_{2}\left(
x\right) \mathrm{e}^{Df_{1}\left( x\right) },
\end{equation*}%
for some positive constant $D\in \mathbb{R}^{+}$, where $f_{1}$ and $f_{2}$
are the functions defined by Condition \ref{Assumptionsasympt} (vi).
\end{theorem}

\begin{proof}
By Lemmata \ref{Lemma-self-adjoint1 copy(1)} and \ref{UinftydomH} together
with Theorem \ref{Uinfinitime}, $\mathbb{F}_{\infty ,s}$ is well-defined on $%
\mathcal{D}_{\infty }$ and, by Theorem \ref{Gluing} together with the
triangle inequality, we compute that, for any $s,t\in \mathbb{R}_{0}^{+}$
and $\varphi \in \mathcal{D}_{\infty }$, 
\begin{equation}
\left\Vert \left( \mathrm{H}_{\infty }-\mathbb{F}_{\infty ,s}\right) \varphi
\right\Vert _{\mathfrak{F}}\leq \left\Vert \left( \mathrm{H}_{\infty }-%
\mathrm{H}_{t}\right) \varphi \right\Vert _{\mathfrak{F}}+\left\Vert \left( 
\mathbb{F}_{\infty ,s}-\mathbb{F}_{t,s}\right) \varphi \right\Vert _{%
\mathfrak{F}}.  \label{eq:Masterestimate}
\end{equation}%
Lemma \ref{strongConv} together with the inclusion $\mathcal{D}_{\infty
}\subseteq \mathcal{D}(\mathrm{N})$ yield 
\begin{equation}
\lim_{t\rightarrow \infty }\left\Vert \left( \mathrm{H}_{t}-\mathrm{H}%
_{\infty }\right) \varphi \right\Vert _{\mathfrak{F}}=0,\qquad \varphi \in 
\mathcal{D}_{\infty }.  \label{eq:Masterestimate1}
\end{equation}%
In addition, similar to (\ref{eq:Grosseestimate}), by the triangle
inequality,%
\begin{eqnarray}
\left\Vert \left( \mathbb{F}_{\infty ,s}-\mathbb{F}_{t,s}\right) \varphi
\right\Vert _{\mathfrak{F}} &\leq &\left\Vert \left( \mathrm{U}_{\infty ,s}-%
\mathrm{U}_{t,s}\right) \mathrm{H}_{s}\mathrm{U}_{s,\infty }\varphi
\right\Vert _{\mathfrak{F}}+\left\Vert \mathrm{H}_{s}\left( \mathrm{U}%
_{s,\infty }-\mathrm{U}_{s,t}\right) \varphi \right\Vert _{\mathfrak{F}} 
\notag \\
&&+\int_{t}^{\infty }\left\Vert \mathrm{W}_{\tau }\mathrm{U}_{\tau ,\infty
}\varphi \right\Vert _{\mathfrak{F}}\mathrm{d}\tau +\int_{s}^{\infty
}\left\Vert \mathrm{W}_{\tau }\left( \mathrm{U}_{\tau ,\infty }-\mathrm{U}%
_{\tau ,t}\right) \varphi \right\Vert _{\mathfrak{F}}\mathrm{d}\tau  \notag
\\
&&\int_{s}^{\infty }\left\Vert \left( \mathrm{U}_{\infty ,\tau }-\mathrm{U}%
_{t,\tau }\right) \mathrm{W}_{\tau }\mathrm{U}_{\tau ,\infty }\varphi
\right\Vert _{\mathfrak{F}}\mathrm{d}\tau  \label{ssdsdsd}
\end{eqnarray}%
for any $\varphi \in \mathcal{D}_{\infty }$ and $s,t\in \mathbb{R}_{0}^{+}$.
Through the limits of (\ref{Uinftystronglimit}) and (\ref{limit F2})--(\ref%
{limit F5}) together with Lemma \ref{UinftydomH}, we deduce from (\ref%
{ssdsdsd}) that, for any $\varphi \in \mathcal{D}_{\infty }$ and $s\in 
\mathbb{R}_{0}^{+}$,%
\begin{equation}
\lim_{t\rightarrow \infty }\left\Vert \left( \mathbb{F}_{\infty ,s}-\mathbb{F%
}_{t,s}\right) \varphi \right\Vert _{\mathfrak{F}}\leq \lim_{t\rightarrow
\infty }\int_{s}^{\infty }\left\Vert \left( \mathrm{U}_{\infty ,\tau }-%
\mathrm{U}_{t,\tau }\right) \mathrm{W}_{\tau }\mathrm{U}_{\tau ,\infty
}\varphi \right\Vert _{\mathfrak{F}}\mathrm{d}\tau .  \notag
\end{equation}%
This upper bound vanishes because of the limit (\ref{Uinftystronglimit}),
Equation (\ref{EstimDomconv4}), Condition \ref{Assumptionsasympt} (vi) and
Lebesgue's dominated convergence theorem. It follows that, for any $\varphi
\in \mathcal{D}_{\infty }$ and $s\in \mathbb{R}_{0}^{+}$, 
\begin{equation}
\lim_{t\rightarrow \infty }\left\Vert \left( \mathbb{F}_{\infty ,s}-\mathbb{F%
}_{t,s}\right) \varphi \right\Vert _{\mathfrak{F}}=0.
\label{eq:Masterestimate2}
\end{equation}%
Inequality (\ref{eq:Masterestimate}) together with the limits (\ref%
{eq:Masterestimate1}) and (\ref{eq:Masterestimate2}) then yields 
\begin{equation}
\mathrm{H}_{\infty }\varphi =\mathbb{F}_{\infty ,s}\varphi ,\qquad \varphi
\in \mathcal{D}_{\infty },\ s\in \mathbb{R}_{0}^{+}.
\label{eq:Presqueresult}
\end{equation}%
In other words, the restriction of $\mathbb{F}_{\infty ,s}$ to $\mathcal{D}%
_{\infty }$ coincides exactly with that of $\mathrm{H}_{\infty }$. Since $%
\mathcal{D}_{\infty }$ is a core for $\mathrm{H}_{\infty }$, the closure of $%
\mathbb{F}_{\infty ,s}|_{\mathcal{D}_{\infty }}$ is precisely given by $%
\mathrm{H}_{\infty }$. Consequently, (\ref{eq:Presqueresult}) implies that 
\begin{equation}
\mathrm{H}_{\infty }=\mathbb{F}_{\infty ,s},\qquad s\in \mathbb{R}_{0}^{+},
\label{first assertion}
\end{equation}%
on the domain $\mathcal{D}\left( \mathrm{H}_{\infty }\right) =\mathcal{D}(%
\mathrm{H}_{0})$ (see Lemma \ref{DomHconst}). By (\ref{Defforwardinfty}),
this last equality is nothing else than the first assertion of the theorem.

Now, applying Equation (\ref{first assertion}) at $s=0$ together with the
triangle inequality, Lemma \ref{Lemma-self-adjoint1 copy(1)}, Theorem \ref%
{Uinfinitime} (iv) and Condition \ref{Assumptionsasympt} (vi), we obtain the
following upper bound:%
\begin{equation}
\left\Vert \left( \mathrm{H}_{\infty }-\mathrm{U}_{\infty ,0}\mathrm{H}_{0}%
\mathrm{U}_{0,\infty }\right) \left( \mathrm{N}+\mathbf{1}\right) ^{-\frac{3%
}{2}}\right\Vert _{\mathrm{op}}\lesssim \mathrm{exp}\left( D\int_{0}^{\infty
}\left\Vert X_{\tau }\right\Vert _{\mathrm{op},2}\mathrm{d}\tau \right)
\int_{0}^{\infty }\left\Vert \left[ \mu _{\tau },X_{\tau }\right]
\right\Vert _{\mathrm{op},2}\mathrm{d}\tau  \label{majdiffdiago1}
\end{equation}%
for some positive constant $D\in \mathbb{R}^{+}$. If $[\mu
_{k,0},X_{k,t}]=[\mu _{k,0},X_{k,t}^{\ast }]=0$ for any $k\in \mathbb{N}$
such that $X_{k,0}\neq 0$, then, bearing in mind Condition \ref%
{AbstractAssumptions} (i) and using the sub-multiplicativity of the norm $%
\Vert \cdot \Vert _{\mathrm{op},2}$ and the triangle inequality, we get the
bound 
\begin{equation}
\left\Vert \left[ \mu _{t},X_{t}\right] \right\Vert _{\mathrm{op},2}\leq
8\left\Vert X_{t}\right\Vert _{\mathrm{op},2}\int_{0}^{\infty }\left\Vert
X_{\tau }\right\Vert _{\mathrm{op},2}^{2}\mathrm{d}\tau
\label{majdiffdiago2}
\end{equation}%
for all $t\in \mathbb{R}_{0}^{+}$. Therefore, the last inequality of the
theorem follows by combining (\ref{majdiffdiago1})--(\ref{majdiffdiago2})
with Condition \ref{Assumptionsasympt} (vi).
\end{proof}

\subsection{Displacement Transformation}

We conclude our technical analysis with a closer look at the family $(%
\mathrm{U}_{t,s})_{s,t\in \lbrack 0,\infty ]}$ of unitary transformations
that generates the Hamiltonian flow $(\mathrm{H}_{t})_{t\in \lbrack 0,\infty
]}$. See Theorems \ref{propagators} and \ref{Uinfinitime} for the unitary
transformation as well as Theorems \ref{Gluing} and \ref{Hinftyunitequiv}
for the Hamiltonian flow. More specifically, we show that the unitary
transformations acting on $\mathfrak{F}\doteq \mathfrak{H}\otimes \mathcal{F}%
_{+}$ do \emph{not} actually implement a displacement transformation of the
form 
\begin{equation*}
\mathbf{1}\otimes a_{k}\mapsto \mathbf{1}\otimes a_{k}+g_{k}\otimes \mathbf{1%
}_{\mathcal{F}_{+}},\qquad k\in \mathbb{N},
\end{equation*}%
on the bosonic annihilation operators, for a given family $(g_{k})_{k\in 
\mathbb{N}}\subseteq \mathcal{B}(\mathfrak{H})$ of operators. In general,
such a property holds only approximately:

\begin{proposition}[Unitary transformation of annihilation operators at
finite time]
\label{TransfoDisp}\mbox{}\newline
Assume Condition \ref{AbstractAssumptions}. For each $k\in \mathbb{N}$ and $%
t\in \mathbb{R}_{0}^{+}$, there exists a closed symmetric operator $\mathcal{%
R}_{k,t}$ on $\mathcal{D}(\mathrm{N}^{1/2})$ such that the identity 
\begin{equation*}
\mathrm{U}_{t,0}\left( \mathbf{1}\otimes a_{k}\right) \mathrm{U}_{0,t}=%
\mathbf{1}\otimes a_{k}+\mathbf{g}_{k,t}\otimes \mathbf{1}_{\mathcal{F}_{+}}+%
\mathcal{R}_{k,t}
\end{equation*}%
holds on $\mathcal{D}(\mathrm{N}^{1/2})$. Here, the displacement coefficient
is given by%
\begin{equation}
\mathbf{g}_{k,t}\doteq \left( \delta _{X_{k,0},0}-1\right)
\int_{0}^{t}X_{k,\tau }^{\ast }\mathrm{d}\tau \in \mathcal{B}(\mathfrak{h}%
_{k})\equiv \mathcal{B}(\mathfrak{h}_{k})^{\otimes },  \label{displacement1}
\end{equation}%
where $\delta _{i,j}$ denotes the Kronecker delta, and the remainder
satisfies the bound 
\begin{equation}
\left\Vert \mathcal{R}_{k,t}\left( \mathrm{N}+\mathbf{1}\right) ^{-\frac{1}{2%
}}\right\Vert _{\mathrm{op}}\leq \left( \mathrm{e}^{D\int_{0}^{t}\left\Vert
X_{\tau }\right\Vert _{\mathrm{op},2}\mathrm{d}\tau }-1\right)
\int_{0}^{t}\left\Vert X_{k,\tau }\right\Vert _{\mathrm{op}}\mathrm{d}\tau ,
\label{displacement1bis}
\end{equation}%
for a time-independent constant $D\in \mathbb{R}^{+}$.
\end{proposition}

\begin{proof}
Recall that we adopt the shorthand $\mathbf{1}\equiv \mathbf{1}_{\mathcal{F}%
_{+}}\equiv \mathbf{1}_{\mathfrak{F}}$ and $\mathbf{1}\otimes a_{k}\equiv
a_{k}$. Fix once and for all $k\in \mathbb{N}$. Note also that $X_{k,0}=0$
yields $X_{k,t}=0$ for all $t\in \mathbb{R}_{0}^{+}$, by Condition \ref%
{AbstractAssumptions} (i). Define the operator 
\begin{equation}
\Tilde{a}_{k,t}\doteq a_{k}+\mathbf{g}_{k,t}\otimes \mathbf{1},\qquad t\in 
\mathbb{R}_{0}^{+},  \label{defatilde}
\end{equation}%
$\mathbf{g}_{k,t}\in \mathcal{B}(\mathfrak{H})$ being defined by (\ref%
{displacement1}), under Condition \ref{AbstractAssumptions}. The operators $%
\Tilde{a}_{k,t}$ are well-defined on $\mathcal{D}(\mathrm{N}^{1/2})$ for all 
$t\in \mathbb{R}_{0}^{+}$, because 
\begin{equation*}
\left\Vert \Tilde{a}_{k,t}(\mathrm{N}+\mathbf{1})^{-\frac{1}{2}}\right\Vert
_{\mathrm{op}}\leq 1+\int_{0}^{t}\left\Vert X_{k,\tau }\right\Vert _{\mathrm{%
op}}\mathrm{d}\tau <\infty \ ,\qquad t\in \mathbb{R}_{0}^{+},
\end{equation*}%
thanks to Condition \ref{AbstractAssumptions} (i) and the triangle
inequality. If $X_{k,0}\neq 0$ then, for any $t\in \mathbb{R}_{0}^{+}$ and
all $\epsilon \in \lbrack -t,\infty )$, 
\begin{equation*}
\epsilon ^{-1}\left( \Tilde{a}_{k,t+\epsilon }-\Tilde{a}_{k,t}\right)
+X_{k,t}^{\ast }\otimes \mathbf{1}_{\mathcal{F}_{+}}=\left( -\epsilon
^{-1}\int_{t}^{t+\epsilon }X_{k,\tau }^{\ast }\mathrm{d}\tau +X_{k,t}^{\ast
}\right) \otimes \mathbf{1}.
\end{equation*}%
(The difference $\Tilde{a}_{k,t+\epsilon }-\Tilde{a}_{k,t}$ clearly extends
to a bounded operator on $\mathfrak{F}$.) By the Lebesgue differentiation
theorem used on the above Riemann integral and Condition \ref%
{AbstractAssumptions} (i), one thus checks that 
\begin{equation}
\partial _{t}\Tilde{a}_{k,t}\doteq \lim_{\epsilon \rightarrow 0}\epsilon
^{-1}\left( \Tilde{a}_{k,t+\epsilon }-\Tilde{a}_{k,t}\right) =\left( \delta
_{X_{k,0},0}-1\right) \left( X_{k,t}^{\ast }\otimes \mathbf{1}\right)
,\qquad t\in \mathbb{R}_{0}^{+},  \label{sdsdsdsdssdsdsdsds}
\end{equation}%
where the limit is defined via the norm topology of $\mathcal{B}(\mathfrak{F}%
)$. Observe from the CCR (\ref{CCR}) and (\ref{def_Gt}) that, for any $t\in 
\mathbb{R}_{0}^{+}$, 
\begin{equation*}
\left[ \mathrm{N},\Tilde{a}_{k,t}\right] =\left[ \mathrm{N},a_{k,t}\right] =%
\left[ a_{k}^{\ast },a_{k}\right] a_{k}=-a_{k}
\end{equation*}%
and 
\begin{equation*}
i\left[ \mathrm{G}_{t},\Tilde{a}_{k,t}\right] =-X_{k,t}^{\ast }\otimes \left[
a_{k}^{\ast },a_{k}\right] +i\left[ \mathrm{G}_{t},\mathbf{g}_{k,t}\otimes 
\mathbf{1}\right] =X_{k,t}^{\ast }\otimes \mathbf{1}+i\left[ \mathrm{G}_{t},%
\mathbf{g}_{k,t}\otimes \mathbf{1}\right] .
\end{equation*}%
In particular, 
\begin{equation}
\left\Vert \left( \mathrm{N}+\mathbf{1}\right) \Tilde{a}_{k,t}\left( \mathrm{%
N}+\mathbf{1}\right) ^{-2}\right\Vert _{\mathrm{op}}\leq \left\Vert \Tilde{a}%
_{k,t}\left( \mathrm{N}+\mathbf{1}\right) ^{-1}\right\Vert _{\mathrm{op}%
}+\left\Vert \left[ \mathrm{N},\Tilde{a}_{k,t}\right] \left( \mathrm{N}+%
\mathbf{1}\right) ^{-2}\right\Vert _{\mathrm{op}}<\infty .
\label{atildepreserveN2}
\end{equation}%
Therefore, using (\ref{sdsdsdsdssdsdsdsds})--(\ref{atildepreserveN2}) and
Theorem \ref{propagators} we compute that, for any $t\in \mathbb{R}_{0}^{+}$%
, 
\begin{equation}
\partial _{t}\left( \mathrm{U}_{0,t}\Tilde{a}_{k,t}\mathrm{U}_{t,0}\varphi
\right) =\mathrm{U}_{0,t}\left( i\left[ \mathrm{G}_{t},\mathbf{g}%
_{k,t}\otimes \mathbf{1}\right] \right) \mathrm{U}_{t,0}\varphi ,\qquad
\varphi \in \mathcal{D}(\mathrm{N}^{2}).  \label{deriveUaU}
\end{equation}%
By integration, we then verify that, for any $t\in \mathbb{R}_{0}^{+}$, 
\begin{equation}
\mathrm{U}_{0,t}\Tilde{a}_{k,t}\mathrm{U}_{t,0}\varphi =\Tilde{a}%
_{k,t}\varphi +\int_{0}^{t}\mathrm{U}_{0,\tau }i\left[ \mathrm{G}_{\tau },%
\mathbf{g}_{k,\tau }\otimes \mathbf{1}\right] \mathrm{U}_{\tau ,0}\varphi 
\mathrm{d}\tau ,\qquad \varphi \in \mathcal{D}(\mathrm{N}^{2}),
\label{sdsssdsd}
\end{equation}%
where the last integral is understood as a Riemann integral on $\mathfrak{F}$%
. Define the operators 
\begin{equation}
\mathcal{R}_{k,t}\doteq \int_{0}^{t}\mathrm{U}_{0,\tau }i\left[ \mathrm{G}%
_{\tau },\mathbf{g}_{k,\tau }\otimes \mathbf{1}\right] \mathrm{U}_{\tau ,0}%
\mathrm{d}\tau ,\qquad t\in \mathbb{R}_{0}^{+},  \label{jk0}
\end{equation}%
on the domain $\mathcal{D}(\mathrm{N}^{2})$, via Theorem \ref{propagators}
and Lemma \ref{TechnicalIntegralWandRWR}. They are indeed well-defined on $%
\mathcal{D}(\mathrm{N}^{2})\subseteq \mathcal{D}(\mathrm{N}^{1/2})$ because,
for some sufficiently large constant $D\in \mathbb{R}^{+}$, 
\begin{eqnarray}
\int_{0}^{t}\Vert \mathrm{U}_{0,\tau }i[\mathrm{G}_{\tau },\mathbf{g}%
_{k,\tau }\otimes \mathbf{1}]\mathrm{U}_{\tau ,0}(\mathrm{N}+\mathbf{1})^{-%
\frac{1}{2}}\Vert _{\mathrm{op}}\mathrm{d}\tau &\leq &\int_{0}^{t}\Vert
X_{k,\tau }\Vert _{\mathrm{op}}\mathrm{d}\tau \int_{0}^{t}D\Vert X_{\tau
}\Vert _{\mathrm{op},2}\mathrm{e}^{D\int_{0}^{\tau }\Vert X_{s}\Vert _{%
\mathrm{op},2}\mathrm{d}s}\mathrm{d}\tau  \notag \\
&=&\left( \mathrm{e}^{D\int_{0}^{t}\Vert X_{\tau }\Vert _{\mathrm{op},2}%
\mathrm{d}\tau }-1\right) \int_{0}^{t}\Vert X_{k,\tau }\Vert _{\mathrm{op}}\ 
\mathrm{d}\tau ,  \label{dfdfdfdfdfdf}
\end{eqnarray}%
thanks to Theorem \ref{propagators}. Being symmetric and densely defined,
the operator is closable, and we identify $\mathcal{R}_{k,t}$ with an
arbitrary choice among its closed extensions. E.g., $\mathcal{R}_{k,t}\equiv 
\mathcal{R}_{k,t}^{\ast \ast }$. In any case, $\mathcal{D}(\mathrm{N}%
^{1/2})\subseteq \mathcal{D}(\mathcal{R}_{k,t})$, because of (\ref%
{dfdfdfdfdfdf}). $\mathcal{D}(\mathrm{N}^{2})$ is a common core for the
closed operators $\tilde{a}_{t}$, $\mathcal{R}_{k,t}$ and $\mathrm{U}_{0,t}%
\Tilde{a}_{k,t}\mathrm{U}_{t,0}$. The last observation can be deduced from
the unitarity of $\mathrm{U}_{t,s}$ for $s,t\in \mathbb{R}_{0}^{+}$, and
Theorem \ref{propagators} (iv) for $n=4$ together with the fact that $%
\mathcal{D}(\mathrm{N}^{2})$ is a core for $\tilde{a}_{t}$. Using standard
continuity arguments we can extend (\ref{sdsssdsd}) to all $\varphi \in 
\mathcal{D}(\mathrm{N}^{1/2})$. The bound given in (\ref{displacement1})
results from (\ref{dfdfdfdfdfdf}). We omit the details.
\end{proof}

\begin{proposition}[Unitary transformation of annihilation operators at
infinite time]
\label{TransfoDisplimit}\mbox{}\newline
Assume Condition \ref{AbstractAssumptions}--\ref{Assumptionsasympt}. For
each $k\in \mathbb{N}$ and $t\in \mathbb{R}_{0}^{+}$, there exists a closed
symmetric operator $\mathcal{R}_{k,t}$ on $\mathcal{D}(\mathrm{N}^{1/2})$
such that the identity 
\begin{equation*}
\mathrm{U}_{\infty ,0}\left( \mathbf{1}\otimes a_{k}\right) \mathrm{U}%
_{0,\infty }=\mathbf{1}\otimes a_{k}+\mathbf{g}_{k,\infty }\otimes \mathbf{1}%
_{\mathcal{F}_{+}}+\mathcal{R}_{k,\infty }
\end{equation*}%
holds on $\mathcal{D}(\mathrm{N}^{1/2})$. Here, the displacement coefficient
is given by%
\begin{equation*}
\mathbf{g}_{k,\infty }\doteq \left( \delta _{X_{k,0},0}-1\right)
\int_{0}^{\infty }X_{k,\tau }^{\ast }\mathrm{d}\tau \in \mathcal{B}(%
\mathfrak{h}_{k})\equiv \mathcal{B}(\mathfrak{h}_{k})^{\otimes },
\end{equation*}%
where $\delta _{i,j}$ denotes the Kronecker delta, and the remainder
satisfies the bound 
\begin{equation*}
\left\Vert \mathcal{R}_{k,\infty }\left( \mathrm{N}+\mathbf{1}\right) ^{-%
\frac{1}{2}}\right\Vert _{\mathrm{op}}\leq \left( \mathrm{e}%
^{D\int_{0}^{\infty }\left\Vert X_{\tau }\right\Vert _{\mathrm{op},2}\mathrm{%
d}\tau }-1\right) \int_{0}^{\infty }\left\Vert X_{k,\tau }\right\Vert _{%
\mathrm{op},2}\mathrm{d}\tau
\end{equation*}%
for a time-independent constant $D\in \mathbb{R}^{+}$.
\end{proposition}

\begin{proof}
Again, $\mathbf{1}\equiv \mathbf{1}_{\mathcal{F}_{+}}\equiv \mathbf{1}_{%
\mathfrak{F}}$ and $\mathbf{1}\otimes a_{k}\equiv a_{k}$. Condition \ref%
{Assumptionsasympt} (vi) ensures the existence of the operator $\mathbf{g}%
_{k,\infty }\in \mathcal{B}(\mathfrak{H})$ for any $k\in \mathbb{N}$ with%
\begin{equation}
\lim_{t\rightarrow \infty }\left\Vert \mathbf{g}_{k,\infty }-\mathbf{g}%
_{k,t}\right\Vert _{\mathrm{op}}\leq \lim_{t\rightarrow \infty
}\int_{t}^{\infty }\left\Vert X_{k,\tau }\right\Vert _{\mathrm{op}}\mathrm{d}%
\tau =0.  \label{smallgconv}
\end{equation}%
Observe next that 
\begin{equation}
\lim_{t\rightarrow \infty }\mathrm{U}_{t,0}a_{k}\mathrm{U}_{0,t}\varphi =%
\mathrm{U}_{\infty ,0}a_{k}\mathrm{U}_{0,\infty }\varphi ,\qquad \varphi \in 
\mathcal{D}(\mathrm{N}^{1/2}),  \label{jk00}
\end{equation}%
by combining (\ref{Uinftystronglimit}) and (\ref{limitUpreservedomNbis})
with the upper bound%
\begin{eqnarray*}
\left\Vert \left( \mathrm{U}_{\infty ,0}a_{k}\mathrm{U}_{0,\infty }-\mathrm{U%
}_{t,0}a_{k}\mathrm{U}_{0,t}\right) \varphi \right\Vert _{\mathfrak{F}}
&\leq &\left\Vert \left( \mathrm{U}_{\infty ,0}-\mathrm{U}_{t,0}\right) a_{k}%
\mathrm{U}_{0,\infty }\varphi \right\Vert _{\mathfrak{F}} \\
&&+\left\Vert a_{k}\left( \mathrm{N}+\mathbf{1}\right) ^{-\frac{1}{2}%
}\right\Vert _{\mathrm{op}}\left\Vert \left( \mathrm{N}+\mathbf{1}\right) ^{%
\frac{1}{2}}\left( \mathrm{U}_{0,\infty }-\mathrm{U}_{0,t}\right) \varphi
\right\Vert _{\mathfrak{F}}.
\end{eqnarray*}%
Similar to (\ref{jk0}), define the operator 
\begin{equation}
\mathcal{R}_{k,\infty }\doteq \int_{0}^{\infty }\mathrm{U}_{0,\tau }i\left[ 
\mathrm{G}_{\tau },\mathbf{g}_{k,\tau }\otimes \mathbf{1}\right] \mathrm{U}%
_{\tau ,0}\mathrm{d}\tau  \label{jk0bis}
\end{equation}%
on the domain $\mathcal{D}(\mathrm{N}^{2})$, via Theorem \ref{propagators}
and Lemma \ref{TechnicalIntegralWandRWR copy(1)}. It is well-defined
because, similar to (\ref{dfdfdfdfdfdf}), for any $t\in \mathbb{R}_{0}^{+}$, 
\begin{equation}
\int_{t}^{\infty }\left\Vert \mathrm{U}_{0,\tau }i\left[ \mathrm{G}_{\tau },%
\mathbf{g}_{k,\tau }\otimes \mathbf{1}\right] \mathrm{U}_{\tau ,0}\left( 
\mathrm{N}+\mathbf{1}\right) ^{-\frac{1}{2}}\right\Vert _{\mathrm{op}}%
\mathrm{d}\tau \leq \left( \mathrm{e}^{D\int_{0}^{\infty }\left\Vert X_{\tau
}\right\Vert _{\mathrm{op},2}\mathrm{d}\tau }-1\right) \int_{t}^{\infty
}\left\Vert X_{k,\tau }\right\Vert _{\mathrm{op}}\mathrm{d}\tau  \label{jk1}
\end{equation}%
for some sufficiently large constant $D\in \mathbb{R}^{+}$. In particular,
for any $\varphi \in \mathcal{D}(\mathrm{N}^{1/2})$, 
\begin{equation}
\lim_{t\rightarrow \infty }\left\Vert \int_{t}^{\infty }\mathrm{U}_{0,\tau }i%
\left[ \mathrm{G}_{\tau },\mathbf{g}_{k,\tau }\otimes \mathbf{1}\right] 
\mathrm{U}_{\tau ,0}\mathrm{d}\tau \varphi \right\Vert _{\mathfrak{F}}=0,
\label{jk2}
\end{equation}%
which, combined with (\ref{jk0}), (\ref{smallgconv})--(\ref{jk1}) and
Proposition \ref{TransfoDisp}, implies the assertion.
\end{proof}

\noindent \textit{Acknowledgments:} This work is supported by the Basque
Government through the BERC 2022-2025 program, as well as the following
grants:

\begin{itemize}
\item Grant PID2024-156184NB-I00 funded by MICIU/AEI/10.13039/501100011033
and cofunded by the European Union.

\item Grant PID2020-112948GB-I00 funded by MCIN/AEI/10.13039/501100011033 and by "ERDF A way of making Europe".

\item FAPESP grants 2025/12824-5 and 2026/04204-0.
\end{itemize}

\end{document}